\documentclass[11pt,a4paper,twoside%,fleqn
]{report}
\usepackage[T1]{fontenc}

\usepackage{graphicx}
\usepackage{feynmp}
\usepackage{ifthen}                   % needed for Feyman-Slash macro
\usepackage{exscale}                  % correct scaling of math-symbols
\usepackage[german,american]{babel}   % switch between languages
\usepackage[latin1]{inputenc}
\usepackage[intlimits]{amsmath}       % improved mathematical typesetting
\usepackage{amsfonts}
\usepackage{amssymb,amsthm}
\usepackage{fancyhdr}                 % user-specified headings
\usepackage{calc}
\usepackage{epsfig}                   % needed to include eps-figures
\usepackage{subeqn}
\usepackage{wick}
\usepackage{a4wide}
\usepackage{epsfig}
\usepackage{theorem}
\usepackage{enumerate}
\usepackage{bbm}
\usepackage{bm}
\usepackage{units}                    % for nicefrac
\usepackage[a4paper,linkcolor=blue,citecolor=green]{hyperref}          % use dvips -z diss.dvi and then
\input{diss.sty}

\begin{document}
\pagenumbering{arabic}         % page numbers agree with those shown
                               % by a pdf viewer

\thispagestyle{empty}

%%%%%%%%%%%%%%%%%%%%%%%%%%%%%%%%%%%%%%%%%%%%%%%%%%%%%%%%%%%%%%%%%%%%%%%%%%%%%%%

\selectlanguage{american}

\begin{titlepage}
 \renewcommand{\baselinestretch}{1.3}
 \centering

%    \Huge\bfseries\sffamily
%    QCD \\

 \vfill

    \Huge
    {\bf Nonperturbative aspects  \\
    of Yang--Mills theory} \\

 \vfill

    \large
    {\bf{Dissertation}}

    zur Erlangung des Grades \\
    eines Doktors der Naturwissenschaften  \\

 \vfill
    der Fakult\"at f\"ur Physik  \\
    der Eberhard-Karls-Universit\"at zu T\"ubingen  \\
 \vfill

    vorgelegt von  \\
 \Large
    {\textsc{Wolfgang Schleifenbaum}}  \\
\vspace{0.2cm}
 \large
  aus Stuttgart  \\

 \vfill

\large
    2008

\end{titlepage}

\clearpage%{\pagestyle{empty}\cleardoublepage}
\pagestyle{empty}

%\cleardoublepage
%\clearpage{\pagestyle{empty}\cleardoublepage}

\vspace*{18cm}
\begin{tabular}{ll}
 \bfseries Tag der m\"undlichen Pr\"ufung :  &  28.05.2008 \\
 \bfseries Dekan :                         & Prof.\ Dr.\ N.\ Schopohl  \\
 \bfseries Erster Berichterstatter :       & Prof.\ Dr.\ H.\ Reinhardt  \\
 \bfseries Zweiter Berichterstatter :      & Prof.\ Dr.\ h.c.\ mult.\
 A.\ F\"a\ss{}ler
\end{tabular}

%\cleardoublepage

%\pagestyle{fancy}
\setlength{\oddsidemargin}{22pt} % *** ZENTRIERT MIT 4pt !! ***
\setlength{\evensidemargin}{\paperwidth-\textwidth-\oddsidemargin-144pt}
\thispagestyle{empty}
\vspace*{8cm}
\begin{center}
{\it Meinem Sohn Benjamin}
\end{center}
\clearpage{\pagestyle{empty}\cleardoublepage}

%\thispagestyle{empty}

%%%%%%%%%%%%%%%%%%%%%%%%%%%%%%%%%%%%%%%%%%%%%%%%%%%%%%%%%%%%%%%%%%%%%%%%%%%%%%%

\selectlanguage{german}

\chapter*{Zusammenfassung}

Das Thema dieser Dissertation ist die Starke Wechselwirkung zwischen
Gluonen und Quarks mit Schwerpunkt auf den nicht st\"orungstheoretischen
Aspekten des Gluonensektors. Es werden %Gitterrechnungen gegen\"ubergestellt, werden
hier Kontinuumsmethoden verwendet, um insbesondere das Ph\"anomen des
Farbeinschlusses zu untersuchen. Der Farbeinschluss, welcher die
Detektion der elementaren Quarks und Gluonen als freie Teilchen
verhindert, verlangt ein Verst\"andnis der langreichweitigen
Wechselwirkungen. In der St\"orungstheorie k\"onnen nur
kurzreichweitige Korrelationen verl\"asslich beschrieben werden. Ein
nicht st\"orungstheoretischer Zugang ist durch das Dyson--Schwinger--Integralgleichungssystem gegeben, das alle Greenfunktionen miteinander
verkn\"upft. Eine L\"osung f\"ur den Gluonpropagator wird im
asymptotisch infraroten und ultravioletten Grenzwert erzielt.

In Kapitel \ref{ch:YMconstr} werden redundante Freiheitsgrade
der Yang--Mills-Eichtheorie durch Fixierung der Weyl-- und
Coulombeichung vor der Quantisierung entfernt. Die Quantisierung mit
Zwangsbedingungen unter Verwendung von Dirac--Klammern wird explizit
ausgef\"uhrt. Als Resultat erh\"alt man den Yang--Mills--Hamiltonoperator.

Der asymptotische Infrarotlimes der Korrelationsfunktionen in
Coulombeichung wird in Kapitel \ref{ghostdom} im Rahmen des
Gribov--Zwanziger--Szenarios f\"ur den Farbeinschluss analytisch untersucht. Das
Coulombpotential zwischen schweren Quarks als Teil des
Yang--Mills--Hamiltonoperators wird in diesem Limes berechnet. Die
\"Ubertragung der L\"osungen in Coulombeichung auf den Infrarotlimes
der Landau\-eichung wird diskutiert.

Der hergeleitete Hamiltonoperator erm\"oglicht die Bestimmung des
Vakuumwellenfunktionals mithilfe des Variationsprinzips in Kapitel
\ref{VarVac}. Numerische L\"osungen der Propagatoren in diesem
Vakuumzustand werden besprochen, und es wird aufgezeigt, dass der
vorhergesagte Infrarotlimes tats\"achlich realisiert ist. Die Diskussion
wird auf Vertexfunktionen erweitert. Des Weiteren wird der Einfluss der
N\"aherungsmethoden auf die L\"osungen untersucht.

Kapitel \ref{UVchap} ist vornehmlich dem Ultraviolettverhalten
der Propagatoren gewidmet. Die Behandlung erfolgt sowohl in der
Coulomb- als auch in der Landaueichung. Ein nicht
st\"orungstheoretischer laufender Kopplungsparameter wird definiert und
berechnet. Der ultraviolette Teil der Variationsl\"osungen aus
Kapitel \ref{VarVac} wird mit den Forderungen der St\"orungstheorie
verglichen. 

In Kapitel \ref{external} wird die R\"uckkopplung des Gluonensektors
auf die Anwesenheit \"au\ss{}erer Ladungen (schwerer Quarks)
behandelt. Zu diesem Zweck werden koh\"arente Gluonenanregungen
vorgeschlagen, welche der Anwesenheit \"au\ss{}erer Ladungen
Rechnung tragen. Weitere Alternativen zur Behandlung dieser
Problemstellung werden besprochen.

%\cleardoublepage
\clearpage{\pagestyle{empty}\cleardoublepage}

%%%%%%%%%%%%%%%%%%%%%%%%%%%%%%%%%%%%%%%%%%%%%%%%%%%%%%%%%%%%%%%%%%%%%%%%%%%%%%%
%\thispagestyle{empty}
\selectlanguage{american}

\chapter*{Abstract}

The subject of this thesis is the theory of strong interactions of
quarks and gluons, with particular emphasis on nonperturbative
aspects of the gluon sector. %Complementary to lattice studies,
Continuum methods are used to investigate in particular the
confinement phenomenon. Confinement---which states that the elementary quarks and gluons cannot be
detected as free particles---requires an understanding of large-scale
correlations. In perturbation theory, only short-range correlations can
be reliably described. A nonperturbative approach is given by the set
of integral Dyson--Schwinger equations involving all Green
functions of the theory. A solution for the gluon propagator is
obtained in the infrared and ultraviolet asymptotic limits.

In chapter \ref{ch:YMconstr}, redundant degrees of freedom
of the Yang--Mills gauge theory are removed by fixing the Weyl and Coulomb gauge prior to
quantization. The constrained quantization in the Dirac bracket
formalism is then performed explicitly to produce the quantized Yang--Mills
Hamiltonian. 

The asymptotic infrared limits of Coulomb gauge correlation functions are
studied analytically in chapter \ref{ghostdom} in the framework of the
Gribov--Zwanziger confinement scenario. The Coulomb potential between heavy quarks
as part of the Yang--Mills Hamiltonian is calculated in this limit. A
connection between the infrared limits of Coulomb and Landau gauge is
established. % It
% is discussed how to short-cut from the Coulomb gauge solutions to the infrared asymptotic solution for the
% Landau gauge propagators. 

The Hamiltonian derived paves the way in chapter \ref{VarVac} for
finding the Coulomb gauge vacuum wave functional by means of the
variational principle. Numerical solutions for the propagators in this
vacuum state are discussed and seen to reproduce the anticipated infrared
limit. The discussion is extended to the vertex functions. The effect
of the approximations on the results is examined.

Chapter \ref{UVchap} is mainly devoted to the ultraviolet behavior of
the propagators. The discussion is issued in both Coulomb and Landau
gauge. A nonperturbative running coupling is defined and calculated. 
The ultraviolet tails of the variational solutions from chapter \ref{VarVac} are compared to
 the behavior demanded by perturbation theory.

In chapter \ref{external}, the back reaction of the gluon sector on
the presence of external charges (heavy quarks) is explored. To
this end, coherent excitations of gluonic modes are suggested to
account for the presence of quarks. Further alternatives for the
discussion of this issue are put forward. % Squeezed states are proposed as an alternative.

\clearpage{\pagestyle{empty}\cleardoublepage}

\tableofcontents
\clearpage{\pagestyle{fancy}\cleardoublepage}
\pagestyle{fancy}
\chapter{Yang--Mills theory as a constrained dynamical system}
\label{ch:YMconstr}
In the development of quantum electrodynamics (QED) \cite{Wey29}, local gauge
invariance of the matter fields served as a guiding principle and it successfully led to a highly
accurate description of electromagnetic forces \cite{Peskin}. The
fundamental requirement of locality necessitates the existence of bosonic gauge fields which
mediate the forces between the fermions. An impediment to the
canonical quantization of such a dynamical system is the
existence of redundant degrees of freedom. This obstacle can be overcome for the electromagnetic theory,
thanks to the linear structure of its equations of motion. Quantum
chromodynamics (QCD), on the other hand, bears non-linearities that make the treatment far more complicated. Since
QCD is the non-abelian generalization of QED, the general structure of
a gauge theory is inherited and along with it its difficulties. It will
be shown in this chapter how the redundant degrees of freedom can
be related to constraints in the classical phase space, and how
quantization of such a dynamical system can be pursued. After a brief
definition of Yang--Mills (YM) theories, amongst which QCD is the
${\mathit{SU(3)}}$ version, we first examine constrained systems in the context of classical mechanics
to pave a way for the terminology. Second, constrained
quantization is exhibited for electrodynamics as an illuminating
example. Finally, we demonstrate how to systematically follow the
steps of constrained quantization in the difficult case of Yang--Mills
theory in Weyl and Coulomb gauge. The explicit form of the quantized
Hamiltonian operator will thus be derived in a quantization approach
that differs from the canonical method.

\section{Definition of Yang--Mills theory}
\label{YMdef}
After a first sporadic attempt by Klein as early as 1939 to generalize the electromagnetic
theory imposing non-abelian group structure 
\cite{Kle38}, it took another 15 years until Yang and Mills were able to provide the
full account of $\mathit{SU(2)}$ non-abelian gauge theory \cite{YanMil54}. Its
generalization to $\mathit{SU(N_c)}$ (with $N_c=3$ realized in QCD) is straightforward, a clear
introduction to the currently accepted definition of YM theory can be
found, e.g., in Ref.\ \cite{Jac80}. There are some good text books on
this topic, e.g.\ Ref.\ \cite{ItzZub}, yet the main elements of YM
theory are outlined here to set up a notation for this thesis.

The basic idea, following electromagnetism as a guiding example, is to cancel the change of the Dirac Lagrangian 
\begin{eqnarray}
  \label{DiracL}
  \cL_{D}=\overline{\psi}\left(i\gamma^\mu\del_\mu-m\right)\psi\; 
\end{eqnarray}
under a local symmetry transformation of its fermion fields
$\psi_a(x)$ in the intrinsic space\footnote{The roman color indices are used
  as subscript or superscript interchangeably.}
\begin{equation}
  \label{psitrafo}
  \psi_a(x)\rarr \psi^{U}_a(x)=U_{ab}(x)\psi_b(x)
\end{equation}
by the introduction of a gauge field $A^a_\mu(x)$. To achieve this,
the derivative $\del_\mu=\frac{\del}{\del x^\mu}$  in the Lagrangian (\ref{DiracL}) is replaced by 
\begin{eqnarray}
  \label{Dfund}
  D_\mu:=\del_\mu+gA_\mu
\end{eqnarray}
which transforms homogeneously, $D_\mu\rarr UD_\mu U^\dagger$, whereas
the matrix-valued gauge fields $A_\mu$ transform inhomogeneously, according to
\begin{equation}
  \label{Atrafo}
  A_\mu(x)\equiv A^a_\mu(x)T^a \rarr A^U_\mu(x)=
  U(x)A_\mu(x)U^\dagger(x) + \frac{1}{g}U(x)\del_\mu U^\dagger(x)\; .
\end{equation}
The subsequent interaction of gauge and matter fields is controlled by
the coupling constant $g$.
Originally, this procedure coined the term ``minimal
substitution''. Nowadays, with a deeper understanding of the
geometrical aspects which allow close analogies to general relativity
\cite{DeW67}, we call $D_\mu$  the ``covariant derivative'', defined in
Eq.\ (\ref{Dfund}) as acting on a field in the fundamental
representation, such as the quark field $\psi^a(x)$.

The unitary group elements $U(x)\in \mathit{SU(N_c)}$,
\begin{equation}
  \label{Udef}
  U(x)=\exp\left(-\alpha^a(x)T^a\right)\; ,\quad \alpha\in\R\; ,
\end{equation}
are continuously generated by $N_c^2-1$ antihermitian
matrices $T_a=-T_a^\dagger$ that obey the Lie algebra
\begin{eqnarray}
  \label{LieAl}
  [T^a,T^b]=f^{abc}T^c
\end{eqnarray}
with the structure constants $f^{abc}$. For the quarks to transform as
$3$-vectors in the fundamental representation of $\mathit{SU(3)}$, one uses the
well-known $3\times 3$ Gell--Mann matrices for $iT^a$. The gauge fields
$A_\mu$, on the other hand, live in a $(N_c^2-1)$-dimensional vector
space spanned by the generators $T^a$ with the additional operation of
commutation (\ref{LieAl}). By an infinitesimal expansion of (\ref{Atrafo}) in
$\alpha$ using (\ref{LieAl}), one can see that $A_\mu^a$ must
transform in the adjoint representation of $\mathit{SU(N_c)}$,
\begin{eqnarray}
  \label{deltaA}
  (A^U)^{a}_\mu=A_\mu^a+(\hat
  T^{b})^{ac}A_\mu^b\alpha^c+\frac{1}{g}\del_\mu\alpha^a+\cO(\alpha^2)\; ,
\end{eqnarray}
where we defined
\begin{eqnarray}
  \label{That}
  (\hat T^b)^{ac}:=f^{abc}\; .
\end{eqnarray}
Any linear combination of the matrices $\hat T^a$ with field operators
shall be denoted by
a caret, e.g.\ $\hat A_\mu\equiv A_\mu^a \hat T^a$, as opposed to the
matrices $A_\mu$ introduced in Eq.\ (\ref{Atrafo}). The covariant derivative $\hat
D_\mu^{ab}$ acting on fields in the adjoint representation is defined by
\begin{eqnarray}
  \label{Dadj}
  \hat D^{ab}_\mu=\delta^{ab}\del_\mu+g\hat A_\mu^{ab}\; .
\end{eqnarray}
Knowing the transformation properties of the gauge field (\ref{Atrafo}), one may
construct from the gauge covariant field strength tensor
\begin{eqnarray}
  \label{Fmunu}
  F_{\mu\nu}:=\frac{1}{g}[D_\mu,D_\nu]=\del_\mu A_\nu - \del_\nu A_\mu
  + g [A_\mu , A_\nu]\; , \quad F_{\mu\nu} \rarr U F_{\mu\nu}U^\dagger
\end{eqnarray}
a gauge invariant contribution to the Lagrangian density that is
purely gluonic. It is the first term in the \emph{Yang--Mills Lagrangian}
\begin{equation}
  \label{YMLagr}
  \cL_{YM}=\frac{1}{2}\tr
  F_{\mu\nu}(x)F^{\mu\nu}(x)+gj_\mu^a(x)A_a^{\mu}(x)\; ,
\end{equation}
whereas the second term represents the coupling to
the quark N\"other current of the color symmetry,
\begin{eqnarray}
  \label{Noether}
  j_\mu^a(x)=\overline{\psi}(x)\gamma_\mu iT^a\psi(x)\; ,
\end{eqnarray}
here treated as a classical external field. This thesis
shall mainly be concerned with the dynamics of the Yang--Mills
Lagrangian (\ref{YMLagr}) and disregard the influence of dynamical
quarks. It is hypothesized that unquenching has a small effect on the
gauge field sector.\footnote{The nonperturbative confinement
  phenomenon is known to exist in the absence of flavors, $N_f=0$, from lattice
  calculations \cite{Bal00}. In the perturbative regime, a small number of quark
  flavors ($N_f<16$) will not remove the property of asymptotic freedom. Recent
  continuum studies \cite{FisWatCas05} as well as lattice calculations
  \cite{Dav+04} indicate
  that the effect of unquenching on hadronic observables 
is small.} The color trace in Eq.\ (\ref{YMLagr}) with the convention
$\tr(T^aT^b)=-\frac{1}{2}\delta^{ab}$ guarantees that the gluonic
energy is positive definite, cf.\ (\ref{classHam}).

With the action principle, the Lagrangian formalism provides a
possibility to derive equations of motion where Lorentz invariance is
manifest. Applying a functional derivative w.r.t.\ the gauge field to
the Yang--Mills action
\begin{eqnarray}
  \label{YMaction}
  S=\int d^4x \cL_{YM}(x)
\end{eqnarray}
one finds\footnote{Note that $F_a^{\mu\nu}$ transforms in the adjoint
  representation, therefore it is natural that the covariant derivative  $\hat
  D_\mu$ is found here.}
\begin{eqnarray}
  \label{EOM1}
  \hat D_\mu^{ab}F_b^{\mu\nu}=-gj_a^{\nu}\; 
\end{eqnarray}
and, using the Bianchi identity
$[D_\mu,F_{\nu\rho}]+cycl. perm.=0$ ,
\begin{eqnarray}
  \label{EOM2}
  \frac{1}{2}\epsilon^{\mu\nu\rho\sigma} \hat D_\mu^{ab}F_{\rho\sigma}^{a}=0\; .
\end{eqnarray}

In order to derive a Hamiltonian, one introduces conjugate momenta
\begin{eqnarray}
  \label{Pidef}
  \Pi^a_\mu(x)=\frac{\delta \cL}{\delta\del_0 A_a^{\mu}(x)}=F^a_{\mu 0}(x)
\end{eqnarray}
and readily notes that due to the antisymmetry of $F^{\mu\nu}$
\begin{eqnarray}
  \label{primconstr}
  \Pi_0^a(x)=0\; .
\end{eqnarray}
Eq.\ (\ref{primconstr}) is understood as a constraint on the phase
space variables and will be shown to complicate the quantization
process. The Legendre transform gives the Hamiltonian $H$,
\begin{eqnarray}
  \label{classHam}
  H&=&\int d^3x\,  \left(\Pi_\mu^a\del_0
    A_a^{\mu}-\cL_{YM}\right)\nn\\
  &=&\int d^3x\,  \left( \Pi_a^{k}F^{a}_{0k}+\Pi_a^{k}\hat
    D^{ab}_kA_0^b+\frac{1}{2}F_{k0}^a F_a^{k0}+\frac{1}{4}F_{ij}^aF_a^{ij}-gj_a^{\mu}A_\mu^a \right)\nn\\
  &=&\int d^3x\,  \left(
    -\frac{1}{2}\Pi_a^k\Pi_k^a+\frac{1}{4}F_{ij}^aF_a^{ij}-A_0^a\hat
    D^{ab}_k\Pi^k_b -gA_0^a \rhoext^a-gA_k^aj^k_a\right)\nn\\
  &=&\frac{1}{2}\int d^3x\,  \left({\bf \Pi}_a^2+{\bf B}_a^2  \right) -
  \int d^3x\,  A_0^a\cG^a + g\int d^3x\, {\bf A}^a\cdot{\bf j}^a\; ,
\end{eqnarray}
with the abbreviations for the external color charge density
$\rhoext^a=j_0^a$, the color magnetic field
$B^a_k=-\frac{1}{2}\epsilon_{ijk}F^a_{ij}$ and 
\begin{eqnarray}
  \label{Gaussdef}
  \cG^a(x)=\hat D^{ab}_k(x)\Pi_b^k(x)+g\rhoext^a(x)\; .
\end{eqnarray}
In the Lagrangian equations of motion we recognize that the Gauss law
can be written as $\cG^a=0$, see Eq.\ (\ref{EOM1}) for
$\nu=0$.\footnote{In electrodynamics, one defines the electric field
  as $E^k=-\Pi^k$ and thus gains the familiar form of the Maxwell equation
  ${\bf \nabla}\cdot{\bf E}=e\rho$.} However, in the Hamiltonian approach, the Gauss law cannot be
found as an equation of motion but must be implemented as a
constraint, as will be discussed below.

The quantization of a gauge theory is a delicate task, already in the
abelian case. Since the gauge transformation indicates that the field
variables $A_k^a(x)$ contain unphysical degrees of freedom, it is not
clear a priori whether the canonical quantization leads to the correct
results, i.e. whether it contains the classical theory in the limit
$\hbar\rarr 0$.  It is discussed in some detail in the following sections how a gauge
theory can be understood as a constrained system and how to go about
with its quantization.  

Although in the Lagrangian formalism Lorentz invariance can be
made manifest, it is sometimes favorable to use the Hamiltonian
formalism instead. The path integral method provides an elegant way of
quantizing a theory covariantly in the Lagrangian formalism. However, it is rather unsuitable to
calculate, e.g., the Balmer formula for hydrogen, one of the first exercises
in quantum mechanics \cite{DurKle79}. The Hamiltonian approach to the quantum
theory of electromagnetism and YM theory shall be the main focus of
this work. Once the conjugate momenta are defined and one
passes over to the Hamiltonian, Lorentz covariance is lost since
a reference frame has been chosen. Nevertheless, Lorentz \emph{invariance} is
maintained. This can be understood, e.g., by explicit calculation in Coulomb
gauge for electromagnetism \cite{BriGoo67} or by criteria for commutation relations in
Coulomb gauge quantum Yang--Mills theory \cite{Sch62}.

\section{Constrained dynamics}
\label{Diracconcept}
In this section we shall discuss the importance of constraints of a
dynamical system for its time evolution and for its quantization
procedure. The basic idea was developed by Dirac \cite{Dir50,Dir58} and very
pedagogically presented in his Yeshiva lectures \cite{Dir64}. Dirac
and many of his successors tried to maintain a high level of
generality in favor of the applicability to any given system, say
electrodynamics or even general relativity \cite{Car01}. However, it turned out
that exceptions to the quite general claims can be found. Therefore,
only the concepts relevant to Yang--Mills theory will be introduced
here. For additional information, see e.g.\
\cite{HenTei92,Sun82,HanRegTei76}.

Consider a classical mechanical system described by a Lagrangian
$L(q_i,\dot{q}_j)$. The Legendre transformation to a Hamiltonian
$H(q_i,p_j)$ with conjugate momenta $p_k$ can be achieved if we can solve
the equation
\begin{equation}
  \label{pdef}
  p_k:=\frac{\del L}{\del \dot{q}_k}
\end{equation}
for the velocities $\dot{q}_k$, which is locally guaranteed by a regular Hessian
\begin{equation}
  \label{Hessian}
  \det\left(\frac{\del^2L}{\del \dot{q}_i \del  \dot{q}_j}\right)\neq 0\; .
\end{equation}
The phase space $\Gamma$ is then defined as a set of independent elements
$\{q_i,p_j\}$. In the case of a singular Lagrangian where Eq.\
(\ref{Hessian}) does not hold, some of the momenta $p_k$ are
dependent. Moreover, additional dependences among the $q_k$ and $p_k$
may arise from the Hamiltonian equations of motion. Generally,
one may formulate all dependences as a number of constraints
\begin{equation}
  \label{phiconstr}
  \varphi_m(q,p)\approx 0\; .
\end{equation}
by which the dimension of phase space is reduced. Below, the symbol
'$\approx$' is explained. The constraints
(\ref{phiconstr}) then hold in a subset $\Gamma_R\subset\Gamma$ of
the original phase space. The Hamiltonian $H$ can be supplemented by
the constraints (\ref{phiconstr}) with
some arbitrary Lagrange multipliers $v_m$,
\begin{equation}
  \label{Htot}
  H_T=H+v_m\varphi_m
\end{equation}
We refer to $H_T$ as the ``total Hamiltonian''. The common variational calculus with constraints gives rise to
Hamiltonian equations of motion in $\Gamma_R$,
\begin{subequations}
\label{HEOM}
\begin{align}
  \label{HamEOM}
  \dot{q}_k&=\frac{\del H}{\del p_k}+v_m\frac{\del \varphi_m}{\del
    p_k}\,\approx\,\{q,H_T\}\\
 \dot{p}_k&=-\frac{\del H}{\del q_k}-v_m\frac{\del \varphi_m}{\del q_k}\,\approx\,\{p,H_T\}
\end{align}
\end{subequations}
Above, we have made use of the Poisson bracket for any functions $f(q,p)$
and $g(q,p)$,
\begin{equation}
  \label{Poissdef}
  \{f,g\}=\frac{\del f}{\del q_k}\frac{\del g}{\del p_k}-\frac{\del f}{\del p_k}\frac{\del g}{\del q_k}
\end{equation}
Note that there is a '$\approx$' symbol in Eq.\ (\ref{HEOM}), a notation due to Dirac which emphasizes that one is to first
calculate the Poisson brackets and then set the constraints
(\ref{phiconstr}) to zero. These kinds of relations are termed ``weak
equations'' and are specific to the use of Poisson brackets. It may be
stressed here that the canonical quantization with such weak equations
leads to the circumstance that some quantum constraints cannot be
imposed as operator identities but only as projections on states. Below, we
shall introduce generalized Poisson brackets that make the usage of
weak equations redundant.

The time evolution of any given phase space function $g(q,p)$ that
does not explicitly depend on time is given
by the Poisson bracket in a concise way:
\begin{equation}
  \label{anyEOM}
  \dot{g}\approx\{g,H_T\}\; .
\end{equation}
The total Hamiltonian $H_T$ obviously serves as the generator of time
evolution. However, the multiplier functions $v_m$ contained in it
leave an arbitrariness. Given some initial conditions, say
$g(0)=g_0$, $g(t)$ at some finite time $t$ is not unique
since we can choose the $v_m$ to be one value or another. After an infinitesimally small time $\delta t$, the function
$g$ will yield
\begin{equation}
  \label{gofdt}
  g(\delta t)\approx g_0+\{g_0,H\}\delta t+v_m\{g_0,\varphi_m\}\delta t\; .
\end{equation}
Imagine we take two different sets of multiplier functions $v'_m$ and
$v''_m$. Comparing these two arbitrary choices to calculate $g'(\delta
t)$ and $g''(\delta t)$, using (\ref{gofdt}), leads to
\begin{equation}
  \label{deltag}
  g''(\delta t)\approx g'(\delta t)+\epsilon_m\{g_0,\varphi_m\}
\end{equation}
with $\epsilon_m=(v''_m-v'_m)\delta t$. Evidently, the two physically
equivalent values $g'(\delta t)$ and $g''(\delta t)$ are related by a
term generated by the constraints $\varphi_m$. This suggests that the
physical state remains unchanged under transformations of the kind
\begin{eqnarray}
  \label{symmtrans}
  g(q,p)\rarr g^\epsilon (q,p)=g(q,p)+\epsilon_m\{g,\varphi_m\}
\end{eqnarray}
However, we have to be
careful with the interpretation, as shown below.

The constraints (\ref{phiconstr}) need to be checked for consistency,
i.e.\ they should hold at all times. In view of Eq.\ (\ref{anyEOM})
this requirement gives rise to the conditions 
\begin{equation}
  \label{phiconsis}
  \dot{\varphi}_k=\{\varphi_k,H\}+\{\varphi_k,\varphi_m\}v_m\approx 0
\end{equation}
which can actually fix the multiplier functions $v_m$ so long as the
matrix
\begin{equation}
  \label{Cdef}
  C_{km}:=\{\varphi_k,\varphi_m\}
\end{equation}
is invertible. We now distinguish \emph{first-class} and
\emph{second-class} constraints \cite{Dir64}. By definition, a
first-class constraint has vanishing Poisson brackets with all other
constraints. With first-class constraints present, the matrix
$(C_{km})$ in Eq.\ (\ref{Cdef}) becomes singular, $\det\left(C_{km}\right)\approx 0$. In the
absence of first-class constraints, all constraints are termed
second-class, and one can show \cite{Sen76} that
\begin{eqnarray}
 \label{2ndclass}
\det\left(C_{km}\right)\napprox 0\; .
\end{eqnarray}
If Eq.\ (\ref{2ndclass}) holds and only second-class constraints are present, we
can fix the multiplier functions $v_m$ using the consistency
conditions (\ref{phiconsis}) by
\begin{equation}
  \label{fixv}
  v_k=-C^{-1}_{km}\{\varphi_m,H\}
\end{equation}
and derive that $g(t)$ with second-class
constraints satisfies the equation of motion
\begin{equation}
  \label{EOMwD}
  \dot{g}=\{g,H\}-\{g,\varphi_k\}C^{-1}_{km}\{\varphi_m,H\}\equiv\{g,H\}_D\; .
\end{equation}
Here, the \emph{Dirac bracket} was introduced,
\begin{equation}
  \label{DBdef}
  {\{f,g\}}_D:=\{f,g\}-\{f,\varphi_k\}C^{-1}_{km}\{\varphi_m,g\}\; .
\end{equation}
It is clear from Eq.\ (\ref{EOMwD}) that the time evolution of the
initial state $g_0$ is unique. The Dirac bracket provides a formalism
that yields equations of motion with (second-class)
constraints. Furthermore, the constraints (\ref{phiconstr}) which are 
regarded as weak equations w.r.t.\ Poisson brackets, can be used as
strong (ordinary) equations if one uses the Dirac bracket instead,
since for any function $f$
\begin{equation}
  \label{strongphi}
  \{\varphi_k,f\}_D= \{\varphi_k,f\} - \{\varphi_k,\varphi_j\}C^{-1}_{jm}\{\varphi_m,f\}=0\; .
\end{equation}
These ideas led Dirac to propose a quantization technique for
constrained dynamical systems, known as \emph{constrained
  quantization}\footnote{It is also associated with the lengthy
  expression ``reduced phase
  space quantization''.}, making use of the
Dirac brackets in the fundamental commutation relations of operators
$\hat q_i$ and $\hat p_i$,
\begin{equation}
  \label{DiracQua}
  i\hbar\{q_i,p_j\}_D\rarr[\hat{q}_i,\hat{p}_j]\; .
\end{equation}
While the Dirac brackets has most of the features of the Poisson
bracket---e.g.\ antisymmetry, linearity, Jacobi identity, product rule---the fundamental Dirac brackets are generally not as simple as the
Poisson brackets,
$\{q_i,p_j\}_D\neq\delta_{ij}$. In linear theories, such as QED
(discussed in the next section) this does not pose a problem to the
quantization prescription (\ref{DiracQua}) since the
fundamental Dirac brackets do not involve field variables. However, in
the non-linear YM theory, constrained quantization is more
difficult. This will be dealt with in section \ref{DiracYM}.

Let us come back to the claim that constraints generate
transformations among physically equivalent states, see Eq.\
(\ref{symmtrans}). For second-class constraints, this can certainly not
be correct, since all multiplier functions are fixed, as we have seen
in Eq.\ (\ref{fixv}). To understand this caveat, note that the
transformations (\ref{symmtrans}) generated by second-class constraints kick the system
out of the reduced phase space $\Gamma_R$ since for $g=\varphi_k$
\begin{equation}
  \label{kickout}
  \varphi''_k=\varphi'_k+\{\varphi_k,\varphi_m\}\epsilon_m\napprox 0\; .
\end{equation}
That is, the transformations (\ref{symmtrans}) generated by
second-class constraints are not symmetries of the theory. 

\begin{figure}
  \centering
  \includegraphics[scale=1.0]{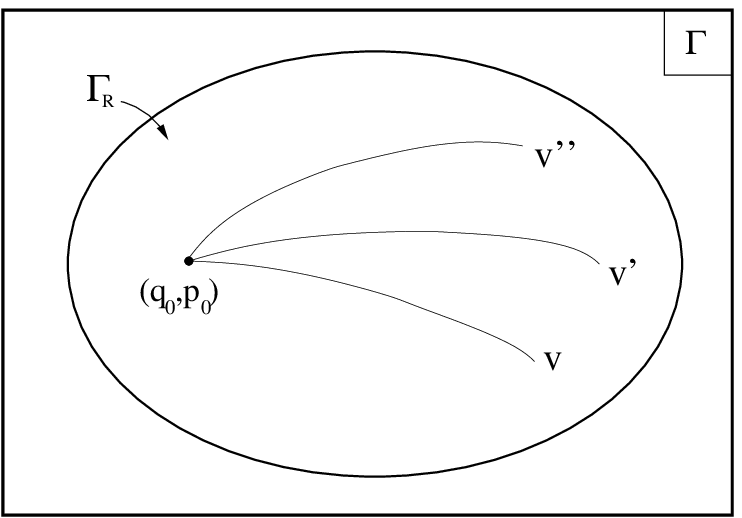}
  \includegraphics[scale=1.0]{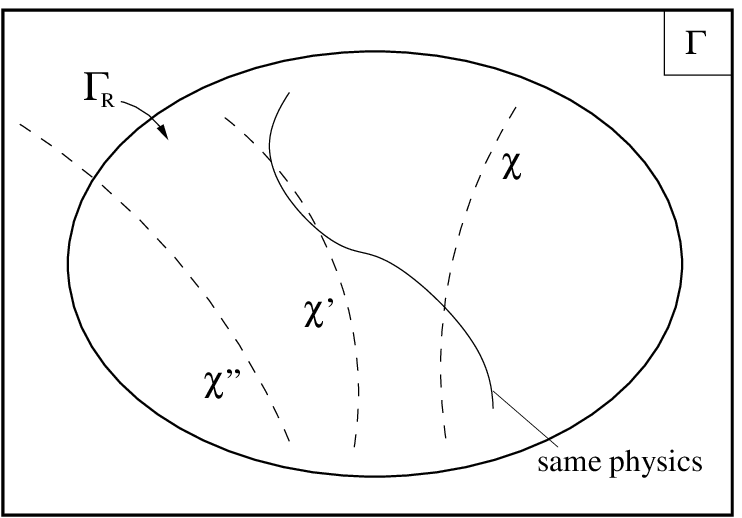}
  \caption{{\it Left:} Ambiguity in time evolution of an initial point
    $(q_0,p_0)$ in the reduced phase space $\Gamma_R$. Different
    choices ($v$, $v'$, $v''$) of the arbitrary multiplier functions
    lead to different trajectories in $\Gamma_R$. {\it Right:} At a
    fixed time, the gauge orbit (solid line) represents physically
    equivalent states. The gauge fixing condition $\chi$ is
    admissible, whereas $\chi '$ fails to be unique and $\chi ''$ is
    not attainable by a gauge transformation.}
  \label{phasespace}
\end{figure}

On the other hand, if we consider first-class constraints only, then
the Poisson brackets all vanish, $\{\varphi_i,\varphi_j\}=0$, and the transformations (\ref{symmtrans}) are indeed
such that the system remains in the reduced phase space $\Gamma_R$
where the constraints are satisfied. They give rise to an equivalence
class\footnote{The equivalence relation $\sim$ is here represented by
  the transformations infinitesimally defined in Eq.\ (\ref{symmtrans}).}
\begin{eqnarray}
  \label{equivcl}
  \left[(q,p)\right]=\left\{\left. (q^\epsilon,p^\epsilon)\in\Gamma_R
      \:\right|\:
    (q,p)\sim(q^\epsilon,p^\epsilon)\; ,\;\;
    \epsilon\in\R\right\}
\end{eqnarray}
%    q_k^\epsilon=q_k+\epsilon_m\{q_k,\varphi_m\}\; ,
%    p_k^\epsilon=p_k+\epsilon_m\{p_k,\varphi_m\}\; , 
of phase space variables that all correspond to the same physical
state. Any set of $(q_i,p_j)$ uniquely determines the state but the reverse is not true, see Fig.\
\ref{phasespace}. These considerations infer that the \emph{first class constraints are the generators
of symmetries of the theory}.

It should be noted here that the above statement, known as
``Dirac's conjecture'' is not a rigorous theorem. Counter examples can
be constructed \cite{Caw79}, and it takes further classifications of
constraints (``primary'' and ``secondary'') to identify the exact set
of generators of symmetry transformations in general. However, all physical
applications agree with Dirac's conjecture and the issue is somewhat
academic. In particular, the gauge transformations of QED and YM theory
are generated by first-class constraints (see below). In this context,
the equivalence class (\ref{equivcl}) is also referred to as the \emph{gauge orbit}.

% Although this statement is entirely correct for QED and YM
% theory, counter examples can be constructed nevertheless
% \cite{Caw79}. Further classifications (primary and secondary constraints)
% aid to rigorously find all generators of the symmetry
% transformations, but for the present purposes the above definitions
% suffice.

The quantization in the presence of first-class constraints is
performed either in the path-integral formalism \cite{Fad70}, or by a
projection on the physical Hilbert subspace \cite{Jac80,ChrLee80}. Alternatively, one can fix the
gauge on the classical level and then quantize with the Dirac
prescription (\ref{DiracQua}). When fixing the gauge, one effectively
picks out a single representative from the equivalence class
(\ref{equivcl}) generated by the first-class
constraints. In practice, this is achieved by imposing supplementary
constraints (``gauge conditions'') 
\begin{equation}
  \label{chidef}
  \chi_n\approx 0
\end{equation}
that turn the
first-class constraints $\varphi_m$ into second-class ones and hence obey
\begin{equation}
  \label{uniquegauge}
  \det(\{\chi_n,\varphi_m\})\neq 0\; .
\end{equation}
The constraints $\varphi_m$ that formerly were first-class thus cease to cause an ambiguous time evolution of the
system. With all constraints being second-class, all multiplier
functions are fixed and one can use the Dirac bracket to proceed with
the quantization.

In order to arrive at a ``physical gauge'' where all unphysical
degrees of freedom are eliminated, the number of gauge conditions must
equal the number of first-class constraints, hence the matrix that
appears in Eq.\ (\ref{uniquegauge}) is quadratic. In principle,
the gauge conditions (\ref{chidef}) are quite arbitrary, except for the requirements of
uniqueness and attainability. Uniqueness can be established locally as
follows: Consider in Fig.\ \ref{phasespace} (right panel) the gauge orbit at a fixed time $t_f$ and
assume without loss of generality that the point $(q^f,p^f)$ satisfies
$\chi_n(q^f,p^f)=0$. The transformation
\begin{eqnarray}
  \label{chitrafo}
  \chi_n\rarr \chi_n^\epsilon=\chi_n+\epsilon_m\{\chi_n,\varphi_m\}
\end{eqnarray}
within the gauge orbit is not to yield any further
solutions $\chi_n^\epsilon\approx 0$ unless it is the identity
transformation, $\epsilon=0$. Otherwise, the gauge fixing condition specifies (at least) two elements of the
gauge orbit, see $\chi '$ in Fig.\ \ref{phasespace}. With the condition
(\ref{uniquegauge}) on the $\chi_n$, one can see from Eq.\
(\ref{chitrafo}) immediately that $\epsilon=0$ is the only solution to
$\chi^\epsilon_n\approx 0$ and therefore the point $(q^f,p^f)$ is unique.\footnote{The
global issues of uniqueness will be discussed below in the context of
YM theory.}
Moreover, a sensible gauge condition has to be
attainable by a gauge transformation. Otherwise, we would
have the situation in which the gauge fixing condition does not intersect
with the gauge orbit at all, see $\chi ''$ in Fig.\ \ref{phasespace}. 

In summary, the general procedure of quantizing classical mechanics in the
presence of constraints has been discussed. It has not been shown how to find the
entire set (\ref{phiconstr}) of constraints starting from the
Lagrangian. This can be achieved with the so-called Dirac-Bergmann
algorithm \cite{Dir50,Dir58,Ber56}. It incorporates the stationarity of constraints that are
found in the equation of motion or in the mere definition of conjugate
momenta.

\section{Quantum electrodynamics}
\label{QED}
An illuminating example of the formalism exhibited above is the
quantization of electrodynamics in Weyl and Coulomb gauge. There are
no difficulties in promoting the ideas from point mechanics to a field
theory, the discrete indices of the generalized coordinates are merely
replaced by the continuous spacetime dependence.

The electromagnetic theory is the abelian version of the general set
of YM theories described in section \ref{YMdef}, recovered by setting
$f^{abc}=0$ in the Lie algebra (\ref{LieAl}). In the absence of
external charges, one finds from (\ref{YMLagr}) the well-known Lagrangian density
\begin{equation}
  \label{EMLagr}
  \cL=-\frac{1}{4}F_{\mu\nu}F^{\mu\nu}\; , \quad F_{\mu\nu}=\del_\mu
  A_\nu-\del_\nu A_\mu\; .
\end{equation}
As quoted in Eq.\ (\ref{primconstr}), the
conjugate momentum of the $A_0$ field vanishes and we can write down a first
constraint $\varphi_1$ as\footnote{In field theory, an infinite
  number of constraints is specified, one at each spacetime point.}
\begin{equation}
  \label{phi1}
  \varphi_1(x):=\Pi_0(x)=\frac{\delta \cL}{\delta \del_0 A_0(x)}\approx
  0\; .
\end{equation}
The Hamiltonian function depending on the fields $A_0(x)$, $A_k(x)$ and the
conjugate momenta $\Pi^k(x)$ reads
\begin{equation}
  \label{EMHam}
  H=\frac{1}{2}\int d^3x\, \left({\bf \Pi}^2(x)+{\bf B}^2(x)\right)-\int
  d^3x\,  A_0(x)\del_k\Pi^k(x)\; .
\end{equation}
From now on, we write $\Pi_k(x)$ for the components of the
contravariant tensor $\Pi^k(x)=F^{k0}(x)$. This change in
notation will become useful in the nonabelian case where the field
operators also have a color index.
The Hamiltonian (\ref{EMHam}) determines the time evolution via the Poisson brackets. The
fundamental Poisson brackets are given by
\begin{equation}
  \label{FPB}
\left\{A_\mu(\fx,t),\Pi_\nu(\fy,t)\right\}=\delta_{\mu\nu}\delta^{3}(\fx,\fy)
\; .
\end{equation}
Since the constraint (\ref{phi1}) is stationary, 
\begin{equation}
  \label{phi2}
  \varphi_2(x):=\{\Pi_0(x),H_T\}=\{\Pi_0(x),H+\int d^3y \: v_1(y)\Pi_0(y)\}=\del_k\Pi_k(x)\approx 0
\end{equation}
has to hold weakly where $v_1(x)$ is an arbitrary multiplier
function. We recognize Eq.\ (\ref{phi2}) as the Gauss law. Following
the Dirac-Bergmann algorithm, we can check if further constraints
arise from requiring that $\varphi_2$ is constant in time. One finds 
\begin{eqnarray}
  \label{EMexhausted}
  \left\{\varphi_2(x),H+\int d^3y\left(v_1(y)\Pi_0(y)+v_2(y)\del_k\Pi_k(y)\right)\right\}=0
\end{eqnarray}
and thus the complete matrix of constraints $C_{ij}$, see Eq.\
(\ref{Cdef}), is given by $\varphi_1$ and $\varphi_2$ which are obviously first-class, $\{\varphi_1,\varphi_2\}=0$.

As we have shown in the preceding section, the first-class constraints
serve as generators of gauge transformations. Here, it can be
explicitly verified that the transformations $A_\mu\rarr A_\mu+\delta
A_\mu$ with
\begin{subequations}
\label{Hamtraf}
\begin{align}
  \label{EMgt1}
  \delta_{\varphi_1}A_0(x)&=\int d^3x'\, 
  \{A_0(x),\varphi_1(x')\}\alpha_1(x')=\alpha_1(x)\; ,\quad
  \delta_{\varphi_1}A_k(x)=0 \\
  \label{EMgt2}
  \delta_{\varphi_2}A_k(x)&=\int d^3x'\,
  \{A_k(x),\varphi_2(x')\}\alpha_2(x')=-\del_k\alpha_2(x)\; ,\quad \delta_{\varphi_2}A_0(x)=0
 \end{align}
\end{subequations}
are symmetries of the Lagrangian (\ref{EMLagr}). The general form (\ref{Atrafo}) of
gauge transformations in the Lagrangian formalism, with generator
$T=i\id$ of the group $\mathit{U(1)}$ and gauge coupling $g=e$, reads
\begin{equation}
  \label{EMginv}
  A_\mu(x)\rarr A_\mu^\alpha(x)=A_\mu(x)+\frac{1}{e}\del_\mu\alpha(x)\; .
\end{equation}
If we wish to identity an $\alpha(x)$ in Eq.\ (\ref{EMginv}) that
corresponds to the transformations (\ref{Hamtraf}),
it is recognized that Eq.\ (\ref{EMgt1})
corresponds to spatially independent gauge transformations $\alpha(t)$
with $\alpha_1(t)=\del_0\alpha(t)/e$ whereas Eq.\ (\ref{EMgt2}) corresponds to
time-independent gauge transformations with $\alpha(\fx)=-\alpha_2(\fx)/e$. 

In the quantization procedure, the problem arises that due to the
gauge invariance (\ref{Hamtraf}) there are more variables than physical degrees of
freedom. The constraints (\ref{phi1}) and (\ref{phi2}) then should be
taken into account in the Dirac formalism. Before we do so, it is
briefly sketched here how the canonical quantization is usually pursued with
Poisson brackets. With the choice of Weyl gauge, $A_0(x)=0$, the first
constraint, $\Pi_0(x)\approx 0$, does not cause any difficulties for 
the quantization of the time-components of the fields. The canonical Poisson brackets are used to
promote the spatial classical fields to operator-valued fields that obey the
equal-time commutation relations
\begin{equation}
  \label{EMcancom}
  [A_i(\fx),\Pi_j(\fy)]=i\delta_{ij}\delta^3(\fx,\fy)\; .
\end{equation}
Since the Gauss law (\ref{phi2}) is an equation that holds only weakly
on the classical level, it is not a surprise that it leads to
a contradiction with the commutation relations,
\begin{equation}
  \label{GaussContra}
  0=[A_k(\fx),\del_j^{\fy}\Pi_j(\fy)]=\del_j^{\fy}[A_k(\fx),\Pi_j(\fy)]=i\del_k^{\fy}\delta^3(\fx,\fy)\neq 0\; .
\end{equation}
The requirement that the Gauss law holds as an operator identity is
therefore abandoned and one usually restricts the Hilbert space to
the kernel of the Gauss law operator $\cG$,
\begin{equation}
  \label{EMGaussProj}
  \partial_k^{\fy}\Pi_k(\fy)\ket{\Psi}=0\; ,
\end{equation}
%There was some debate in the literature about the consistency of this procedure \cite{}.

Let us mention here that the quantum theory is described in the
Schr\"odinger picture where the field operators are independent of time.

Constrained quantization has the generic feature that the constraints can be
imposed as strong equations once the Poisson brackets are replaced by
the Dirac brackets. In the so quantized theory, the constraints hold
as operator identities and can be used to eliminate unphysical degrees
of freedom. Before the Dirac brackets can be introduced, all
first-class constraints have to be degraded to second class by
imposing gauge fixing conditions. We now choose Weyl and Coulomb gauge to render the constraints
(\ref{phi1}) and (\ref{phi2}) second class,
\begin{subequations}
\begin{align}
  \label{EMWeyl}
  \chi_1(x)&=A_0(x)\approx 0 \\
  \label{EMCoul}
  \chi_2(x)&=\del_k A_k(x)\approx 0
\end{align}
\end{subequations}
The Weyl or ``temporal'' gauge condition (\ref{EMWeyl}) fixes the
gauge transformations (\ref{EMgt1}) generated by the first-class
constraint $\Pi_0\approx 0$, whereas the Coulomb gauge
condition (\ref{EMCoul}) fixes the gauge
transformations (\ref{EMgt2}) generated by the first-class constraint
$\del_k\Pi_k(x)\approx 0$ which is the Gauss law. Thus, the gauge is fixed
completely\footnote{\label{foot:global} Up to global gauge transformations.} on the
classical level and we will refer to the joint conditions
(\ref{EMWeyl}) and (\ref{EMCoul}) as the \emph{temporal Coulomb gauge}.
% and gain a non-singular anti-symmetric matrix of second-class constraints,
% \begin{equation}
%   \label{Cij}
%   (C_{ij})(\fx,\fy)=\left(
%     \begin{array}{cccc}
%       \{\phi_1(\fx),\phi_1(\fy)\} & \{\phi_1(\fx),\phi_2(\fy)\} &
%       \{\phi_1(\fx),\chi_1(\fy)\} & \{\phi_1(\fx),\chi_2(\fy)\} \\
%       \{\phi_2(\fx),\phi_1(\fy)\} & \{\phi_2(\fx),\phi_2(\fy)\} &
%       \{\phi_2(\fx),\chi_1(\fy)\} & \{\phi_2(\fx),\chi_2(\fy)\} \\
%       \{\chi_1(\fx),\phi_1(\fy)\} & \{\chi_1(\fx),\phi_2(\fy)\} &
%       \{\chi_1(\fx),\chi_1(\fy)\} & \{\chi_1(\fx),\chi_2(\fy)\} \\
%       \{\chi_2(\fx),\phi_1(\fy)\} & \{\chi_2(\fx),\phi_2(\fy)\} &
%       \{\chi_2(\fx),\chi_1(\fy)\} & \{\chi_2(\fx),\chi_2(\fy)\} 
%     \end{array}
%     \right)
% \end{equation}

All (second-class) constraints are now collected in a vector
$\phi=\left(\varphi_1,\varphi_2,\chi_1,\chi_2\right)$, and we redefine
the matrix (\ref{Cdef}) by
$C_{ij}(\fx,\fy)=\{\phi_i(\fx),\phi_j(\fy)\}$ which can be evaluated
for a fixed time $t$ using the fundamental
Poisson brackets (\ref{FPB}) to yield
\begin{equation}
    (C_{ij})(\fx,\fy) = \left(
      \begin{array}{cccc}
        0 & 0 & -\delta^3(\fx,\fy) & 0 \\
        0 & 0 & 0 & \del^2\delta^3(\fx,\fy) \\
        \delta^3(\fx,\fy) & 0 & 0 & 0 \\
        0 & -\del^2\delta^3(\fx,\fy) & 0 & 0
      \end{array}
      \right)\; ,
\end{equation}
Since $C_{ij}$ is regular by construction, we can also write down its inverse,
\begin{eqnarray}
  \label{C-1ij}
  (C^{-1}_{ij})(\fx,\fy)&=&\left(
      \begin{array}{cccc}
        0 & 0 & +\delta^3(\fx,\fy) & 0 \\
        0 & 0 & 0 & \frac{1}{4\pi|\fx-\fy|} \\
        -\delta^3(\fx,\fy) & 0 & 0 & 0 \\
        0 & -\frac{1}{4\pi|\fx-\fy|} & 0 & 0
      \end{array}
      \right)\; .
\end{eqnarray}
The fundamental Dirac brackets can be calculated straightforwardly
from the definition (\ref{DBdef}),
\begin{eqnarray}
  \label{EMFDB}
  \{A_i(\fx),\Pi_j(\fy)\}_D&=&\{A_i(\fx),\Pi_j(\fy)\}\nn\\
&&\hspace{0.4cm}-\int d^3x'd^3y'\{A_i(\fx),\phi_m(\fx')\}C^{-1}_{mn}(\fx',\fy')\{\phi_n(\fy'),\Pi_j(\fy)\} \nn\\
&=&\delta_{ij}\delta^3(\fx,\fy)-\int d^3x'd^3y'\del_i^{\fx'}\delta(\fx,\fx')\frac{1}{4\pi|\fx'-\fy'|}\del_j^{\fy'}\delta(\fy',\fy)\nn\\
&=&\delta_{ij}\delta^3(\fx,\fy)-\del_i^{\fx}\del_j^{\fy}\frac{1}{4\pi|\fx-\fy|}=:t_{ij}(\fx)\delta^3(\fx,\fy)=:t_{ij}(\fx,\fy)
\end{eqnarray}
where the last line defines the transverse projector
$t_{ij}(\fx)$. The definition of the matrix-valued distribution $t_{ij}(\fx,\fy)$ will
be useful as well.

Quantization is achieved by the prescription\footnote{The
  field operators are denoted by the same symbols as their classical
  counterparts. It should be clear from the context which objects the
  symbols refer to.}
\begin{eqnarray}
  \label{EMfuncomrel}
  {\{A_i(\fx),\Pi_j(\fy)\}_D}=t_{ij}(\fx,\fy)\quad\rarr\quad [{A}_i(\fx),{\Pi}_j(\fy)]=it_{ij}(\fx,\fy)
\end{eqnarray}
which provides information on how the momentum operator $\Pi_k(\fx)$ acts on the wave
functional $\braket{A}{\Psi}=\Psi[A]$ in the coordinate
representation. Yet, not all components of $\Pi_k(\fx)$ are determined
by Eq.\ (\ref{EMfuncomrel}) since the
transverse projector has zero modes,
$\del^i_{\fx}t_{ij}(\fx,\fy)=\del^j_{\fy}t_{ij}(\fx,\fy)=0$. To be specific, the longitudinal components of the
field operators are left unconstrained by Eq.\
(\ref{EMfuncomrel}). We split the field operators into longitudinal and
transversal components, $A_i=A_i^\perp+A_i^\parallel$ with $A_i^\perp(\fx)=t_{ij}(\fx)A_j(\fx)$ (and for
$\Pi$ accordingly), and find from  Eq.\
(\ref{EMfuncomrel}) that
\begin{eqnarray}
  \label{EMtrcomrel}
  [A_i^\perp(\fx),\Pi_j^\perp(\fy)]=it_{ij}(\fx,\fy)\; .
\end{eqnarray}
The additional information needed for $A_k^\parallel$ and $\Pi_k^\parallel$ comes from the
constraints themselves, Eqs.\ (\ref{phi2}) and (\ref{EMCoul}), that now hold as operator
identities. The longitudinal projections  by
$\ell_{ij}:=\delta_{ij}-t_{ij}$ of both field operators
are identically zero,
\begin{align}
  \label{EMAlong}
  {A}_i^\parallel(\fx)&\equiv \ell_{ij}(\fx)A_j(\fx)=0\\
  \label{EMPilong}
  \Pi_i^\parallel(\fx)&\equiv\ell_{ij}(\fx)\Pi_j(\fx)=0\; 
\end{align}
Here, no contradiction
to the commutation relations (\ref{EMfuncomrel}) arises,
\begin{equation}
  \label{GaussNoContra}
  0=[A_k(\fx),\del_j^{\fy}\Pi_j(\fy)]=\del_j^{\fy}[A_k(\fx),\Pi_j(\fy)]=i\del_j^{\fy}t_{kj}(\fx,\fy)= 0\; ,
\end{equation}
unlike in the
quantization procedure with Poisson brackets, cf.\ Eq.\
(\ref{GaussContra}). The transverse components of the momentum
operator is found in view of Eq.\ (\ref{EMtrcomrel}) to be
\begin{eqnarray}
  \label{Pitr}
  {\Pi}^\perp_k(\fx)=t_{kj}(\fx)\frac{\delta}{i\delta{A}^\perp_j(\fx)}\; .
\end{eqnarray}
Quantization is thus complete and one may find from (\ref{EMHam}) the gauge-fixed
quantum Hamiltonian 
\begin{eqnarray}
  \label{EMqHam}
  {H}[A^\perp,\Pi^\perp]=\frac{1}{2}\int d^3x\,\left({{\bf \Pi}^\perp}^2+{\bf
      B}^2[{A}^\perp]\right)\; .
\end{eqnarray}
After all, we have arrived at a Hamiltonian expressed in terms of
gauge-invariant variables. Note that $A_k^\perp$ is left unchanged by
gauge transformations (\ref{EMginv}) and $\Pi_k=F_{k0}$ transforms
homogeneously in view of (\ref{Fmunu}) and hence transforms trivially in the
abelian theory.

\section{Constrained quantization of Yang--Mills theory}
\label{DiracYM}
The transfer of the techniques used in the previous section to non-abelian theories is aggravated by non-linearities
in the equations of motions. Nevertheless, constrained quantization of
YM theory in the temporal Coulomb gauge is feasible. This section is
devoted to presenting the steps of calculation that lead to the
gauge-fixed Hamiltonian operator of YM theory with static external
quark fields. 

Usually, the canonical quantization is performed in the
Weyl gauge where the spatial components of the gauge field can be
treated as a set of Cartesian coordinates, and after quantization a
coordinate transformation to curvilinear coordinates in the Coulomb
gauge is undertaken \cite{ChrLee80,CreMuzTud78,GerSak78,CheTsa87}. 
This corresponds to an incomplete gauge fixing before quantization and
is known to bring about a number of difficulties \cite{Lan89}. For instance,
if only the Weyl gauge, $A_0^a=0$, is enforced, the
Gauss law cannot be used as an operator equation and has to be imposed as a projection onto a physical Hilbert
space \cite{Bjo80,Jac80}, see also \cite{Rei97}. The physical states are not normalizable
 as an artefact of the residual gauge invariance
 \cite{RosTes84}. Moreover, the commutator $\left[ \cG^a,\cG^b \right]$ may be
 anomalously broken which obscures the projection on states
 \cite{FadSha86,FadSla91}. Another difficulty in fixing only the Weyl
 gauge is that the perturbative gauge field propagator is
plagued by unphysical poles \cite{CarCurMen82}. 

In the present thesis, the gauge shall be completely fixed\footnote{See footnote
  \ref{foot:global}.} already at the
classical level by imposing the temporal Coulomb gauge conditions
\begin{subequations}
\label{tempCoul}
\begin{align}
  \label{Weyl}
  A_0^a(x)&\approx 0 \\
  \label{Coulomb}
  \hat D_k^{ab}(x)A_k^b(x)&=\del_k A^a_k(x)\approx 0
\end{align}
\end{subequations}
and only afterwards, the quantization will be issued with the Dirac
brackets. This choice was also discussed in
\cite{LerMicRos87,GirRot84}. Complete gauge fixing, such as the temporal Coulomb gauge (\ref{tempCoul}), has the advantage that it
can be chosen such as to remove all unphysical degrees of freedom. The
Gauss law then holds as an operator identity. In exchange, the
algebra of the quantization procedure is slightly more involved. 

Analogously to electrodynamics, the first constraint to momentum
phase-space is given by the time component of the momentum operator
\begin{eqnarray}
  \label{Phi1}
    \varphi_1^a(x):=\Pi_0^a(x)=\frac{\delta \cL_{YM}}{\delta \del_0 A_0^a(x)}\approx
  0\; .
\end{eqnarray}
The Hamiltonian (\ref{classHam}) with ${\bf j}^a=0$ for simplicity,
\begin{eqnarray}
  \label{YMHam}
    H=\frac{1}{2}\int d^3x \left(\Pi_k^a\Pi_k^a+B_k^aB_k^a\right)-\int d^3x\,A^a_0\cG^a
\end{eqnarray}
serves as the generator of time translations.\footnote{In the same way
as in the previous section, we write $\Pi_k^a(x)$, meaning $\Pi^{ka}(x)$.} The stationarity of the
constraint (\ref{Phi1}) gives a further constraint,
\begin{eqnarray}
  \label{Phi2}
  \varphi_2^a(x):=\{\varphi_1^a(x),H_T\} = \cG^a(x)=\hat{D}^{ab}_k(x)\Pi_k^b(x)+g\rhoext^a(x)\approx 0\; ,
\end{eqnarray}
where the equal-time fundamental Poisson brackets were employed,
\begin{eqnarray}
  \label{FPBcolor}
  \left\{A_\mu^a(\fx,t),\Pi_\nu^b(\fy,t)\right\}=\delta_{\mu\nu}\delta^{ab}\delta^{3}(\fx,\fy)  \; .
\end{eqnarray}
The constraint $\varphi_2$ is found to be stationary
using the well-known identity \cite{FadSla91}
\begin{eqnarray}
  \label{GPB}
  \{\cG^a(\fx,t),\cG^b(\fy,t)\}=gf^{abc}\cG^c(\fx,t)\delta^3(\fx,\fy)\approx 0\; .
\end{eqnarray}
The above quantity weakly vanishes\footnote{A theorem due to Dirac
  \cite{Dir50} ensures that a Poisson bracket of two first-class
  constraints is again first-class.} in view of Eq.\ (\ref{Phi2}) and
therefore the constraints $\varphi_1$ and $\varphi_2$ in
Eqs. (\ref{Phi1}) and (\ref{Phi2}) form the complete set of
constraints. The latter are first-class and are turned into second-class
by the gauge fixing condition
(\ref{tempCoul}). 

Before proceeding to the calculation of the Dirac
brackets, let us introduce a  convenient short-hand notation. The
coordinate dependence is absorbed into the color index which is thus
understood as a collective index. Vectors in Lorentz space
  keep their Lorentz indices to avoid an over-estranged notation. All fields are considered at a fixed time. Repeated indices are summed over,
i.e.\ for the color indices this implies an integration over the
implicit coordinate dependence. 

For instance,
\begin{eqnarray}
  \label{shorthand}
  A^a_k(\fx)\equiv A^a_k\; ,\quad \Pi^a_k(\fx)\equiv \Pi^a_k\; .
\end{eqnarray}
In addition, derivatives are written as matrices in coordinate and color space,
\begin{eqnarray}
  \label{shorthand2}
  \del_k^{\fx}\delta^3(\fx,\fy)\delta^{ab}\equiv\del_k^{ab}\; ,\quad
  \hat D_k^{ab}(\fx)\delta^3(\fx,\fy)\equiv \hat D^{ab}_k\; .
\end{eqnarray}
Note that these matrices are antisymmetric under the exchange of the
collective indices,
\begin{eqnarray}
  \label{antisymmderiv}
  \del_k^{ab}=-\del_k^{ba}\; ,\quad \hat D^{ab}_k=-\hat D^{ba}_k\; .
\end{eqnarray}
Also note the symmetric matrices
\begin{eqnarray}
  \label{symmmat}
   \delta^{ab}\delta^3(\fx,\fy)\equiv\delta^{ab}\; ,\quad
   \delta^{ab}t_{ij}(\fx,\fy)\equiv t_{ij}^{ab}\; , \quad
   \delta^{ab}\ell_{ij}(\fx,\fy)\equiv \ell_{ij}^{ab}\; .
\end{eqnarray}

The constraints along with the gauge-fixing conditions are collected
in a vector $\phi$ with elements
\begin{subequations}
  \label{collectphi}
  \begin{align}
    \phi_1^a &= \varphi_1^a=\Pi_0^a\,\approx\, 0\\
    \phi_2^a &= \varphi_2^a=\cG^a\,\approx\, 0\\
    \phi_3^a &= \chi_1^a=A_0^a\,\approx\, 0\\
    \phi_4^a &= \chi_2^a=\del_k^{ab}A_k^b\,\approx\, 0
  \end{align}
\end{subequations}
to
give a regular constraint matrix $C^{ab}=\left(\{\phi_m^a,\phi_n^b\}\right)$.
The relevant Poisson brackets for its calculation yield
\begin{subequations}
\label{relevPB}
\begin{align}
  \label{relevPB1}
  \{A_0^a,\Pi_0^b\}&=\delta^{ab}\\
  \label{relevPB2}
  \{A_i^a,\Pi_j^b\}&=\delta_{ij}\delta^{ab}\\
  \label{relevPB3}
  \{A_k^a,\phi_2^b\}&=\hat D_j^{bc}\{A_k^a,\Pi_j^c\}=-\hat D_k^{ab}\\
  \label{relevPB4}
  \{\phi_4^a,\Pi_k^b\}&=\del_j^{ac}\{A_j^c,\Pi_k^b\}=\del_k^{ab}\\
  \label{relevPB5}
  \{\phi_2^a,\phi_4^b\}&=\hat D_i^{ac}\del_j^{bd}\{\Pi_i^c,A_j^d\}=\hat D_i^{ac}\del_i^{cb}=:-[G^{-1}]^{ab}
\end{align}
\end{subequations}
In the last line we have defined the non-local matrix $G^{-1}$ which is
symmetric in view of the antisymmetric matrices in Eq.\
(\ref{antisymmderiv}). Furthermore, due to the Coulomb gauge condition
(\ref{Coulomb}), it obeys the relation
\begin{eqnarray}
  \label{Gsymm}
  [G^{-1}]^{ab}=-\hat D_k^{ac}\del_k^{cb}=-\del_k^{ac}\hat D_k^{cb}
\end{eqnarray}
and its inverse $G$ is defined by
\begin{eqnarray}
  \label{Ginvdef}
  G^{ac}[G^{-1}]^{cb}=[G^{-1}]^{ac}G^{cb}=\delta^{ab}\; .
\end{eqnarray}
All (weakly) non-zero matrix elements $(C^{ab})_{mn}$ are now given by
Eqs.\ (\ref{relevPB1}) and (\ref{relevPB5}) so that the constraint matrix reads
\begin{eqnarray}
  \label{constrmat}
  C^{ab}\,\approx\,\left(
  \begin{array}{cccc}
    0 & 0 & -\delta^{ab} & 0\\
    0 & 0 & 0 & -[G^{-1}]^{ab} \\
    \delta^{ab} &0&0&0\\
    0&[G^{-1}]^{ab}&0&0
  \end{array}
\right)
\end{eqnarray}
and its inverse is given by
\begin{eqnarray}
  \label{C-1}
  [C^{-1}]^{ab}\,\approx\,\left(
  \begin{array}{cccc}
    0 & 0 & \delta^{ab} & 0\\
    0 & 0 & 0 & G^{ab} \\
    -\delta^{ab} &0&0&0\\
    0&-G^{ab}&0&0
  \end{array}
\right)\; .
\end{eqnarray}
At this point, let us emphasize that the matrix $G$, which will be the
classical counterpart of the ghost operator, can only be
made explicit by an infinite series expansion in the coupling
constant $g$.\footnote{On top of that, zero modes of the matrix
  $G^{-1}$ infer that $G$ exists only in subspace of $\Gamma_R$. This
  will be discussed in chapter \ref{ghostdom}.} This has led some authors to refrain from the constrained
quantization with such a constraint matrix, see Refs.\
\cite{Sun82,KugoBook,HalRen03}. In Refs.\ \cite{Tud79,GirRot82}, a Hamiltonian
operator is obtained but one is left with operator ordering ambiguities. In other gauges, where $A_0\neq 0$, the expression for $C^{ab}$ becomes
even more complicated \cite{HanRegTei76}. Our approach aims at a
comparison to the Christ-Lee Hamiltonian \cite{ChrLee80} in the
temporal Coulomb gauge. Therefore, let us press on with the formal expression
for $G$ defined by (\ref{Ginvdef}) and with the inverse constraint matrix
(\ref{C-1}).

The fundamental Dirac brackets are defined by
\begin{eqnarray}
  \label{fundDBdef}
  \{A_\mu^a,\Pi_\nu^b\}_D&=&\{A_\mu^a,\Pi_\nu^b\}-\{A_\mu^a,\phi_{m}^{a'}\}([C^{-1}]^{a'b'})_{mn}\{\phi_n^{b'},\Pi_\nu^b\}\; 
\end{eqnarray}
and, using Eq.\ (\ref{relevPB}), are calculated to yield
\begin{align}
  \label{DBtriv}
  \{A_0^a,\Pi_k^b\}_D&=\{A_0^a,\Pi_0^b\}_D=\{A_k^a,\Pi_0^b\}_D=0\\
  \label{DBij}
  \{A_i^a,\Pi_j^b\}_D&=\delta_{ij}\delta^{ab}-(-\hat D^{ac}_i)G^{cd}\del^{db}_j=\delta_{ij}\delta^{ab}+\hat D^{ac}_iG^{cd}\del^{db}_j=:T_{ij}^{ab}
\end{align}
It is clear that the Dirac brackets (\ref{DBtriv}) must vanish since, by
construction, the Dirac bracket of a constraint with any phase space
function vanishes, recall Eq.\ (\ref{strongphi}). In particular, the Gauss law functional $\cG^a=\phi_2^a$ has
vanishing Dirac bracket with any function $\cO[A,\Pi]$,
\begin{eqnarray}
  \label{DBGauss}
  \{\cG^a,\cO^b\}_D=0\; ,
\end{eqnarray}
as one may easily convince oneself, cf.\ Eq.\ (\ref{strongphi}). All second-class constraints
equations hold strongly.
The non-symmetric matrix $T^{ab}_{ij}$ occurring in Eq.\ (\ref{DBij}) can be
understood in some sense as a non-abelian generalization of the transverse projector $t_{ij}$.\footnote{Letting $\hat D\rarr\del$, one regains the familiar
expression (\ref{EMFDB}). Note the sign in $G=-(\hat D\del)^{-1}$.} It
obviously has the projector property $T^2=T$, and the projector
orthogonal to it is given by $L:=\id-T$. We thus have
\begin{eqnarray}
  \label{LandT}
  L_{ij}^{ab}=-\hat D_i^{ac}G^{cd}\del_j^{db}\; ,\quad
  T_{ij}^{ab}=\delta_{ij}^{ab}+\hat D_i^{ac}G^{cd}\del_j^{db}\; ,
  \quad TL=LT=0\; .
\end{eqnarray}
Just like for $t_{ij}$,
it is important to note that $T_{ij}^{ab}$ has zero modes,
\begin{eqnarray}
  \label{Tzeros}
  \del_i^{ab}T_{ij}^{bc}=T_{ij}^{ab}\hat D_j^{bc}=0\; .
\end{eqnarray}

We are now ready to pass to the quantum theory with the prescription
\begin{eqnarray}
  \label{funcomrel}
i\{A_\mu^a,\Pi_\nu^b\}_D\rarr [{A}_\mu^a,{\Pi}_\nu^b]\; .  
\end{eqnarray}
It is
understood that all fields are operators henceforth. The second-class
constraints (\ref{collectphi}) now hold strongly and can be imposed as
operator identities,
\begin{eqnarray}
  \label{opid}
  A_0^a=\Pi_0^a=\cG^a=A_k^{\parallel a}=0\; .
\end{eqnarray}
In order to arrive at a representation for the momentum operator
$\Pi_k^a$ that is to obey
\begin{eqnarray}
\label{comrel}
[A_i^{\perp a},\Pi_j^b]=iT_{ij}^{ab}\; ,
\end{eqnarray}
it is helpful to split it into the orthogonal projections (from the
right side) of $T$ and
$L$,
\begin{eqnarray}
  \label{Pisplit}
  \Pi_k^a=\Pi_k^{Ta}+\Pi_k^{La}\;
  ,\qquad\Pi_k^{Ta}=\Pi_j^bT_{jk}^{ba}\; ,\quad
  \Pi_k^{La}=\Pi_j^bL_{jk}^{ba}\; .
\end{eqnarray}
The commutation relations (\ref{comrel}) are thus decomposed into
two separate ones,
\begin{eqnarray}
  \label{comrelL}
  [A_i^{\perp a},\Pi_j^{Lb}]&=&0\\
 \label{comrelT}
  [A_i^{\perp a},\Pi_j^{Tb}]&=&iT_{ij}^{ab}
\end{eqnarray}
and a situation similar to the abelian case, cf.\ section \ref{QED},
occurs. Recall that the longitudinal component $\Pi^\parallel$ obeyed a
trivial commutation relation equivalent to $\Pi^L$ in (\ref{comrelL}) and had to be determined by the (strong)
Gauss law (\ref{EMPilong}). The transverse component $\Pi^\perp$ obeyed $[A_i^\perp,\Pi_j^\perp]=t_{ij}$
which has the easily found solution (\ref{Pitr}). Here, the
``non-abelian projector'' $T$ (\ref{LandT}) in $[A^\perp,\Pi^T]=iT$ contains field
operators $A^\perp$. Nevertheless, a
representation of $\Pi_k^{Ta}$ can be found to be
\begin{eqnarray}
  \label{PiT}
  \Pi_k^{Ta}=T^{ba}_{jk}\frac{\delta}{i\delta A_j^{\perp
      b}}=T^{ba}_{jk}\Pi_j^{\perp b}\; ,
\end{eqnarray}
as verified by plugging Eq.\ (\ref{PiT}) into Eq.\
(\ref{comrelT}). Note here that $T$ is not symmetric.

The component $\Pi^L$ is determined by the Gauss law (\ref{Phi2}) which yields
\begin{eqnarray}
  \label{Gauss+PiL}
  \hat D_k^{ab}\Pi_k^b=\hat D_k^{ab}T_{jk}^{cb}\Pi_j^{\perp c}+\hat
  D_k^{ab}\Pi_j^cL_{jk}^{cb}=\hat
  D_k^{ab}\Pi_k^{Lb}=-g\rho_{\textrm{ext}}^a\; .
\end{eqnarray}
where the contribution of $\Pi^T$ vanishes in view of Eq.\ (\ref{Tzeros}). To solve
Eq.\ (\ref{Gauss+PiL}) for $\Pi^L$, one may use the Helmholtz  theorem
to write $\Pi^L$ as the gradient field of some field operator $\Phi$ since
\begin{eqnarray}
  \label{PiLgrad}
  \left({\boldsymbol{\del}}\times{\bf
      \Pi}^L\right)^a_k=\epsilon_{ijk}\del_i^{ab}\Pi_j^{Lb}=-\epsilon_{ijk}\del_i^{ab}\Pi_n^e\hat D_n^{ed}G^{dc}\del_j^{cb}=0\; .
\end{eqnarray}
We can then write $\Pi_k^{La}=-\del_k^{ab}\Phi^b$ and solve
the Gauss law (\ref{Gauss+PiL}) by
$\Phi^a=-gG^{ab}\rho_{\textrm{ext}}^b$. The operator $\Pi^L$ then is explicitly
given as a functional of field operators $A^\perp$,
\begin{eqnarray}
  \label{PiL}
  \Pi_k^{La}[A^\perp]=g\del_k^{ab}G^{bc}[A^\perp]\rho_{\textrm{ext}}^c\; .
\end{eqnarray}
One may subsequently verify the commutation relation (\ref{comrelL}).

Before we go on to derive the Hamiltonian operator from the classical
one (\ref{YMHam}) by promoting the fields $A_k^a$ and $\Pi_k^a$ to their
operator-valued counterparts, operator ordering is briefly
discussed, cf.\ Ref.\ \cite{ChrLee80}. Gauge fixing gives rise to non-trivial factors in the
Hamiltonian, as can be seen in the path integral formalism. Using the
the Faddeev--Popov trick \cite{FadPop67}, the kinetic energy $E_k$
yields\footnote{The result is reminiscent of the Laplace--Beltrami
  operator $\Delta$ with a non-trivial metric $g_{\mu\nu}$ and its determinant $g=\det
  (g_{\mu\nu})$, see Ref.\ \cite{Nak90},
  \begin{equation}
    \label{LaplaceBeltrami}
    \Delta=-\frac{1}{\sqrt{g}}\del_\mu\sqrt{g}g^{\mu\nu}\del_\nu\; ,
  \end{equation}
as pointed out in Refs.\ \cite{ChrLee80, Kuc86}. The Weyl gauge operators
are then in Cartesian coordinates, whereas the Coulomb gauge operators
correspond to curvilinear ones. In contrast to Ref.\ \cite{ChrLee80}, we
here have no Cartesian coordinates to start with, since the temporal
Coulomb gauge was already fixed on the classical level.}
\begin{eqnarray}
  \label{eq:pathint}
  E_k&=&\braket{\Psi}{\Psi}^{-1}\frac{1}{2}\bra{\Psi}\Pi^a_k\Pi^a_k\ket{\Psi}\nn\\
&=&\braket{\Psi}{\Psi}^{-1}\frac{1}{2}\int\cD A\:\Psi^*[A]\Pi^a_k\Pi^a_k\Psi[A]\nn\\
&=&\braket{\Psi}{\Psi}^{-1}\frac{1}{2}\int\cD A\: (\Pi_k^a\Psi)^*[A](\Pi_k^a\Psi)[A]\nn\\
&=&\frac{1}{2}\int\cD A^\perp\cJ[A^\perp] \left((\Pi_k^{Ta}+\Pi_k^{La})\Psi\right)^*[A^\perp]\left((\Pi_k^{Ta}+\Pi_k^{La})\Psi\right)[A^\perp]\nn\\
&=&\frac{1}{2}\int\cD A^\perp\cJ[A^\perp] \Psi^*[A^\perp]
\left(\frac{1}{\cJ[A^\perp]} \left({\Pi_k^{Ta}}^\dagger+{\Pi_k^{La}}^\dagger\right)
\cJ[A^\perp](\Pi_k^{Ta}+\Pi_k^{La})
\right) \Psi[A^\perp]\; ,\nn\\
\end{eqnarray}
where the state norm $\braket{\Psi}{\Psi}$ was chosen such as to cancel the
group volume, and 
\begin{equation}
  \label{FPdet}
  \cJ[A^\perp]=\textrm{Det}(G^{-1})[A^\perp]
\end{equation}
is the Faddeev--Popov determinant. The last line in Eq.\
(\ref{eq:pathint}) is obtained by taking the
hermitian adjoint of the product $\cJ(\Pi^T+\Pi^L)$ with respect to the
inner product in functional space when bringing the $\Psi^\ast$ to the left. One can read off from Eq.\
(\ref{eq:pathint}) the form of the kinetic part of the Hamiltonian
operator
\begin{eqnarray}
  \label{Hk}
   H_k[A^\perp,\Pi^T,\Pi^L]=\frac{1}{2}\cJ^{-1}[A^\perp]\left({\Pi^{Ta}_k}^\dagger+{\Pi_k^{La}}^\dagger\right)\cJ[A^\perp]
\left(\Pi^{Ta}_k+\Pi_k^{La}\right)\; .
\end{eqnarray}
The decomposition of the momentum operator $\Pi=\Pi^T+\Pi^L$ with the
expressions (\ref{PiT}) and (\ref{PiL}) can now be inserted into (\ref{Hk}), giving rise to four terms. Let us consider
these separately. One term reads
\begin{eqnarray}
  \label{HkLL}
  H_k^{LL}&=&\frac{1}{2}\:\cJ^{-1}{\Pi_k^{La}}^\dagger\cJ\Pi_k^{La}\nn\\
&=&\frac{1}{2}\:\cJ^{-1}\left(g\del_k^{ab}G^{bc}\rho_{\textrm{ext}}^c\right)\cJ\left(g\del_k^{ad}G^{de}\rho_{\textrm{ext}}^e\right)\nn\\
&=&\frac{g^2}{2}\:\rho_{\textrm{ext}}^cG^{cb}\left(-\del_k^{ba}\del_k^{ad}\right)G^{de}\rho_{\textrm{ext}}^e\nn\\
&=&\frac{g^2}{2}\:\rho_{\textrm{ext}}^cF^{ce}\rho_{\textrm{ext}}^e
\end{eqnarray}
and is recognized as the Coulomb interaction of external charges via
the Coulomb Green function $F:=G(-\del^2)G$. Using the relation
$t_{ij}^{ab}T_{jk}^{bc}\del_k^{cd}=-t_{ij}^{ab}L_{jk}^{bc}\del_k^{cd}$
as well as the identification of the dynamical gluonic charge
$\rhodyn$ in\footnote{With the conventions used here, $\rhodyn^a =
  \hat A_k^{\perp ab}\Pi_k^{\perp b} = -\hat{\bf A}^{\perp
    ab}\cdot{\bf \Pi}^b$. Thus, one arrives at $\del_k\Pi_k^a=-g(\rhoext^a+\rhodyn^a)$.}
\begin{equation}
  \label{rhodyndef}
\hat D_k^{ab}\Pi_k^{\perp b}=\Pi_k^{\perp b}\hat D_k^{ab}=g\hat A_k^{\perp ab}\Pi_k^{\perp
  b}=:g\rhodyn^a 
\end{equation}
we further find
\begin{eqnarray}
  \label{HkLT}
    H_k^{LT}&=&\frac{1}{2}\:\frac{1}{\cJ}{\Pi_k^{La}}^\dagger\cJ\Pi_k^{Ta}\nn\\
    &=&\frac{1}{2}\:g\del_k^{ab}G^{bc}\rho_{\textrm{ext}}^c T_{mk}^{da}\Pi_m^{\perp d}\nn\\
    &=&\frac{g}{2}\:\rhoext^cG^{cb}\left(-\del_k^{ba}\right)\hat D_m^{dd'}G^{d'a'}\del_k^{a'a}\Pi_m^{\perp d}\nn\\
    &=&\frac{g}{2}\:\rhoext^cG^{cb}\left(-\del_k^{ba}\del_k^{aa'}\right)G^{a'd'}\hat D_m^{d'd}\Pi_m^{\perp d}\nn\\
    &=&\frac{g^2}{2}\:\rhoext^cF^{cd'}\rhodyn^{d'}
\end{eqnarray}
and 
\begin{eqnarray}
  \label{HkTL}
      H_k^{TL}&=&\frac{1}{2}\:\frac{1}{\cJ}{\Pi_k^{Ta}}^\dagger\cJ\Pi_k^{La}\nn\\
      &=&\frac{1}{2}\:\frac{1}{\cJ}\Pi_m^{\perp b}T_{mk}^{ba}\cJ\:g\del_k^{ac}G^{cd}\rho_{\textrm{ext}}^d\nn\\
      &=&\frac{g}{2}\:\frac{1}{\cJ}\Pi_m^{\perp b}\hat D_m^{bb'}\cJ G^{b'a'}\del_k^{a'a}\del_k^{ac}G^{cd}\rhoext^d\nn\\
      &=&\frac{g^2}{2}\:\frac{1}{\cJ}\rhodyn^{b'}\cJ F^{b'd}\rhoext^d\; ,
\end{eqnarray}
two contributions that account for the Coulomb interaction of
dynamical and external charges. Using the identity
$t_{m'm}^{b'b}T_{mk}^{ba}T_{nk}^{ca}t_{nn'}^{cc'}=t_{m'n'}^{b'c'}+t_{m'm}^{b'b}L_{mk}^{ba}L_{nk}^{ca}t_{nn'}^{cc'}$
we get two further  contributions to $H_k$,
\begin{eqnarray}
  \label{HkTT}
        H_k^{TT}&=&\frac{1}{2}\:\frac{1}{\cJ}{\Pi_k^{Ta}}^\dagger\cJ\Pi_k^{Ta}\nn\\
        &=& \frac{1}{2}\:\frac{1}{\cJ}\Pi_m^{\perp
        b}T_{mk}^{ba}\cJ T_{nk}^{ca}\Pi_n^{\perp c}\nn\\
      &=&\frac{1}{2}\:\frac{1}{\cJ}\Pi_k^{\perp a}\cJ\Pi_k^{\perp a} +
      \frac{1}{2}\:\frac{1}{\cJ}\Pi_m^{\perp
        b}\hat D_m^{bb'}G^{b'a'}\del_k^{a'a}\cJ
      \hat D_n^{cc'}G^{c'a''}\del_k^{a''a}\Pi_n^{\perp c}\nn\\
&=&\frac{1}{2}\:\frac{1}{\cJ}\Pi_k^{\perp a}\cJ\Pi_k^{\perp a} +
\frac{g^2}{2}\:\frac{1}{\cJ}\rhodyn^{b'}\cJ
G^{b'a}\left(-\del_k^{a'a}\del_k^{aa''}\right)G^{a''c'}\rhodyn^{c'}\nn\\
&=&\frac{1}{2}\:\frac{1}{\cJ}\Pi_k^{\perp a}\cJ\Pi_k^{\perp a} +
\frac{g^2}{2}\:\frac{1}{\cJ}\rhodyn^{b'}\cJ F^{b'c'}\rhodyn^{c'}\; ,
\end{eqnarray}
namely the kinetic energy of transverse gluons as well as the Coulomb
interaction of dynamical charges. The magnetic potential operator
$H_p$ can be obtained straightforwardly from its classical counterpart in Eq.\ (\ref{YMHam}) since
it only comprises the coordinates $A_k^a$. It simply yields
\begin{eqnarray}
  \label{Hp}
  H_p=\frac{1}{2}B_k^a[A^\perp]B_k^a[A^\perp]\; .
\end{eqnarray}
Adding the contributions (\ref{HkLL}),
(\ref{HkLT}), (\ref{HkTL}), (\ref{HkTT}) and (\ref{Hp}), we finally arrive at the
Hamiltonian operator in temporal Coulomb gauge. In explicit notation, it reads
\begin{equation}
  \label{CLHam}
     \boxed{\begin{gathered}
   H[A^\perp,\Pi^\perp]=\frac{1}{2}\int d^3x\,\frac{1}{\cJ}\Pi_k^{\perp
    a}(\fx)\cJ\Pi_k^{\perp a}(\fx)+\frac{1}{2}\int d^3 x B_k^{\perp
    a}(\fx)B_k^{\perp a}(\fx)\\
\qquad +\frac{g^2}{2}\int d^3x\:
  d^3y\frac{1}{\cJ}\rho^a(\fx)\cJ\: F^{ab}(\fx,\fy)\rho^b(\fy)\; .
     \end{gathered}}
   \end{equation}
Here, we have defined the total charge density 
\begin{equation}
  \label{rhodef}
  \rho^a(\fx):=\rhodyn^a(\fx)+\rhoext^a(\fx)=\hat
  A^{ab}_k(\fx)\Pi^b_k(\fx)+\psi^\dagger(\fx)iT^a\psi(\fx)\; .
\end{equation}
For future reference, let us write down the explicit form of the
Coulomb operator $F$,
\begin{equation}
  \label{Fdef}
  F^{ab}(\fx,\fy)=\int d^3x'\:G^{ac}(\fx,\fx')\left(-\del^2\right)G^{cb}(\fx',\fy)
\end{equation}
and of the inverse ghost operator
\begin{equation}
  \label{G-1def}
  [G^{-1}]^{ab}(\fx,\fy)=-\delta^{ab}\del^2\delta^3(\fx,\fy)-g\hat
  A_k^{ab}(\fx)\del_k^{\fx}\delta^3(\fx,\fy)\; .
\end{equation}
Christ and Lee \cite{ChrLee80} derived the same result as in Eq.\
(\ref{CLHam}) using a different approach. They quantized the Cartesian
coordinates in Weyl gauge and only subsequently transformed into the
Coulomb gauge, keeping the Gauss law as a constraint on the wave
functional. Here, we were able to reinforce the Gauss law as a operator
identity, having used Dirac's concept of constrained quantization. It
is reassuring to see that the two methods produce the same result for
the Yang--Mills energy spectrum,
keeping in mind that in general such an equivalence cannot be
established \cite{Kuc86,RomTat89,Sch90,Lol90,Sha00}.

\chapter{Infrared ghost dominance}
\label{ghostdom}

The complex structure of the Yang--Mills Hamiltonian (\ref{CLHam})
derived in the previous section does not allow for a rigorous
calculation of the energy spectrum or the Green
functions. Perturbation theory in the gauge coupling $g$ provides a
possibility to describe interactions with high momentum transfer in QCD
where asymptotic freedom guarantees that $g$ can be treated as a small
parameter. The infrared sector, however, is not accessible by
perturbative methods since we know from the renormalization group that
$g$ increases as we go to lower energies. An understanding of long-range interactions
of quarks and gluons thus calls for nonperturbative methods. \emph{Color
confinement}, the experimental evidence that quarks and gluons have
never been detected as asymptotic states and must be confined into color
singlets, is to date one of the most challenging topics in
theoretical physics. 

By means of lattice calculations, it has
been possible to penetrate the infrared nonperturbative sector of QCD
and uncover a confining potential between (static) quarks
\cite{Gre03,Bal00}. At present, however,
available lattice sizes do not suffice to describe the Green
functions in the deep infrared \cite{Fis+07,FisAlkRei02}. Despite the
ever-growing computational power, a numerical simulation of
confinement on the lattice will always fall short of proving it%, due
                                %to the weakly defined continuum limit
. Nevertheless, the achievements
of the lattice community with calculations that are particularly reliable in the
intermediate momentum regime, are indispensable for the mutual dialogue
with the continuum approach. A joint theoretical investigation of QCD
phenomena is a promising project for progressive research.

The continuum approach has the intriguing feature that
the asymptotic infrared limit can be studied analytically. In the last decade,
a new understanding of infrared QCD has arisen from studying
continuum Yang--Mills (YM) theory via Dyson--Schwinger equations. The Landau gauge has the
advantage of being covariant, allowing for straightforward
perturbative calculations, and it therefore encouraged many to
intensive investigation of the infrared properties of YM
theory \cite{SmeAlkHau97,SmeHauAlk98,Fis06}. In Coulomb gauge, non-covariance brings about severe
technical difficulties which are only recently on the verge of being
overcome \cite{WatRei06,WatRei07a,WatRei07b}. Nevertheless, the Coulomb gauge might be the more
efficient choice to identify the nonabelian degrees of freedom. It is well-known that screening and
anti-screening contributions to the interquark potential are neatly
separated in Coulomb gauge perturbation theory \cite{Khr70}. As for the infrared
domain, the Gribov--Zwanziger scenario serves as a transparent
confinement mechanism \cite{Gri78,Zwa97}.

In this chapter, the Gribov--Zwanziger scenario shall be introduced and
analyzed in the temporal Coulomb gauge. It will be thus possible to
provide information on the asymptotic infrared behavior of the Green
functions and recover a linearly rising potential between static color
charges, i.e.\ heavy quark confinement. A full numerical calculation
in chapter \ref{VarVac}
will be seen to
reproduce the infrared asymptotics given here
analytically. Moreover, the infrared asymptotics of Landau gauge can
be retrieved from a generalization of the results in the temporal Coulomb gauge.

\section{Gribov--Zwanziger scenario of confinement}
\label{GribovZwanziger}

Having outlined the issues of redundant degrees of freedom in section
\ref{Diracconcept}, a discussion of gauge fixing in the context of
Yang--Mills theory may follow naturally. In particular, we have touched
upon the issue of uniqueness of a gauge fixing condition
condition 
\begin{eqnarray}
  \label{chi0}
  \chi^a[A]=0 \; .
\end{eqnarray}
Uniqueness was shown to be locally guaranteed by the criterion
(\ref{uniquegauge}) which reads for the classical YM theory
\begin{eqnarray}
  \label{uniqueYM}
  \Det\left(\{ \chi^a,\cG^b \}\right) \napprox 0 \; .
\end{eqnarray}
Locally means here that only a small neighborhood of the point $\cA$ in
configuration space that satisfies (\ref{chi0}) is considered. An
infinitesimal gauge transformation according to Eq.\ (\ref{symmtrans}),
\begin{eqnarray}
  \label{infinitrafo}
  \chi^a[\cA]\rarr \chi^a[\cA]+\left. \{ \chi^a, \cG^b
    \}\right|_{A=\cA}\epsilon^b\; ,
\end{eqnarray}
shows that $\cA$ is the unique solution to (\ref{chi0}) only if
Eq.\ (\ref{uniqueYM}) holds. Let us now turn to the quantum theory in
the Schr\"odinger picture. Gribov \cite{Gri78} discovered that
if the Coulomb gauge condition $\chi^a(\fx)=\del_kA^a_k(\fx)=0$ is chosen,
uniqueness of its solution $\cA$ is lost if the
matrix $G^{-1}$ with components 
\begin{equation}
  \label{FPop}
  \left[G^{-1}\right]^{ab}(\fx,\fy)=\left(-\del^2\delta^{ab}-g\hat A_k^{ab}\del_k^{\fx}\right)\delta^3(\fx,\fy)
\end{equation}
develops zero modes. From the explicit form of $G^{-1}$ this is seen to happen for large magnitudes of $gA$, i.e.\ in the
nonperturbative domain. The field configuration $\cA$ can then shown to
be connected by a gauge transformation to one or more $\cA^U$, called  \emph{Gribov copies}, that also
obey the Coulomb gauge condition. Gribov copies lead to incorrect
results in the calculation of expectation values due to over-counting
of physically equivalent field configurations. %Moreover, it was
                                %argued in \cite{} that the existence
                                %of Gribov copies imply a breakdown of
                                %color charges in QCD. 
The criterion for avoiding Gribov copies is the quantum analogue of
Eq.\ (\ref{uniqueYM}), noting that $G^{-1}=\left\{\del_kA_k,\cG\right\}$, and reads
\begin{equation}
  \label{Jneq0}
  \cJ[A]=\Det\left(G^{-1}\right)\neq 0\; .
\end{equation}
It is illuminating to see how the above uniqueness condition follows
from a direct calculation. Consider a gauge transformation $\cA^U$ where
$U(\fx)=\e^{-\alpha(\fx)}\approx 1-\alpha(\fx)$
% with $[(\del_k\alpha),\alpha]=0$, as
% realized in infinitesimal transformations 
is very close to unity. The field
$\cA^U$ is a Gribov copy of $\cA$ if% \footnote{For
%   $[(\del_k\alpha),\alpha]=0$, we can write $\del_kU^\dagger=(\del_k\alpha)U^\dagger=U^\dagger\del_k\alpha$.}
\begin{eqnarray}
  \label{globalGribov}
  0&=&\del_k\cA_k^U=\del_kU\cA_kU^\dagger+\frac{1}{g}\del_kU\del_kU^\dagger\nn\\
&=&U\left(-\left(\del_k\alpha\right)\cA_k+\cA_k\left(\del_k\alpha\right)-\frac{1}{g}\left(\del_k\alpha\right)\left(\del_k\alpha\right)+\frac{1}{g}\left(\del^2\alpha\right)+\frac{1}{g}\left(\del_k\alpha\right)\left(\del_k\alpha\right)\right)U^\dagger\nn\\
&=&\frac{1}{g}U\left(\left(\del^2\alpha\right)+g\left[\cA_k,\left(\del_k\alpha\right)\right]\right)U^\dagger\nn\\
&=&\frac{1}{g}UT^a\left(\left(\delta^{ab}\del^2+g\hat
    \cA^{ab}_k\del_k\right)\alpha^b\right)U^\dagger\; .
\end{eqnarray}
Since the matrices $U$ are regular and $T^a$ are linearly independent,
Eq.\ (\ref{globalGribov}) can be written as
\begin{eqnarray}
  \label{whyFMR}
  \int d^3y  \left[G^{-1}\right]^{ab}(\fx,\fy)\alpha^b(\fy)=0\; .
\end{eqnarray}
Thus, if the condition (\ref{Jneq0}) is fulfilled, then
$\alpha^a(\fx)=0$ and $\cA^U=\cA$, and the solution $\cA$ of the gauge fixing
condition (\ref{chi0}) is unique.

\begin{figure}
  \centering
  \inc[scale=0.55]{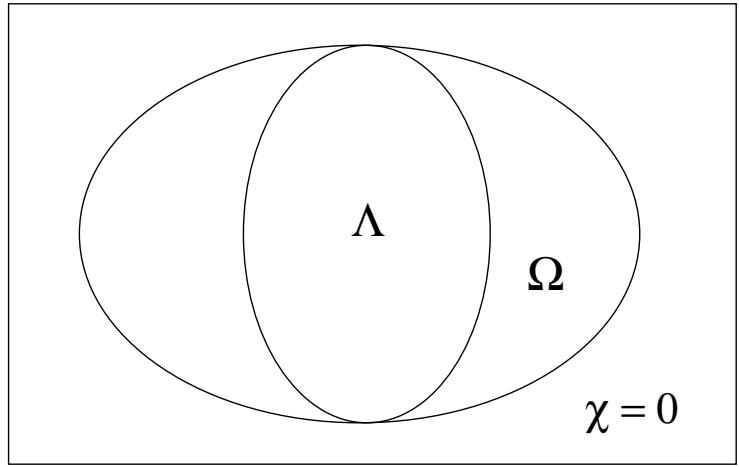}
  \caption{Configuration space and its restriction to the gauge fixing condition $\chi=0$, the (first) Gribov region $\Omega$ and the fundamental modular region $\Lambda$. }
  \label{fig:Gribov}
\end{figure}

The configuration space
is therefore decomposed into the so-called \emph{Gribov regions},
alternating in sign of the Faddeev--Popov determinant $\cJ$. By definition, the first Gribov region $\Omega$ contains the
perturbative vacuum, $gA=0$, and has a positive definite Faddeev--Popov determinant,
\begin{equation}
  \label{1stGribov}
  \Omega=\left\{A\: | \:  \del_kA_k=0\; ,\;\cJ[A]>0\right\}\; .
\end{equation}
In order to avoid Gribov copies, it is necessary to restrict the
configuration space to the first Gribov region $\Omega$, bounded by
the Gribov horizon $\del\Omega$, see Fig.\ \ref{fig:Gribov}. Practically,
it is not straightforward to implement this restriction. One might
naturally ask whether a gauge fixing condition exists, different from
the Coulomb gauge, for which the Faddeev--Popov determinant is manifestly
positive and the Gribov problem does not arise. However, a theorem due to
Singer \cite{Sin78} states that (on a three- or four-dimensional compact manifold) there exists no
gauge fixing condition that does not have the Gribov problem.

A restriction to $\Omega$ can be realized by an action principle, as noted by
Polyakov\footnote{See footnote in Ref.\ \cite{Gri77}.}. It requires that the
$L^2$ norm of the gauge field $A_k(\fx)$ be minimal along its orbit,
parametrised by the gauge transformations \mbox{$U(\fx)=\e^{-\alpha(\fx)}$},
\begin{eqnarray}
  \label{L2}
   F_A[U]&:=&\int d^3x\, \tr \left(A_k^U(\fx)\right)^2\nn\\
%&=&  \int d^3x\, \tr \left(A_k(\fx)+\frac{1}{g}U(\fx)\del_kU^\dagger(\fx)\right)^2\nn\\
&=&  F_A[1]-\frac{2}{g}\int
d^3x\, \tr\left(\alpha(\fx)\del_kA_k(\fx)\right)+\nn\\
&&\qquad\frac{1}{g^2}\int d^3xd^3y
\tr\left(\alpha^\dagger(\fx) G^{-1}[A](\fx,\fy)\alpha(\fy)\right)\; .
\end{eqnarray}
The stationarity condition sets $A=\cA$ and the minimum condition
yields $\cJ[\cA]>0$. Action principles are a common technique to implement
gauge fixing on the lattice \cite{ManOgi86}.

However, the restriction to $\Omega$ is not
sufficient to exclude gauge copies. 
Various examples \cite{JacMuzReb78,Hen79} indicate that there are Gribov copies inside
$\Omega$. A proof of existence was given by Ref.\ \cite{SemFra82}%and
                                %further established in \cite{}
, see also Ref.\ \cite{DelZwa91}. These remnant Gribov copies show that
the minima of $F_A[U]$ in Eq.\ (\ref{L2}) are degenerate. Lattice
calculations confirm that there are indeed Gribov copies inside
$\Omega$ \cite{ManOgi90,Mar93,Qua+07}. 
A further restriction of configuration space is needed to completely
exclude Gribov copies. %was advocated by Zwanziger \cite{}. 
The \emph{fundamental
  modular region} $\Lambda$ is defined to be the set of unique\footnote{Up to
  global gauge transformations.}
absolute minima of the functional (\ref{L2}) among gauge orbits,
\begin{equation}
  \label{defLambda}
  \Lambda:=\left\{A:F_A[1]\leq F_A[U]\;\;\forall U \right\}\; .
\end{equation}
It was shown that the interior of $\Lambda$ is indeed free of
Gribov copies whereas the boundary $\del\Lambda$ still contains Gribov
copies \cite{Baa92}. The gauge condition specified by Eq.\ (\ref{defLambda}) is also termed
``minimal Coulomb gauge'' \cite{Zwa98}. Although conceptually vital,
the fundamental modular region as yet lacks the utility for explicit
calculations. Little is known about the boundary
$\del\Lambda$. %although some of its properties were gained in Refs.\
               %\cite{}. 
Fortunately, it was shown in Ref.\ \cite{Zwa03} by means of
stochastic quantization that despite the Gribov copies
inside $\Omega$, expectation values are computed correctly if the
integration domain is $\Omega$ instead of $\Lambda$. This is due to
the finding \cite{Zwa94} that the dominant field configurations lie on
$\del\Omega\cap\del\Lambda$, the common boundary of $\Omega$ and $\Lambda$.\footnote{In this context, 
let us mention that the field configurations on
$\del\Omega\cap\del\Lambda$ can be identified as center
vortices. These field
configurations were found to drive the confining mechanism on the lattice in Coulomb gauge \cite{GreOle03} as well as in
Landau gauge \cite{GatLanRei04}.} We shall
therefore restrict the configuration space to $\Omega$ in the
following. Gauge fields $A$ are always transverse and the
superscript $\perp$ is abandoned for brevity.

The restriction to a compact region in configuration space, be it
$\Omega$ or $\Lambda$, has important consequences for the infrared
sector of the theory. Due to asymptotic freedom, the ultraviolet is
not affected by the horizon since when the coupling becomes small, all
relevant configurations are in the vicinity of $gA=0$. It was shown
that this point always has a finite distance from the horizon
\cite{DelZwa89}. The nonperturbative infrared sector, on the other hand, can
in principle be governed by any region within $\Omega$. From
statistical mechanics, we know that in a compact sphere with radius $r$ of high
dimension $N$, the probability distribution is concentrated at the
boundary, due to the ``entropy factor'' $r^{N-1}dr$. If Yang--Mills theory
is regularized on the lattice, Zwanziger argued \cite{Zwa97} that the
situation is comparable to the latter example, the dimensionality $N$ even diverges in the
continuum limit. In the asymptotic infrared, the momentum space ghost
propagator
\begin{eqnarray}
  \label{DGdef}
  D_G(k):=\frac{1}{N_c^2-1}\delta^{ab}\int d^3x\,  \lla G[A]\rra^{ab}(\fx,\fy) \e^{-i\fk\cdot(\fx-\fy)}\; ,\quad k=|\fk|
\end{eqnarray}
is therefore expected to diverge strongly,
\begin{equation}
  \label{horizondef}
  \lim_{k\rarr 0}\left[ k^2 D_G(k) \right]^{-1} = 0\; ,
\end{equation}
due to divergent eigenvalues of the operator $G$ on the
Gribov horizon. Eq.\ (\ref{horizondef}) is known as the {\it horizon
  condition}. Gribov discussed perturbatively in Ref.\
\cite{Gri78} that only for $k=0$ the ghost propagator $D_G(k)$ can have a pole. Zwanziger's horizon condition
(\ref{horizondef}), on the other hand, is a nonperturbative
statement. Since the Coulomb potential $V_C(k)$ between external
color charges is proportional to
\begin{equation}
  \label{V_Cdef}
\int d^3x\,  \lla G(-\del^2)G\rra^{ab}(\fx,\fy)\e^{-ik\cdot (\fx-\fy)}\; ,
\end{equation}
 the infrared enhancement of $D_G(k)$ from
effects on the Gribov horizon might be the driving mechanism for
$V_C(k)$ to diverge as $k^{-4}$ and for a linear confining  potential
between heavy quarks to emerge. This notion is known as the
\emph{Gribov--Zwanziger scenario} of confinement in the Coulomb gauge.

\section{Stochastic vacuum in Coulomb gauge}
\label{stochvac}
Vacuum expectation values that are solely dominated by entropy in
configuration space are intriguingly simple. According to the
Gribov--Zwanziger scenario, infrared correlation functions of Coulomb
gauge YM theory fall under this category and hence can asymptotically be described by a vacuum wave functional
$\Psi_0[A]$ stochastically distributed in configuration space. Whilst the full structure of a vacuum expectation value, in
particular the perturbative regime, is sensitive to a non-trivial
probability distribution $|\Psi[A]|^2$, the infrared limit is fully
determined by the geometry in phase space, and one may write
$\Psi_0[A]=1$ to find correct results \cite{Zwa03a}. This choice of the wave
functional is here referred to as the \emph{stochastic vacuum} and suggests
the following schematic prescription for the asymptotic evaluation of
Coulomb gauge expectation values of any operator $\cO(k)$, depending
on momentum $k$,
\begin{equation}
  \label{SVexpval}
  \bra{\Psi}\cO(k)\ket{\Psi}=\int_\Omega  \cD A\, \cJ\Psi^*\cO(k)\Psi\;\stackrel{k\rarr
    0}{\rarr}\;\frac{1}{\mathit{Vol}(\Omega)}\int_\Omega \cD A\,  \cJ\cO=\bra{\Psi_0}\cO\ket{\Psi_0}\; .
\end{equation}
We introduced a factor $\mathit{Vol}(\Omega)=\int_\Omega \cD A\, \cJ[A]$ to normalize the wave functional,
$\braket{\Psi_0}{\Psi_0}=1$, which is possible in the compact region $\Omega$. A legitimate question to ask is whether the
factor of the Faddeev--Popov determinant $\cJ$ does not spoil the
argument that the probability distribution is strongly enhanced at the
Gribov horizon. Recall that the Gribov horizon $\del\Omega$ was defined as the
boundary of $\Omega$ where the Faddeev--Popov matrix $G^{-1}$ picks up
vanishing eigenvalues. Hence, the factor $\cJ=\Det G^{-1}$ will
\emph{suppress} the probability distribution on
$\del\Omega$. Nevertheless, it is not clear a priori whether this
suppression is strong enough to spoil the enhancement of entropy and
we shall put forth as a working hypothesis that it does
not.\footnote{As a simple example \cite{Zwa91}, consider a $d$-dimensional
  sphere with $0<r<1$ and let the probability distribution $p(r)$ be damped at
  the boundary by a factor $(1-r)^n$ with order $n(d)$. The entropy factor
  from the integral measure, $dr r^{d-1}$, enhances the probability at
  the boundary. For large dimensions $d$, $p(r)=r^{d-1}(1-r)^n$ is then sharply
  peaked at $r=1-\frac{n}{d}$, provided $\frac{n}{d}\ll 1$, despite
  the damping factor.} On the
lattice, the density of eigenmodes of the Faddeev--Popov operator
is under current investigation \cite{GreOleZwa04}.

Although the prescription (\ref{SVexpval}) for infrared correlations
can be motivated from the Gribov--Zwanziger scenario, it cannot be
taken for granted that the solutions thus obtained are mathematically unique on the one hand, or physical on
the other. To further contemplate Eq.\ (\ref{SVexpval}), note that
$\Psi_0[A]=1$ is the correct wave functional in $1+1$ dimensional
Yang--Mills theory \cite{Raj88,HetHos89}. However, the entropy argument is lacking in
$1+1$ dimensions where the gauge field becomes a compact quantum
mechanical variable. In $3+1$ dimensions, the dynamics might be
substantially different. After all, we will try to find the vacuum wave
functional in $3+1$ dimensions that describes not only the infrared limit but also
ultraviolet correlations. In the 
Hamiltonian formalism, the Ritz-Rayleigh variational method gives the
possibility to introduce a variational parameter into the wave
functional and determine it by minimizing the energy% \cite{Sch62a}
. This method will be discussed in chapter \ref{VarVac} to
discuss the full momentum dependence of the Green functions. It will
turn out that the full solutions  agree with the
solution in the stochastic vacuum for asymptotically small
momenta \cite{FeuRei04,FeuRei04a,ReiFeu04,EppReiSch07}. Turning the argument around, these findings support the Gribov--Zwanziger
scenario and the notion of a stochastically distributed probability
distribution $|\Psi[A]|^2$ in the infrared. The investigation of the
stochastic vacuum pursued in this chapter is quite different from a
variational approach in that it lacks equations of motion arising from
minimizing the energy. However, we may extract 
information from the geometry in configuration space. In the following
section, it will be shown how the boundary conditions of the Gribov region
give rise to Dyson--Schwinger integral equations that determine the
properties of the Green functions.

\section{Dyson--Schwinger equations}
\label{DSEs}

A convenient tool for the evaluation of expectation values are
generating functionals. In particular, the Dyson--Schwinger equations
that intercorrelate all Green functions can be straightforwardly
derived from the path integral representation of a generating
functional. The classification of Green functions into \emph{full}, \emph{connected}
and \emph{proper} Green functions is common and the same applies for
generating functionals \cite{ItzZub}. Full Green functions corresponding
to the stochastic vacuum $\Psi_0$ can be found by
\begin{equation}
  \label{Greenany}
  \lla \cO[A] \rra \equiv \bra{\Psi_0}\cO[A]\ket{\Psi_0}=\frac{1}{Z[j]}\left.\cO\left[\frac{\delta}{\delta j}\right] Z[j]\right|_{j=0}
\end{equation}
if we define the generating functional $Z[j]$ of full Green functions
by
\begin{equation}
  \label{Zdef}
  Z[j]=\int \cD A\, \cJ[A]\e^{j\cdot A}\; ,\quad j\cdot A\equiv \int d^3x\, 
  A_i^a(\fx)t_{ik}(\fx)j_k^a(\fx)\; .
\end{equation}
It is easily verified that Eq.\ (\ref{Greenany}) indeed reproduces
the expectation values in the stochastic vacuum, see Eq.\
(\ref{SVexpval}). Dyson--Schwinger equations follow from a quite simple
statement, namely that the (path) integral of a total derivative can
be expressed by boundary terms. This idea is applied to
the path integration in configuration space, bounded by the Gribov
horizon. The gauge-fixing procedure endowed us the Faddeev--Popov
determinant $\cJ$ in the path integral, see Eq.\ (\ref{FPop}). One
is thus led to the identity
\begin{equation}
  \label{DSE}
  0=\int \cD A\, \frac{\delta}{\delta A_j^b(\fy)}\cJ[A]\e^{j\cdot A}=\int
  \cD A\, \cJ[A]\left(\frac{\delta\ln\cJ[A]}{\delta A_j^b(\fy)}+t_{jm}(\fy)j_m^b(\fy)\right)\e^{j\cdot A}
\end{equation}
since at the Gribov horizon the Faddeev--Popov determinant vanishes by definition,
$\left.\cJ[A]\right|_{\del\Omega}=0$. Applying to Eq.\ (\ref{DSE}) a
functional derivative $\delta/\delta j_i^a(\fx)$, dividing by $Z[j]$ and setting sources to zero, $j=0$, gives
\begin{equation}
  \label{DSE1}
  t_{ij}^{ab}(\fx,\fy)=-\lla A_i^a(\fx)\frac{\delta\ln\cJ[A]}{\delta
    A_j^b(\fy)}\rra\; .
\end{equation}
In order to make sense of the above expression, first note that\footnote{It is convenient to denote by
  $\cO$ a matrix with elements $\cO^{ab}(\fx,\fy)=\bra{\fx,
    a}\cO\ket{\fy,b}$ in color and (continuous) coordinate space. The
  trace ``$\Tr$'' sums up diagonal elements in both the latter
  spaces. For notation, see also appendix \ref{AppA}.} \cite{FeuRei04}
\begin{equation}
\label{DSE1a}
  \frac{\delta\ln\cJ[A]}{\delta A_i^a(\fx)}=\Tr\frac{\delta\ln
    G^{-1}[A]}{\delta
    A_i^a(\fx)}=-\Tr\left(G[A]\Gamma_i^{0,a}(\fx)\right)\; ,
\end{equation}
 where $\Gamma_i^{0,a}(\fx)$ is the tree-level ghost-gluon vertex,
 here matrix-valued with components
  \begin{equation}
    \label{GGAdef}
    \big(\Gamma_i^{0,a}(\fx)\big
    )^{bc}(\fy,\fz)=-\frac{\delta\left[G^{-1}\right]^{bc}(\fy,\fz)}{\delta A_i^a(\fx)}=g\big (\hat T^a\big )^{bc}t_{ik}(\fx)\delta^3(\fx,\fy)\del_k^{\fy}\delta^3(\fy,\fz)\; .
  \end{equation}
Second, plugging (\ref{DSE1a}) into (\ref{DSE1}), the
object $\lla AG\rra$ can be identified as the dressed connected ghost-gluon vertex that
may be decomposed by\footnote{If the generating functional is
  Gaussian, Wick's theorem infers the factorization in the first line
  of Eq.\ (\ref{GGAconn}). Without the use of
  Wick's theorem, as for the stochastic vacuum, one may use functional methods~\cite{Zwa03}.}
\begin{eqnarray}
  \label{GGAconn}
  \lla A_i^a(\fx) G[A] \rra &=&  \int d^3y 
\lla A_i^a(\fx) A_j^b(\fy)\rra 
\lla \frac{\delta G[A]}{\delta A_j^b(\fy)} \rra
\nn\\
&=& \int d^3y 
\lla A_i^a(\fx) A_j^b(\fy)\rra 
\lla G[A] \Gamma_j^{0,b}(\fy) G[A] \rra \nn\\
&=&\int d^3y D_{ij}^{ab}(\fx,\fy)D_G\Gamma_j^b(\fy)D_G
\end{eqnarray}
into the proper ghost-gluon vertex $\Gamma_k^a$ with propagators attached, namely the gluon propagator 
\begin{eqnarray}
  \label{Dijdef}
  D_{ij}^{ab}(\fx,\fy):=\lla A_i^a(\fx)A_j^b(\fy)\rra\; ,
\end{eqnarray}
and the (here matrix-valued) ghost propagator $D_G$, with matrix
elements
\begin{eqnarray}
  \label{DGdefmom}
  D_G^{ab}(\fx,\fy)=\lla G[A]^{ab}(\fx,\fy)\rra\; .
\end{eqnarray}

The present calculations circumvent the common introduction of ghost
fields by interpreting $\cJ[A]$ as a Gaussian integral over Grassmann
variables. In fact, the introduction of Grassmann fields and their
sources spoils the argument that boundary terms from the Gribov
horizon vanish \cite{Zwa03}. It is therefore advisory to stick to
the functional determinant $\cJ[A]$ and not to introduce ghost sources. Nevertheless, we refer to $D_G$ as the ``ghost'' propagator and
$\Gamma_k$ as the ``ghost-gluon'' vertex.

With the above definitions, the Dyson--Schwinger equation (\ref{DSE1}) becomes
\begin{eqnarray}
  \label{DSE2}
  t_{ij}^{ab}(\fx,\fy)&=&\Tr\left(\lla
    A_i^a(\fx)G[A]\rra\Gamma_j^{0,b}(\fy)\right)
\nn\\
    &=&\int d^3z
    D_{ik}^{ac}(\fx,\fz)\Tr\left(D_G\Gamma_k^c(\fz)D_G\Gamma_j^{0,b}(\fy)\right)\; .
\end{eqnarray}
It would be pleasant to invert the gluon propagator in Eq.\
(\ref{DSE2}), but the transverse projector $t_{ij}$ is singular. Yet, without loss of generality, we can define the matrix $D_A$ by
$D_{ij}^{ab}(\fx,\fy)=t_{ij}(\fx)D_A^{ab}(\fx,\fy)$ and its inverse,
$D_A^{-1}D_A=\id$. The translationally
invariant function
\begin{equation}
  \label{D-1def}
  \big [D^{-1}_{ij}\big ]^{ab}(\fx,\fy):=t_{ij}(\fx)\left[
    D_A^{-1}\right]^{ab}(\fx,\fy)
\end{equation}
is referred to as the inverse gluon propagator.\footnote{Note that $D_{ij}^{-1}$ is
inverse to $D_{ij}$ only in the transverse subspace of the full
Lorentz space,
\begin{eqnarray}
  \label{D-1D=1}
  \int d^3z
  \left[D^{-1}_{ik}\right]^{ac}(\fx,\fz)\left[D_{kj}\right]^{cb}(\fz,\fy)=t^{ab}_{ij}(\fx,\fy)\; .
\end{eqnarray}}
 One can now deduce from Eq.\ (\ref{DSE2}) the coordinate space gluon DSE in its
most useful form,
\begin{equation}
  \label{DSEgluon}
  \big [D_{ij}^{-1}\big ]^{ab}(\fx,\fy)=\Tr\left(D_G\Gamma_i^a(\fx)D_G\Gamma_j^{0,b}(\fy)\right)\; .
\end{equation}

\begin{figure}
\centering
% \begin{eqnarray*}
% D_G^{-1}\quad =\quad\parbox{20mm}{\includegraphics{eps/fullghost}}^{-1}&=&\quad\parbox{20mm}{\includegraphics{eps/bareghost}}^{-1}\quad-\quad\parbox{35\unitlength}{\includegraphics[scale=1.2]{eps/ghostSE}}\\
% D_{ij}^{-1}\quad =\quad\parbox{20mm}{\includegraphics{eps/fullgluon}}^{-1}&=&\quad\parbox{35\unitlength}{\includegraphics[scale=1.2]{eps/gluonghostloop}}\\
% \end{eqnarray*}
\inc[scale=1.0, bb= 150 600 460 700, clip=]{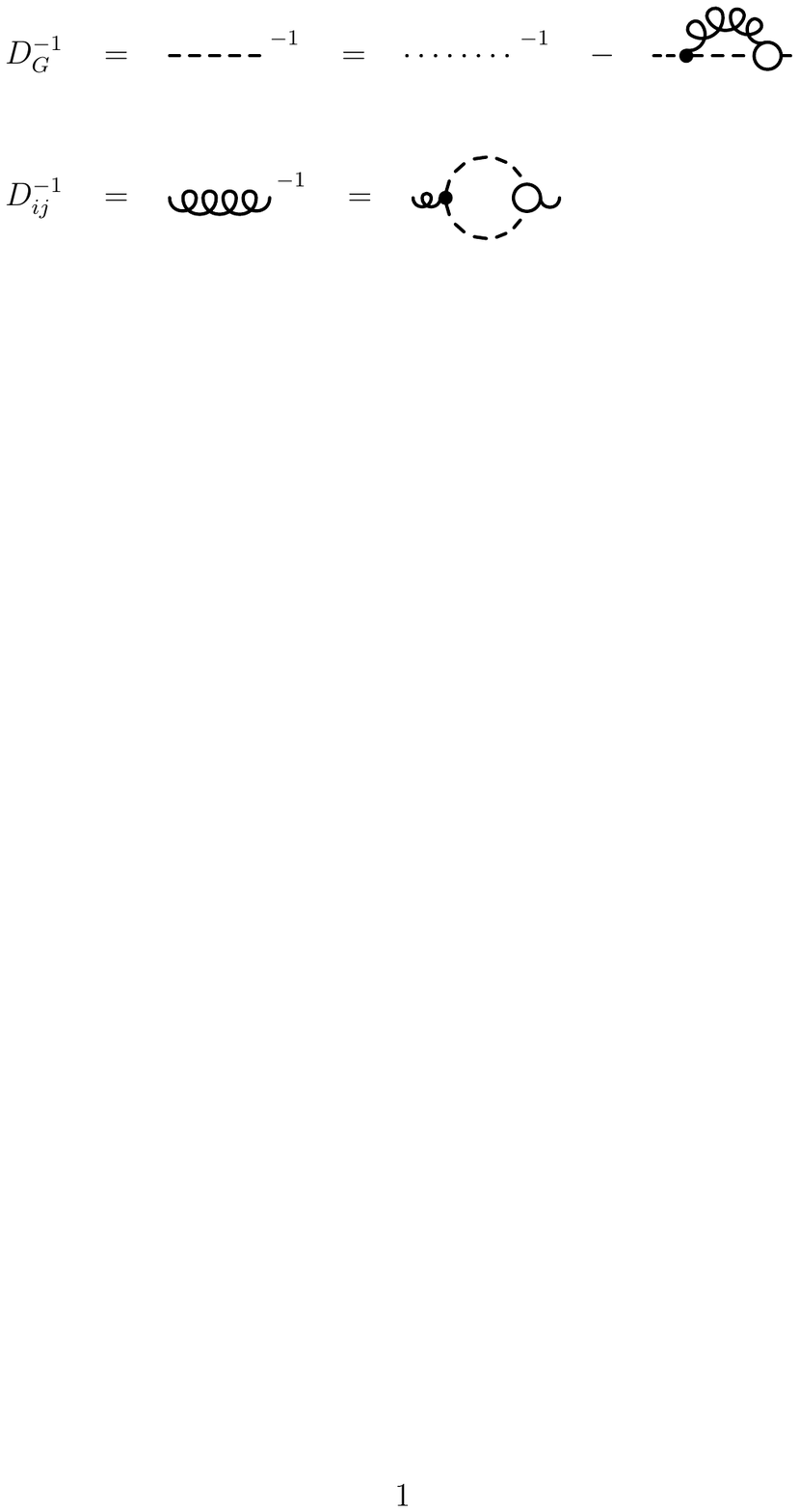}
\caption{The set of Dyson--Schwinger equations in the stochastic
  vacuum. 
Dashed lines represent interacting connected ghost propagators whereas the
dotted line is the tree-level ghost propagator. Curly lines represent
the gluon propagator. Empty blobs stand for proper ghost-gluon
vertices and dots represent tree-level ghost-gluon vertices.}
\label{figDSE}
\end{figure}

In Fig.\ \ref{figDSE}, the gluon DSE (\ref{DSEgluon}) is interpreted
diagrammatically. By the loop diagram on the r.h.s., the so-called
\emph{ghost loop}, the gluon propagator $D_{ij}$ is correlated with the ghost
propagator $D_G$. The latter needs to be determined by its own DSE. To
derive it \cite{FeuRei04}, first note that
\begin{equation}
  \label{G-1mitGamma}
  G^{-1}=G_0^{-1}-\int d^3x\,  A_k^a(\fx)\Gamma_k^{0,a}(\fx)\; .
\end{equation}
Multiplying the above equation with $G_0$ from the left and $G$ from
the right gives
\begin{equation}
  \label{ghostop}
  G=G_0+G_0\int d^3x\,  A_k^a(\fx)\Gamma_k^{0,a}(\fx)G\; .
\end{equation}
Taking the expectation value of (\ref{ghostop}) and using the identity (\ref{GGAconn}) for
the connected ghost-gluon vertex yields
\begin{equation}
  \label{propDSEghost}
  D_G=G_0+G_0\Sigma D_G
\end{equation}
with the ghost self energy
\begin{equation}
  \label{Sigma}
  \Sigma^{ab}(\fx,\fy)=\int
  d^3[x'y'uv]\big [\Gamma_i^{0,c}(\fu)\big
  ]^{aa'}(\fx,\fx')D_{ij}^{cd}(\fu,\fv)D_G^{a'b'}(\fx',\fy')\big
  [\Gamma_j^d(\fv)\big ]^{b'b}(\fy',\fy)\; .
\end{equation}
A multiplication with $G_0^{-1}$ from the left and $D_G^{-1}$ from the
right gives the final form of the ghost DSE in coordinate space,
\begin{equation}
  \label{DSEghost}
  \left[D_G^{-1}\right]^{ab}(\fx,\fy)=\left[G_0^{-1}\right]^{ab}(\fx,\fy)-\Sigma^{ab}(\fx,\fy)\; ,
\end{equation}
shown in Fig.\ \ref{figDSE}. Let us emphasize that Eq.\ (\ref{DSEghost}) was not derived from the path integral, in
contrast to the gluon DSE (\ref{DSEgluon}). The ``operator method'' using Eq.\
(\ref{ghostop}) offers an alternative to the conventional path integral
method, without the introduction of ghost sources. Restriction to the Gribov region
$\Omega$ is ensured by cutting off the path integral at $\del\Omega$,
the operator method formally also accounts for this restriction since
the operator $G$ in Eq.\ (\ref{ghostop}) is not defined on $\del\Omega$.

Having at hand the gluon DSE (\ref{DSEgluon}) and the ghost DSE (\ref{DSEghost})
shown in Fig.\ \ref{figDSE}, one may now try and solve for the two-point
Green functions (``propagators'') $D_G$ and $D_{ij}$. However, the proper ghost-gluon vertex
that appears in both DSEs couples the propagators to higher $n$-point
functions, thus giving rise to the typical infinite tower of coupled
integral equations. After establishing a connection to the Landau
gauge, an approximation for the ghost-gluon vertex will be motivated
in section \ref{secGGA} which makes a solution for $D_G$ and $D_{ij}$ feasible.

\section{Connection to Landau gauge}
\label{toLandau}
Linear covariant gauges are a natural choice for perturbative calculations in
the Lagrangian formalism. With Lorentz invariance being manifest and
after Wick rotation to Euclidean space, 
perturbative loop integrals can be computed with standard methods
\cite{Smirnov}. 
However, this is at the cost of losing a positive definite metric in Hilbert
space. One cannot simultaneously maintain locality, positivity and
manifest Lorentz invariance \cite{Yndurain}. 
A Hamiltonian approach is thus obscured and the
potential between static quarks as it occurs in the Coulomb gauge
Hamiltonian is not accessible in the Landau gauge. The confinement
problem must therefore be studied by alternative means, e.g.\ the
Kugo--Ojima confinement criterion \cite{KugOji79,Kug95}. 

Usually, the class of linear covariant gauge
conditions $\chi^a(x)=\del^\mu A_\mu^a(x)-B^a(x)\approx 0$, is implemented in the path
integral by means of the Faddeev--Popov method%\cite{FadPop67}
. The arbitrary function
$B^a(x)$ is integrated out with a Gaussian weighting of width $\xi$,
which effectively smears of the Lorenz gauge condition,
  \begin{equation}
    \label{Landaugauge}
    \delta\left(\del^\mu A_\mu^a\right)\rarr \e^{-\frac{1}{2\xi}\left(\del^\mu A_\mu^a\right)^2}
  \end{equation}
controlled by an arbitrary parameter $\xi$. Apart from the
Faddeev--Popov determinant $\cJ=\Det\cM$, with 
\begin{equation}
  \label{FPpathint}
  \cM^{ab}(x,y)=\frac{\delta\left(\del^\mu
      {A^U}_\mu^a(x)\right)}{\delta \alpha^b(y)}\; ,
\end{equation}
the YM Lagrangian is thus supplemented by a gauge fixing term
$\cL_{gf}=-\frac{1}{2\xi}\left(\del^\mu A_\mu^a\right)^2$.
The Landau gauge is then obtained in the limit $\xi\rarr 0$. 
From gauge invariance, a Slavnov--Taylor identity can be derived
\cite{Sla72} that
states that the longitudinal part of the gluon propagator remains
unchanged by radiative corrections,
\begin{equation}
  \label{STI}
  \ell^{\mu\nu}(k)D_{\mu\nu}(k)=\xi\; .
\end{equation}
It is
important to realize that during calculations, $\xi$ is kept non-zero,
so that the gluon propagator has an inverse. The fields $A_\mu^a(x)$ are
therefore not transverse, only the Landau gauge correlation functions
are. In particular, the ghost-gluon vertex in Landau gauge does not
have a transverse projector attached to its gluon leg, since the
transversality is imposed ``off-shell''. This is in contrast to the
constrained quantization in temporal Coulomb
gauge, where the transversality condition is enforced ``on-shell'' and
the operators $A^a_i(\fx)$ are manifestly transverse.\footnote{The
  terminology using ``on-shell'' and ``off-shell'' to describe the
  gauge-fixing technique is encountered, e.g., in Ref.\ \cite{Zwa03}.}

The Gribov problem affects the configuration space in Landau gauge
just like it does for Coulomb gauge. In fact, the Faddeev--Popov gauge
fixing procedure in Landau gauge gives rise to the Faddeev--Popov
determinant 
\begin{equation}
  \label{JLandau}
  \cJ[A]=\Det\left(-\delta^{ab}\del_\mu\del^\mu-g\hat A_\mu^{ab}\del^\mu\right)
\end{equation}
and one cannot miss to acknowledge the similarity to the one in
Coulomb gauge. It was pointed out by Gribov in his seminal paper
\cite{Gri77} that if Wick rotation is employed, the Landau gauge
condition becomes equivalent to the Coulomb gauge condition. The
dimension $D=3+1$ of Minkowski spacetime infers that $d=4$ after Wick
rotation, as opposed to $d=3$ in the temporal Coulomb gauge.

Recall the Gribov--Zwanziger scenario in the Coulomb gauge. The entropy
argument for infrared dominant field configurations at the Gribov
horizon along with the horizon condition (\ref{horizondef}) for the ghost
propagator applies to the Landau gauge in Euclidean space
just the same \cite{Zwa94}. One may therefore expect that the infrared
limit in Landau gauge yields the same
set of integral equations as for the Coulomb gauge, with the
dimension shifted to $d=4$. To see how it comes about, consider the
expression for a Landau gauge-fixed expectation value,
\begin{equation}
  \label{expLandau}
  \lla \cO[A] \rra = \int \cD A\,  \cJ[A] \e^{-\int d^4x_E \cL_{YM}(x_E)}
\end{equation}
where the weight of field configurations is determined by the
Euclidean Yang--Mills action. Shifting the field variable by $A\rarr
gA$, the weight acquires the form 
\begin{equation}
  \label{shiftweight}
  \e^{-\int d^4x_E \cL_{YM}(x_E)}\rarr \e^{-\frac{1}{g^2}\int d^4x_E \cL_{YM}(x_E)}
\end{equation}
and one recognizes that to zeroth order of the \emph{strong coupling expansion} in
$\frac{1}{g^2}\ll 1$, the evaluation of expectation values depends only
on the Faddeev--Popov determinant \cite{Zwa03} and is therefore
equivalent to the stochastic vacuum (\ref{SVexpval}), with the difference that the
Faddeev--Popov determinant in (\ref{JLandau}) is to be computed in $d=4$ dimensional
Euclidean Lorentz space. 
% For an attempt of perturbation theory in
% $\frac{1}{g}$, see Ref.\ \cite{Man94}.

The full structure of Landau gauge Dyson--Schwinger equations was
studied intensively in the past decade \cite{AlkSme00,Fis06}. It turned out that the
infrared sector of the solutions is asymptotically determined by the
set of equations shown in Fig.\ \ref{figDSE}. Apparently, the strong
coupling expansion in the Landau gauge Lagrangian formalism accounts for the correct infrared
physics, while in the Coulomb gauge Hamiltonian formalism the stochastic vacuum is a simple but
sufficient state to describe infrared correlations. The approximation
needed to arrive at a solution to the coupled Dyson--Schwinger
equations shown in Fig.\ \ref{figDSE} concerns the ghost-gluon
vertex. This will be the topic of the next section.

\section{Ghost-gluon vertex}
\label{secGGA}

In gauges where the gluon propagation is transverse, such as Coulomb
and Landau gauge, the ghost-gluon vertex has the feature of nonrenormalization. In
this context, one can motivate that the vertex is approximately
tree-level for all momentum configurations. The solutions to all Green
functions will qualitatively depend on the simple structure of the
ghost-gluon vertex and therefore this section is devoted to assess the tree-level approximation. 

\begin{figure}
  \centering
  \inc[bb= 146 611 458 716,clip=]{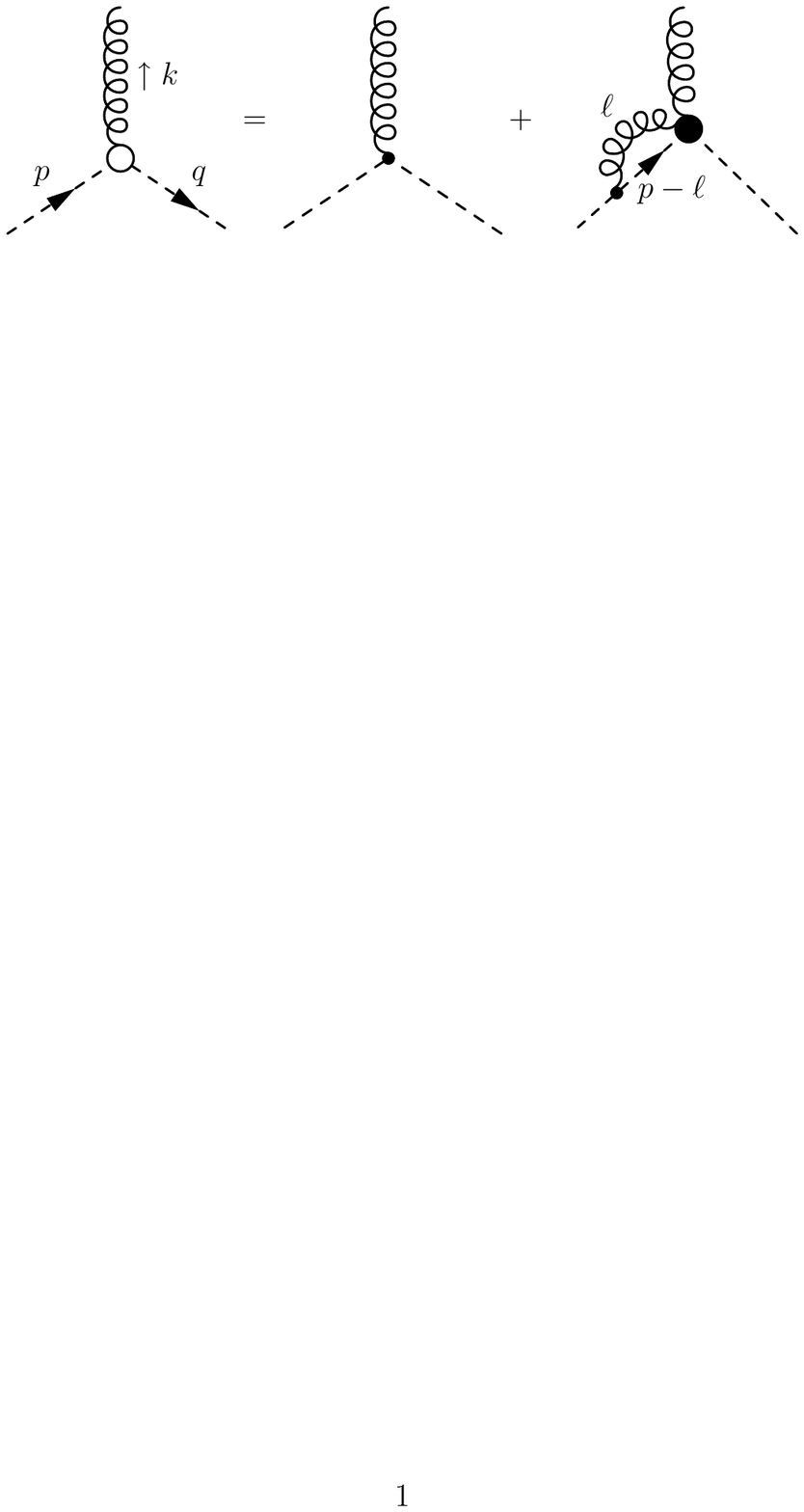}
  \caption{Dyson--Schwinger equation for the ghost-gluon vertex. The
    l.h.s.\ represents the proper ghost-gluon vertex, with definitions
  of the momenta $k$, $q$ and $p=q+k$. The filled blob on the r.h.s.\
  corresponds to a four-point function, see text.}
  \label{figGGADSE}
\end{figure}

As far as renormalization is concerned, the ghost-gluon vertex in
Coulomb or Landau gauge is quite an unusual Green function. Whereas
one usually encounters ultraviolet divergences in a perturbative
expansion of a Green function, the ghost-gluon vertex appears to be a
finite function, to all orders of perturbation theory. This can be
understood by considering a Dyson--Schwinger equation for this
vertex. In the Landau gauge, the ghost-gluon vertex Dyson--Schwinger
equation was derived in Ref.\ \cite{Sch+05} and can be represented in
momentum space as shown in Fig.\ \ref{figGGADSE}. The proper ghost-gluon vertex $\Gamma_\mu(k;q,p)$ is related to its
tree-level counterpart $\Gamma_\mu^{0}(q)$ and a connected four-point function $\Gamma_{\mu\nu}(k,\ell;q,p)$ that will
be explicitly defined in section \ref{vertices}. Here, it shall suffice to
recognize the generic structure of the loop diagram,
\begin{equation}
  \label{GGADSE}
  \Gamma_\mu(k;q,p)=\Gamma_\mu^0(q)+\int \dbar^4\ell\,\Gamma_\rho^0(p-\ell)D_{\rho\nu}(\ell)D_G(\ell-p)\Gamma_{\nu\mu}(k,-\ell;\ell-p,q)\; .
\end{equation}
Taylor anticipated in Ref.\ \cite{Tay71} the structure of the
Dyson--Schwinger equation (\ref{GGADSE}) and perceived that the
contribution of $\Gamma_\mu^0(p-\ell)$ to the loop integration can be factored
out due to the transversality of the gluon
propagator. In momentum space, with the color structure suppressed,
the contraction of the gluon propagator $D_{\mu\nu}$  with the tree-level
ghost-gluon vertex, $\Gamma^0_\mu(q)=igq_\mu$, yields

%,\footnote{In momentum space, the color structure is
%  suppressed. For conventions, see appendix LABEL.}
\begin{equation}
  \label{Taylor}
 \Gamma_\rho^0(p-\ell) D_{\rho\mu}(\ell)=\Gamma_\rho^0(p)D_{\rho\mu}(\ell)\; .
\end{equation}
Hence, the degree of divergence of the $\ell$-integration is diminished. Landau gauge
perturbation theory explicitly confirms that the divergent part of
the $g^3$ contribution to the vertex is proportional to $\xi$, i.e.\ it
vanishes in the Landau gauge limit, see e.g.\ \cite{DavOslTar96}. Taylor further argued that if the
loop diagram does not develop any poles in the infrared, one can take
the infrared limit of the incoming ghost momentum $p\rarr 0$ and find
from Eq.\ (\ref{GGADSE})
\begin{equation}
  \label{GGAp0}
  \lim_{p\rarr 0}\Gamma_\mu(k;q,p)=\Gamma_\mu^0(q)\; .
\end{equation}

\begin{figure}
  \centering
  \includegraphics[scale=0.8]{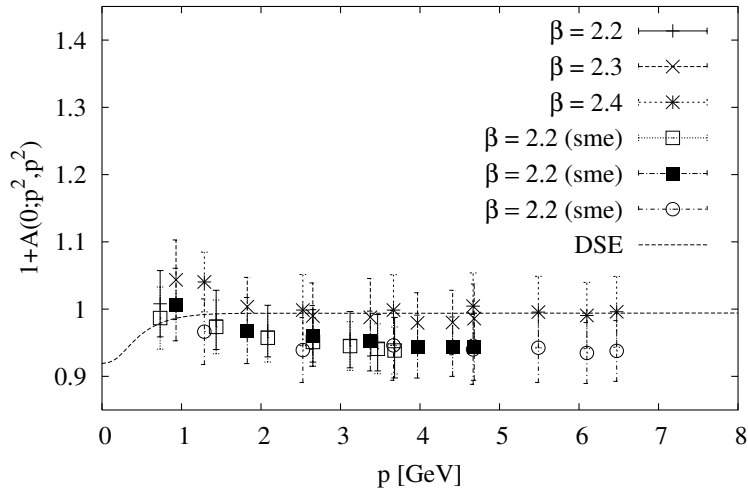}
  \caption{Projection of the ghost-gluon vertex in $3+1$ dimensional
    Landau gauge as a function of one
    momentum, taken from Ref.\ \cite{Sch+05} (for details, see there). Both the DSE result and the
    lattice data show that the ghost-gluon vertex is basically at tree-level.}
  \label{figGGAnonrel}
\end{figure}

For the labelling of momenta, see Fig.\ \ref{figGGADSE}. Since one particular momentum configuration, given by Eq.\
(\ref{GGAp0}), is found where the ghost-gluon vertex is finite, its
multiplicative renormalization constant $\tilde{Z}_1$, being momentum
independent, has to be finite, $\tilde{Z}_1<\infty$. In an appropriate
renormalization scheme, $\tilde{Z}_1$ can defined to be
trivial, $\tilde{Z}_1=1$. Multiplicative renormalization of the
bare ghost-gluon vertex $\Gamma_B$, understood here as a vector, is defined by
\begin{equation}
  \label{Z1def}
  \Gamma_B(k,q,p;\Lambda)=\tilde{Z}_1(\mu,g,\xi,\Lambda) \Gamma(k,q,p;\mu)
\end{equation}
and relates $\Gamma_B$ to the renormalized vertex
$\Gamma$. %\footnote{Additionally, the function $\tilde{Z}_1$ depends on
%the coupling $g$ and the gauge parameter $\xi$ which need to be
%renormalized themselves.} 
Since in the Landau gauge limit, $\xi\rarr 0$, one has $\tilde{Z}_1<\infty$, we can drop all dependences on the regulator
$\Lambda$. The dependence of $\tilde{Z}_1$ on the dimensionful renormalization
scale $\mu$ must drop out since $g$ and $\tilde{Z}_1$ are both
dimensionless. The number $\tilde{Z}_1$ is then defined by a renormalization
prescription, i.e.\ a certain momentum configuration $P:=\{k_r^2,q_r^2,p_r^2\}$ is
chosen for which $\Gamma(k_r,q_r,p_r)=\Gamma^0(\mu)$. In order to
avoid cuts in the complex plane, $P$ is usually chosen in the
perturbative regime. Nevertheless, the renormalization prescription
with $P=\{\mu^2,\mu^2,0\}$  is
a convenient one since Eq.\ (\ref{GGAp0}) then
infers that $\tilde{Z}_1=1$. Marciano and Pagels \cite{MarPag78}
pointed out that for the symmetric point $P=\{\mu^2,\mu^2,\mu^2\}$ with $\mu^2\neq 0$, one
generally has $\tilde{Z}_1\neq 1$ (cf.\ \cite{LerSme02}). The symmetric point with $\mu^2=0$,
however, does not necessarily give $\tilde{Z}_1=1$ since the infrared
limits of momenta are not interchangeable, as argued below in section \ref{vertices}. In
Ref.\ \cite{CucMenMih04}, the choice $P=\{0,\mu^2,\mu^2\}$ was
advocated. In the following, we will set $\tZ_1=1$, but come back to
it in chapter \ref{UVchap}.
%In the latter renormalization prescription, we will calculate $\tilde{Z}_1$ analytically in section \ref{vertices}.

% It is not surprising that Green functions depend on the
% renormalization prescription. In finite order perturbative expansions, even
% the gauge coupling is renormalization prescription dependent
% \cite{CelGon79} and leads to scale ambiguities \cite{BroLepMac83}.%%,Ste81,,SolShi97}. 

Due to the nonrenormalization of the ghost-gluon vertex and the
related property (\ref{GGAp0}), recent Landau gauge DSE studies used a
tree-level ghost-gluon vertex for all momentum configurations
\cite{AlkSme00,Fis06}. Of course, this approximation needs to be
assessed further. It was
discussed in Ref.\ \cite{Sch+05} that with the tree-level vertex as a starting
point, the iteration of the vertex DSE (\ref{GGADSE}) will not lead too far away
from the tree-level value. The changes stay in the range of
$10\%-20\%$. Furthermore, lattice calculations \cite{CucMenMih04,Ste+05,Maa07} confirmed that the vertex
is basically tree-level. In Fig.\ \ref{figGGAnonrel}, the DSE results
along with the lattice results for the vertex are shown. Thus, strong support is
available that the perturbative arguments of Taylor, stated above, hold
in the nonperturbative regime. 

In Coulomb gauge, the argumentation for a tree-level ghost-gluon vertex is
similar. Although the perturbative calculations in Coulomb gauge are
far from trivial \cite{WatRei06, WatRei07a,WatRei07b}, the DSE for the
ghost-gluon vertex can be shown to reproduce the Landau gauge struture
depicted in Fig.\ \ref{figGGADSE} \cite{WatRei06}. The gluon propagator is transverse as
well and therefore Taylor's reasoning for nonrenormalization can also
be applied to the Coulomb gauge. Within a set of interpolating gauges,
the nonrenormalization of the ghost-gluon vertex can be established as
well, considering the Coulomb gauge limit \cite{FisZwa05}. The Landau
gauge calculations shown in Fig.\ \ref{figGGAnonrel} were also carried
out in $d=3$ spatial dimensions and the results were qualitatively the
same \cite{Sch+05}. Therefore,
it seems to be an appropriate approximation to the Coulomb gauge DSEs
to use a tree-level vertex instead of a proper vertex, analogously to
the Landau gauge,
\begin{eqnarray}
  \label{GGAapprox}
  \Gamma_k(\fx)\approx\Gamma_k^0(\fx)\; .
\end{eqnarray}

% SonstNoch:
% \begin{itemize}
% \item iteration of tree-level
% \item constant dressing
% \item other renormalization prescr $\rarr Z_1\neq 1$ 
% \item other power laws in [LerSme02]
% \item lattice mit Bild
% \item Boucaud
% \end{itemize}

\section{Analytical solution for propagators}
\label{rechnung}

The approximation scheme (\ref{GGAapprox}) exhibited in the previous section enables us
to find a solution to the coupled set of Dyson--Schwinger equations. With
the Coulomb and Landau gauge differing in the number $d$ of spatial
dimensions only, it is convenient to keep $d$ as an unspecified
parameter. The gluon DSE (\ref{DSEgluon}) can be transformed into momentum space
defining the scalar gluon propagator $D_A(k)$ by\footnote{Note that in
momentum space, the matrix notation of coordinate space is no longer
used. Moreover, we simplify the notation henceforth by abandoning
the bold-face typesetting for spatial vectors. Latin Lorentz indices now range
from $1$ to $d$.}
\begin{equation}
  \label{DAmomdef}
  \delta^d(k-p)\delta^{ab}D_A(k)=\int d^d[xy] D_A^{ab}(x,y)\e^{-ik\cdot
    x+ip\cdot y}\; .
\end{equation}
Translational invariance allows us to write $D_A(k)$ as a function of one
variable which makes momentum space calculations convenient. The color
structure of propagators is chosen to be diagonal, i.e.\ $D_G^{ab}(x,y)=D_G(x,y)\delta^{ab}$. In the
approximation scheme with a tree-level ghost-gluon vertex, it is
verified straightforwardly in a perturbative expansion to all orders (``rainbow
expansion'') that color-diagonal propagators solve the set of integral equations. 
Using Eq.\ (\ref{DGdef}) for the definition of the momentum space ghost
propagator $D_G(k)$, Fourier transformation of  the gluon DSE
(\ref{DSEgluon}) expressed in terms of
$D_{ij}^{-1}(k):=t_{ij}(k)D_A^{-1}(k)$ yields
\begin{equation}
  \label{gluonDSEuncontract}
  D_{ij}^{-1}(k)=-N_c\int\dbar^d\ell\, \Gamma_i^0(k,\ell)\Gamma_j^0(k,\ell-k)D_G(\ell)D_G(\ell-k)\; .
\end{equation}
Here, the tree-level ghost-gluon vertex in momentum space is derived
by Fourier transform from Eq.\ (\ref{GGAdef}),
\begin{align}
  \label{GGAdefmom}
\int d^d[xyz]
  \big (\Gamma_k^{0,a}(x)\big )^{bc}(y,z)\e^{-ik\cdot x-iq\cdot
    y+ip\cdot z}&=  \big(\hat
    T^{a}\big )^{bc}\delta^d(p-q-k)\Gamma_k^0(k,q)\nn\\
\Rightarrow \Gamma_k^0(k,q)&=igt_{kj}(k)q_j\; .
\end{align}
The color trace in (\ref{gluonDSEuncontract}) was computed using
$\tr\big (\hat T^a\hat
  T^b\big )=-N_c\delta^{ab}$, see appendix \ref{AppA}. Eq.\ (\ref{gluonDSEuncontract}) is seen
to be manifestly transverse. This follows from the fact that in the
derivation of the DSEs in section \ref{DSEs} the Coulomb gauge
condition was imposed  ``on-shell''. We shall see in section \ref{uniqueness} that the
``off-shell'' gauge condition leads to a (slightly) different gluon
DSE. In Eq.\ (\ref{gluonDSEuncontract}), we can take the trace in Lorentz space to find, using
$t_{ii}(k)=d-1$ and Eq.\ (\ref{GGAdefmom}) and writing $\hat k=k/|k|$,
\begin{equation}
  \label{gluonDSE}
  D_A^{-1}(k)=g^2N_c\frac{1}{(d-1)}\int
  \dbar^d\ell\:\ell^2\left(1-(\hat \ell\cdot\hat k)^2\right)D_G(\ell)D_G(\ell-k)\; .
\end{equation}
The ghost DSE (\ref{DSEghost}) can be Fourier transformed in the same
manner. With a tree-level ghost-gluon vertex, it reads in momentum space
\begin{equation}
  \label{ghostDSE}
   D_G^{-1}(k)=k^2-g^2N_ck^2\int \dbar^d\ell \: \left(1-(\hat{\ell}\cdot \hat{k})^2\right)D_A(\ell)D_G(\ell-k)\: .
\end{equation}
Eq.\ (\ref{ghostDSE}) can be put in a form that explicitly satisfies the horizon
condition (\ref{horizondef}) by subtracting $k^{-2}D_G^{-1}(k)$ at zero momentum. This subtraction
does not introduce an extra parameter since by the horizon condition (\ref{horizondef}), $(k^2D_G(k))^{-1}|_{k=0}=0$,
and we get
\begin{equation}
  \label{ghostDSEren}
  \left(k^2D_G(k)\right)^{-1}=g^2N_c\int  \dbar^d\ell \: D_A(\ell)\left(1-(\hat{\ell}\cdot \hat{k})^2\right)\left(D_G(\ell)-D_G(\ell-k)\right)
\end{equation}
for the ghost DSE with an integral that converges more strongly in the
ultraviolet than the one in Eq.\ (\ref{ghostDSE}). In the ultraviolet sector,
the horizon condition can be used to remove ultraviolet divergences in
the ghost DSE. %%dont forget to note in the UV chapter that Z_3tilde
               %%starts with g^2 and contradicts PT!

The set of equations (\ref{gluonDSE}) and (\ref{ghostDSEren}) was
first solved without approximations in Refs.\ \cite{LerSme02,Zwa02} by a power law
ansatz. The derivation and the solution is reviewed here, and some subtleties are pointed
out. 

With power law ans\"atze for the propagators,
\begin{eqnarray}
  \label{plawAnsatz}
  D_A(k)=\frac{A}{(k^2)^{1+\alpha_A}}\; ,\quad D_G(k)=\frac{B}{(k^2)^{1+\alpha_G}}
\end{eqnarray}
the integrals in (\ref{gluonDSE}) and (\ref{ghostDSEren}) turn into
linear combinations of a particularly simple kind of integrals we
refer to as \emph{two-point integrals},
\begin{eqnarray}
  \label{2ptdef}
 \Xi_m(\alpha,\beta):=\int \dbar^d\ell\:\frac{(\ell\cdot
   k)^m}{(\ell^{2})^{\alpha}((\ell-k)^{2})^{\beta}}\; ,\quad \alpha
 ,\beta\in\mathbb{R}\, ,\,\, m\in\mathbb{N}\; .
\end{eqnarray}
In appendix \ref{app:2ptint}, explicit solutions for the two-point
integrals (\ref{2ptdef}) are given and their convergence properties are
discussed. The gluon DSE (\ref{gluonDSE}) may be
rewritten in terms of two-point integrals as
\begin{eqnarray}
  \label{DSE2ptA}
  \left(k^2\right)^{1+\alpha_A}&=&\frac{g^2N_cAB^2}{d-1}\left(\Xi_0(\alpha_G,1+\alpha_G)-\frac{1}{k^2}\Xi_2(1+\alpha_G,1+\alpha_G)\right)
\end{eqnarray}
and it converges in the ultraviolet (as $\ell\rarr\infty$) for
$1+2\alpha_G>\frac{d}{2}$. Infrared divergences are absent so long as
$\alpha_G<\frac{d}{2}$. 

In the ghost DSE (\ref{ghostDSEren}), the horizon condition allowed
for a subtraction which in turn improves the UV convergence. The
two-point integral occurring in the ``unrenormalized'' DSE
(\ref{ghostDSE}) is divergent, but it can be
unambiguously split into a momentum-independent divergent part and a
regular part, see appendix \ref{app:2ptint}. The divergent part is subtracted using the horizon
condition in (\ref{ghostDSEren}) and the formulae (\ref{2ptints}) for
$\Xi_m(\alpha,\beta)$ yield the correct values for the regular part. Thus,
\begin{eqnarray}
  \label{DSE2ptG}
   \left(k^2\right)^{\alpha_G}=-g^2N_cAB^2\left(\Xi_0(1+\alpha_Z,1+\alpha_G)-\frac{1}{k^2}\Xi_2(2+\alpha_Z,1+\alpha_G)\right)_{reg.}
\end{eqnarray}
where the remark ``reg.'' indicates that the regular part of
$\Xi_m(\alpha,\beta)$ is taken. The integrals in (\ref{DSE2ptG}) are then
UV-convergent if $\alpha_G<1$. In the infrared, poles are avoided for
$-\frac{1}{2}<\alpha_G<\frac{d}{2}$. Only values for $\alpha_A$ and
$\alpha_G$ are considered for which the integrals converge. 

Solving a well-defined (convergent) set of integral equations, we can
rescale the integration variable in (\ref{2ptdef}) by
$\ell\rarr\lambda\ell$, and readily find that 
$\Xi_m(\alpha,\beta)=I_m(\alpha,\beta)(k^2)^{d/2+m-\alpha-\beta}$ with
some dimensionless functions $I_m(\alpha,\beta)$. These power law
solutions can be plugged into the DSEs (\ref{gluonDSE}) and
(\ref{ghostDSEren}) and a relation between the exponents $\alpha_A$ and
$\alpha_G$ is gained,
\begin{eqnarray}
  \label{sumrule}
  \alpha_A+2\alpha_G=\frac{d-4}{2}\; .
\end{eqnarray}
Eq.\ (\ref{sumrule}) is referred to as the ``\emph{sum rule} (for power
law exponents)''. 

Let us emphasize that the sum rule (\ref{sumrule}) can be directly
traced back to the nonrenormalization of the ghost-gluon vertex
\cite{SchLedRei06}. Following Ref.\ \cite{WatAlk01}, a quite general
ansatz for the proper ghost-gluon vertex is
\bee
\label{GGAvertex}
\Gamma_k(k;q,p)=igt_{kj}(k)q_j\sum_aC_a\Big (\frac{k}{\sigma}\Big )^{l_a}\left(\frac{q}{\sigma}\right)^{m_a}\left(\frac{p}{\sigma}\right)^{n_a}\: ,
\eeq
where the constraint $l_i+m_i+n_i=0, \forall i,$ guarantees the independence of the renormalization scale $\sigma$, i.e.\ nonrenormalization of the vertex. It is readily shown that the sum rule (\ref{sumrule}) is not affected by a dressing of the ghost-gluon vertex such as (\ref{GGAvertex}), since it turns into 
\bee
\alpha_A+2\alpha_G=\frac{d-4}{2}+\sum_i(l_i+m_i+n_i)\: .
\eeq
Any consequence of the sum rule is therefore understood as due to the nonrenormalization of the ghost-gluon vertex.

By virtue of the sum rule (\ref{sumrule}), one of the exponents
$\alpha_A$ and $\alpha_G$ can be eliminated. We choose to define
\begin{eqnarray}
  \label{kappadef}
  \kappa:=\alpha_G
\end{eqnarray}
and express $\alpha_A$ in terms of $\kappa$.

The integrals in (\ref{gluonDSE}) and (\ref{ghostDSEren}) can now be
written down concisely, using the formulae in (\ref{2ptints}),
\begin{subequations}
\label{DSEIAIG}
  \begin{eqnarray}
    \label{DSEIA}
    (k^2)^{d/2-1-2\kappa}&=&g^2N_cAB^2I_A(\kappa)(k^2)^{d/2-1-2\kappa}\\
    \label{DSEIG}
    (k^2)^\kappa&=&g^2N_cAB^2I_G(\kappa)(k^2)^\kappa
  \end{eqnarray}
\end{subequations}
where the dimensionless functions $I_A$ and $I_G$ were introduced,
\begin{subequations}
  \label{IZIGdef}
\begin{eqnarray}
\label{IAdef}
I_A(\kappa)&=&\frac{1}{2(4\pi)^{d/2}}\,\frac{{\Gamma(\frac{d}{2} - \kappa )}^2\,\Gamma(1 - \frac{d}{2} + 2\,\kappa )}
  {\Gamma(d - 2\,\kappa )\,{\Gamma(1 + \kappa )}^2}\; ,\\
\label{IGdef}
  I_G(\kappa)&=&- \frac{4^\kappa (d-1)}{(4\pi)^{d/2+1/2}} \,\frac{\Gamma(\frac{d}{2} - \kappa )\,\Gamma(-\kappa )\,
      \Gamma(\frac{1}{2} + \kappa )}{\Gamma(\frac{d}{2} - 2\,\kappa
      )\,\Gamma(1 + \frac{d}{2} + \kappa )}\; .
\end{eqnarray}
\end{subequations}
Eq.\ (\ref{DSEIAIG}) requires that
\begin{eqnarray}
  \label{coeffrule}
g^2N_cAB^2I_A=1=g^2N_cAB^2I_G\; .   
\end{eqnarray}
The infrared exponent $\kappa(d)$ is
therefore implicitly defined by
\begin{eqnarray}
  \label{kappavond}
  I_{id}:=\frac{I_G(\kappa)}{I_A(\kappa)}\stackrel{!}{=}1\; .
\end{eqnarray}

\begin{figure}
\centering
\includegraphics[scale=1.4]{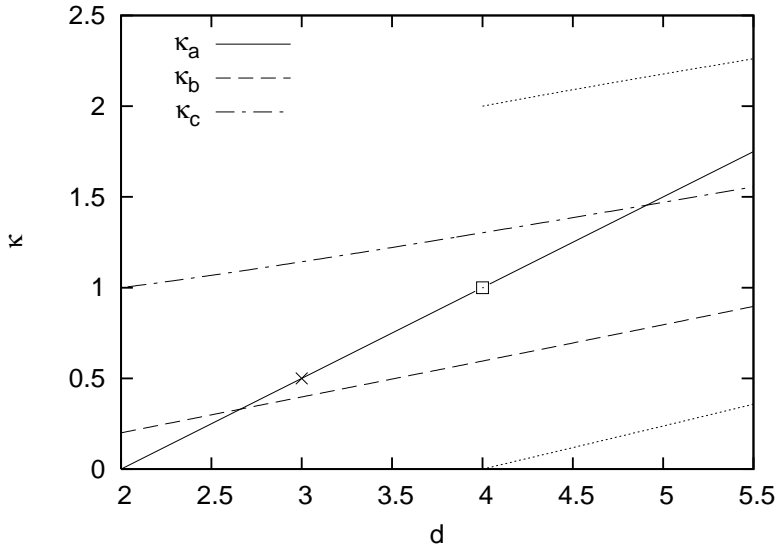}
\caption{Solutions for $\kappa(d)$. The box
  indicates an exceptional point, see text. The cross marks the
  solution favorable for reasons given below. The dotted lines are
  extra solutions that occur for $d>4$. Note that $\kappa$ is
  restricted to the range (\ref{kappaconverge}).}
\label{kvond}
\end{figure}

Above, convergence of the integrals in the DSEs was discussed. In terms of the
exponent $\kappa$, the conditions for convergence can be
re-expressed. However,
recall that the ultraviolet behavior of $D_A(k)$ and $D_G(k)$ is not
governed by the stochastic vacuum but by perturbation theory. The power
law behavior (\ref{plawAnsatz}) is just regarded as a possible
infrared limit that will actually be realized (see chapter \ref{VarVac}). Hence, the
restriction on the infrared exponents by analyzing the UV convergence
is spurious. Discarding UV criteria for both $I_A(k)$ and $I_G(k)$, we
find from the condition that no divergences appear in the infrared
integration domain the relation $-\frac{1}{2}<\kappa<\frac{d}{2}$. Moreover, the horizon condition
demands $\kappa>0$. Altogether, we have
\begin{eqnarray}
  \label{kappaconverge}
  0<\kappa<\frac{d}{2}
\end{eqnarray}
and in this range solutions to (\ref{kappavond}) are
investigated.\footnote{According to the principle of locality, the
  Green functions should be tempered distributions which  restricts
  $\kappa$ further. We do not focus on this aspect here.} Manipulating the gamma functions in (\ref{IZIGdef}), one finds the compact expression
\begin{eqnarray}
  \label{Iid}
  I_{id}(d,\kappa)=\frac{\sin(\frac{\pi}{2}(d-4\kappa))}{\sin(\pi\kappa)}\frac{(d-1)\Gamma(1+2\kappa)\Gamma(d-2\kappa)}{\Gamma(1+\kappa+d/2)\Gamma(d/2-\kappa)}\stackrel{!}{=}1
\end{eqnarray}
for finding $\kappa(d)$. The solutions are shown in Fig.\
\ref{kvond}. These are exhaustive for $\kappa$ in the range (\ref{kappaconverge}) and
complete a recent calculation \cite{Hub+07}.
For a given Euclidean dimension $d$, several solutions for $\kappa$
are available. One of them can be found by algebraic manipulations,
\begin{eqnarray}
  \label{kappa2}
  \kappa_a(d)=\frac{d}{2}-1\; ,
\end{eqnarray}
as verified by plugging $\kappa_a(d)$ into Eq.\ (\ref{Iid}). The
other (irrational) solutions $\kappa_b(d)$ and $\kappa_c(d)$ were determined numerically. 

There are
some exceptional points in the $(d,\kappa)$ diagram that are only
solutions to Eq.\ (\ref{Iid}) if approached in a certain direction. One of
them is the point $\binom{d=4}{\kappa=1}$ that solves Eq.\ (\ref{Iid}) if
approached along the line (\ref{kappa2}). This peculiarity
appears for \emph{even} dimensions $d$ and \emph{integer} values for $\kappa$. Setting
\bee
d&=& 2m-2\lambda\epsilon \\
\kappa&=& n+\epsilon
\eeq
with $m,n\in\bN$, we approach in the limit $\epsilon\rarr 0$ the
values for even $d$ and integer $\kappa$. The parameter $\lambda\in\R$
controls the slope $-\frac{1}{2\lambda}$ of the line on which the point $\binom{2m}{n}$ is approached in a
$(d,\kappa)$-diagram. Choosing $n<m$ guarantees the integrability of
infrared poles in the ghost loop integral, see Eq.\ (\ref{kappaconverge}). Treating the $\frac{0}{0}$ expressions that
occur in (\ref{Iid}) carefully, we find
\bee
\label{Iidmn}
\lim_{\epsilon\rarr
  0}I_{id}(2m-2\lambda\epsilon,n+\epsilon)=(-1)^{1+m+n}(m-n)_{m-n}\frac{(2m-1)(2n)!}{(m+n)!}(2+\lambda)
\eeq
where $(a)_n=a(a+1)\dots(a+n-1)$ is the Pochhammer
symbol. Setting
$\lambda=0$ corresponds to keeping a fixed dimension throughout the
calculation. From Eq.\ (\ref{Iidmn}) it can be deduced that for no values
of $m,n$ the equation $I_{id}=1$ (\ref{Iid}) is solved if
$\lambda=0$. Hence the failure of trying to find the point
$\binom{d=4}{\kappa=1}$ as a Landau
gauge solution \cite{FisAlk02},
with arguments in Refs.\ \cite{AtkBlo98a,Zwa02} that this solution
should exist. For the set of points with $n=m-1$, i.e.\
\bee
\label{excpts}
\binom{d}{\kappa}\in\left\{\binom{2}{0},\, \binom{4}{1},\,
  \binom{6}{2},\,\dots \right\}\; ,
\eeq
we find from (\ref{Iidmn}) the simple result
\bee
\lim_{\epsilon\rarr
  0}I_{id}(2m-2\lambda\epsilon,m-1+\epsilon)=2+\lambda\, .
\eeq
Tuning $\lambda=-1$, the points (\ref{excpts}) are solutions to
(\ref{Iid}), at least in principle. One may adopt the point of view that
the dimension $d$ is fixed and the points in Fig.\ \ref{kvond} can
only be approached along vertical lines. In fact, rendering $d$ a
variable is here not understood in the context of dimensional
regularization where divergences are thus regularized. Rather, $d$ is
here a fixed parameter for which (\ref{Iid}) needs to be
solved. The exceptional points (\ref{excpts}) are therefore discarded.

As seen in Fig.\ \ref{kvond}, there are several solutions for $\kappa$ as a function of $d$. How many solutions are found for a given dimension? For even dimensions $d$, Eq.\ (\ref{Iid}) can be turned into a
polynomial equation in $\kappa$ of degree $d+1$ which has the
$\frac{d}{2}+1$ integer solutions $\kappa\in\{0,0,1,2,\dots
,\frac{d}{2}-1\}$ for $\lambda$ given by Eq.\ (\ref{Iidmn}). The
double root at $\kappa=0$ is excluded by the horizon condition. Thus,
these integer solutions lie below (or on) the branch (\ref{kappa2}) looking somewhat
odd in Fig.\ \ref{kvond}. There are further non-integer solutions for even
dimensions, $\frac{d}{2}$ at the most being real.

\begin{figure}
  \centering
  \includegraphics[scale=1.2]{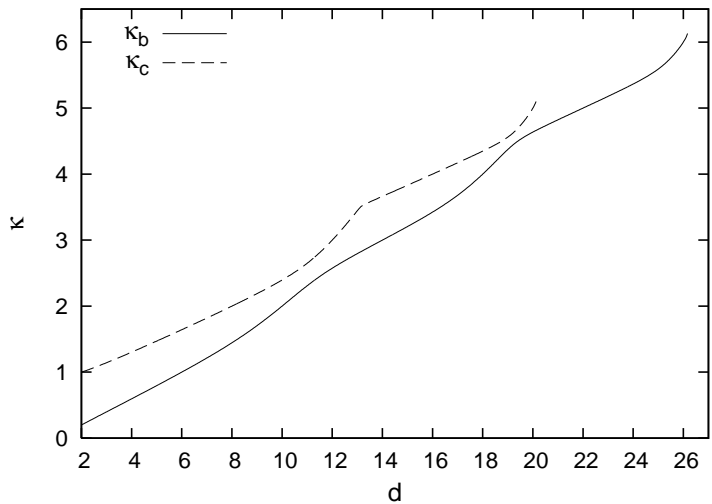}
\caption{Two solution branches of $\kappa(d)$ and their behavior for
  large dimensions $d$. The upper solution becomes complex at
  $d\approx 20.15$, the lower one at $d\approx 26.18$.}
  \label{highd}
\end{figure}

Apart from the branch $\kappa_a(d)$ in Eq.\ (\ref{kappa2}), there are
two other solutions $\kappa_b(d)$ and $\kappa_c(d)$ in
the range in $d\in (2,4]$, see Fig.\ \ref{kvond}. Let us investigate the continuity of these
three solution branches as functions of $d$, cf.\ \cite{Zwa03a}. Here,
we do not consider the limit $d\to 1$ since the $1+1$
dimensional theory does not have any degrees of freedom and power law
solutions might not be appropriate. Instead, we
discuss the high-$d$ range. The Gribov--Zwanziger scenario should work
particularly well for higher-dimensional YM theory where entropy is
even more enhanced at the boundary of configuration space. First of all, the
solution $\kappa_a(d)$ holds for all $d$. The two other solutions in
$\kappa_b(d)$ and $\kappa_c(d)$ in the interval $2<d\leq 4$ connect to the curves shown in Fig.\ \ref{highd}
or higher $d$. The upper solution approximately yields $\kappa_c=\frac{d}{6}+\frac{2}{3}$ for
$2<d\leq 8$ and oscillates around $\kappa=\frac{d}{4}$ for $8\leq d
\leq 20$. It ceases to be real for $d\approx 20.15$ . The lower solution matches $\kappa_b=\frac{d}{5}-\frac{1}{5}$
for $2<d\leq 6$, in agreement with Ref.\ \cite{Zwa03a}, but oscillates around $\kappa=\frac{d}{4}-\frac{1}{2}$
for $6\leq d \leq 26$ and becomes complex at $d\approx 26.18$. The
continuity of the solutions for $\kappa$ as functions of $d$ is
therefore strongest for the solution $\kappa_a(d)$ which holds for all
$d$.

A restriction on the set of solutions in Fig.\ \ref{kvond}
comes from dealing with only power laws for the propagators, due to
the ansatz (\ref{plawAnsatz}). The two-point integrals in (\ref{2ptdef}) cease to
be power laws if $\frac{d}{2}+m-\alpha-\beta=0$. Actually, they cease
to exist, due to a logarithmic ultraviolet divergence. We have argued
that these ultraviolet divergences are subtracted by some 
renormalization scheme in the perturbative sector. One might therefore
suggest to consider instead the integrals $\Xi_m(\alpha,\beta)$ subtracted at
$\mu$ and then approach $\frac{d}{2}+m-\alpha-\beta=0$. For instance,
if $m=0$, then
\begin{eqnarray}
  \label{logsol}
  \lim_{\alpha\rarr\frac{d}{2}-\beta} \left(\Xi_0(\alpha,\beta)-\left.\Xi_0(\alpha,\beta)\right|_{q=\mu}
\right)
&=&  \frac{1}{(4\pi)^{\frac{d}{2}}}\frac{1}{\Gamma(\frac{d}{2})}\lim_{\epsilon\rarr
    0}\Gamma(\epsilon)\left[(k^2)^{-\epsilon}-(\mu^2)^{-\epsilon}\right]\nn\\&=&
  \frac{1}{(4\pi)^{\frac{d}{2}}}\frac{1}{\Gamma(\frac{d}{2})}\ln\frac{\mu^2}{k^2}\; .
\end{eqnarray}
Evidently, after subtraction of the ultraviolet divergence, the
solution of the two-point integral (with exponent zero) is a
logarithm. These functions call for a separate treatment. The allowed
values of $\kappa$ (\ref{kappaconverge}) are therefore further restricted by
\begin{eqnarray}
  \label{nologs}
  \kappa\neq\frac{d}{4}-\frac{1}{2}
\end{eqnarray}
to avoid a ghost loop that gives a logarithm. The lower solution
branch $\kappa_b(d)$ in Fig.\ \ref{highd} oscillates around the forbidden values
(\ref{nologs}) for high $d$, exactly yielding (\ref{nologs}) for
$d=6,10,14,18,\dots$. A discussion on a
possible extension to solutions with logarithms will follow in section
\ref{uniqueness}. 

\begin{table}
  \centering
  \begin{tabular}{c|c|c|c|}
    $d$ & $\kappa$ & $D_G$ & $D_A$\\
    \hline
    $2$ & $\frac{1}{5}$ & $\sim\frac{1}{k^{2.400}}$ & $\sim k^{0.800}$
    \\ 
    \hline
    $3$ & $0.398$ & $\sim\frac{1}{k^{2.795}}$ & $\sim k^{0.590}$ \\
    $3$ & $\frac{1}{2}$ & $\sim\frac{1}{k^3}$ & $\sim k$ \\
    $3$ & $1.143$ & $\sim\frac{1}{k^{4.285}}$ & $\sim k^{3.570}$ \\
    \hline
    $4$ & $0.595$ & $\sim\frac{1}{k^{3.191}}$ & $\sim k^{0.381}$\\
    $4$ & $1.303$ & $\sim\frac{1}{k^{4.605}}$ & $\sim k^{3.210}$
  \end{tabular}
  \caption{Power law solutions for propagators in the stochastic vacuum.}
  \label{tab:sol}
\end{table}

To summarize this section, we solved the momentum space
Dyson--Schwinger equations of the stochastic vacuum with a tree-level
ghost-gluon vertex by power law ans\"atze and thoroughly
discussed the validity of the solutions. For the propagators in the temporal
Coulomb and the Landau gauge, with $d=3$ and $d=4$, resp., the
solutions are shown in Table \ref{tab:sol}. In $2+1$ dimensional YM
theory, the temporal Coulomb gauge corresponds to $d=2$ and the Landau
gauge to $d=3$.

% Further investigations in \cite{LerSme02} showed that the value for $\kappa$, determined by Eq.\ (\ref{detkappa}), only slightly depends on the values of $\{l_i,m_i,n_i\}$. 

% Since neither the DSE studies \cite{dipl_Schleifenbaum,Sch+05} nor the lattice calculations \cite{CucMenMih04, Ste+05} show any infrared divergences, the dressing function of the ghost-gluon vertex must be some finite function. To investigate the consequences of a finite dressing function of the ghost-gluon vertex, let us assume, for simplicity, that it is given by a finite constant,
% \bee
% \label{constGGA}
% \Gamma_\mu(k;q,p)=C\Gamma^{(0)}_\mu(q)\; ,
% \eeq
% The constant $C$ can be determined by self-consistingly solving the DSE for the ghost-gluon vertex in the infrared gluon limit, see section \ref{}. 
% where $\Gamma^{(0)}_\mu(q)=igq_\mu$ is the bare ghost-gluon vertex. Then, the infrared analysis of the propagators can be performed in the same way as above. The ghost self-energy and the ghost loop are both multiplied by the constant $C$. In Eq.\ (\ref{detkappa}), this constant appears on both sides to one power and thus trivially cancels. Therefore, a constant dressing of the ghost-gluon vertex is completely irrelevant for the infrared behaviour of the propagators.

\vfill

\section{Linear confining potential}
\label{confinement}
The Hamiltonian approach in the temporal Coulomb gauge allows for a
direct calculation of the potential energy between two static
quarks. Let the external charge density be given by two point charges,
separated by the distance $r$,
\begin{equation}
  \label{rhom}
  \rhoext^a(x)=\delta^{a3}\left(\delta^{(3)}(x-r/2)-\delta^{(3)}(x+r/2)\right)\; .
\end{equation}
The color vector $\rhoext(x)$ was here chosen in the $3$-direction of $SU(2)$,
for convenience. From the Hamiltonian (\ref{CLHam}) the $|x-y|=r$ dependent
color Coulomb potential $V_C(r)$ can be identified by
\begin{equation}
  \label{potential}
  V_C(r)=\frac{g^2}{2}\int d^3[xy]\rhoext^a(x)\:\lla F^{ab}(x,y) \rra\:
  \rhoext^b(y)\; .
\end{equation}
%The factor$\delta^3(0)$ is the volume of space, coming from the
%translational invariance of the vacuum expectation value. 
With Eq.\ (\ref{rhom}), the Coulomb potential $V_C(r)$ can be written as 
\begin{equation}
  \label{VCmeson}
  V_C(r)=g^2\lla F^{33}(0)-F^{33}(r)\rra\; ,
\end{equation}
with the self-energy of the quarks, $\lla F^{33}(0)\rra\equiv\lla
F^{33}(x,x)\rra$. 

An
approximation is needed to express $V_C(r)$ in terms of the known
Green function $D_G(k)$. In arriving at the solutions for the
propagators in the previous section, the ghost-gluon vertex was
approximated to be at tree-level. An approximation for writing
$V_C(r)$ in terms of $D_G$ that is equivalent to the vertex
approximation would be desirable. Let us recall that factorizing the
expectation value $\lla G \Gamma^0 G \rra \approx \lla G\rra\Gamma^0\lla
G\rra$ implies $\Gamma\approx\Gamma^0$, see Eq.\ (\ref{GGAconn}). In the same manner, we factorize
\begin{eqnarray}
  \label{Ffact}
  \lla F^{ab}(x,y) \rra &=&
\int d^3z \lla G^{ad}(x,z)(-\del^2) G^{db}(z,y)\rra\nn\\
&\approx& \int d^3z \lla G^{ad}(x,z)\rra (-\del^2)\lla
G^{db}(z,y)\rra\; .
\end{eqnarray}
Checking the validity of this approximation is an issue dealt with in
section \ref{fsection}.
The expression (\ref{VCmeson}) thus turns into
\begin{eqnarray}
  \label{V_C}
  V_C(r)=g^2\int \dbar^3k\, k^2 D_G^2(k)\left(1-\e^{ik\cdot r}\right)\; .
\end{eqnarray}
Plugging in $D_G(k)=B/(k^2)^{1+\kappa}$ from (\ref{plawAnsatz}) permits an explicit
evaluation of $V_C(r)$,\footnote{In the first line, the pole at $k=0$
  is regularized by integrating only from $0<\epsilon\ll 1$ to
  $\infty$ and letting $\epsilon\rarr 0$ at the end. In the second
  line, partial integration is used repeatedly, and in the last line
  an integral representation of the gamma function in Ref.\ \cite{Gradshteyn}, formula $3.761.4$ was used. }
\begin{eqnarray}
  \label{V_Csol}
  V_C(r)&=&\frac{g^2B^2}{2\pi^2}\int_0^\infty dk\,
  k^{-4\kappa}\left(1-\frac{\sin(kr)}{kr}\right)\nn\\
  &=&\frac{g^2B^2}{2\pi^2}r^{4\kappa-1}
\left\{
    \left(
      \frac{1}{4\kappa-1}-\frac{1}{4\kappa}-\frac{1}{4\kappa(4\kappa-1)}
    \right)
        \frac{1}{\epsilon^{4\kappa-1}}
    +\frac{1}{4\kappa(4\kappa-1)}\int_0^\infty dx\, \frac{\sin x}{x^{4\kappa-1}}
\right\} \nn\\
  &=&\frac{g^2B^2}{2\pi^2}\Gamma(-4\kappa)\sin\left(2\kappa\pi\right)r^{4\kappa-1}
\end{eqnarray}
for values $\frac{1}{4}<\kappa<\frac{3}{4}$ where this integral
exists. In the latter interval, two solutions of the DSEs exist, see Table \ref{tab:sol}. One
of them, found on the branch $\kappa_a(d)$ in Fig.\ \ref{kvond}, yields
$\kappa_a=\frac{1}{2}$ and thus a heavy quark potential $V_C(r)$ that rises
linearly. From Eq.\ (\ref{V_Csol}) one gets for $\kappa=\kappa_a$
\begin{eqnarray}
  \label{V_Clin}
  V_C^{lin}(r)=\frac{g^2B^2}{8\pi}r\; .
\end{eqnarray}
By the above linear potential, heavy quarks are held together by a
constant force given by the {\it Coulomb string tension}
$\sigma_C=g^2\frac{B^2}{8\pi}$. Thus, heavy quarks are confined. 

Let us reflect on how we obtained such a heavy
quark potential. It is known that $V_C(r)$ cannot rise stronger than
linearly \cite{Sei78,Bac86}, but it could saturate for large $r$ (see
further discussions in the chapters below). The
linearity of the potential agrees exactly with lattice calculations (see
e.g.\ Ref.\ \cite{Bal00}). Hence, support is given for the validity of the
result (\ref{V_Clin}). % The spinning top model \cite{Gre03} is a simple
% possible idea how this confining potential may be thought of. On the
% other hand, it
It is enlightening to survey the origin of this outcome. The
Gribov--Zwanziger scenario was exhibited in section \ref{GribovZwanziger}. Gribov copies necessitate the restriction of configuration 
space and Zwanziger's entropy arguments demand the horizon 
condition. The simplest possible state, the stochastic vacuum, is
deemed sufficient to describe the infrared properties of the theory. 
Dyson--Schwinger equations comprise the information on the Green
functions and with a few approximations, we thus arrived at the linear
potential (\ref{V_Clin}). One might be astonished by the simplicity of
these ideas. In particular, discarding the entire Yang--Mills action $S_{YM}$
(setting $\Psi[A]=1$) and still producing sensible results seems
puzzling. The only physical content in the generating functional
\begin{eqnarray}
  \label{Zagain}
     Z[j]=\frac{1}{\mathit{Vol}(\Omega)}\int_\Omega \cD A\, \cJ[A]\e^{j\cdot A}
\end{eqnarray}
that drives the (infrared) dynamics comes from the gauge fixing
procedure.\footnote{Note that the Faddeev--Popov determinant is gauge invariant.} This notion goes by the name \emph{infrared ghost dominance}. Merely the Faddeev--Popov kernel $\cM$ (\ref{FPpathint}) and its boundary
conditions on the Gribov horizon $\del\Omega$ are sufficient to give rise
to a linear confinement potential. The crucial infrared properties are
thus governed by entropy in configuration space. 
% On top of that, the Faddeev--Popov determinant $\cJ[A]$ is gauge
% invariant. \dots

A caveat in the above reasoning for quark confinement in the Coulomb
gauge is that the Coulomb potential $V_C(r)$ is only an upper bound to
the gauge-invariant potential $V_W(r)$ defined by the Wilson loop \cite{Zwa03b}. The
Coulomb string tension is therefore larger or equal to the string
tension from the Wilson loop, $\sigma_C\geq\sigma_W$. Hence, it cannot
serve as an order parameter for the deconfinement transition. The
mechanism that lowers the Coulomb string tension to its physical value
is expected to be due to constituent gluons along the color flux
tube. So far, the vacuum Green functions are unaware of the presence
of external charges. It will be the subject of chapter \ref{external} to investigate the
effect of the back reaction of the external color charges onto the
gauge field sector.

Some further critical remarks are at order.
\begin{itemize}
\item \emph{Perturbative sector}. If the solutions in the stochastic
  vacuum are independent of $S_{YM}$, how do they connect to the
  ultraviolet regime?
\item \emph{Approximations}. How can we estimate the effect of the
  approximations made in deriving the Green functions?
\item \emph{Uniqueness}. Are the power law solutions found unique?
  Which one of them is realized in physics?
\end{itemize}
In the following it will be attempted to give answers to the questions
listed above. The latter item is dedicated to the next section.

\section{On the uniqueness of the solution}
\label{uniqueness}

Whether the solutions for the propagators found in section \ref{rechnung} can
be regarded as unique solutions to the infrared sector of YM theory is
an important issue addressed in this section. A thorough discussion of
uniqueness would probably have to begin with the perturbative sector
and any approximations should be avoided. Here, we shall turn to a much
simpler task. In view of the approximated DSEs (\ref{gluonDSE}) and (\ref{ghostDSEren}), it is
investigated if any of the solutions listed in Table \ref{tab:sol} can
be considered unique.

\subsection*{A free parameter}
First of all, let us start with a generalization of the DSEs.
In the context of renormalization in the perturbative regime,
a local action is indispensable and the
transversality condition on the gauge field, be it Coulomb gauge or
Landau gauge, needs to be taken ``off-shell''%, i.e.\ with the aid of
                                %a Nakanishi-Lautrup multiplier field
                                %\cite{}
. A crucial difference to
the ``on-shell'' formulation is that the tree-level Green functions
are slightly different. In particular, the ghost-gluon vertex does not
come with a transversal projector attached to its gluon leg. Instead
of $\Gamma_k^0(k,q)$, the local formulation yields $\Gamma_k^0(q)$ with
\begin{eqnarray}
  \label{GGAtloff}
  \Gamma_k^0(q)=igq_k \quad \Leftrightarrow \quad
  \Gamma_k^0(k,q)=igt_{kj}(k)q_j\; .
\end{eqnarray}
Consequently, in the local formulation not all Feynman graphs are transverse in the gluon
momenta. For instance, the $\cO(g^2)$ gluon propagator in Landau gauge
is known to be comprised by a gluon loop, a ghost loop and a tadpole
term. Only the sum of these diagrams yields a transverse gluon
propagator. This subtlety slightly alters the DSEs derived in section
\ref{rechnung}. We do not repeat the derivation here, but simply state the
results,
\begin{subequations}
  \label{DSEoff}
\begin{eqnarray}
\label{gluonDSEoffij}
    D_{ij}^{-1}(k)&=&-N_c\int\dbar^d\ell\, \Gamma_i^0(\ell)\Gamma_j^0(\ell-k)D_G(\ell)D_G(\ell-k)\\
\label{ghostDSEoff}
    D_G^{-1}(k)&=&k^2-g^2N_ck^2\int \dbar^d\ell \: \left(1-(\hat{\ell}\cdot \hat{k})^2\right)D_A(\ell)D_G(\ell-k)\; ,
\end{eqnarray}
\end{subequations}
as opposed to Eqs.\ (\ref{gluonDSEuncontract}) and (\ref{ghostDSE}). Here, the proper
ghost-gluon vertices were again replaced by the tree-level 
ones. The ghost DSE stays the same. In the gluon
DSE (\ref{gluonDSEoffij}), on the other hand,
note the tiny difference that the ghost-gluon vertices $\Gamma_k^0(q)$
are the ones defined in Eq.\ (\ref{GGAtloff}). Although the
nonperturbative ghost loop
on the r.h.s.\ of Eq.\ (\ref{gluonDSEoffij}) should be transverse with
a proper ghost-gluon vertex,\footnote{If the ghost loop dominates in
  the infrared, it must itself be transverse.} it is generally not transverse with the
tree-level ghost-gluon vertices (\ref{GGAtloff}). Transversality is
lost due to the approximation of the vertex. To control the
longitudinal contributions, contract Eq.\ (\ref{gluonDSEoffij}) with
the tensor
\begin{eqnarray}
  \label{Rtens}
  R_{ij}^\zeta(k):=\delta_{ij}-\zeta\ell_{ij}(k)\; .
\end{eqnarray}
The factor $\zeta\ell_{ij}(k)$ would give no contribution if both left
and right hand side of  (\ref{gluonDSEoffij}) were transverse. The
sensitivity of the solution with respect to $\zeta$ thus measures the
violation of transversality and the quality of the approximation
used. 

Before we use the tensor $R_{ij}^\zeta$, let us calculate the inverse
gluon propagator in Eq.\ (\ref{gluonDSEoffij}) explicitly with the
ansatz (\ref{plawAnsatz}) for the ghost propagator
$D_G(k)$,\footnote{The tensor integral may be expanded into scalar
  integrals by means of the Passarino--Veltman algorithm \cite{PasVel79}.}
\begin{eqnarray}
  \label{CurvTens}
  B^{-2}D_{ij}^{-1}(k)&=&g^2N_c\int \dbar^d\ell\, \frac{\ell_i(\ell-k)_j}{(\ell^2)^{1+\kappa}((\ell-k)^2)^{1+\kappa}} \nn\\
&=&g^2N_c\frac{1}{(4\pi)^{d/2}}(k^2)^{d/2-1-2\kappa}
\left\{
\frac{\Gamma(d/2-\kappa)^2\Gamma(2\kappa+1-d/2)}{2\Gamma(1+\kappa)^2\Gamma(d-2\kappa)}\delta_{ij}
\right. \nn\\
&& - 
\left.
\frac{\Gamma(d/2-\kappa)\Gamma(d/2-1-\kappa)\Gamma(2\kappa-d/2+2)}{\Gamma(1+\kappa)^2\Gamma(d-1-2\kappa)}\ell_{ij}(k)
\right.\nn\\
&&\qquad +
\left.
\frac{\Gamma(d/2+1-\kappa)\Gamma(d/2-1-\kappa)\Gamma(2\kappa-d/2+2)}{\Gamma(1+\kappa)^2\Gamma(d-2\kappa)}\ell_{ij}(k)
\right\}\nn\\
&=&I_A\left[\:\delta_{ij}-(4\kappa-d+2)\:\ell_{ij}(k)\:\right](k^2)^{d/2-1-2\kappa}\; ,
\end{eqnarray}
with the function $I_A$ given by Eq.\ (\ref{IAdef}). The Lorentz structure of
$D_{ij}^{-1}(k)$ is most easily recognized by the last line in
(\ref{CurvTens}). It reproduces a familiar result from
perturbation theory ($\kappa=0$): the ghost loop has longitudinal
components. Furthermore, one can see that quite generally the ghost
loop is not transverse. Transversality is maintained if and only if the
infrared exponent $\kappa$ satisfies
\begin{eqnarray}
  \label{kappatransv}
  \kappa=\frac{d-1}{4}\; ,
\end{eqnarray}
as seen directly in Eq.\ (\ref{CurvTens}). For other values of
$\kappa$, projections of the gluon DSE (\ref{gluonDSEoffij}) with the tensor
$R_{ij}^\zeta(k)$ in Eq.\ (\ref{Rtens}) will depend on the value of
$\zeta$. In this case, the computation of the DSEs can be repeated to
give a condition $I_{id}(\kappa,\zeta)=1$, analogously to Eq.\
(\ref{kappavond}). The latter condition yields a solution $\kappa(d,\zeta)$ that is not
unique in the sense that it depends on $\zeta$. This dependence is due
to the approximation of the ghost-gluon vertex and must be
spurious. In principle, $\zeta$ can be regarded as a free parameter. A
good approximation will give a weak $\zeta$-dependence.

\begin{figure}
\centering
\includegraphics[scale=1.3]{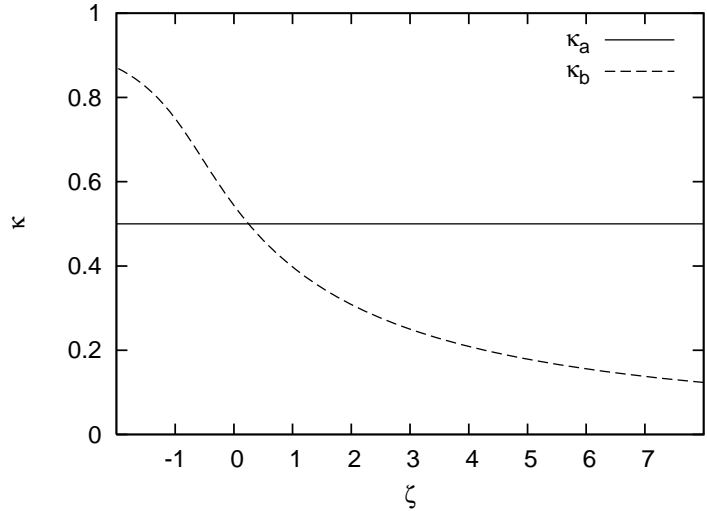}
\caption{The function $\kappa(d=3,\zeta)$. Due to the approximation of
  the ghost-gluon vertex, the infrared
  exponent $\kappa$ depends on the parameter $\zeta$, except for $\kappa_a=1/2$.}
\label{kvonz}
\end{figure}

We now focus on $d=3$ where one can derive that \cite{Maa+04}
\bee
\label{eins}
I_{id}(\kappa,\zeta)=\frac{32(\kappa-1)\kappa\cos(2\pi\kappa)}{(1+2\kappa)(3+2\kappa)(2(1-\kappa)-\zeta(1-2\kappa))\sin(\pi\kappa)^2}=1
\eeq
determines the value of $\kappa$, spuriously dependent on $\zeta$. For
$\zeta=1$, the two solutions $\kappa_a=\frac{1}{2}$ and
$\kappa_b=0.398$ from Table \ref{tab:sol} are recovered. In Fig.\
\ref{kvonz}, it is shown that other solutions for $\kappa$ can be found if
$\zeta\neq 1$. The important feature to notice here is that while
$\kappa_b$ depends on $\zeta$, $\kappa_a$ is
$\zeta$-independent. This can be seen immediately from
(\ref{kappatransv}) by noting that $\kappa_a=\frac{1}{2}$ yields a
transverse ghost loop.
Although favored in Refs.\ \cite{Maa+04,Hub+07} for $3$-dimensional
Landau gauge, the solution $\kappa_b$ is quite sensitive to
approximations in the off-shell formulations, and hence might be unphysical.
For the solution $\kappa_a=\frac{1}{2}$, the approximations made are
more trustworthy and it allows any value of $\zeta$. Further arguments
in favor of $\kappa_a$ will be given in section \ref{vertices}, in the context of
well-defined infrared limits of the ghost-gluon vertex. Here, let us
finish the discussion by noting that it is $\kappa_a$ that naturally
yields a linearly confining potential in the Coulomb gauge.

In the Landau gauge ($d=4$), the solutions listed in Table \ref{tab:sol} are all
sensitive to the value of $\zeta$ and none of the solutions is
characteristic in this respect. The accepted value for the infrared
exponent, is $\kappa=0.595$, as reviewed in Ref.\ \cite{Fis06}.

\subsection*{Logarithmic degeneracy}
The infrared power law ans\"atze made so far yielded integrals that can be generically represented by 
\bee
\label{defI}
I(\alpha;p^2):=\int d^d\ell\,  (\ell^2)^\alpha
\phi(\ell-p)=I(\alpha)(p^2)^{\alpha+\varphi}\; ,\quad  p\rarr 0
\eeq
for some suitable function $\phi(\ell^2)$. The solutions to the integral in Eq.\ (\ref{defI}) with $\phi(\ell^2)$
being a power law were discussed, and it is clear how to find the
exponent $\varphi$ that depends on the dimension $d$ of the integral
and the function $\phi$. The dimensionless numbers $I(\alpha)$ can be
computed with the methods given in Ref.\ \cite{SchLedRei06}. Each of the
DSEs are of the form
\bee
\label{logDSE}
(p^2)^\beta=I(\alpha;p^2)\: .
\eeq
Matching both sides of Eq.\ (\ref{logDSE}), we get the sum rule $\beta=\alpha+\varphi$ on the one hand and the condition $I(\alpha)=1$ on the other hand. This determines $\alpha$ as well as $\beta$.

In order to incorporate logarithms into the infrared ans\"atze such as
$(\ell^2)^\alpha \ln^n(\ell^2)$ we note that simple powers can serve as generating functions of the logarithm to some power $n\in\mathbb{Z}$, that is,
\bee
\label{gener}
(p^2)^\alpha\ln^np^2=\left\{
\begin{aligned}
\frac{\partial^n}{\partial\alpha^n}(p^2)^\alpha & \quad n>0 \\
\underbrace{\int\dots\int d\alpha}_{|n|}\: (p^2)^\alpha & \quad n<0 \quad .
\end{aligned} 
\right.
\eeq
As long as the generators of the logarithm, i.e.\ the derivatives or the $\alpha$-integrals, may be interchanged with the loop integration of $d^d\ell$ we can compute
\bee 
\label{Jdef}
J(\alpha,n;p^2) :=\int d^d\ell\,  (\ell^2)^\alpha \ln^n\ell^2
\phi(\ell-p)\; ,\quad p\rarr 0
\eeq
by applying these generators to Eq.\ (\ref{defI}).
E.g., for $n=1$ we simply have to write
\bee
J(\alpha,1;p^2)&=&\int d^d\ell\,  (\ell^2)^\alpha \ln\ell^2 \phi(\ell-p)=\int d^d\ell\,  \frac{\partial}{\partial\alpha}(\ell^2)^\alpha \phi(\ell-p)\nn\\
&=& \frac{\partial}{\partial\alpha} I(\alpha;p^2)= \left(\frac{\partial I(\alpha)}{\partial\alpha}+I(\alpha)\ln p^2\right)(p^2)^{\alpha+\varphi} \nn\\
&\stackrel{p\rarr 0}{\rarr}& \ln p^2\: I(\alpha;p^2) 
\eeq
In the limit $p\rarr 0$, the logarithm dominates any constant and thus only the above term remains. For any $n\in\mathbb{N}$, the result yields
\bee 
J(\alpha,n;p^2)=(p^2)^{\alpha+\varphi}\sum_{k=0}^n\binom{n}{k}\frac{\partial^kI(\alpha)}{\partial\alpha^k}\ln^{n-k}p^2
\eeq
Equivalently to the case $n=1$, one gets the infrared limit
\bee
\lim_{p\rarr 0}J(\alpha,n;p^2)=\ln^np^2\: I(\alpha;p^2)\: \quad ,\: n\in\mathbb{N}\: .
\eeq

Let us turn to $n<0$. From the prescription (\ref{gener}) one can derive the formula for $J$ by repeated partial integration. To prove the result thus obtained, it is easier to turn the integral form of Eq. (\ref{gener}) into a differential equation,
\bee
\label{DGLneg}
I(\alpha;p^2)=\frac{\partial^{|n|}}{\partial\alpha^{|n|}}J(\alpha,n;p^2)\quad ,\: n<0\: .
\eeq
For $n=-1$, we get
\bee
\label{minus1}
J(\alpha,-1;p^2)=\frac{(p^2)^{\alpha+\varphi}}{\ln p^2}\sum_{k=0}^\infty\frac{\partial^kI(\alpha)}{\partial\alpha^k}\frac{(-1)^k}{\ln^kp^2}+C(p^2)
\eeq
with a constant $C(p^2)$ that does not depend on $\alpha$.

{\it Proof}:
\bee
\frac{\partial}{\partial\alpha}J(\alpha,-1;p^2)&=&\frac{(p^2)^{\alpha+\varphi}}{\ln p^2}\sum_{k=0}^\infty\left(\frac{\partial^kI(\alpha)}{\partial\alpha^k}\frac{(-1)^k}{\ln^{k-1}p^2}+\frac{\partial^{k+1}I(\alpha)}{\partial\alpha^{k+1}}\frac{(-1)^{k}}{\ln^kp^2}
\right)\nn\\
&=&(p^2)^{\alpha+\varphi}\sum_{k=0}^\infty\left(\frac{\partial^kI(\alpha)}{\partial\alpha^k}\frac{(-1)^k}{\ln^{k}p^2}-\frac{\partial^{k+1}I(\alpha)}{\partial\alpha^{k+1}}\frac{(-1)^{k+1}}{\ln^{k+1}p^2}\right)
\nn\\
&=& (p^2)^{\alpha+\varphi} I(\alpha)=I(\alpha;p^2)\: .
\eeq
Note in Eq.\ (\ref{minus1}) that for $p\rarr 0$ only the term with $k=0$ is relevant. The arbitrary integration constants $C(p^2)$ can always be chosen such that they are subleading in the infrared. The result (\ref{minus1}) can now be generalized to any $n<0$. By induction, one can show that 
\bee
\label{minus}
J(\alpha,n;p^2)=\frac{(p^2)^{\alpha+\varphi}}{\ln^{|n|} p^2}\prod_{j=1}^{|n|}\sum_{k_j=0}^\infty\frac{\partial^{\sum_i^{|n|} k_i}I(\alpha)}{\partial\alpha^{\sum_i^{|n|} k_i}}\frac{(-1)^{\sum_i^{|n|} k_i}}{\ln^{\sum_i^{|n|} k_i}p^2}+\cO(\alpha^{|n-1|})\quad ,\: n<0
\eeq
satisfies the differential equation (\ref{DGLneg}). The infrared limit yields
\bee
\lim_{p\rarr 0}J(\alpha,n;p^2)=\frac{I(\alpha;p^2)}{\ln^{|n|}p^2}\; , \quad  n\in\mathbb{Z}^-\; .
\eeq
In summary, we have shown that in the infrared limit for any
$n\in\mathbb{Z}$ the logarithms in the integrals $J(\alpha,n;p^2)$ in
Eq.\ (\ref{Jdef}) can be removed from under the integral and placed in front of the integral, evaluated at the external momentum $p^2$, i.e.\
\bee
\label{bottomline}
\lim_{p\rarr 0}J(\alpha,n;p^2)=\ln^np^2\:I(\alpha;p^2)\; ,\quad  n\in\mathbb{Z} \; .
\eeq
Therefore, if an integral equation of the DSE type, as given in Eq.\
(\ref{logDSE}), is solved by plain power laws in the infrared, an
ansatz of the kind $(\ell^2)^\alpha \ln^n(\ell^2)$ with
$n\in\mathbb{Z}$ will also be a solution. To see that, note that the
logarithm will appear on both sides of Eq.\ (\ref{logDSE}) and the
determining equation $I(\alpha)=1$ for the infrared exponent $\alpha$
is left unchanged. In this sense, there is a degeneracy in the
infrared power law solutions that are insensitive to
logarithms. 

\begin{figure}
  \centering
  \hspace{-0.0cm}\includegraphics[scale=0.95]{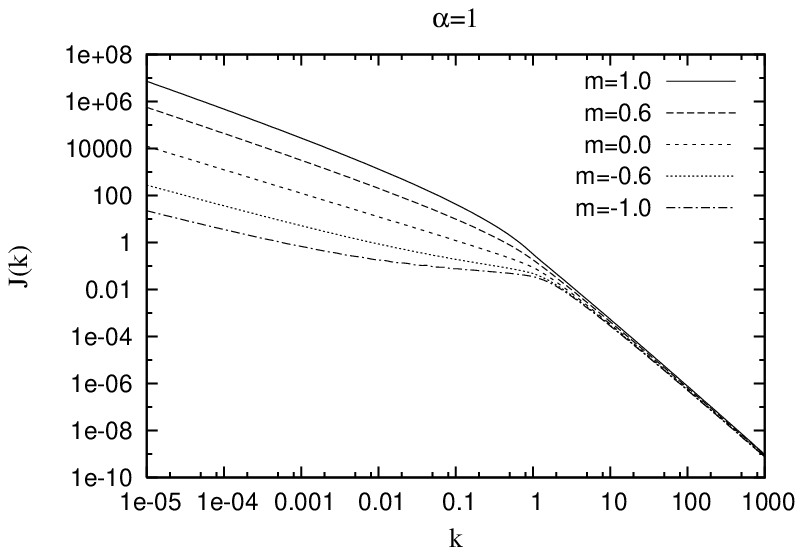}\includegraphics[scale=0.95]{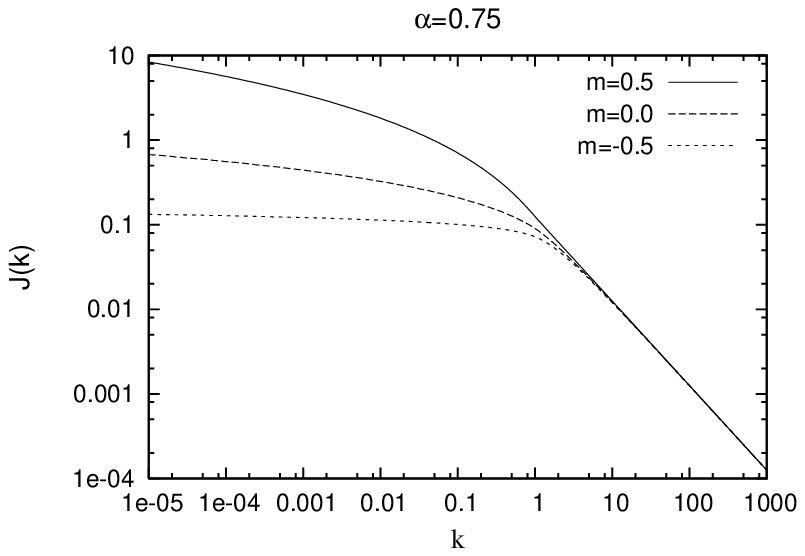}
  \caption{\emph{Left}: Logarithmic dressings of a power law  with
    $\alpha=1$ in the infrared. The curves confirm that in the infrared limit, the
    logarithm simply factors out, according to Eq.\
    (\ref{conj}). \emph{Right}: Here, $\alpha=0.75$. The upper curve with $m=0.5$ yields
    $\ln^2k^2$, the middle curve with $m=0$ yields $\ln k^2$ and the
    lower curve with $m=-0.5$ gives $\ln\ln k^2$ in the infrared.}
\label{logs}
\end{figure}

However, it is not clear how to rigorously generalize these results to
$n\in\mathbb{R}$. There are methods to define fractional derivatives
and fractional integration \cite{Frac} and a careful treatment might
recover the result (\ref{bottomline}) for arbitrary
$n\in\mathbb{R}$. This is something that would be worth
looking into. We conjecture here that in the infrared limit,
\begin{eqnarray}
  \label{conj}
  \lim_{k\to 0}\int \dbar^d q \frac{\ln^m
    q^2\;\ln^n(q-k)^2}{(q^2)^\alpha ((q-k)^2)^\beta}=\ln^{m+n}k^2\;\Xi_0(\alpha,\beta)
\end{eqnarray}
where $\Xi_0(\alpha,\beta)$ are the functions defined in (\ref{2ptdef}), basically
the l.h.s. of (\ref{conj}) without the logarithms. Hence, Eq.\
(\ref{conj}) states that in the infrared limit the logarithms factor out.
For the time being, it shall suffice to corroborate this
conjecture by a few modest numerical investigations. In Fig.\
\ref{logs}, we have implemented the integrals
\begin{eqnarray}
  \label{toyint}
  J(k):=\int \dbar^3 q\: \phi(q)\phi(q-k)\to \ln^{2m}k^2 \frac{I(\alpha) }{k^{4\alpha-3}}
\end{eqnarray}
with the simple functions
\begin{eqnarray}
  \label{toyfunc}
  \phi(k)=\left\{
  \begin{array}{cc}
    \frac{\ln^mk^2}{(k^2)^\alpha} & \quad\textrm{for}\;\; k<1\\
    \frac{1}{k^3} & \quad \textrm{for}\;\; k>1
  \end{array}
  \right.
\end{eqnarray}
for several values of $\alpha$ and $m$. In the left panel of Fig.\
\ref{logs}, the infrared limit indicated in Eq.\ (\ref{toyint}) with
$I(\alpha)=\Gamma^2(3/2-\alpha)\Gamma(2\alpha-3/2)/\left(8\pi^{3/2}\Gamma^2(\alpha)\Gamma(3-2\alpha)\right)$ following from Eq.\ (\ref{eq:2pt}) is
found exactly. This supports the hypothesis (\ref{conj}). 

In Ref.\ \cite{Kon02}, the possibility of logarithmic corrections to
the power law solutions was discussed using the angular
approximation. It is easily verified that the angular approximation is
unable to reproduce reliable results for the class of two-point
integrals we are dealing with. Moreover, the hypothesis (\ref{conj})
needs a more refined treatment than the angular approximation can
provide. The denial of logarithmic corrections in Ref.\ \cite{Kon02}
is therefore likely to be due to the angular approximation.

A further
point is made in the right panel of Fig.\ \ref{logs}. If the power
law exponent $\alpha$ in the integrand of Eq.\ (\ref{toyint}) is such that the integral is
dimensionless, i.e.\ $\alpha=\frac{3}{4}$ with $m=0$, the results will be a
logarithm, rather than $k^0=const$. If furthermore $m\neq 0$, the
result is $\ln^{1+2m}k^2$. This can be seen in the two upper curves of
the plot. The lower curve with the
choice $m=-\frac{1}{2}$ renders the exponent of the logarithm
zero. The asymptotic result found numerically is the function $\ln\ln k^2$.
With zero exponents of power laws incrementing the logarithm power,
and with zero logarithm power incrementing the $\ln\ln k^2$ power, it
is intuitive to expect this logic to continue to $\ln\ln\ln k^2$,
$\ln\ln\ln\ln k^2$ and so on. Let us have these general ideas be
followed up with a specific example. Consider an iteration of the DSEs in $d=3$ with $\kappa=\frac{1}{4}$
as a starting point, i.e.\ $D_G^{(0)}(k)\sim 1/(k^2)^{5/4}$. The ghost loop then produces, after UV
subtraction, a logarithm in the IR, cf.\ Eq.\
(\ref{logsol}). Consequently, the IR gluon propagator
yields $D_A^{(1)}(k)\sim 1/(\ln k^2)$. Plugged into the ghost DSE, the
log sum rule  (\ref{conj}) yields that $D_G^{(1)}(k)\sim (\ln k^2)/(k^2)^{5/4}$. Every
iteration increments the logarithm power of the ghost propagator, so
after the $n$th iteration, $D_G^{(n)}(k)\sim (\ln^n k^2)/(k^2)^{5/4}$. The
choice $\kappa=\frac{1}{4}$ must therefore be excluded, see Eq.\ (\ref{nologs}).

However, as long as such subtleties are avoided, the solutions for the
ghost and gluon propagators can be appended by
\begin{eqnarray}
  \label{logsols}
  D_A(k)&=&\frac{A}{(k^2)^{1+\alpha_A}\ln^\gamma\left(k^2/\mu^2\right)}\nn\\
  D_G(k)&=&\frac{B}{(k^2)^{1+\alpha_G}\ln^\delta\left(k^2/\mu^2\right)}
\end{eqnarray}
with the conditions (\ref{sumrule}) and (\ref{kappavond}) from the power law analysis and the log sum
rule
\begin{eqnarray}
  \label{logsumrule}
  \gamma+2\delta=0
\end{eqnarray}
from Eq.\ (\ref{conj}). Due to the factoring out of the logarithms in the
infrared limit, no changes are expected for the value of
$\kappa=\alpha_G$. Neither are any qualitative changes expected if logarithms
are present in the infrared, for they are always subdominant to power laws.

To summarize this section, the power solutions found in sections
\ref{rechnung} are not necessarily unique. First of all, a free
parameter can be dialled in the off-shell formulation to yield different
values of the infrared exponents. Only the one solution,
$\kappa_a=\frac{1}{2}$ for $d=3$ (see Fig.\ \ref{kvond}), is insensitive to this parameter. Secondly,
the amendment of logarithms to the power law solutions give one
possibility to generalize the solutions. Logarithms will be of
particular interest in the discussion of the ultraviolet behavior of
the Green functions. That will be the topic of chapter~\ref{UVchap}.

\vfill

\chapter{Variational solution for the vacuum state}
\label{VarVac}

While in the previous chapter the stochastic vacuum state was
investigated to derive the infrared properties of the theory, this
chapter is devoted to determine the vacuum wave functional in the
temporal Coulomb gauge by
variational methods. A Gaussian type of ansatz wave functional is put forward to
calculate the energy and minimize it with respect to the parameters of
the ansatz. 
The solutions for the Green functions thus obtained turn
out to approach the power law solutions for $d=3$ of the previous
chapter in the infrared. In the ultraviolet, the propagators are those
of free particles, up to logarithmic corrections.

By virtue of the (time-independent) Yang--Mills
Schr\"odinger equation, additional information is available from the
spectral properties of the Hamiltonian, as opposed to the exclusive
consideration of Dyson--Schwinger equations. In this respect, the
Hamiltonian approach is superior to the Lagrangian approach,
utilizing the principle of minimal energy. In quantum field theory, 
variational methods have enjoyed considerable attention
\cite{Sch62a,Wangerooge87}. However, critical remarks were issued by
Feynman. In the 1987 Wangerooge conference,
Feynman claimed that variational methods were ``no damn good at all''
in quantum field theory \cite{Wangerooge87}, essentially for the following reason. The only wave functionals that can be used
reasonably, are Gaussians. Any corrections to the Gaussians can only
be calculated numerically, involving intolerable errors. The
inaccuracies in the ultraviolet sector spoil the reliability of
infrared quantities, especially in non-linear theories such as QCD
where the UV and IR modes mix. Therefore, if Gaussian wave
functionals are not good enough, there is no chance in finding
reasonable results with the variational method, according to
Feynman. The calculations presented in this chapter are, of course,
overshadowed by the latter remarks. Nevertheless, on a qualitative
level, the results turn out immune to the criticism. First of all,
a generalization of Gaussian wave functionals is feasible and will be
exhibited in the following section. Secondly, the Gribov--Zwanziger
scenario infers that 
the gauge-fixed configuration space itself governs the infrared,
regardless of the ultraviolet accuracy of the wave functional.
The crucial infrared
properties do not follow from the variational minimization of the
vacuum energy, but can be extracted directly from the path integration
inside the Gribov horizon, along with the horizon condition. We are
therefore confident that the infrared power laws do not suffer from any
other approximations than those discussed in the previous chapter. 

It is well-known from nuclear physics that the ground state energy is quite insensitive to errors in the wave function. The accuracy
of the wave function is tested by the calculation of transition
amplitudes, not by the vacuum state itself. In the same sense, one may expect
that the Yang--Mills vacuum energy density is not too far off using the
crude approximation of a Gaussian vacuum wave functional. Other
quantities, such as the vertices, may be more sensitive to changes in
the wave functional. These assertions will be discussed in the course
of this chapter.

\vfill
%\pagebreak

\section{Gaussian types of wave functionals}
\label{sectionpsi}
Gaussian types of Coulomb gauge wave functionals are put forward in
this section. They were first
proposed in Refs.\ \cite{ReiFeu04,FeuRei04},
\begin{eqnarray}
\label{Psi}
\Psi_\lambda [A] =\cN\cJ^{-\lambda}[A]\exp\left[-\frac{1}{2}\int
  d^3[xy]A_i^a(x)\omega(x,y)A_i^a(y)\right]\; ,
\end{eqnarray}
where  $\omega (x, y)$ is a variational kernel and $\cN$ normalizes
the wave functional, $\braket{\Psi}{\Psi}=1$. [Note that henceforth, we refrain from using bold-faced symbols for
$3$-vectors, unless necessary for clarity.] The factor of the
Faddeev--Popov determinant with the real exponent $\lambda$ allows for
a further enhancement (if $\lambda>0$) of the probability density near
the Gribov horizon. One specific choice, $\lambda=\frac{1}{2}$,
resembles the usage of a ``radial wave function'', familiar from the calculation of the hydrogen atom. In this case, the evaluation of
expectation values becomes straightforward because the generating
functional of full Green functions,
\bee
\label{ZCoulomb}
Z_\lambda [j] & = & \langle {\Psi} | \exp \left[ { \int d^3 x j^a_k ({ x}) A^a_k 
({ x})} \right] | {\Psi} \rangle
\nonumber\\
& = & {\cal N}^2 \int \mathcal{D} A \cJ^{1-2\lambda}[A]\exp \left[ - \int d^3[xy]A_k^a(x)\omega(x,y)A_k^a(y) +  \int
  d^3 x j^a_k ({ x}) A^a_k ({ x}) \right] \; ,\nn\\ 
\eeq
is actually a Gaussian path integral for $\lambda=\frac{1}{2}$. Before
we write down the result, let us try to understand the meaning of the
parameter $\lambda$.

A Dyson--Schwinger equation (DSE) for the gluon propagator can be written down for the path
integral in Eq.\ (\ref{ZCoulomb}) in order to relate $\omega$ and
$\lambda$ to the gluon propagator $D_{ij}$,
\begin{eqnarray}
  \label{Dandw}
  0&=&\cN^2\int \cD A\,  \frac{\delta}{\delta A_i^a(x)}\cJ^{1-2\lambda}
\nn\\&&\quad\exp \left[ - \int d^3 [x'y'] A^c_k(x')\omega(x',y')A_k^c(y') +  \int
d^3 x' j^c_k ({ x'}) A^c_k ({ x'}) \right]\nn\\
&=&\cN^2\int
\cD A\, \cJ^{1-2\lambda}\left((1-2\lambda)\frac{\delta\ln\cJ}{\delta
    A_i^a(x)}-2\int d^3z\:\omega(x,z)A_i^a(z)+t_{im}(x)j_m(x)\right)
\nn\\ &&
\quad\exp \left[ - \int d^3 [x'y'] A^c_k(x')\omega(x',y')A_k^c(y') +  \int
d^3 x' j^c_k ({ x'}) A^c_k ({ x'}) \right]\; .
\end{eqnarray}
Taking a derivative w.r.t.\ $j_j^b(y)$ and setting sources to zero
yields
\begin{eqnarray}
  \label{Dandw2}
  0=(1-2\lambda)
\lla
\frac{\delta\ln\cJ}{\delta A_i^a(x)}A_j^b(y)
\rra_\lambda
-2\int d^3z\:\omega(x,z)
\lla A_i^a(z)A_j^b(y)\rra_\lambda
+t_{ij}^{ab}(x,y)\; .
\end{eqnarray}
Setting $\lambda=\frac{1}{2}$ in the above equation, the gluon propagator 
is seen to be determined by the variational kernel $\omega(x,y)$,
\begin{eqnarray}
  \label{Dhalf}
 \lla
 A_i^a(x)A_j^b(y)\rra_{\frac{1}{2}}=\frac{1}{2}\delta^{ab}t_{ij}(x)\omega^{-1}(x,y)\; .
\end{eqnarray}
For $\lambda\neq\frac{1}{2}$, we rewrite the first term in Eq.\
(\ref{Dandw2}) by partial integration,
\begin{eqnarray}
  \label{Dandw3}
 \lla \frac{\delta\ln\cJ}{\delta A_i^a(x)}A_j^b(y)\rra_\lambda&\nn\\
 &&\hspace{-2.9cm}=\cN^2\int
\cD A\, \cJ^{1-2\lambda}\frac{\delta\ln\cJ}{\delta
  A_i^a(x)}\left(-\frac{1}{2}\int d^3z
  \omega^{-1}(y,z)\frac{\delta}{\delta A_j^b(z)}\right)
 \nn\\
&&\quad\exp \left[ - \int d^3 [x'y'] A^c_k(x')\omega(x',y')A_k^c(y')
\right]\nn\\
&&\hspace{-2.9cm}=\int d^3z\left(
\lla 
\frac{\delta^2\ln\cJ}{\delta A_i^a(x) A_j^b(z)}
\rra_\lambda+(1-2\lambda)
\lla
\frac{\delta\ln\cJ}{\delta A_i^a(x)}
\frac{\delta\ln\cJ}{\delta
  A_j^b(z)}\rra_\lambda
\right)
\frac{1}{2}\:\omega^{-1}(z,y)
\; .
\end{eqnarray}
We now introduce an approximation for the evaluation of Green
functions, referred to as the \emph{loop expansion}. Whereas its common
definition concerns an expansion in $\hbar$, we here use the term
\emph{loop expansion} for ordering diagrams by the number of loops,
integrating nonperturbative ghost or gluon propagators. Applied to
Eq.\ (\ref{Dandw3}), it can be realized that the first term comprises one loop,
and the second term comprises two loops.\footnote{Each functional
  trace ``$\Tr$'', as found in $\ln\cJ=\Tr\ln G^{-1}$, gives rise to a
  loop.}
In the one-loop expansion, we may write Eq.\ (\ref{Dandw2}) as
\begin{eqnarray}
  \label{Dandw4}
  0&=&(1-2\lambda)
\int d^3z\lla 
\frac{\delta^2\ln\cJ}{\delta A_i^a(x) A_j^b(z)}
\rra_\lambda
\frac{1}{2}\omega^{-1}(z,y)\nn\\
&&-2\int d^3z\:\omega(x,z)
\lla A_k^a(z)A_j^b(y)\rra_\lambda
+t_{ij}^{ab}(x,y)\; .
\end{eqnarray}
We recognize in Eq.\ (\ref{Dandw4}) the ghost loop, here denoted by $\chi_{ij}$ with
\begin{eqnarray}
  \label{chidef2}
  \chi_{ij}^{ab}(x,y)=-\frac{1}{2}\lla 
\frac{\delta^2\ln\cJ}{\delta A_i^a(x) A_j^b(y)}
\rra_\lambda\; .
\end{eqnarray}
Since by the quantity $\chi_{ij}$ the non-trivial metric of the gauge-fixed
variables is expressed, the momentum space quantity
\begin{eqnarray}
  \label{chiscadef}
  \chi(k)=\frac{1}{(d-1)(N_c^2-1)}t_{ij}(x)\delta^{ab}\int d^3x\, \chi_{ij}^{ab}(x,y)\e^{-ik\cdot(x-y)}
\end{eqnarray}
is also referred to as the \emph{curvature} \cite{FeuRei04}, with the spatial dimension $d=3$.
Introducing for the gluon propagator the function $\Omega(x,y)$ by
\begin{eqnarray}
  \label{Omegadef}
  \lla A_i^a(x) A_j^b(y)\rra_\lambda =
  \frac{1}{2}t_{ij}(x)\Omega^{-1}(x,y)\; ,
\end{eqnarray}
and employing the one-loop approximation, Eq.\ (\ref{Dandw4}) can be found to yield
\begin{eqnarray}
  \label{Dandw5}
  \Omega(x,y)=\omega(x,y)+(1-2\lambda)\chi(x,y)\; .
\end{eqnarray}

Recall from the last chapter that the DSEs, following directly from
the boundary conditions on the path integration within the Gribov
horizon, fully determine the Green functions of the
theory. In the variational approach, on the other hand, the Green
functions have a parametric dependence on the kernel $\omega$ and on
the parameter $\lambda$. The DSE (\ref{Dandw5}) merely tells us how the gluon
propagator is related to the variational kernel $\omega$. The equation that actually determines $\omega$ is the
\Sch equation, $H\ket{\Psi}=E\ket{\Psi}$, that yields for the vacuum
state a variational equation of the Rayleigh-Ritz type,
\begin{eqnarray}
  \label{Scheq}
  \lla H \rra \to \textrm{min}\; .
\end{eqnarray}
The \Sch equation is not strictly a DSE, regarding its
origin. Nevertheless, we will often refer to the \Sch equation as a DSE
since, after all, it has the same form of a non-linear integral
equation.

One infers from Eq.\ (\ref{Dandw5}) that for the evaluation of the
gluon propagator to one-loop order, the Faddeev--Popov determinant in (\ref{ZCoulomb}) can be replaced by a
Gaussian,
\begin{eqnarray}
  \label{Jforchi}
  J=\exp\left[{-\int d^3[xy]A_i^a(x)\chi_{ij}^{ab}(x,y)A_j^b(y)}\right]\; ,
\end{eqnarray}
and it suffices to use the following generating functional
\begin{eqnarray}
  \label{Zexpl}
  Z_\lambda [j] &=& \int \cD A \exp \left[ - \int d^3 [xy] A_k^a(x)\Omega(x,y)A_k^a(y) +  \int
d^3 x j^a_k ({ x}) A^a_k ({ x}) \right]\nn\\
&=&\exp \left[\frac{1}{4} \int d^3 [xy]
  j_m^a(x)t_{mn}(x)\Omega(x,y)j_n^a(y) \right]\; .
\end{eqnarray}
From the relation
\begin{eqnarray}
  \label{Wick}
  \lla \cO[A] \rra_\lambda = \cO\left[\frac{\delta}{\delta j}\right] Z_\lambda[j]
\end{eqnarray}
one can verify that the gluon propagator yields Eq.\ (\ref{Dandw5}).

It was shown in Ref.\ \cite{ReiFeu04} that for the calculation of
expectation values in the \Sch equation to one-loop order, the
Faddeev--Popov determinant can be replaced according to Eq.\
(\ref{Jforchi}). Thus, the ``Gaussian type'' of wave functionals
(\ref{Psi}) are indeed Gaussian to the level of approximation. The
functional form of Wick's theorem \cite{ZinnJustin} thus becomes
applicable and results in the concise prescription (\ref{Wick}) to
evaluate expectation values. With $\omega$ replaced by $\Omega$ in
Eq.\ (\ref{Zexpl}), the \Sch equation minimizes the energy with
respect to the gluon propagator (\ref{Omegadef}) and the value of $\lambda$ is therefore
irrelevant \cite{ReiFeu04}. We will make the choice
$\lambda=\frac{1}{2}$ to solve the \Sch equation following Ref.\
\cite{FeuRei04}. Let us point out here that the restriction to the
Gribov region is abandoned in the evaluation of the Gaussian path integrals,
see Eq.\ (\ref{Zexpl}). It is therefore of particular interest to compare
the results to lattice calculations.

With a Gaussian wave functional at hand,
the gauge fields can be transformed into the particle representation
and interpreted as particles with energy modes $\omega(k)$. Since the
function $\omega(k)$ turns out be a non-trivial dispersion
relation, gluons are interpreted as quasi-particles. This will be
further discussed in chapter \ref{external}.

\section{Minimizing the energy density}
The vacuum energy density of Yang--Mills theory with vanishing external
charges, $\rhoext^a(x)=0$, was calculated in the state (\ref{Psi})
with $\lambda=\frac{1}{2}$ to two-loop order (in the energy) in Ref.\
  \cite{FeuRei04}. Subsequently, the Yang--Mills \Sch equation was used
  to determine the vacuum state by minimizing the energy density
  w.r.t.\ the kernel $\omega(k)$. This leads to the
  equation\footnote{In this section, all expectation values are taken
    in the vacuum state (\ref{Psi})
with $\lambda=\frac{1}{2}$.}
  \begin{eqnarray}
    \label{gapdef}
    \frac{\delta}{\delta\omega}\lla H\rra=0
  \end{eqnarray}
which is basically the only equation that follows from minimizing the
energy. Since it gives rise to a dispersion relation with a mass gap,
we refer to Eq.\ (\ref{gapdef}) as the \emph{gap equation}. The
gap equation has quite a complex structure, involving the ghost
propagator in particular, and we have to resort to auxiliary equations
in order to calculate the expectation values. One of
these equations is the ghost DSE derived in chapter \ref{ghostdom} and another
one accounts for the factorization of the Coulomb potential, see
below. The entire calculation of Ref.\ \cite{FeuRei04} shall not be
repeated here, but only the essential steps are outlined and some
subtleties are pointed out.

For the expectation value $E=\lla H[A,\Pi]\rra$ with the Yang--Mills
Hamiltonian $H[A,\Pi]$ given by Eq.\ (\ref{CLHam}), to be calculable by the
prescription (\ref{Wick}), the momentum operators $\Pi$ are eliminated by
action on the wave functional $\Psi[A]=\braket{\Psi}{A}$. We introduce by 
\begin{eqnarray}
  \label{PiOnPsi}
  \sqrt{\cJ}\Pi_k^a(x)\frac{1}{\sqrt{\cJ}}\left(\sqrt{\cJ}\Psi[A]\right)&=&\left(\frac{\delta}{i\delta A_k^a(x)}-\frac{1}{2i}\frac{\delta\ln\cJ}{\delta A_k^a(x)}\right)\cN\e^{-\frac{1}{2}\int A\omega A}\nn\\
&=&iQ_k^a(x)\left(\sqrt{\cJ}\Psi[A]\right)
\end{eqnarray}
the quantity
\begin{eqnarray}
  \label{Qdef}
  Q_k^a(x)=\int
  d^3y\,\omega(x,y)A_k^a(y)-\frac{1}{2}t_{kj}(x)\Tr\left(G\Gamma_k^{0,a}(x)\right)\; .
\end{eqnarray}
One can then show that for the given wave functional
$\Psi_\frac{1}{2}[A]$,
\begin{eqnarray}
  \label{Hvev}
  E=\lla H[A,\Pi]\rra = \lla H[A,Q[A]] \rra
\end{eqnarray}
and
% One may thus write
% \begin{eqnarray}
%   \label{Hvev}
%   E&=&\int \cD A \cJ \Psi[A]^* H[A,\Pi] \Psi[A]\nn\\
% &=&\int \cD A \left(\sqrt{\cJ}\Psi[A]\right)^*
% H[A,\sqrt{\cJ}\Pi\frac{1}{\sqrt{\cJ}}]\left(\sqrt{\cJ}\Psi[A]\right)\nn\\
% &=&\int \cD A \left(\sqrt{\cJ}\Psi[A]\right)^*
% H[A,Q[A]]\left(\sqrt{\cJ}\Psi[A]\right)\nn\\
% &=& \lla H[A,Q[A]] \rra
% \end{eqnarray}
Wick's theorem (\ref{Wick}) becomes applicable. The vacuum energy $E=E_k+E_p+E_C$ can be
computed \cite{FeuRei04} and it yields the kinetic energy 
\begin{eqnarray}
  \label{Ek}
  E_k=\frac{N^2_c - 1}{2}  (2\pi)^3\delta^3(0) \int \dbar^3 k \frac{ \left[ \omega ({ k}) - \chi
({ k}) \right]^2}{\omega ({ k})}\; ,
\end{eqnarray}
the magnetic potential
\begin{eqnarray}
  \label{Ep}
  E_p&=&\frac{N^2_c - 1}{2} (2\pi)^3\delta^3(0) \int \dbar^3 k
  \frac{{ k}^2}{\omega ({ k})}\; \nn\\
&&\quad+ \; \frac{N_c \left( N^2_c - 1
    \right) }{16} g^2  (2\pi)^3\delta^3(0) \int \dbar^3 k\: \dbar^3 k'
  \frac{1}{\omega ({ k}) \omega ({ k}')}  
\left( 3 - (\hat{{ k}} \cdot \hat{{ k}'})^2 \right)\; ,
\end{eqnarray}
and the Coulomb potential  
\begin{eqnarray}
  \label{EC}
  E_C &=& g^2\frac{N_c (N^2_c - 1)}{8}  (2\pi)^3\delta^{3}(0) \int
  \dbar^3 k\:  \dbar^3 k' 
\left( 1 + (\hat{{ k}} \cdot \hat{{ k}'})^2 \right) 
\frac{d ({ k} -{ k}')^2 f({ k} - { k}')}{({ k} - { k}')^2}\nonumber\\
&& \quad\frac{\left( \left[ \omega ({ k}) - \chi ({ k}) \right] - 
\left[ \omega ({ k}') - \chi ({ k}') \right] \right)^2}{\omega
({ k}) \omega ({ k}')}\; .
\end{eqnarray}
These terms arise from the kinetic, magnetic, and Coulomb parts of the
Hamiltonian, listed in that order in Eq.\ (\ref{CLHam}). The overall volume
factor of $(2\pi)^3\delta^3(0)$ is due to the translational invariance in the
absence of localized external charge distributions. In the above
expressions, two form factors were introduced. The ghost form factor
$d(k)$ measures the deviation of the ghost propagator $D_G(k)$ from
tree-level,\footnote{Note that this definition is different from Ref.\
\cite{FeuRei04}. Here, at tree-level $d(k)=1$.}
\begin{eqnarray}
  \label{ddef}
  D_G(k)=\frac{d(k)}{k^2}\; ,
\end{eqnarray}
and the Coulomb form factor $f(k)$ measures the deviation from the
factorization in the expectation value of the Coulomb operator $F$ (cf.\ Eq.\ (\ref{Ffact})),
\begin{eqnarray}
  \label{fdef}
  \lla F(k)\rra = \frac{d^2(k)f(k)}{k^2}\; .
\end{eqnarray}
From the definition (\ref{chiscadef}), the curvature $\chi(k)$ is found to define
the following integral,
\begin{eqnarray}
  \label{chiDSE}
  \chi(k)=g^2\frac{N_c}{4}\int \dbar^3\ell
\left(
1-(\hat{\ell}\cdot\hat{k})^2\right)
\frac{d(\ell)d(\ell-k)}{(\ell-k)^2}\; .
\end{eqnarray}
The gap equation can be derived from Eq.\ (\ref{gapdef}) using
\begin{eqnarray}
  \label{gapvariation}
  \frac{\delta E}{\delta \omega (k)}  = \frac{N^2_C - 1}{2} {\delta}^3
  (0) \frac{1}{\omega^2 (k)} \left[ - k^2 + \omega^2 (k) - \chi^2 (k) - I^0_\omega -
I_\omega (k) \right]\; ,
\end{eqnarray}
 to yield in
momentum space \cite{FeuRei04}
\begin{eqnarray}
  \label{gapeq}
  \omega^2(k)=k^2+\chi^2(k)+I_\omega(k)+I_\omega^0
\end{eqnarray}
with the abbreviations
\begin{subequations}
\begin{eqnarray}
  \label{Iw0def}
  I^0_\omega & = & g^2\frac{N_C}{4} \int\dbar^3 \ell
\lk 3 - (\hat{{ k}}\cdot \hat{{ \ell}})^2 \rk \frac{1}{\omega ({
\ell})} \\
\label{Iwdef}
I_\omega ({ k}) & = & g^2\frac{N_C}{4} \int \dbar^3 \ell \lk 1 +
(\hat{{ k}} \cdot \hat{{ \ell}})^2
\rk 
\frac{d ({ k} -{ \ell})^2 f ({ k} - { \ell})}{({ k} - { \ell})^2} 
\nn\\&&\qquad\qquad\frac{\left[ \omega ({ \ell}) - \chi ({ \ell}) + \chi
({ k}) \right]^2 - \omega ({ k})^2}{\omega ({ \ell})} 
\end{eqnarray}
\end{subequations}
In addition, the ghost from factor $d(k)$ obeys the ghost DSE
(\ref{ghostDSE}),
\begin{eqnarray}
  \label{dDSE}
  \frac{1}{d(k)}=1-g^2\frac{N_c}{2}\int \dbar^3\ell
\left(
1-(\hat{\ell}\cdot\hat{k})^2\right)
\frac{d(\ell-k)}{(\ell-k)^2\omega(\ell)}
\end{eqnarray}
and the Coulomb form factor obeys another integral equation that will
be discussed further in section \ref{fsection},
\begin{eqnarray}
  \label{fDSE}
  f(k)=1+g^2\frac{N_c}{2}\int \dbar^3\ell
\left(
1-(\hat{\ell}\cdot\hat{k})^2\right)
\frac{d^2(\ell-k)f(\ell-k)}{(\ell-k)^2\omega(\ell)}
\end{eqnarray}
The equations (\ref{gapeq}), (\ref{dDSE}) and (\ref{fDSE}) are here
collectively called the Dyson--Schwinger equations (DSEs) of our
approach to Coulomb gauge Yang--Mills theory. The approximations used
in deriving the DSEs include the
tree-level approximation for the ghost-gluon vertex and the one-loop
expansion in the gap equation.

\section{Renormalization}

Each one of the Dyson--Schwinger equations, the one for the ghost form factor $d(k)$, the 
Coulomb form factor DSE for $f(k)$, the gap equation for $\omega(k)$
and also the curvature integral $\chi(k)$ require subtraction of the
UV divergences. This can be seen by plugging in the ultraviolet
behavior of the form factors known from perturbation theory. As
discussed in more detail in chapter \ref{UVchap}, the ultraviolet asymptotic
propagators should attain tree-level up to logarithmic
corrections. The tree-level values for the ghost form factor as well as the Coulomb form factor
are simply unity. Whereas the latter are genuinely instantaneous in our approach, the
gluon propagator may be regarded as the equal-time part of the
tree-level propagator in Euclidean spacetime,
\begin{eqnarray}
  \label{Dequaltime}
  D_{ij}^0(\fk,\Delta t=0)=\int_{-\infty}^{\infty}\dbar k_0\frac{t_{ij}(\fk)}{k_0^2+\fk^2}=\frac{t_{ij}(\fk)}{2|\fk|}
\end{eqnarray}
and a first guess\footnote{Recall that $D_{ij}(k)=\frac{1}{2}t_{ij}(k)\omega^{-1}(k)$.} for $\omega(k)$ in the UV therefore is
$\omega(k\to\infty)=k$, yielding the dispersion relation of a free
massless particle. By simple power counting in the ultraviolet, the integrals
in (\ref{dDSE}) and (\ref{fDSE}) are found to be logarithmically
divergent. The gap equation (\ref{gapeq}) and the curvature integral
(\ref{chiDSE}), on the other hand, contain power divergences. 

Renormalization, the concept of redefining the parameters of a theory
such that expectation values are free of infinities, can be introduced
multiplicatively in perturbation theory \cite{Collins}. That is, the local field
operators are multiplied by renormalization constants such as to remove
the divergences occurring in the perturbative
expansion. In the Lagrangian language, multiplicative renormalization
may be described by appropriate counter terms in the Lagrangian that
respect its symmetries. The arbitrariness in defining the finite part of a
divergent quantity can be controlled by the renormalization group. 
Nonperturbatively, a systematic concept of removing
divergences is lacking. This also applies to the present
nonperturbative approach to Coulomb gauge YM theory. In Ref.\ \cite{Epp+07},
counter terms to the Hamiltonian were introduced which are seen---a
posteriori---to remove the divergences in the DSEs. Below, we will
follow the renormalization procedure given in Ref.\
\cite{Epp+07}. Thus, some additional parameters are introduced into
the theory. It will be seen that, at least for the asymptotic IR and
UV behavior, the solutions are independent of these parameters.

\subsection*{Renormalization of the Faddeev--Popov determinant}
The curvature integral (\ref{chiDSE}) is linearly divergent. From the
identity $\cJ=\exp(-\int A\chi A)$, see Eq.\ (\ref{Jforchi}), valid to one-loop order, it is
evident that the divergences in $\chi(k)$ can be traced back to the
Faddeev--Popov determinant. Eq.\ (\ref{Jforchi}) suggests that the counter terms
required to renormalize the Faddeev--Popov determinant, or more precisely its
logarithm, have to be of the form $\sim \int A A \hk $.
Since the log of the Faddeev--Popov determinant can be regarded as
part of the ``action'', we renormalize the Faddeev--Popov determinant as
\begin{eqnarray}
\label{12}
J \to J \cdot \Delta J=\exp \left[ \Tr \ln G^{-1} + C_\chi (\Lambda)
  \int d^3x\,  A_k^a (x)A_k^a(x) \right]
\end{eqnarray}
or by using the representation Eq.\ (\ref{Jforchi}), we obtain
\be
\label{13}
J \cdot \Delta J = \exp \left[ - \int A \lk \chi - C_\chi (\Lambda) \rk A \right] 
\hk.
\ee
Obviously, the counter term $C_\chi (\Lambda)$ has to be chosen to
eliminate the ultraviolet divergent part of the curvature $\chi (k)$. 
Thus the renormalization condition reads in momentum space
\be
\label{14}
\chi (k) - C_\chi (\Lambda) = \hk \mbox{finite} 
\hk .
\ee
As usual, there is some freedom in choosing the finite constants of the right hand
side of Eq.\ (\ref{14}). In principle, we could just eliminate the ultraviolet
divergent part of $\chi(k)$ by appropriately choosing 
 the counter
term $C_\chi (\Lambda)$. 
However,  it is more convenient to choose $C_\chi
(\Lambda)$ to be the curvature at some renormalization scale $\mu$, 
resulting in
the renormalization condition 
\be
\label{**}
C_\chi (\Lambda) = \chi (\mu)
\ee
 and in the finite
renormalized curvature
\be
\label{15}
\bar{\chi} (k) = \chi (k) - \chi (\mu) \hk .
\ee
It is easy to check that this quantity is indeed ultraviolet finite and
obviously it satisfies the condition
\be
\label{16}
\bar{\chi} (k = \mu) = 0 \hk .
\ee
By adopting the renormalization condition Eq.\ (\ref{**}),
the renormalized quantity $\bar{\chi} (k)$ in Eq.\ (\ref{15}) depends on the so far
arbitrary scale $\mu$. By choosing the renormalization condition Eq.\ (\ref{16}) this
renormalization scale becomes a parameter of our ``model'', since it defines the
infrared content of the curvature $\chi (k)$ kept in the
renormalization process. For instance, choosing $\mu = 0$, the whole infrared divergent
part of the curvature is chopped off. The numerical calculations below
use a finite $\mu>0$.

In Ref.\ \cite{ReiEpp07}, a different renormalization
condition was chosen, keeping from the ultraviolet divergent quantity $\chi (\mu)$ the
finite part $\chi' (\mu)$. This amounts to setting
\be
\label{20}
C_\chi (\Lambda) = \chi (\mu) - \chi' (\mu)\; ,
\ee
which results in the renormalized curvature
\be
\label{19}
\chi (k) = \bar{\chi} (k) + \chi' (\mu) \; .
\ee
Only the divergent part of the curvature is subtracted, keeping
fully its finite part. Here, we have an extra parameter $\chi' (\mu)$ of the
theory and the renormalization scale $\mu$ is not related to the zero of the
renormalized quantity ${\chi} (k)$ in Eq.\ (\ref{19}). The latter
method of removing divergences may be looked upon as an alternative,
the former (\ref{15}) will be used below. Both methods can be
accounted for by the specific choice of the counter term $\Delta\cJ$
to the Faddeev--Popov determinant.

\subsection*{Counter terms to the Hamiltonian}

In order to eliminate the divergences in the gap equation, it turns
out the counter terms 
\begin{eqnarray}
\label{1}
\Delta H = C_0 (\Lambda) \int d^3x\,  A_k^a(x)A_k^a(x) + iC_1 (\Lambda) \int d^3x\,   A_k^a(x) \Pi_k^a(x) 
\end{eqnarray}
should be added to the Hamiltonian.
Here the coefficients $C_0(\Lambda)$ and $C_1(\Lambda)$ depend on the momentum cutoff
$\Lambda$ and have to be adjusted so that the 
UV singularities in the gap equation
disappear. Note that these coefficients multiply ultralocal operators which
are singular in quantum field theory. For the wave functional at
hand, the expectation value of the counter term
Eq.\ (\ref{1})  is given by
\be
\label{2}
\Delta E = C_0 (\Lambda) \frac{1}{2} t_{ii} \delta^{aa} \int d^3x\, 
\omega^{- 1} (x,x) -  C_1 (\Lambda) \int d^3x\,  \lla A_k^a(x) Q_k^a(x)
\rra\; .
\ee
With the identity
\begin{eqnarray}
\label{4}
\lla A_k^a(x) Q_k^a(x) \rra = \int d^3 x'
\omega^{- 1} (x, x') \lk \omega (x', x) - \chi (x', x) \rk \delta^{a a} 
\end{eqnarray}
shown in Ref.\ \cite{FeuRei04}, Eq.\ (\ref{2}) can be written in momentum
space as
\begin{eqnarray}
\label{5}
\Delta E = \lk N^2_C - 1 \rk (2\pi)^3\delta^3(0) \left[ C_0 (\Lambda) \int
  \dbar^3k \frac{1}{\omega (k)} - C_1 \int \dbar^3 k \frac{\omega (k) - \chi
(k)}{\omega (k)} \right] 
\hk .
\end{eqnarray} 
Taking the variation of this expression with respect to $\omega (k)$, we obtain
\begin{eqnarray}
\label{6}
\frac{\delta \Delta E}{\delta \omega (k)}= \lk N^2_C - 1 \rk \delta^3(0) \left[ -C_0 (\Lambda) \frac{1}{\omega^2 (k)} - C_1 (\Lambda) \frac{\chi
(k)}{\omega^2 (k)} \right] \hk .
\end{eqnarray}
Note that $\chi (k)$ depends only on the ghost propagator but not on the gluon
energy $\omega$, at least as long as $\chi (k)$ is not yet the self-consistent
solution. Adding the counter terms from Eq.\ (\ref{5}) to the energy
in the gap equation (\ref{gapvariation}), one finds
\begin{eqnarray} 
\label{8}
\omega^2 (k) = k^2 +\chi^2 (k)+ I^0_\omega + I_\omega (k) - 2 C_0 (\Lambda) -
2 C_1 (\Lambda) \chi (k)\; .
\end{eqnarray}

\subsection*{Renormalization of the gap equation}
We now turn to the renormalization of the gap equation  (\ref{8}) 
which already includes the counter terms. After the renormalization of the
Faddeev--Popov determinant, the curvature $\chi (k)$ can be replaced by the
renormalized one $\bar{\chi} (k)$, Eq.\ (\ref{15}). Note that the Coulomb integral
$I_\omega (k)$ 
does not depend on any constant part of the curvature, see Eq.\
(\ref{Iwdef}), so that $\chi (k)$ could
have been replaced right away by the finite quantity $\bar{\chi} (k)$. Replacing
$\chi (k)$ by $\bar{\chi} (k)$ and using the relation
\be
\label{21}
I_\omega (k) = I^{(2)}_\omega (k) + 2 \bar{\chi} (k) I^{(1)}_\omega (k)\; ,
\ee
the gap equation in Eq.\ (\ref{8}) becomes 
\begin{eqnarray}
\label{22}
\omega^2 (k)- \bar{\chi}^2 (k) = k^2 + I^0_\omega + I^{(2)}_\omega (k) - 2 C_0
(\Lambda) + 2 \bar{\chi} (k) \lk I^{(1)}_\omega(k) - C_1 (\Lambda) \rk \; .
\end{eqnarray}
The integrals $I^{(n = 1, 2)}_\omega (k)$ are the linearly $(n = 1)$ and
quadratically $(n = 2)$ UV divergent parts of $I_\omega(k)$ in Eq.\
(\ref{Iwdef}), see also Ref.\ \cite{FeuRei04}.
The integral $I^0_\omega$ in Eq.\ (\ref{Iw0def}) is also quadratically
divergent but independent of the external momentum. We can therefore eliminate
all UV-divergences by choosing $C_0 (\Lambda) \sim \Lambda^2$ and $C_1
(\Lambda) \sim \Lambda$. In principle, the coefficients of the counter
terms $C_0 (\Lambda)$ and $C_1(\Lambda)$ have to be chosen to
eliminate the UV divergent parts of the quantities
appearing in the gap equation. This means that we should choose these infinite
coefficients as
\be
\label{26}
\lk I^0_\omega + I^{(2)}_\omega (k) \rk_{\textrm{UV divergent part}} - 2 C_0
(\Lambda) & = & 0 \\ 
\label{27} 
\left. I^{(1)}_\omega  (k) \right|_{\textrm{UV-divergent part}} -  C_1
 (\Lambda) & = & 0
\ee
Note that the UV divergent parts of $I^{(n = 1, 2)}_\omega (k)$ are by
dimensional arguments independent of the external momentum $k$. Technically it
is more convenient to choose the following alternative renormalization conditions.
 As usual we have the freedom in choosing the
renormalization conditions up to finite constants. Given the fact that
$I^0_\omega$ is independent of the external momentum and the differences 
\be
\label{23}
\Delta I^{(n)}_\omega (k, \nu) = I^{(n)}_\omega (k) - I^{(n)}_\omega (\nu)
\ee
are UV-finite, we can eliminate all UV-divergences by choosing the
renormalization conditions
\begin{subequations}
\label{renprs}
\be
\label{24}
I^0_\omega + I^{(2)}_{\omega} (k = \nu) - 2 C_0 (\Lambda) &  = & 0  \\
\label{25}
I^{(1)}_\omega (k = \nu) - C (\Lambda) & = & 0 \hk ,
\ee
\end{subequations}
where $\nu$ is an arbitrary renormalization scale, which could be chosen to be
the same scale $\mu$ of the renormalization of the curvature, but given the fact
that with the renormalization prescription (\ref{**}), the scale $\mu$ becomes a
physical parameter, the two renormalization parameters $\nu$ and $\mu$ need not
necessarily  be the same. With the renormalization conditions (\ref{renprs}) the renormalized
(finite!) gap equation reads
\be
\label{28}
\omega^2 (k) - \bar{\chi}^2 (k) = k^2 + \Delta I^{(2)}_\omega (k, \nu) + 2
\bar{\chi} (k) \Delta I^{(1)}_\omega (k , \nu) \hk .
\ee
Assuming that the integrals $I^{(n = 1, 2)}_\omega (k)$  are infrared finite 
and
furthermore that the renormalized curvature $\bar{\chi} (k)$
  is infrared divergent,
the infrared limit of the renormalized gap equation is given by
\be
\label{29}
\lim\limits_{k \to 0}
 \lk \omega (k) - \bar{\chi} (k) \rk = \Delta I^{(1)}_\omega (k = 0,\nu)\; .
\ee
The 't Hooft loop, which yields a perimeter law in the confinement
phase, can be explicitly calculated using its representation
found in Ref.\ \cite{Rei02}. In the Gaussian vacuum, it is sensitive to the
quantity (\ref{29}) and it is shown in Refs.\ \cite{ReiEppSch06,ReiEpp07} that only the
choice $\nu =0$ gives the perimeter law. With this choice
the renormalized gap equation becomes
\begin{eqnarray} 
\label{gapren}
\omega^2 (k) - \bar{\chi}^2 (k) = k^2 + \Delta I^{(2)}_\omega (k, 0) + 2
\bar{\chi} (k) \Delta I^{(1)}_\omega (k, 0)\; . 
\end{eqnarray}

\subsection*{Renormalization of the ghost and Coulomb form factor} 

The Dyson--Schwinger equation for the ghost propagator is given by Eq.\
(\ref{dDSE}) and bears a logarithmic divergence in the ultraviolet
part of the integral. By a simple subtraction at $\mu_d$, this
divergence is removed. Hence, 
\begin{equation}
\label{dren} 
\frac{1}{d(k)} = \frac{1}{d(\mu_d)} - \left[g^2\frac{N_c}{2}\int \dbar^3\ell
\left(
1-(\hat{\ell}\cdot\hat{k})^2\right)
\frac{d(\ell-k)}{(\ell-k)^2\omega(\ell)}- (k\leftrightarrow\mu_d)\right]
\end{equation} 
is a finite equation. In principle, one can choose $\mu_d$ different
from $\mu$. In view of the horizon condition, it is convenient to set
$\mu_d=0$ while keeping a finite $\mu$ (in particular for the sake of
the curvature). 

The DSE for the Coulomb form factor is given by Eq.\ (\ref{fDSE}) and
can be subtracted at $\mu_f$ to yield the finite expression
\begin{eqnarray} 
\label{fren} 
f(k)  = f(\mu_f)  +  \left[ \frac{g^2N_c}{2}\int \dbar^3\ell
\left(
1-(\hat{\ell}\cdot\hat{k})^2\right)
\frac{d^2(\ell-k)f(\ell-k)}{(\ell-k)^2\omega(\ell)}-
(k\leftrightarrow\mu_f)\right]\; .
\end{eqnarray}
We choose here $\mu_f=\mu$. 
The coupled DSEs then contain the following 
undetermined parameters: the renormalization scale $\mu$ of the
renormalization of the curvature and the renormalization constant $f(\mu)$ of the Coulomb
form factor. The parameter $d(\mu_d)$ may (or may not) be taken care of by
implementing the horizon condition.

\section{Full numerical solutions}
\label{numsol}

The renormalized and thus finite DSEs (\ref{gapren}, \ref{dren}, \ref{fren}) are cast in a form
tractable for a numerical study. Solving such a coupled set of
integral equations is pursued by iteration. An educated guess for
initial form factors is plugged into the integrals and the results are
recorded by the Chebychev approximation to be processed further. 
For asymptotic values of momentum, it is instructive for the numerics to make use of
the analytical results, cf.\ the methods used in Refs.\ \cite{AtkBlo98,diss_Bloch,FisAlk02,Maa05}. The set of nodes cover a finite momentum
region. In order to extrapolate this to the whole infinite momentum range, the general algebraic
forms obtained analytically serve as ans\"atze. However, contrary to what has been previously done in Dyson--Schwinger studies, the parameters of
these asymptotic forms are still determined numerically. E.g., in the infrared
it was shown analytically that power laws provide a solution. The numerics will have the asymptotic form of the power
laws (\ref{plawAnsatz}) as an input but
determine the various exponents and coefficients as parameters
by nonlinear least-squares fitting. Thus, this numerical method provides a check
on the analytical results of Table \ref{tab:sol} rather than imposing the
latter. 

Previously, a numerical solution of the Coulomb gauge DSEs was found
in Ref.\ \cite{FeuRei04}. The infrared behavior agreed approximately
with the exponent $\kappa=0.398$ of Table \ref{tab:sol}. A heavy quark
potential was obtained that confines but does not rise linearly, as
compared to the Wilson loop on the lattice. Therefore, a string
tension cannot be extracted from these results. After analytical
investigations of the infrared behavior of the Coulomb gauge Green
functions in Ref.\ \cite{SchLedRei06}, it became clear that yet another
solution, with $\kappa=\frac{1}{2}$ should exist. Eventually, improved numerical
methods were able to indeed confirm $\kappa=\frac{1}{2}$ and reveal a strictly
linearly rising potential in Ref.\ \cite{EppReiSch07}. These results
will be exhibited in this section.

The technical details of the numerical calculation are described in
Ref.\ \cite{EppReiSch07}. It shall suffice here to mention the essence
of the set-up. For convenience, the gauge coupling $g$ is set to unity. The functions\footnote{We drop the bar on $\chi(k)$,
  still meaning $\bar{\chi}(k)$ in Eq.\ (\ref{15}).} $\omega(k)$ and
$\chi(k)$ are both described in the far infrared by a power law
$A/k^\alpha$ where $A$ and $\alpha$ are extracted by fitting the solution.
It is possible to gain more information on the intermediate momentum
regime by setting up an infrared expansion of the function
\begin{align}
\label{defnu}
\omega(k)-\chi(k)=c+c_1k^\gamma\, ,\quad k\rightarrow 0\, , \gamma>0\, .
\end{align}
and again determining its parameters by fitting. While both
$\omega(k)$ and $\chi(k)$ are infrared enhanced, we have
$c<\infty$. Recalling that the 't Hooft loop requires for confinement $\nu=0$
in Eq.\ (\ref{29}), we realize that $c=0$.

In the ultraviolet, we use the asymptotic behavior of the form
factors found in Ref.\ \cite{FeuRei04} and make the corresponding
ans\"atze for $k\to\infty$,
\begin{subequations}
\label{UVnumerics}
\begin{eqnarray}
\label{UVw}
\omega(k)&=&k\\ 
\label{UVx}
\chi(k)&\sim& k/\sqrt{\ln(k/m_\chi)} \\
\label{UVd}
d(k)&\sim& 1/\sqrt{\ln(k/m_d)}\\
\label{UVf}
f(k)&\sim& 1/\sqrt{\ln(k/m_f)}
\end{eqnarray}
\end{subequations}
extracting the coefficients as well as the different scale parameters
$m_\chi$, $m_d$, $m_f$ using least-squares fitting. An
ultraviolet behavior different from Eq.\ (\ref{UVnumerics}) will be
proposed in chapter~\ref{UVchap}.

\begin{figure}
\begin{center}
\includegraphics{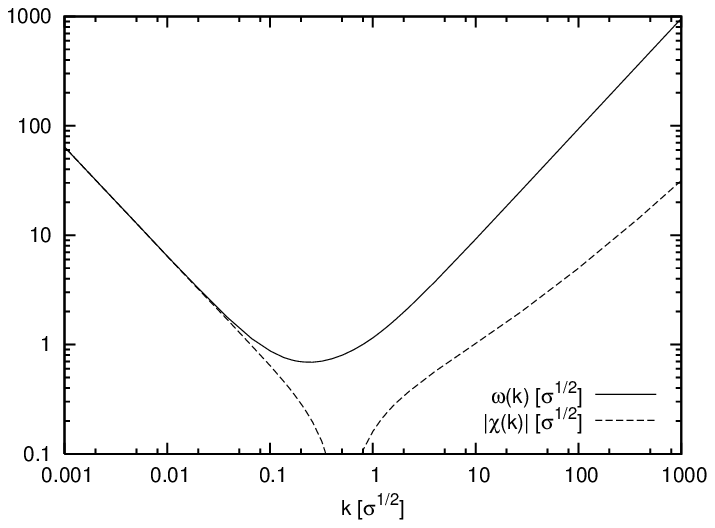}
\includegraphics{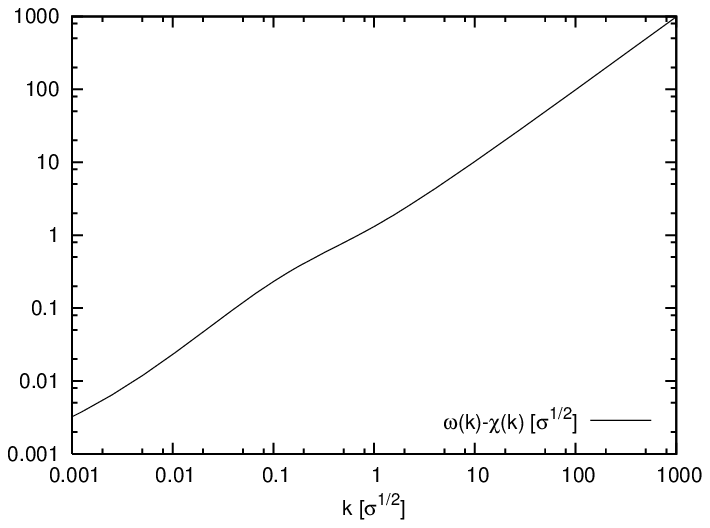}
\end{center}
\caption{Left: The gluon energy $\omega(k)$ and the modulus of
  the scalar curvature $\chi(k)$. Right: The difference
  $\omega(k)-\chi(k)$.
\label{fig-omega-chi}}
\end{figure}

\begin{figure}
\begin{center}
\includegraphics[scale=1.0]{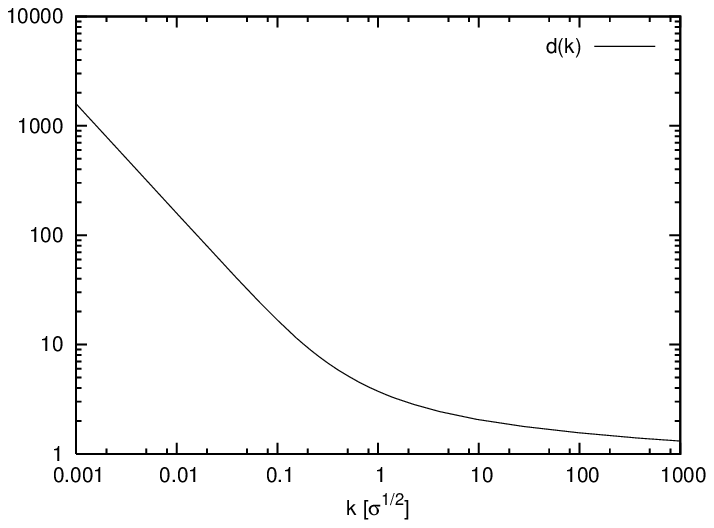}
\includegraphics[scale=1.0]{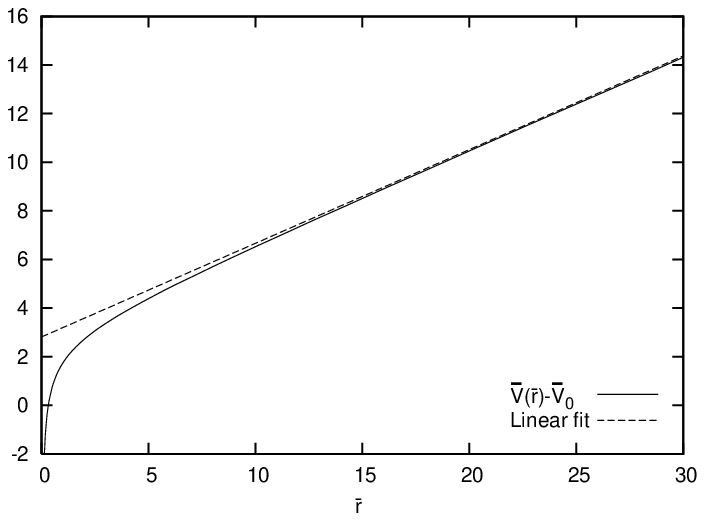}
\end{center}
\caption{Left: The ghost form factor $d(k)$. Right: The heavy quark Coulomb potential $V_C(r)$.}
\label{fig-renorm}
\end{figure}

All numerical plots displayed in this section were calculated by D.\
Epple \cite{EppReiSch07}. 
% Figures \ref{fig-omega-chi} and \ref{fig-renorm} show the
% numerical solution of the coupled Dyson--Schwinger equations for
% the choice $\bar{\omega} (1) = 1.6$. 
As seen from the left panel in
Fig.\ \ref{fig-omega-chi}, both $\omega(k)$ and $\chi(k)$ are enhanced
like $1/k$ in the infrared, as predicted in the stochastic vacuum. The
right panel of Fig.\ \ref{fig-omega-chi} 
shows the function $\omega(k)-\chi(k)$ vanishing for $k\to 0$, in accord with our choice of the renormalization condition $c=0$.
% Figure \ref{fig-renorm} shows the dependency of $\omega(k)$
% and $d(k)$ upon the renormalization constant $\omega(\mu)$. The
% asymptotic behaviour is apparantly insensitive to the value of
% $\omega(\mu)$. There are some quantitative changes in the intermediate
% momentum range, indicating that here the approximations are less trustworthy.
 Fig.\ \ref{fig-renorm} (left panel) shows the infrared enhanced ghost form factor
 $d(k)$. Thus, the functions $d(k)$, $\omega(k)$ and $\chi(k)$ are all
 enhanced as $1/k$ in the infrared.
On the right panel of  Figure \ref{fig-renorm}, the heavy quark Coulomb potential $V_C(r)$, as given by
Eq.\ (\ref{VCmeson}), is shown. An exactly
linearly rising behavior (within an estimated error of less than
one percent) is found, again in agreement with the infrared analysis
of chapter \ref{ghostdom}. The linearly rising potential allows us to fit our
scale from the string tension. 
Lattice calculations \cite{CucZwa03,GreOle03,LanMoy04}
show, however, that the Coulomb string tension $\sigma_C$ is about a factor of
$1.5\ldots 3$ larger than the string tension $\sigma_W$ extracted from the Wilson
loop, in agreement with the analytic result \cite{Zwa03b} that the Coulomb
string tension is an upper bound to the Wilson loop string
tension. Setting $\sigma_C=\sigma_W$ therefore yields a small
quantitative error. However, the \emph{qualitative} features of the results
shown in Figs.\ \ref{fig-omega-chi} and \ref{fig-renorm} are reliable on the one hand, and they
successfully describe the physical nonperturbative phenomenon of
confinement on the other. Quarks are confined by a linearly rising
potential $V_C(r)$, and gluons are confined since the dispersion
relation $\omega(k)$ diverges for $k\to 0$, prohibiting the
propagation of gluons over large distances.

The numerical confirmation of the infrared behavior predicted by the
analytical calculations in the stochastic vacuum, see chapter
\ref{ghostdom}, deserves the following two comments. Firstly, among the
full structure of the gap equation (\ref{gapren}), the infrared leading
contributions were obviously already accounted for by the stochastic
vacuum. Fig.\ \ref{fig-omega-chi} clearly shows that the curvature
$\chi(k)$, by which the gluon propagator in the stochastic vacuum  is expressed over the entire
momentum range, is the infrared asymptotic function of the (inverse)
gluon propagator $\omega(k)$. In retrospect, these findings may be
understood as a confirmation of the Gribov--Zwanziger
scenario. \label{page:asymp}The second comment is more a mathematical one. Given the
full structure of the DSEs, how can an asymptotic solution be
obtained analytically? We make power law ans\"atze for the infrared
behavior and, knowing that in the UV they are not valid, we use them
for integrating the propagators over the entire momentum range. This method was
first used in Landau gauge DSE studies in Ref.\ \cite{AtkBlo98a} and later
led to the now accepted infrared exponents by the calculations in
Refs.\ \cite{LerSme02,Zwa02}. From the experience with the numerical
evaluation of integrals, this procedure seems acceptable, but a
justification would be desirable. In Ref.\ \cite{Fis06} some arguments
are given why infrared asymptotics may be computed in the way
mentioned above. A sketch of a proof was given in the context of the
``infrared integral approximation'' in Ref.\ \cite{SchLedRei06}. It is also
worth mentioning the useful reference \cite{Erdelyi} in the context of
asymptotic expansions. Here, we give a short and rather obvious
explanation why the method is correct.
It follows from the Riemann-Lebesgue lemma that convolution integrals $C(k)$
of the type
\begin{align}
\label{convol}
C(k)&=\int dq\, D_1(q)D_2(q-k)=\int dq \int dx \int dy\, \tilde{D}_1(x) \tilde
{D}_2(y) \e^{-iqx-i(q-k)y}\nn\\
&=\int dx\, \tilde{D}_1(x) \tilde{D}_2(-x) \e^{-ikx}
\end{align}
asymptotically approach the function $C^{as}(k)$ obtained by
replacing in the integrand the infrared
asymptotic functions $D_i^{as}(k)=\lim_{k\to 0}D_i(k)$ and
$\tilde{D}_i^{as}(x)=\lim_{x\to\infty}\tilde{D}_i(x)$ ($i=1,2$),
\begin{align}
\label{convol2}
C^{as}(k)&=\lim_{k\to 0}C(k)=\int dx\, \tilde{D}_1^{as}(x)\tilde{D}_2^{as}(-x) \e^{-ikx}\nn\\
&=\int dq\, D_1^{as}(q)D_2^{as}(q-k)\; .
\end{align}
This argument can be applied to the (renormalized)
DSEs. Replacing the scalar products from the Lorentz structure by
identities such as $2\ell\cdot k=k^2+\ell^2-(\ell-k)^2$, all
(one-loop) DSEs turn
into a sum of convolution integrals of the type $C(k)$ given in Eq.\ (\ref{convol}).

\begin{figure}
  \centering
\includegraphics[scale=1.35]{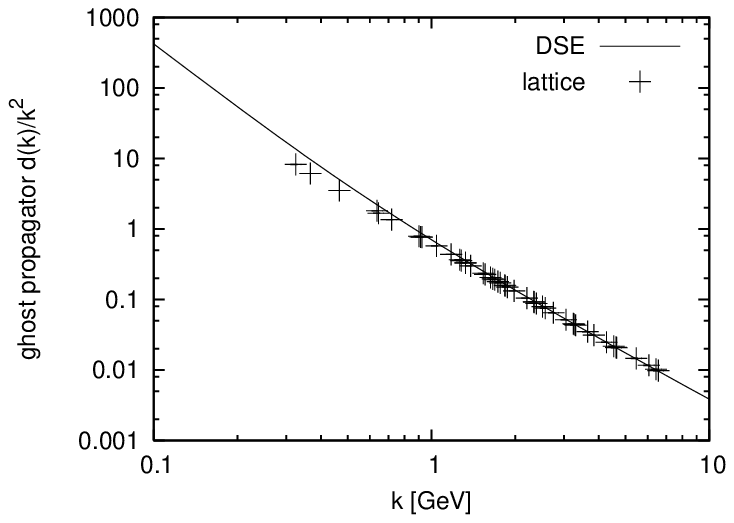} \includegraphics[scale=1.35]{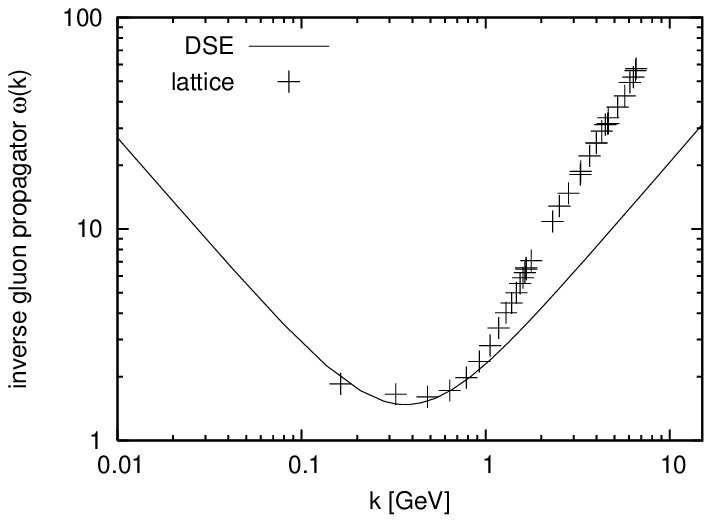}
  \caption{Ghost propagator \cite{privBur} and inverse gluon propagator
    \cite{Qua+07} from lattice
    calculations, in comparison to the continuum results in the
    Hamiltonian approach.}
  \label{lattice}
\end{figure}

Recently, there have been investigations of Coulomb gauge Green
functions on the lattice \cite{CucZwa02,LanMoy04,Qua+07}. In Fig.\ \ref{lattice}, the lattice data for
the ghost propagator and the inverse gluon propagator are shown in
comparison to the continuum solutions. The qualitative infrared
enhancement of the ghost propagator is reproduced by the lattice
calculations. At the same time, the gluon form factor $\omega(k)$
seems to be infrared enhanced as well. It is difficult to
extract reliable data for the infrared sector on the lattice. Yet, the
lattice data for the Coulomb gauge gluon propagator of Ref.\ \cite{CucZwa02} gives $\kappa\approx 0.49$, 
in agreement with the continuum result. In the
ultraviolet the striking feature of Fig.\ \ref{lattice} is that the
lattice data for the function $\omega(k)$ rise
stronger than linearly. This might actually be an indication that the
ultraviolet properties of the continuum solutions need to be
improved. We will come back to that in section \ref{propsinUV}.

Finally, let us mention that in $2+1$ dimensional Coulomb gauge YM
theory, numerical continuum results from the Coulomb gauge Hamiltonian approach were recently obtained
in Ref.\ \cite{FeuRei07}. Similarly to the $d=3$ case, it
is found that all form factors are infrared enhanced, with infrared
exponents that approximately agree with $\kappa=\frac{1}{5}$, the
value given in Table \ref{tab:sol} for $d=2$.

\section{Vertices}
\label{vertices}

The energy density that was minimized in the previous section to give
rise to the vacuum Green functions was shown to be rather insensitive to
the choice of the wave functional, according to the discussion in
section \ref{sectionpsi}. The three-gluon vertex we are turning to in this
section will show a strong dependence on the wave functional, in
particular on the exponent $\lambda$ of the Faddeev--Popov
determinant. Knowing the infrared behavior of the three-gluon vertex,
it will be possible to further investigate the ghost-gluon vertex in
the infrared.

\subsection*{Three-gluon vertex}
\label{sec:3gl}
As pointed out above, it is only the Faddeev--Popov
determinant that influences the infrared behavior of Yang--Mills theory. The
solution obtained for the infrared exponents of the propagators was found to
be independent of the three-gluon vertex, in particular, since it contributes to neither the ghost self-energy term nor the ghost loop contribution to the gluon self-energy. In the Coulomb
gauge, we have seen that any of the vacua  $\Psi_\lambda[A]$ given by Eq.\
(\ref{Psi}) minimizes the energy w.r.t.\ $\lambda$, evaluated to one-loop order in the DSE (two-loop order in the diagrams for the energy). The question is how the three-gluon vertex changes in the infrared for different values of $\lambda$ without resorting to the one-loop approximation.

The full three-gluon vertex is defined as 
\bee
{\left(\Gamma_{\textrm{full}}\right)}^{abc}_{ijk}(x,y,z)=\lla A^a_i(x)A^b_j(y)A^c_k(z) \rra \; .
\eeq
In the particular case of $\lambda=1/2$ it is found that
\bee
\label{vertexzero}
\int\cD A A_i A_j A_k \e^{-\int A\omega A}=0
\eeq
in a symmetric integration domain. Leaving aside a discussion whether
the Gribov region $\Omega$ is symmetric about $A=0$, we here extend
the integration domain to the whole gauge-fixed configuration space,
for practical purposes. Hence, the three-gluon vertex vanishes for
$\psi_{1/2}$ \cite{FeuRei04}. Now consider the case $\lambda \neq
1/2$. To one-loop order in the gap equation we may employ the form
(\ref{Zexpl}) for the generating functional $Z_\lambda[j]$. The three-gluon
vertex then vanishes for any $\lambda$ since the weighting of path
integration remains in a Gaussian form.

\begin{figure}[t]
\centering
\inc[bb = 131 608 477 713,clip=]{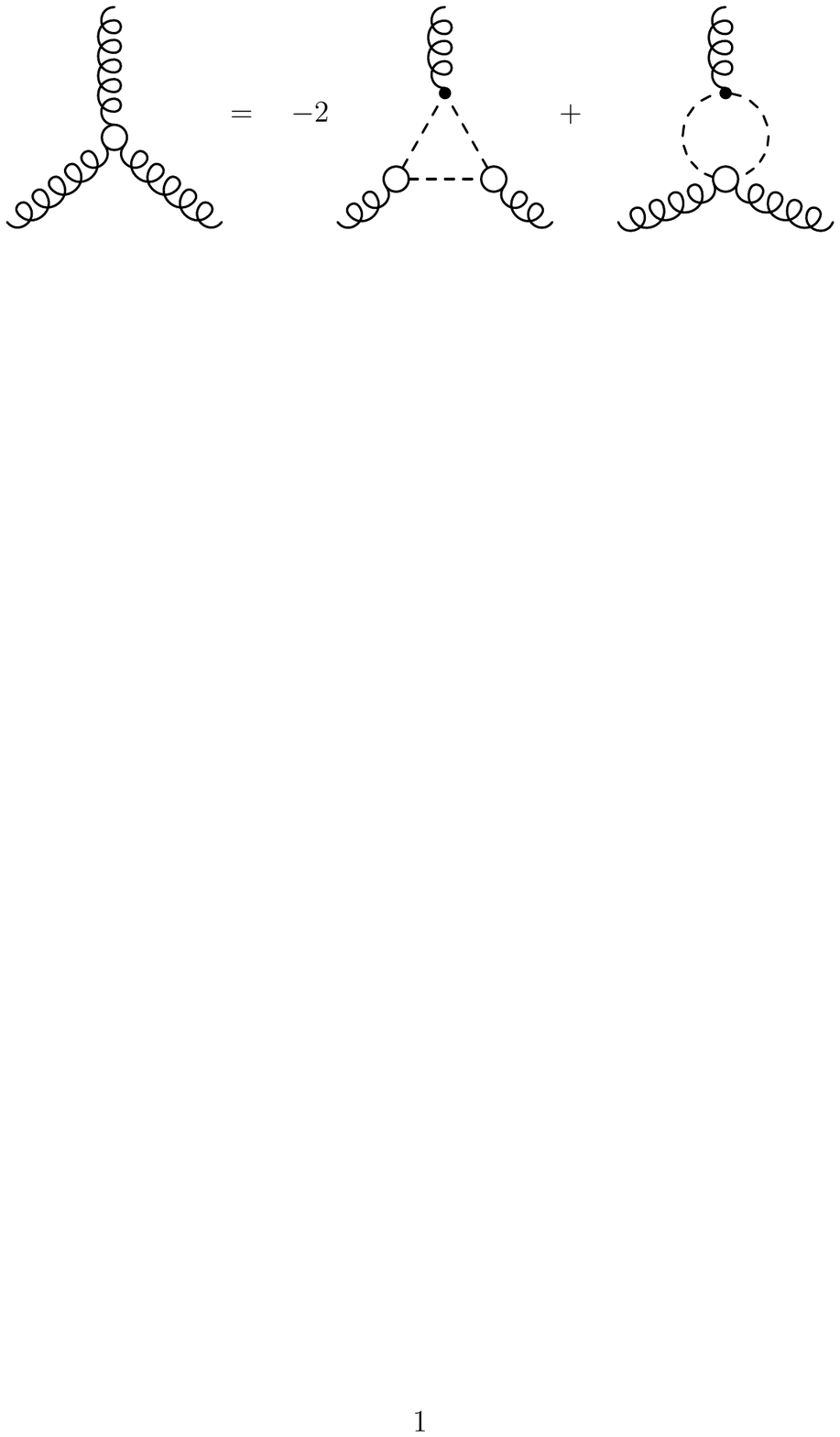}
\caption{The (complete) DSE for the three-gluon vertex derived from the
  generating functional given in Eq.\ (\ref{ZCoulomb}).}
\label{DSE diag}
\end{figure}

On the other hand, without the use of the one-loop approximation,
$\lambda\neq 1/2$ will give a non-zero three-gluon vertex, in contrast
to Eq.\ (\ref{vertexzero}), as will be shown. Thus, the three-gluon
vertex shows great sensitivity to the choice of the vacuum wave
functional $\psi_\lambda$, a behavior not exhibited to one-loop
order. Making the choice $\lambda=0$ permits the standard
representation of the Faddeev--Popov determinant by ghosts. One can
then derive \cite{SchLedRei06} the Dyson--Schwinger equation for the Coulomb gauge
three-gluon vertex. It is depicted diagrammatically in Fig.\ \ref{DSE diag}. Its complete form
comprises a diagram with the unknown two-ghost-two-gluon vertex which is
truncated here. For the following calculation, the ghost-gluon vertex
is used at tree-level. The DSE for the proper three-gluon vertex then reads
\cite{SchLedRei06}
\bee
\label{DSEmomentum}
\hspace{-0.6cm}\Gamma_{ijk}(k_1,k_2,k_3) = N_c \int \dbar^d\ell\; D_G(\ell) D_G(\ell+k_1) D_G(\ell-k_2)
 \Gamma^0_i(\ell) \Gamma^0_j(\ell-k_2)\Gamma_k^0(\ell+k_1)\; ,
\eeq
where the outgoing momenta $k_i$ obey the conservation law 
\begin{equation}
k_1 + k_2 + k_3 = 0 \; .
\end{equation}

\begin{figure}
\centering
\includegraphics[scale=0.39,clip=]{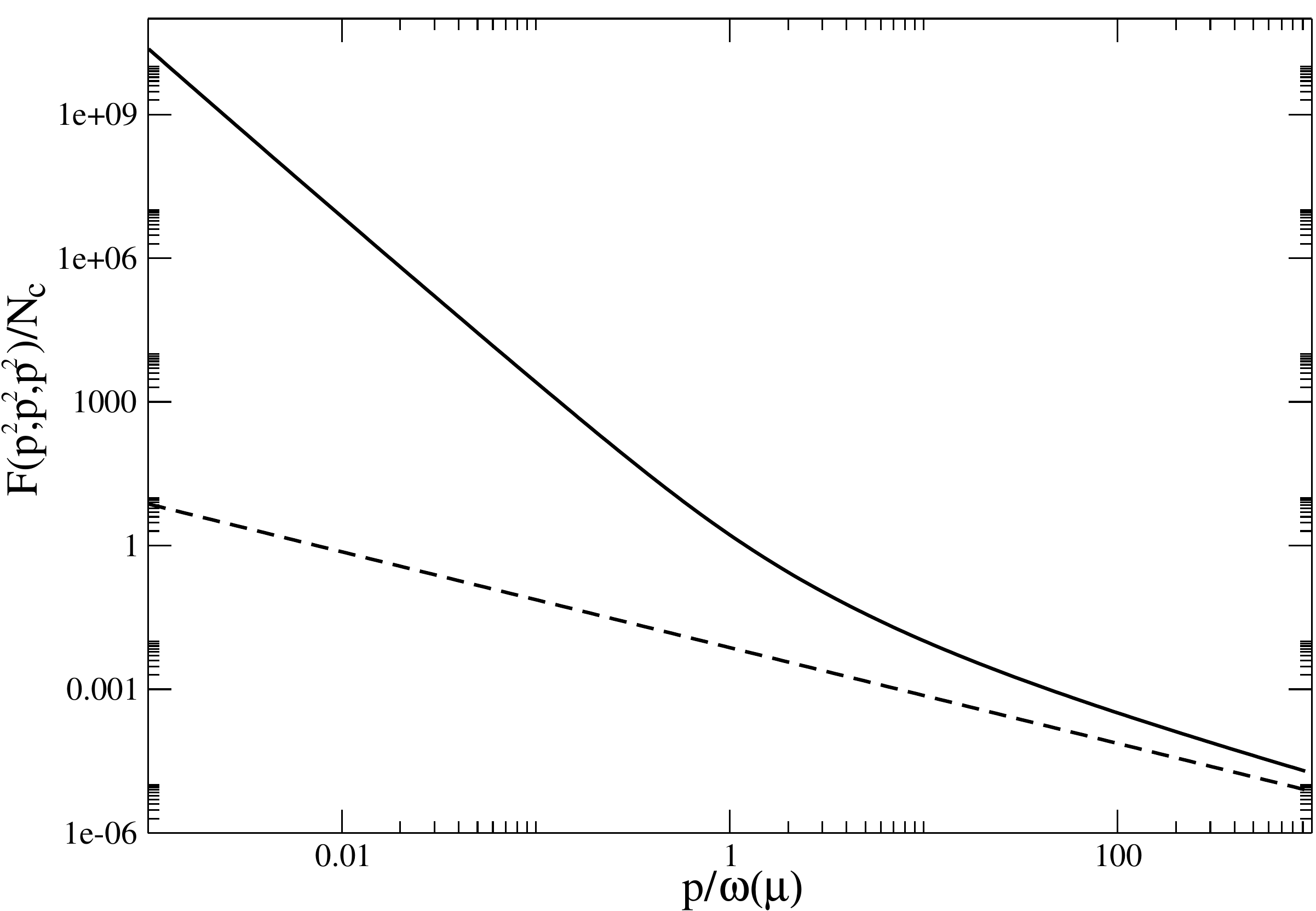}
\caption{Calculation by M.\ Leder, published in \cite{SchLedRei06}. Dressing function of the Coulomb gauge proper three-gluon vertex at the symmetric point. The dashed curve shows in contrast the perturbative case where the propagators in the loop are tree-level, i.e.\ $\kappa=0$.}
\label{fig:ZZZ}
\end{figure}

The vertex given by Eq.\ (\ref{DSEmomentum}) is projected onto the tensor subspace spanned by the tensor components of the tree-level vertex. Due to Bose symmetry, the coefficient functions of these six components are all the same, but their signs alternate as the vertex without the color structure is antisymmetric under gluon exchange. One finds

\begin{equation}\label{tensor}
\begin{aligned}
\Gamma_{ijk}(k_1,k_2,k_3) = & - i(k_2)_i \delta_{jk} F (k_2^2,k_1^2,k_3^2) && + i(k_3)_i \delta_{jk} \; 
F (k_3^2,k_1^2,k_2^2) \\
& + i(k_1)_j \delta_{ik} F (k_1^2,k_2^2,k_3^2) && - i(k_3)_j \delta_{ik} \; F (k_3^2,k_2^2,k_1^2) \\
& - i(k_1)_k \delta_{ij} F (k_1^2,k_3^2,k_2^2) && + i(k_2)_k \delta_{ij} \; F (k_2^2,k_3^2,k_1^2)\; . \\
\end{aligned}
\end{equation} 

Equating Eq.\ (\ref{DSEmomentum}) with (\ref{tensor}) and contracting with these six tensors, yields a set of six linear equations for $F$, the solution of which reads

\begin{multline}\label{formfactor}
F(k_1^2,k_2^2,k_3^2)= \frac{-N_c}{10(k_1^2 k_2^2 - (k_1\cdot k_2)^2)} \int
\dbar^d\ell \: D_G(\ell) D_G(k_3+\ell)D_G(\ell-k_2) \\
\left( (k_1^2 + k_1\cdot k_2)(-2J_2-J_4+3J_5) +(k_2^2 + k_1\cdot k_2)(2J_1-3J_3+J_6)\right)
\end{multline}
 
where

\begin{equation}\label{J}
\begin{aligned}
J_1:=& (k_1\cdot \ell)\ell^2 & J_2:=& (k_2\cdot \ell)\ell^2 & J_3:=& (k_2\cdot \ell)(k_1\cdot \ell)\\
J_4:=& k_2^2\ell^2 & J_5:=& (k_2\cdot \ell)^2 & J_6:=& (k_1\cdot k_2)\ell^2 \; .
\end{aligned}
\end{equation}

The integral (\ref{formfactor}) depends only on the ghost propagator in this
truncation, and despite the infrared enhancement of the latter it is
convergent. The numerical calculation of the form factor $F$ at the symmetric
point,  where $k_1^2=k_2^2=k_3^2=:k^2$, shows a strong infrared enhancement,
see Fig.\ \ref{fig:ZZZ}. For the ghost propagator we used the numerical
results of Ref.\  \cite{FeuRei04} where $\kappa=0.425$. A
fit to the data in Fig.\  \ref{fig:ZZZ} yields an infrared power law, $F(k^2)
\sim (k^2)^{-1.77}$. It is clear from the discussion on page \pageref{page:asymp} that this asymptotic power law can be obtained
analytically. Replacing the integrand in Eq.\
(\ref{formfactor}) by its infrared behavior and shifting $\ell
\rightarrow \lambda \ell$, one finds from the homogeneity that $F(k^2)
\sim (k^2)^{d/2-2-3\kappa}$. Plugging in the value $\kappa=0.425$ gives an
infrared exponent of $-1.775$ which agrees (within errors) with the numerical
result in Fig.\ \ref{fig:ZZZ}. At large momenta the vertex vanishes
which complies with asymptotic freedom since in the Gaussian vacuum there is no tree-level vertex.

The above results for the infrared behavior of the Coulomb gauge three-gluon
vertex can be generalized to any value of $\kappa$ and any
dimension $d$. Therefore, we can also make statements about the Landau
gauge where $d$ represents the dimension of Euclidean spacetime. Noting that $d/2-2-3\kappa=\alpha_A-\alpha_G$, see Eq.\
(\ref{sumrule}), we find

\begin{equation}\label{3gl_powerlaw}
F(k^2) \sim \frac{1}{(k^2)^{\alpha_G-\alpha_A}}\; .
\end{equation}

With the analytical results for $\kappa$ in Coulomb ($d=3$) as well as in
Landau ($d=4$) gauge, the exponents in Eq.\ (\ref{3gl_powerlaw}) have
the numerical values shown in Table \ref{Fresult}.
\begin{table}
  \centering
  \begin{tabular}{c|c|c}
    $d$ & $\kappa$ & $F(k^2)$ \\
    \hline
    $3$ & $0.398$  & $\frac{1}{(k^2)^{1.775}}$ \\
    $3$ & $\frac{1}{2}$ & $\frac{1}{(k^2)^{2}}$ \\
    $4$ & $0.595$ & $\frac{1}{(k^2)^{1.785}}$
  \end{tabular}
\caption{Infrared power laws of the function $F(k^2)$.}
\label{Fresult}
\end{table}
The Landau gauge result agrees exactly with Ref.\ \cite{AlkFisLla04}. 

Another interesting kinematic point is where one of the gluon
momenta, say $k_1$, is set to zero while the others remain finite. A
technical remark: we switch here to the off-shell gauge
condition. Thus, no ambiguities are found when trying to evaluate
transverse projectors at zero momentum. Another technical problem is
the following. Trying to calculate the vertex
with one momentum vanishing from Eq.\ (\ref{tensor}) by setting $k_1=0$, the projections onto the tensor components fail because the determinant of the coefficient matrix that defines the tensor expansion vanishes in this case.\footnote{In this context, one might note that the infrared limit of any tensor integral is non-trivial. Given an integral 
\bee
\label{tensint}
I_{\mu_1\mu_2\dots\mu_M}(\{k^{(i)}\})=\int \dbar^d\ell\:\ell_{\mu_1}\ell_{\mu_2}\dots\ell_{\mu_M}\:f(\ell,\{k^{(i)}\})
\nn
\eeq
we can construct a tensor basis from the external scales $\{k^{(i)}\}$
and Lorentz invariant tensors. According to the Passarino--Veltman
formalism \cite{PasVel79}, the above integral can then be expanded in this basis, which is nothing but solving a set of linear equations for the expansion coefficients in this basis. If one sets up a tensor expansion for finite $\{k^{(i)}\}$ and then tries to perform the infrared limit of a single external momentum, say $k^{(k)}\rarr 0$, the coefficient matrix becomes singular, and the tensor expansion is not well-defined. Instead, one can set $k^{(k)}=0$ from the beginning (if the integral exists here) and construct the tensor basis spanning a vector space which is of a lower dimension than beforehand. The expansion coefficients are then well-defined.} It is advisory to impose $k_1=0$ in the DSE (\ref{DSEmomentum}), 
\bee
\label{ZZZDSEk0full}
\Gamma_{ijk}(0,k,-k)=-ig^3N_c\int \dbar^d\ell\:\ell_i(\ell-k)_j\ell_k D_G^2(\ell)D_G(\ell-k)\: .
\eeq
One can then realize that this integral exists. It can be expanded into a
tensor basis constructed by the only scale $k$ and Lorentz invariant tensors,
i.e.\
$\{k_i\delta_{jk},k_j\delta_{ki},k_k\delta_{ij}\}$.
However, the only component that survives the transverse projections of the
gluon legs of $\Gamma_{ijk}$ with finite momenta, is obviously $k_i\delta_{jk}$. Thus, we can write
\bee
\label{ZZZDSEk0}
\Gamma_{ijk}(0,k,-k)=-ig^3N_ck_i\delta_{jk}\frac{1}{(d-1)k^2}k_m t_{np}(k)\int \dbar^d\ell\:\ell_m\ell_n\ell_k D_G^2(\ell)D_G(\ell-k)+\dots
\eeq
where the ellipsis represents irrelevant tensor components which shall
be discarded henceforth. In the asymptotic infrared, we use the power
laws (\ref{plawAnsatz}) for the propagators and obtain from Eq.\ (\ref{ZZZDSEk0})
\bee
\label{ZZZk0}
\Gamma_{ijk}(0,k,-k)=-ig^3B^3N_ck_i\delta_{jk}\frac{I_3}{(k^2)^{\alpha_G-\alpha_A}}\; , \quad k\rarr 0\: 
\eeq
with 
\begin{eqnarray}
  \label{I3def}
  I_3=&=&\frac{(k^2)^{\alpha_G-\alpha_Z}}{d-1}\left(\Xi_1(1+2\kappa,1+\kappa)-\Xi_3(2+2\kappa,1+\kappa)/k^2\right)\nn\\
&=&\frac{1}{2(4\pi)^{d/2}(d-1)}\frac{\Gamma(\frac{d}{2} - 2\kappa
  )\Gamma(\frac{d}{2} - \kappa )\Gamma(2 - \frac{d}{2} + 3\kappa
  )}{\Gamma(d - 3\kappa )\Gamma(1 + \kappa )\Gamma(2 + 2\kappa )}\; .
\end{eqnarray}

In view of the strong infrared divergence of the three-gluon vertex,
one has to check the ghost dominance in the propagator DSEs. The
infrared power law of the three-gluon vertex (\ref{3gl_powerlaw}),
expresses that the vertex dressing replaces the infrared exponent of a
gluon by that of a ghost propagator, for any dimension $d$. The
infrared hierarchy of terms in the gluon DSE remains untouched, since
even with the dressing of the three-gluon vertex, terms involving it
remain subleading to the ghost-loop in the infrared. E.g., the gluon
loop, which has an infrared exponent of $d/2-2-2\alpha_A$ with a
tree-level three-gluon vertex, attains an infrared power law with the
exponent $d/2-2-\alpha_A-\alpha_G$ if the vertex is dressed. Clearly,
this term is still subleading w.r.t.\ the ghost loop which bears an
infrared exponent of $d/2-2-2\alpha_G$. In the Landau gauge, the
infrared hierarchy was checked systematically by a skeleton expansion in Ref.\ \cite{AlkFisLla04}.

\subsection*{Ghost-gluon vertex}
The ghost-gluon vertex is investigated here with focus on the
(off-shell) Landau gauge. The reason is that in this gauge the corresponding DSE is
available from Ref.\ \cite{Sch+05}. In the Coulomb gauge, with the
choice $\lambda=0$ for the wave functional $\Psi_\lambda[A]$, the
derivation of the DSE should be straightforward, albeit tedious, and
technically equivalent to the three-gluon vertex.

Since neither the DSE studies \cite{Sch+05} nor the lattice
calculations \cite{CucMenMih04, Ste+05, Maa07} show any infrared divergences, the dressing function of the ghost-gluon vertex must be some finite function. To investigate the consequences of a finite dressing function of the ghost-gluon vertex, let us assume, for simplicity, that it is given by a finite constant,
\bee
\label{constGGA}
\Gamma_\mu(k;q,p)=C_0\Gamma^0_\mu(q)\; ,
\eeq
The constant $C_0$ can be determined by self-consistently solving the
DSE for the ghost-gluon vertex in the infrared gluon limit, see
section \ref{rechnung}, 
where $\Gamma^0_\mu(q)=igq_\mu$ is the tree-level ghost-gluon
vertex. Then, the infrared analysis of the propagators can be
performed in the same way as above. The ghost self-energy and the
ghost loop are both multiplied by the constant $C_0$. 
After evaluation of the integrals, this constant appears on both sides
of the equation $C_0I_G=C_0I_A$, cf.\ Eq.\ (\ref{kappavond}), and thus trivially cancels. Therefore, a constant dressing of the ghost-gluon vertex is completely irrelevant for the infrared behavior of the propagators.

The question that arises is if a non-constant dressing of the
ghost-gluon vertex might result in a change for the determining
equation (\ref{kappavond}) of $\kappa$. The investigations in Ref.\ 
\cite{Sch+05} showed that after one iteration step of the ghost-gluon
vertex DSE, the vertex remains approximately tree-level over the whole
momentum range, i.e.\ $C_0\approx 1$. Also, the results in Ref.\ 
\cite{Sch+05} confirmed that for
vanishing incoming ghost momentum $p$, the ghost-vertex becomes tree-level
in the Landau gauge\footnote{This agrees with the corresponding
  Slavnov--Taylor identity in the Landau gauge.}, see Eq.\ (\ref{GGAp0}). Recalling the discussion
in section \ref{secGGA}, this is expected to hold true in the Coulomb gauge. If we discard the irrelevant component of the vertex along the gluon momentum $k$, $\Gamma_\mu(k;q,p)$ also becomes tree-level for vanishing outgoing ghost momentum $q$ \cite{LerSme02,dipl_Schleifenbaum}, i.e.\   
\bee
\label{symm}
\lim_{p\rarr 0}\Gamma_\mu(k;q,p)=\lim_{q\rarr 0}\Gamma_\mu(k;q,p)=\Gamma^0_\mu(q)\; .
\eeq

However, the infrared limits of the ghost and gluon momenta are generally not interchangeable In particular, zero gluon momentum yields a dressing that is different from unity, although quite close to it, as we will see. The following relation shall define $C$,
\bee
\label{Cdef2}
C\:\Gamma^0_\mu(q)\equiv\lim_{p\rarr 0}\left(\lim_{k\rarr 0}\Gamma_\mu(k;q,p)\right)\neq\lim_{k\rarr 0}\left(\lim_{p\rarr 0}\Gamma_\mu(k;q,p)\right)=\Gamma^0_\mu(q) \: , \quad q\rarr 0\: .
\eeq

Does $C\neq 1$ or a non-constant $C$ affect the value for the infrared
exponent $\kappa$? To see this, it is not necessary to get involved in
a numerical calculation but we can argue qualitatively
instead. Consider any loop integral that involves the ghost-gluon
vertex.  Wherever it may appear in the loop diagram, the ghost-gluon
vertex is always attached to ghost propagators. The integrand will be
strongly enhanced for those loop momenta where the ghost propagator
diverges, i.e.\ for $p\rightarrow 0$. Since the gluon propagator, on
the other hand, is finite for all momenta, any infrared singularities
in the integrand of the loop integral can actually be due to the ghost
propagator only. The ghost-gluon vertex does not introduce any
additional singularities, since its dressing function is finite. For
the value of the integral, the singularities in the integrand will
give the dominant contribution.  The only value of the dressing
function of the ghost-gluon vertex that is relevant to the integral,
is then the one where any of the ghost momenta vanish. According to
Eq.\ (\ref{symm}), the vertex is tree-level in these limits. We can
therefore infer that in any loop integral the tree-level ghost-gluon vertex
will yield the correct result. This circumstance can thus be traced
back to the horizon condition and the transversality of the gluon
propagator. In Ref.\ \cite{LerSme02}, infrared divergent dressing functions
of the ghost-gluon vertex were employed and it was shown that $\kappa$
does not vary much.

\begin{figure}[t]
\centering
\ing[bb= 105 600 519 700, clip=]{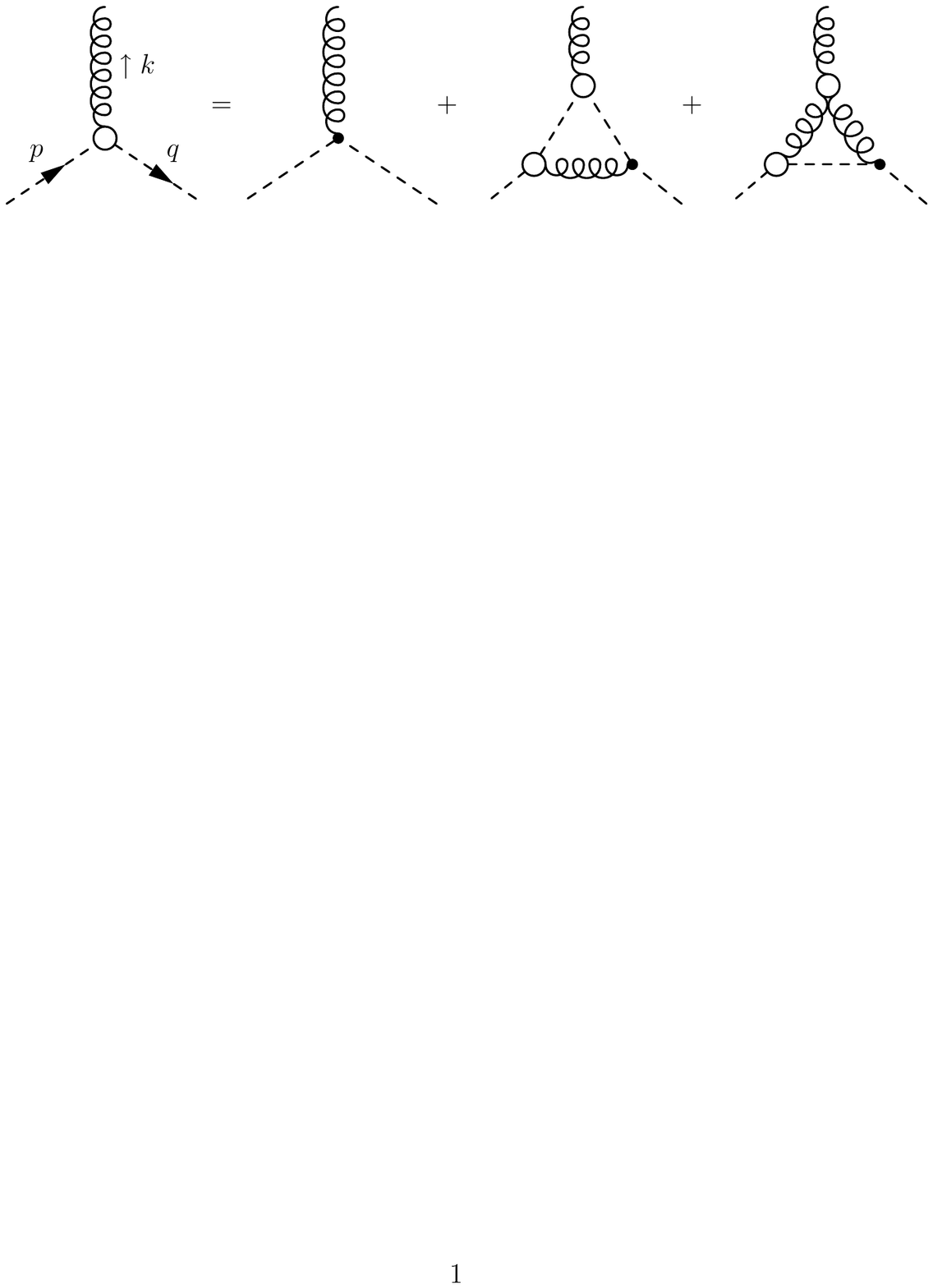}
\caption{The (truncated) Dyson--Schwinger equation for the ghost-gluon vertex.}
\label{fig:GGADSE}
\end{figure}

Nevertheless, the constant $C$, defined by Eq.\ (\ref{Cdef2}) is not
entirely meaningless since the introduction of a running coupling, see
chapter \ref{UVchap} below, makes use of it. One can actually analytically calculate $C$ by means of the DSE for the ghost-gluon vertex \cite{Sch+05}, see Fig.\ \ref{fig:GGADSE},

\bee
\label{GGADSEoff}
\Gamma_\mu(k;q,p)=\Gamma^0_\mu(q)+\Gamma^{(GGA)}_\mu(k;q,p)+\Gamma^{(GAA)}_\mu(k;q,p)\; .
\eeq

Here, $\Gamma^{(GGA)}_\mu$ is a graph with two full ghost and one full gluon propagator in the loop,
\bee
\label{GGA1loopGGA}
\Gamma^{(GGA)}_\mu(k;q,p)&=&-\frac{N_c}{2}\int \dbar^d\ell\: 
\Gamma_\alpha^0(-\ell)D_{\alpha\beta}(\ell-q)\Gamma_\beta(q-\ell;-p,-\ell-k)\nn\\&&\hspace{1cm}D_G(\ell+k)\Gamma_\mu(k;\ell,\ell+k)D_G(\ell)
q_\alpha t_{\alpha\beta}(\ell-q) (\ell+k)_\beta\ell_\mu
\: ,
\eeq
and $\Gamma^{(GAA)}_\mu$ has two gluon and one ghost propagator in the loop, but involves a proper reduced three-gluon vertex $\Gamma_{\mu\nu\rho}$,
\bee
\label{GGA1loopGAA}
\Gamma^{(GAA)}_\mu(k;q,p)&=&-\frac{N_c}{2}\int \dbar^d\ell\: 
\Gamma_\alpha^0(q)D_G(\ell-q)\Gamma_\beta(\ell+k;q-\ell,p)
\nn\\&&\hspace{1cm}D_{\beta\rho}(\ell+k)\Gamma_{\mu\nu\rho}(k;\ell,-\ell-k)D_{\nu\alpha}(\ell)
\: .
\eeq

Comparing Eq.\ (\ref{GGADSEoff}) and Eq.\ (\ref{GGADSE}), the four-point
function $\Gamma_{\mu\nu}(k,\ell;q,p)$ used in section \ref{secGGA}
can be expanded to yield the two graphs $\Gamma^{(GGA)}_\mu$ and
$\Gamma^{(GAA)}_\mu$, as well as a proper four-point function omitted
here. For further details, see Ref.\ \cite{Sch+05}.

Since $\Gamma_\mu(k;q,p)$ exists in the limit $k\rarr 0$ \cite{Sch+05,CucMenMih04,Ste+05}, we set $k=0$ in the integrands which greatly simplifies the tensor structure of Eqs.\ (\ref{GGA1loopGGA}) and (\ref{GGA1loopGAA}). Furthermore, the proper ghost-gluon vertices that appear in the loop integrals are rendered tree-level, as discussed above. We then get
\bee
\label{GGADSE1stk0}
\Gamma^{(GGA)}_\mu(0;q,q)&=&ig^3C\frac{N_c}{2}\int \dbar^d\ell\: 
\ell_\alpha D_{\alpha\beta}(\ell-q)q_\beta\ell_\mu D_G^2(\ell)
\: ,
\eeq
and 
\bee
\label{GGADSE2ndk0}
\Gamma^{(GAA)}_\mu(0;q,q)&=&g^2\frac{N_c}{2}\int \dbar^d\ell\: 
q_\alpha q_\beta D_{\alpha\nu}(\ell)D_{\beta\rho}(\ell)D_G(\ell-q)\Gamma_{\mu\nu\rho}(0;\ell,-\ell)
\: .
\eeq
Naively, we would expect from ghost dominance in the infrared that the
contribution (\ref{GGADSE2ndk0}) is subdominant since it incorporates
only one and not two ghost propagators, like
(\ref{GGADSE1stk0}). Using a tree-level three-gluon vertex, we can
calculate both integrals for $q\rarr 0$ in the infrared integral
approximation and indeed find that the graph (\ref{GGADSE2ndk0}) becomes
negligible. If the dressed three-gluon vertex (\ref{ZZZk0}) is included, the graph
(\ref{GGADSE2ndk0}) has a substantial contribution to this limit of
the ghost-gluon vertex. The calculation is then somewhat more
involved and deferred to appendix \ref{AppC}, but one can
extract from it the value of C solving the infrared limit of the DSE
(\ref{GGADSEoff}) self-consistently. For the various solutions of $\kappa$
found in chapter \ref{rechnung}, the numerical values of $C$ are shown
in Table \ref{Cresult}.\footnote{Note that the results (\ref{Cresult}) are independent of $N_c$. The color trace that occurs in the loop diagrams of Eq.\ (\ref{GGADSEoff}) yields a factor of $N_c/2$, see Eqs.\ (\ref{GGA1loopGGA}) and (\ref{GGA1loopGAA}), but it cancels with the propagator coefficient term $g^2AB^2=1/(I_GN_c)$.}

\begin{table}
  \centering
  \begin{tabular}{c|c|c}
    $d$ & $\kappa$ & $C$ \\
    \hline
     $3$ & $0.398$  & $1.089$ \\
    $3$ & $\frac{1}{2}$ & $1$ \\
    $4$ & $0.595$ & $1.108$ 
 \end{tabular}
  \caption{Solution for the infrared dressing $C$ of the ghost-gluon vertex.}
  \label{Cresult}
\end{table}

It is quite remarkable that in the Coulomb gauge with the solution $\kappa=\frac{1}{2}$, the two non-trivial graphs that appear in the DSE for the ghost-gluon vertex show an exact mutual cancellation in the infrared gluon limit,
\bee
\label{cancel}
\inc[scale=1, bb = 144 610 461 713, clip=]{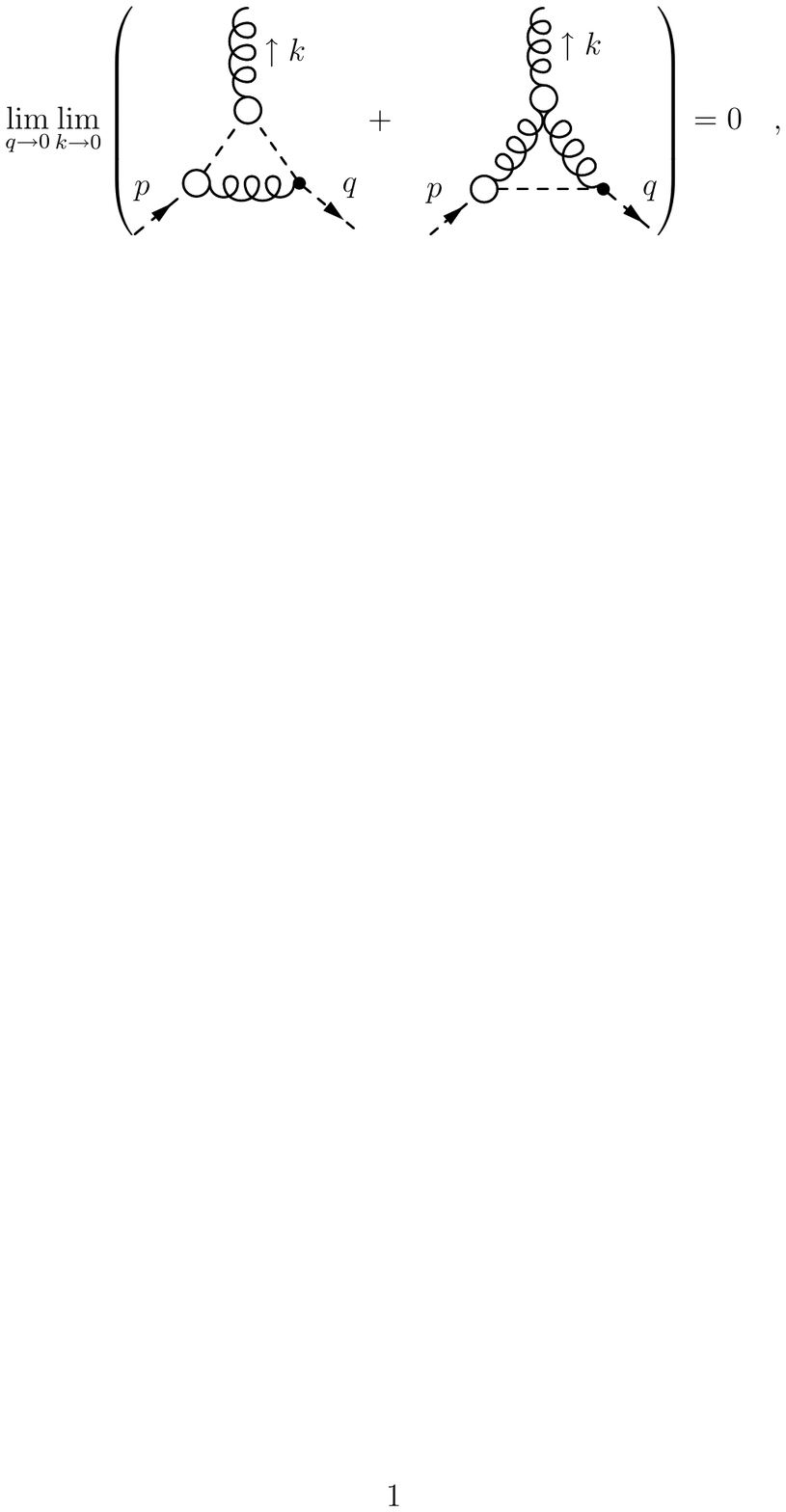}
\eeq
so that the corrections to tree-level vanish and $C=1$. Therefore, the
interchangeability of limits is recovered in this case only and the
ghost-gluon vertex becomes tree-level in all infrared limits.\footnote{An extension to
  ghost-antighost symmetric Landau gauge is straightforward and is
  seen not to alter the results for $C$.} For
other values of $\kappa$ and $d$, the ghost-gluon vertex is not
regular in the infrared. It must be granted that the omission of the four-point function in the
vertex DSE spoils the argument. A skeleton expansion of the
omitted object would be an interesting project. The original
derivation of the Landau gauge solution for $\kappa$ in Ref.\
\cite{LerSme02} relied on the regularity of the ghost-gluon vertex in
the infrared. Their result $\kappa=0.595$ for $d=4$ does not produce a
regular ghost-gluon vertex though, the infrared limits are not
interchangeable. As long as the omitted four-point function has no
effect, this would
contradict the working hypothesis of Ref.\ \cite{LerSme02}.

\section{Coulomb form factor}
\label{fsection}
The success of finding a linear heavy quark potential
$V_C(r\to\infty)\sim r$ in the Coulomb gauge
Hamiltonian approach relies on the approximation made for the
expectation value $\lla F \rra$, see section \ref{confinement}. The
Coulomb form factor $f(k)$ was defined by Eq.\ (\ref{fdef}),
\begin{eqnarray}
  \label{fredef}
  f(k)=\frac{\lla F \rra(k)}{k^2 D_G^2(k)}\; ,
\end{eqnarray}
and here we try and assess the factorization 
$\lla F \rra=k^2D^2_G(k)$, i.e.\ $f(k)=1$. To this end, we employ the operator
identity 
\begin{eqnarray}
  \label{MarSwi}
  F=\frac{\del}{\del g}(gG)
\end{eqnarray}
which follows easily from $g\del G^{-1}/\del g = G^{-1}-G_0^{-1}$. We
refer to the interesting relation (\ref{MarSwi}), first written down in Ref.\
\cite{MarSwi84}, as the {\it Marrero-Swift relation}.
Taking the expectation value of (\ref{MarSwi}) in the Gaussian vacuum
and neglecting contributions of $\frac{\del\omega}{\del g}$, 
\begin{eqnarray}
  \label{fbyd}
  f(k)=-\frac{g^2}{k^2}\frac{\del}{\del g}\frac{1}{gD_G(k)}\; .
\end{eqnarray}
Plugging the ghost DSE (\ref{ghostDSE}) into Eq.\ (\ref{fbyd}), an
integral equation for $f(k)$ is derived,
\begin{eqnarray}
  \label{fDSEderiv}
  f(k)&=&-\frac{g^2}{k^2}\frac{\del}{\del
    g}\left(\frac{k^2}{g}-N_ck^2\int \dbar^d\ell \:
    \left(1-(\hat{\ell}\cdot
      \hat{k})^2\right)D_A(\ell)(gD_G(\ell-k))\right)\\
\label{fDSEs}
  &=& 1+ g^2N_c\int \dbar^3\ell
\left(
1-(\hat{\ell}\cdot\hat{k})^2\right)
D_A(\ell)D_G^2(\ell-k)f(\ell-k)(\ell-k)^2
\end{eqnarray}
which agrees with the DSE (\ref{fDSE}) for $f(k)$ displayed
above. With a power law ansatz for $f(k)$ in the infrared,
\begin{eqnarray}
  \label{fplaw}
  f(k)=\frac{C_f}{(k^2)^{\alpha_f}}\; , \quad k\to 0\; , \quad
  \alpha_f>0\; ,
\end{eqnarray}
we now aim at a solution of the Coulomb form factor DSE (\ref{fDSEs}). For $k\rarr 0$, one finds
\be
\label{DSEf}
(k^2)^{-\alpha_f}=(k^2)^{-\alpha_f} g^2AB^2N_cI_f(\kappa,\alpha_f)
\ee
where
\be
I_f(\kappa,\alpha_f)=\frac{(d-1)\kappa}{(4\pi)^{d/2}}\frac{\Gamma\left(\alpha_f\right)\Gamma\left(2\kappa\right)\Gamma\left(d/2-2\kappa-2\alpha_f\right)}{\Gamma\left(d/2-2\kappa\right)\Gamma\left(d/2-\alpha_f+1\right)\Gamma\left(\alpha_f+2\kappa+1\right)} .
\ee
From Eq.\ (\ref{DSEf}) we can see that 
\be
\label{ABf}
g^2AB^2N_cI_f(\kappa,\alpha_f)=1
\ee
has to be fulfilled. The above relation from the Coulomb form factor
DSE can now be plugged into the coefficient rule (\ref{coeffrule}) from
the ghost DSE to find 
\be
\label{condition}
I_G(\kappa)=I_f(\kappa,\alpha_F) .
\ee

\begin{figure}
\centering
\includegraphics[scale=1.4]{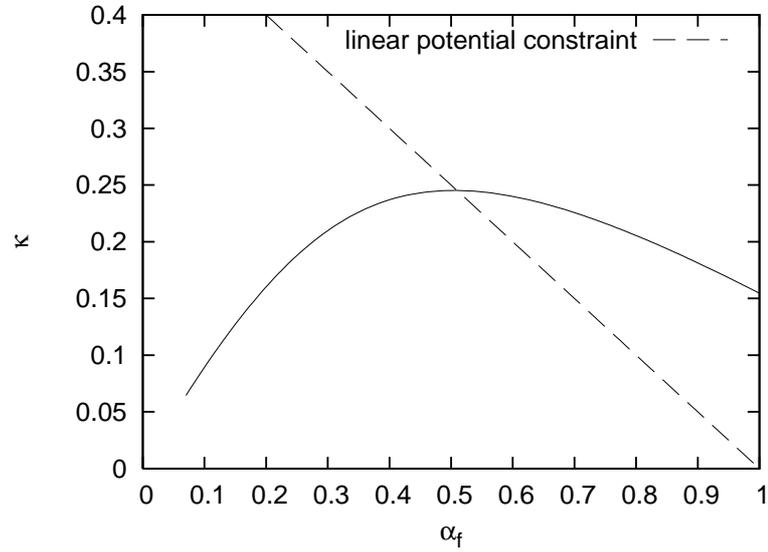} 
\caption{The ghost exponent $\kappa$ as a function of $\alpha_f$ is shown by a solid
line for $3+1$ dimensions. A dashed line indicates the values for which a linear potential will
arise.} 
\label{kvonf}
\end{figure}

We now specify the spatial dimension $d$. For $d=3$, Eq.\ (\ref{condition}) yields
\be
\label{solf}
1= \frac{\left( -1 + 2\,\kappa  \right) \,\cos (\pi \,\left( \alpha_f  + 2\,\kappa  \right) )\,\Gamma(4 - 2\,\alpha_f )\,\Gamma(-2\,\kappa )\,
    \Gamma(1 + 2\,\alpha_f  + 4\,\kappa )\,\sin (\pi \,\alpha_f )}{\pi \,\left( -1 + \alpha_f  \right) \,\left( 3 + 2\,\kappa  \right) \,
    \left( -1 + 2\,\alpha_f  + 4\,\kappa  \right) \,\Gamma(2 + 2\,\kappa )} .
\ee
The numerical solution to this equation gives $\kappa$ as a function of 
$\alpha_f$, as shown in Fig.\ \ref{kvonf}.
One can see immediately that for any value of $\alpha_f$, the ghost exponent $\kappa$ yields
\be
\label{restrict}
\kappa<\frac{1}{4}.
\ee
 Therefore, recalling the results for $\kappa$ in Table \ref{tab:sol}, there exists no value
 of $\alpha_f$ for which all three DSEs are satisfied. 

It is instructive to focus on the case where 
\be
\alpha_f+2\kappa=1
\ee
since this leads directly to a linearly rising potential for static quarks. Plugging this constraint into 
Eq.\ (\ref{solf}), we find
\be
1=\frac{-2\,\kappa \,\left( 3 + 2\,\kappa  \right) \,\cos (2\,\pi \,\kappa )\,\Gamma(-1 - 4\,\kappa )\,\Gamma(1 + 2\,\kappa )}  {\left( -1 + 2\,\kappa  \right) \,\Gamma(-1 - 2\,\kappa )}
\ee
which has the numerical solution $\kappa=0.245227$. Surprisingly, this result is exactly agreed upon 
by lattice calculations \cite{LanMoy04}, where $\kappa=0.245(5)$ was found.
However, one should notice that the lattice calculations carried out so far in
Coulomb gauge and also in Landau gauge use too small lattices to give reliable
results in the infrared.

\begin{figure}
\centering
\includegraphics[scale=1.4]{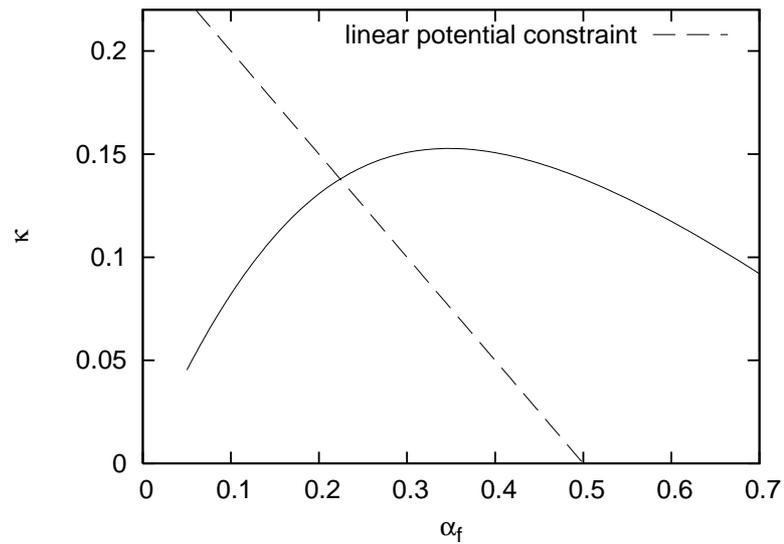}  
\caption{Same as Fig.\ \ref{kvonf}, here in $2+1$ dimensions.} 
\label{kvonf2}
\end{figure}

In $2+1$ dimensional Coulomb gauge, the calculations are equivalent. From a
simultaneous treatment of the ghost DSE and the gluon DSE we find from
the stochastic vacuum results in section \ref{rechnung}
a unique solution for the ghost exponent \cite{Zwa02},
\be
\label{2+1}
\kappa=\frac{1}{5} .
\ee
Note that with angular approximation the result is $\kappa=1/4$. In a
recent publication \cite{FeuRei07}, the value (\ref{2+1}) for $\kappa$ was confirmed numerically. On the other hand, if 
we consider only the ghost DSE and the equation for the
Coulomb form factor for $d=2$, one gets by 
enforcement of the condition Eq.\ (\ref{condition}):
\be
\label{fd2}
\frac{\Gamma(\alpha_f )\,\Gamma(1 - \alpha_f  - 2\,\kappa )\,\Gamma(\kappa )\,\Gamma(2 + \kappa )}
  {\Gamma(2 - \alpha_f )\,{\Gamma(-\kappa )}^2\,\Gamma(1 + \alpha_f  + 2\,\kappa )}=1\,  .
\ee
The numerical solution is shown in Fig.\ \ref{kvonf2}. If we require the potential to be linearly rising, the solution has to obey
$\alpha_f+2\kappa=\frac{1}{2}$.
Eq.\ (\ref{fd2}) then leads to the numerical value of
$\kappa=0.138$. Lattice results do not agree with this value \cite{diss_Moyaerts}, although they do for $3+1$ dimensions.

It is interesting to note that, leaving the DSE for $\omega$ aside, a
restriction on the infrared exponents (\ref{restrict}) arises
from self-consistency of the equations for $d$ and $f$ only. This
restriction also concerns the infrared behavior of $\omega$ via the
sum rule (\ref{sumrule}). Consider a fixed wave functional with a
kernel $\omega$ that yields $\kappa>\frac{1}{4}$ by its infrared behavior. In
principle, it should be possible to calculate $d$ and $f$ for such a
wave functional. However, due to the restriction (\ref{restrict}), no solution can
be found. We therefore conclude that the Eq.\ (\ref{restrict}) must be due to
the approximations made in the DSEs for $d$ and $f$. The relaxation of
the horizon condition allows for a self-consistent solution of all
DSEs where $\omega(k)$, $\chi(k)$, $f(k)$ and $d(k)$ are infrared
finite. This will be shown in section \ref{subcrit}.
However, we consider the infrared enhancement, of $d(k)$ in
particular, as the driving mechanism in the Gribov--Zwanziger
confinement scenario. Therefore, we infer from the above discussion
that the DSE for $f(k)$, as it stands in Eq.\ (\ref{fDSEs}), along
with the resulting restriction (\ref{restrict}) on $\kappa$, must be ignored.

Let us reconsider the Marrero-Swift relation (\ref{MarSwi}) with focus
on the neglect of $\frac{\del\omega}{\del g}$. In the infrared limit,
the ghost propagator becomes independent of the gauge coupling $g$, as
seen in the stochastic vacuum,
\begin{eqnarray}
  \label{gindep}
  \frac{\del}{\del g}\lla G \rra &=&  \frac{\del}{\del g} \frac{\int
    \cD A\, G[gA] \Det G^{-1}[gA]}{\int
    \cD A\,  \Det G^{-1}[gA]} \nn\\
&=&  \frac{\del}{\del g} \frac{\int
    \cD A'\, G[A'] \Det G^{-1}[A']}{\int
    \cD A'\, \Det G^{-1}[A']}= 0\; ,
\end{eqnarray}
where in Eq.\ (\ref{gindep}) the notation reads $G^{-1}[gA]=-\del^2-g\hat A\del$. Using
the above relation and taking a derivative $\del/\del g$ of the ghost DSE (\ref{ghostDSE}) yields
\begin{eqnarray}
  \label{contradict}
  0 = -N_c k^2 \int \dbar^3\ell
\left(
1-(\hat{\ell}\cdot\hat{k})^2\right)
D_G(\ell-k)\frac{\del}{\del g}\left( g^2 D_A(\ell) \right)\; \neq 0\;\; .
\end{eqnarray}
Neglecting $\frac{\del D_A}{\del g}$, one must run into
contradictions. In the expectation value of the Marrero-Swift relation (\ref{MarSwi}), the neglected terms can be
carried along for the stochastic vacuum to give
\begin{eqnarray}
  \label{fullgdel}
  \frac{\del}{\del g}\lla gG \rra &=&  \frac{\del}{\del g} \frac{\int
    DA\, gG[A] \cJ[A]}{\int
    \cD A\,  \cJ[A]} \nn\\
 &=&  \frac{\int \cD A\,  \cJ[A]\frac{\del}{\del g}\left( gG \right) }{\int
   \cD A\,  \cJ[A]} 
+\frac{\int \cD A\,  \cJ[A]\frac{\del\ln\cJ}{\del g} gG}{\int \cD A\,  \cJ[A]} 
-\frac{\int \cD A\,  \cJ[A]\: gG }{\int \cD A\,  \cJ[A]} \frac{\int \cD A\, 
  \cJ[A]\frac{\del\ln\cJ}{\del g} }{\int \cD A\,  \cJ[A]} \nn\\
&=& \lla F\rra + \lla gG\frac{\del\ln\cJ}{\del g} \rra - \lla gG \rra
\lla \frac{\del\ln\cJ}{\del g}\rra\; .
\end{eqnarray}
There are two extra terms that were omitted in Eq.\ (\ref{fbyd}),
having the mathematical structure of a \emph{covariance}. This
covariance also occurs in the Gaussian vacuum. Since with Eq.\
(\ref{gindep}) we have $\frac{\del}{\del g}\lla gG \rra = \lla G \rra$,
the covariance cancels the Coulomb propagator $\lla F \rra $ on the
r.h.s.\ of Eq.\ (\ref{fullgdel}) in favor of a ghost propagator $\lla
G\rra$,
\begin{eqnarray}
  \label{gdelreal}
  \lla F\rra + \lla \frac{\del\ln\cJ}{\del g} gG \rra - \lla gG \rra
\lla \frac{\del\ln\cJ}{\del g}\rra =  \lla G \rra\; .
\end{eqnarray}
The derivation of the Coulomb form factor DSE (\ref{fDSEs}) would
therefore fail if the approximation $\frac{\del \omega}{\del g}\approx
0$ was relaxed. This does not mean that the DSE (\ref{fDSEs}) is wrong
within the chosen approximation scheme. Applying Wick's  theorem to
derive an integral equation for $f(k)$ in the one-loop approximation
also leads to Eq.\ (\ref{fDSEs}). However, the above considerations do
show that, for the asymptotic infrared, the DSE (\ref{fDSEs}) for
$f(k)$ describes a higher-order effect. On the level of approximations
used, we find from Eqs.\ (\ref{fbyd}) and (\ref{gindep}) that 
\begin{eqnarray}
  \label{fisd}
  f(k) = \frac{1}{k^2D_G^2(k)}\frac{\del}{\del g}\left( gD_G
  \right)\stackrel{k\to 0}{\longrightarrow}
  \frac{1}{k^2D_G  (k)}\equiv \frac{1}{d(k)} \; .
\end{eqnarray}
Plugging this into the Coulomb form factor DSE (\ref{fDSE}) yields
\begin{eqnarray}
  \label{contra2}
  d^{-1}(k) = 1+ g^2\frac{N_c}{2}\int \dbar^3\ell
\left(
1-(\hat{\ell}\cdot\hat{k})^2\right)
\frac{d(\ell-k)}{(\ell-k)^2\omega(\ell)}\; , \quad k\to 0
\end{eqnarray}
which explicitly contradicts the ghost DSE (\ref{dDSE}), note the plus
sign. This contradiction is the analogon of Eq.\ (\ref{contradict})
for $\frac{\del D_A}{\del g}= 0$.

In conclusion, the quality of the approximation scheme must be
improved in order to satisfy the DSE for the Coulomb form
factor. Vertex corrections, which effectively increment the loop
order, would be an interesting starting point in that direction.

\begin{figure}
\centering
\includegraphics[scale=0.9]{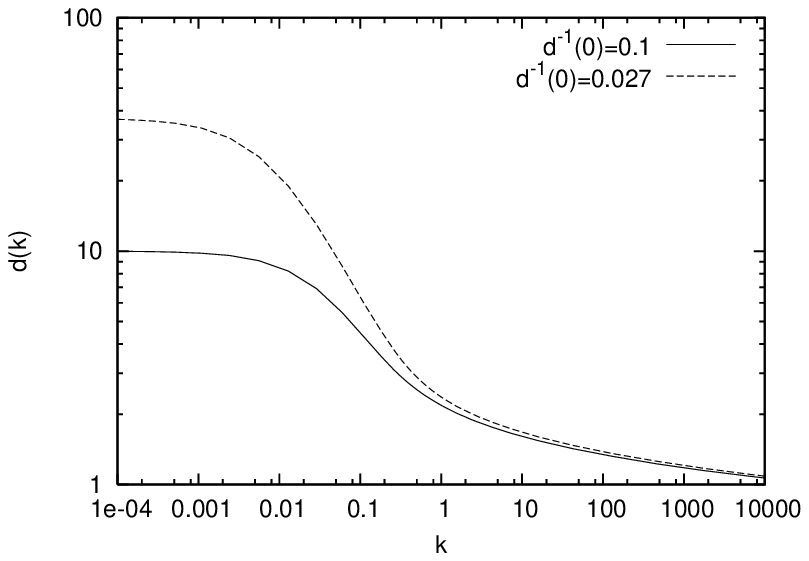} 
\includegraphics[scale=0.9]{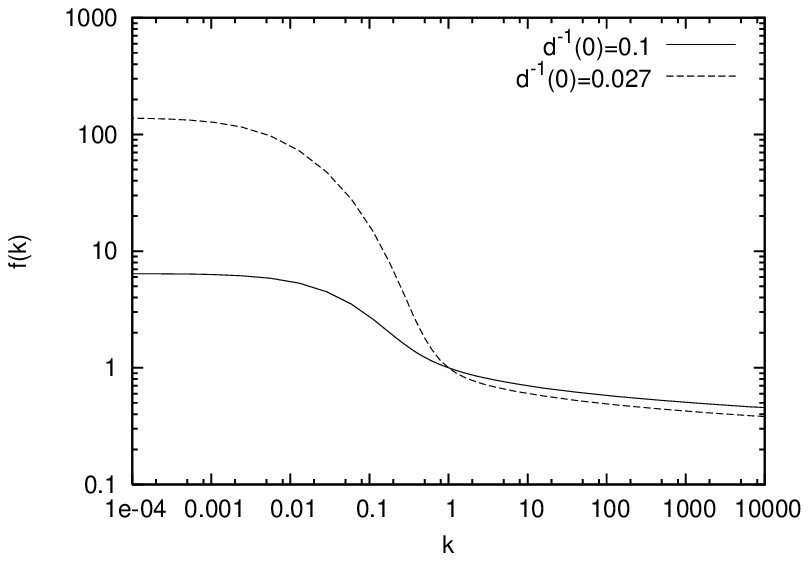} 
\caption{Left: The subcritical ghost form factor $d(k)$. Right: The
  Coulomb form factor $f(k)$. Calculated by D.\ Epple
in Ref.\ \cite{Epp+07}.} 
\label{fig-res-df}
\end{figure}

\section{Subcritical solutions}
\label{subcrit}

\begin{figure}
\centering
\inc[scale=0.9]{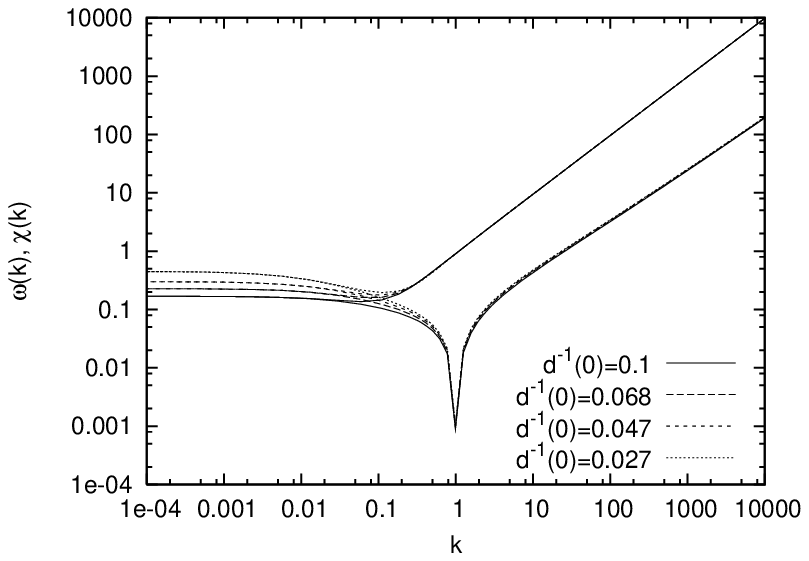}
\inc[scale=0.9]{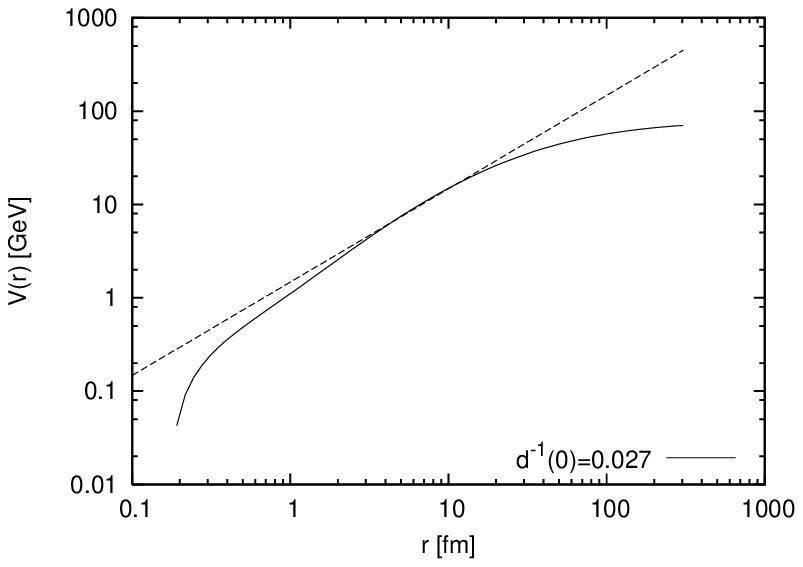}
\caption{Left: The gluon energy $\omega(k)$ and the modulus of the
scalar curvature, $|\chi(k)|$, in the subcritical case. Note that $\chi(k)$ changes sign at
$k=1$. Internal units are used. Right: The heavy quark potential $V_C(r)$ for the subcritical solutions
and a linear fit for the intermediate regime. Calculated by D.\ Epple
in Ref.\ \cite{Epp+07}.
\label{fig-res-omega-chi} } 
\end{figure}

Although it was argued above that in the chosen approximation scheme,
the DSE (\ref{fDSEs}) for the Coulomb form factor $f(k)$ should be
ignored, we here relax the horizon condition to still find a solution
to all DSEs. The renormalization parameter $d_0^{-1}:=d^{-1}(\mu_d=0)$
distinguishes the subcritical solutions for $d_0^{-1}>0$ and the
critical ones for $d_0^{-1}=0$. The subcritical form factors $d(k)$
and $f(k)$, the gap function $\omega(k)$, and the curvature $\chi(k)$
are all infrared finite functions, as the numerical calculation
shows \cite{Epp+07}. In Fig.\ \ref{fig-res-df}, the
functions $d(k)$ and $f(k)$ are shown for $d_0^{-1}=0.1$ and
$d_0^{-1}=0.027$. The functions $\omega(k)$ and $\chi(k)$ are shown in
a common plot in Fig.\ \ref{fig-res-omega-chi}, for several values of
$d_0^{-1}$. These functions solve their respective DSEs
self-consistently. The DSE (\ref{fDSEs}) for $f(k)$ is solved as well
for subcritical $d_0^{-1}$. Lowering the value of $d_0^{-1}$, the
solutions are seen to break down. The smallest possible value for
which all DSEs are solved self-consistently with positive $d(k)$ and
$f(k)$ was found to be $d_0^{-1}\approx 0.02$. In the numerical
calculations, the other renormalization parameters were chosen such
that $\chi(\mu=1)=0$ and $f(\mu=1)=1$. With the subcritical solutions
so obtained, one may calculate the heavy quark potential $V_C(r)$, as
shown in  Fig.\ \ref{fig-res-omega-chi}. For intermediate quark separation $r$, the
potential $V_C(r)$ is approximately linear; fitting the slope to the
lattice string tension, $r$ is found to be in the range of hadronic
phenomenology. However, for $r\to\infty$, the heavy quark potential
becomes questionable since permanent confinement is lost for the
subcritical solutions. In addition, the running coupling $\alpha(k)$
\emph{vanishes} in the infrared, as seen in the forthcoming
chapter. Altogether, the implications of the subcritical solutions for
the infrared physics are not compatible with Coulomb confinement. This
indicates that the horizon condition must not be abandoned. Despite the
inconsistency in the DSE for $f(k)$, which is due to higher-order
effects, we favor the solutions with critical $d_0^{-1}=0$, i.e.\ with
the horizon condition implemented where the functions $\omega(k)$,
$\chi(k)$ and $d(k)$ are infrared enhanced.

\chapter{Running coupling and ultraviolet behavior}
\label{UVchap}

A quantitatively correct treatment of QCD, with the possibility to
describe phenomenological effects, needs to take into account
accurately its scale $\Lambda_{QCD}$. In the absence of flavor
charges, $N_f=0$, this is the only scale of the theory. The value of $\Lambda_{QCD}$
can be related to the perturbative expansion of the (Gell--Mann--Low) beta function,
\begin{eqnarray}
  \label{betafdef}
  \beta(g)=\frac{\del g(\mu)}{\del\ln\mu}=-\beta_0 g^3 - \beta_1 g^5 -
   \beta_2 g^7+\cO(g^9)
\end{eqnarray}
that describes the dependence of the renormalized gauge coupling $g$
on the renormalization scale $\mu$. Solving the differential equation
(\ref{betafdef}) with neglect of $\cO(g^5)$, the gauge coupling is
seen to diverge at the scale
\begin{eqnarray}
  \label{LamQCD}
  \Lambda_{QCD}=\mu\exp\left(-\frac{1}{2\beta_0 g^2(\mu)}\right)\; .
\end{eqnarray}
This breakdown of perturbation theory occurs at a finite
$\Lambda_{QCD}$, if $g^2(\mu)>0$, giving rise to dimensional
transmutation of the dimensionless gauge coupling to the dimensionful
energy scale $\Lambda_{QCD}$. The $\cO(g^3)$
coefficient of the beta function (\ref{betafdef}), $\beta_0$, is a
gauge invariant quantity that is calculated in perturbation theory to
be
\begin{eqnarray}
  \label{beta0}
  \beta_0=\frac{1}{(4\pi)^2}\frac{11N_c}{3}\; .
\end{eqnarray}
Higher order coefficients of the beta function were obtained in cumbersome calculations
\cite{TarVlaZha80,RitVerLar97} and serve for an improved matching of
$\Lambda_{QCD}$ to experiment.

High-energy experiments are able to provide us with the scale
$\Lambda_{QCD}$ immanent in QCD.\footnote{Estimates can also be
  obtained directly by lattice simulations \cite{ProRacSim06}.} A physical process that involves measurable quantities, e.g.\ the mass
of the $Z$ boson, needs to be described by theoretical calculations
which are then compared to the experimental data. The corresponding
scattering matrix is computed by means of perturbative QCD,
preferably in a covariant gauge, in a certain chosen renormalization
scheme. As a result, the running coupling 
\bee\alpha(k)=\frac{g^2(k)}{4\pi}\eeq
can be extracted for the momentum $k$ of the given process. For the
mass $M_Z=91.19\: GeV$ of the Z boson, it is found that
$\alpha(M_Z)=0.1187$ in the $\overline{MS}$ scheme and by the extrapolation
of the perturbative $\alpha(k)$ into the infrared,\footnote{Various
  conventions exist for the definition of the $\beta$ function and its
coefficients of the perturbative expansion. The barred symbols in Eq.\
(\ref{beta3loop}) are related to those of Eq.\ (\ref{betafdef}) by $\beta_0
=\frac{\bar{\beta}_0}{(4\pi)^2}$, $\beta_1 =
\frac{2\bar\beta_1}{(4\pi)^4}$ and $\beta_2 = \frac{\bar\beta_2}{2(4\pi)^6}$.}
\begin{eqnarray}
  \label{beta3loop}
  \alpha(\mu) &=& \frac{4 \pi}{\bar\beta_0 \ln(\mu^2/\Lambda_{QCD}^2)}\left[
1- \frac{2\bar\beta_1}{\bar\beta_0^2} \frac{\ln\left[\ln(\mu^2/\Lambda_{QCD}^2)\right]}{\ln(\mu^2/\Lambda_{QCD}^2)}
+\frac{4\bar\beta_1^2}{\bar\beta_0^4\ln^2(\mu^2/\Lambda_{QCD}^2)} \right.\nonumber\\
&&\hspace*{3cm} \left. \times \left(\left(\ln\left[\ln(\mu^2/\Lambda_{QCD}^2)\right] -\frac{1}{2} \right)^2
+\frac{\bar\beta_0\bar\beta_2}{8\bar\beta_1^2} - \frac{5}{4}\right)\right]
 \; ,
\end{eqnarray}
one finds
%$\alpha(\Lambda_{\overline{MS}})=\infty$ for 
$\Lambda_{\overline{MS}}=217\,$MeV \cite{PDG04}. 
% These values
% are
%, of course, gauge independent. 
%Also, the
%running coupling is a renormalization froup invariant, in the sense
%that ...
They do, however, depend on the renormalization prescription, hence the
subscript on $\Lambda_{\overline{MS}}$, as an artefact of perturbation
theory. A change in the renormalization
prescription can in fact be compensated for by a finite
renormalization group transformation. It is the truncation of the
perturbative series that makes the running coupling scheme dependent \cite{CelGon79}.

The variational solutions presented in the previous chapter only
loosely account for the ultraviolet properties of the
theory and it will be shown that the $\beta$ function turns out to be
slightly incorrect. The first few coefficients of the beta function, in particular
$\beta_0$, record inaccuracies of the solutions and any observable
is immensely sensitive to such errors, as can be seen directly in Eq.\
(\ref{LamQCD}). If the nonperturbative variational solutions are to
give quantitatively sensible results, they have to reproduce $\beta_0$
with high accuracy. In this chapter, we first recall the crucial
properties of the YM coupling constant and then define a
nonperturbative running coupling to come by some information on how 
the Green functions are to behave in the ultraviolet.

\section{Anti-screening of color charges}
\label{antiscreen}

An effect specific to the non-abelian nature of YM theory is the
\emph{anti-screening} of color charges. The interaction of two color charges
on a short distance can be treated by perturbation theory. The
perturbative corrections increase the magnitude of the (renormalized)
color charges, i.e.\ the potential is ``anti-screened'', contrary to
the abelian theory. Anti-screening is a crucial property of YM theory
and intimately connected to asymptotic freedom. 

\begin{figure}
  \centering
  \ing[bb = 123 641 490 700, clip=]{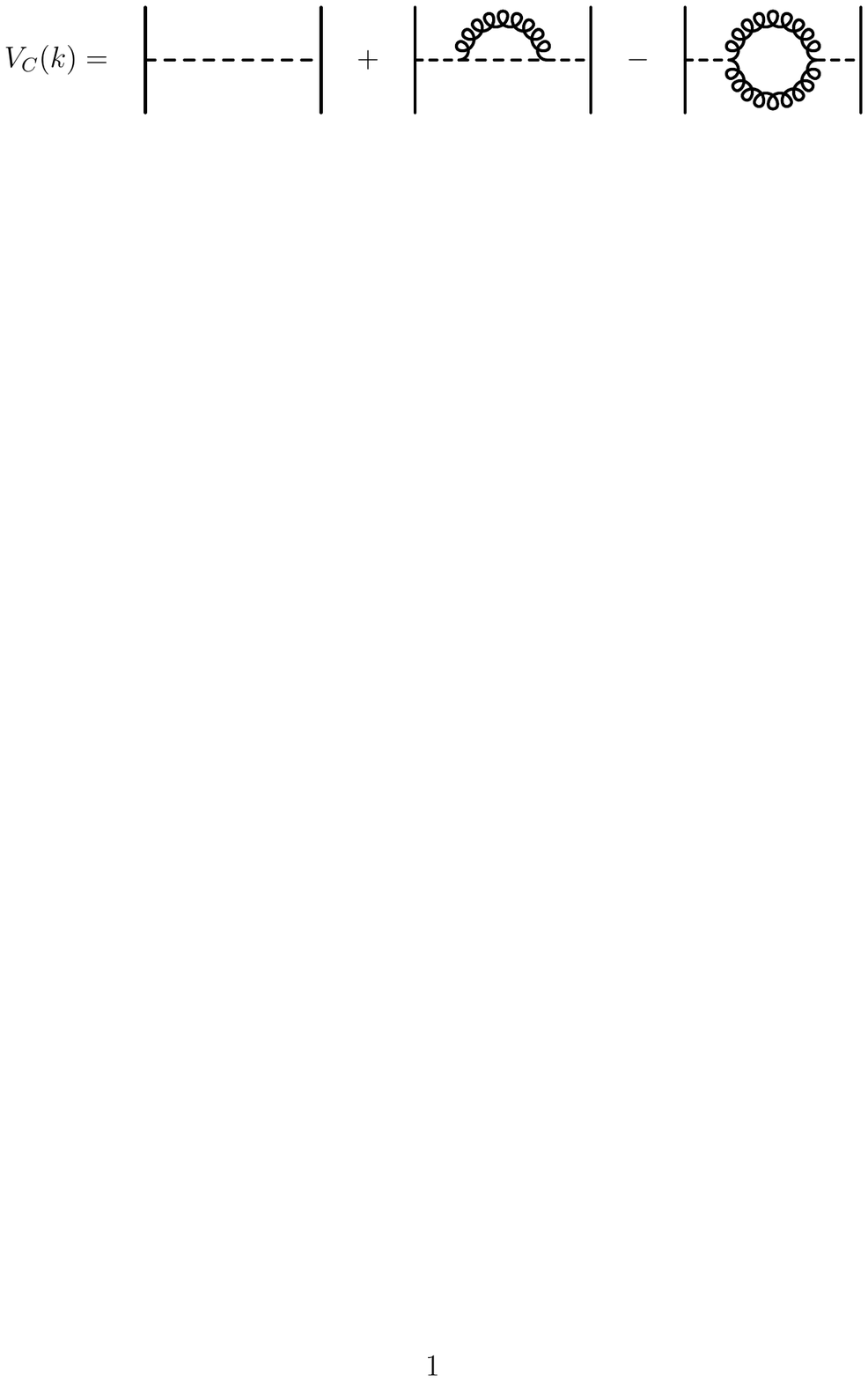}
  \caption{Perturbative screening and anti-screening contributions to
    the color Coulomb potential. The first graph represents the usual
    Coulomb interaction, the second graph shows the genuinely
    nonabelian anti-screening contribution and the last graph
    corresponds to the usual screening term. These graphs are not Feynman diagrams.}
  \label{Khri}
\end{figure}

One may find the potential between two color charges by setting up a basis
of states $\{\ket{n}\}$ in the absence of external charges, then
turning on $\rhoext$, that is $\rho=\rhodyn\to\rhodyn+\rhoext$, and
calculating the correction to the potential energy $E_C$ to second
order in $\rhoext$,\footnote{The Faddeev--Popov determinants $\cJ$ are
  omitted since they do not contribute in the lowest-order perturbation
in $g_B$.}
\begin{eqnarray}
  \label{PTdef}
  E_C
&=&\overbrace{\frac{g_B^2}{2}\int d^3[xy] \bra{0} \rhoext^a(x) F^{ab}(x,y)
    \rhoext^b(y)
    \ket{0}}^{=:E_C^{(+)}}\: +\nn\\
&&\underbrace{-\frac{g_B^2}{2}\int d^3[xy]\sum_n\frac{\left|\bra{0} \rhoext^a(x) F^{ab}(x,y)
    \rhodyn^b(y)
    \ket{n}\right|^2}{E_n}}_{=:E_C^{(-)}}
\; .
\end{eqnarray}
The screening vacuum polarization is of course still
present as the manifestly negative contribution $E_C^{(-)}$ in
(\ref{PTdef}). Its calculation in a perturbative expansion in the bare
gauge coupling $g_B$ to order $\cO(g_B^4)$ is familiar from the vacuum
polarization in the abelian theory \cite{Peskin}. To let two
point-like charges, $\rhoext^a(x)$ as in (\ref{rhom}), approach each
other from an infinite separation to $r$, the energy
$E_C|_{\infty}-E_C|_r = \int \dbar^3k
V_C(k)\e^{ik\cdot r}$ is needed. The momentum space potential
$V_C(k)=V_C^{(+)}(k)+V_C^{(-)}(k)$ comprises the screening contribution
\begin{eqnarray}
  \label{screen}
V_C^{(-)}(k)\approx\frac{g_B^2}{k^2}\left(-\frac{g_B^2}{(4\pi)^2}\frac{C_2}{3}\ln\frac{\Lambda^2}{k^2}\right)\; ,
\end{eqnarray}
where the cut-off $\Lambda$ regularizes the UV divergence and the quadratic Casimir yields here for $\mathit{SU(N_c)}$ the value $C_2=N_c$. 
The screening effect is, however, over-compensated by the contribution
$V_C^{(+)}(k)$. To see that, expand the non-abelian Coulomb operator
around its abelian part,
\begin{eqnarray}
  \label{Fexpand}
  F^{ab}(x,y)=\left( 
   G_0 + 2\: G_0\: g_B\hat A\del\: G_0 + 3\: G_0\: g_B\hat A\del \:
   G_0\: g_B\hat A\del\: G_0+\cO(g_B^3) 
   \right)^{ab}(x,y)\; ,
\end{eqnarray}
and then take the expectation value, see e.g.\ Ref.\ \cite{book_Lee}, where the third term in (\ref{Fexpand}) strengthens the potential,
\begin{eqnarray}
  \label{anti-screen}
  V_C^{(+)}(k) 
   &=&\frac{g_B^2}{k^2}\left(1+\frac{g_B^2}{(4\pi)^2}\frac{12
       C_2}{3}\ln\frac{\Lambda^2}{k^2}\right)\; .
\end{eqnarray}
Note that in the abelian theory only the first term in the
perturbative expansion (\ref{Fexpand}) exists, and thus
$V_C^{(+)}(k)=g_B^2/k^2$. The anti-screening correction in Eq.\ (\ref{anti-screen})
is genuinely non-abelian. In Fig.\ \ref{Khri}, the
neatly separated screening and anti-screening contributions are shown
diagrammatically. The net effect of the $\cO(g^4)$ contributions
increases the potential between static $\mathit{SU(N_c)}$ color charges,
\begin{eqnarray}
  \label{netanti}
   V_C(k) 
   &=&\frac{g_B^2}{k^2}\left(1+\frac{g_B^2}{(4\pi)^2}\frac{11 N_c}{3}\ln\frac{\Lambda^2}{k^2}\right)
\end{eqnarray}
 as first
demonstrated by Khriplovich for $N_c=2$ \cite{Khr70}. 
The net anti-screening correction to the static quark potential can be
used to define the renormalized coupling $g(\mu)$, with the ambiguity
in subtracting the divergence inherent in the renormalization scale
$\mu$. From the so-defined $g(\mu)$, the $\beta$ function may be
calculated via Eq.\ (\ref{betafdef}), reproducing its Nobel-prize
awarded perturbative expansion 
(with the interpretation of asymptotic freedom \cite{GroWil73,Pol73}), as shown in
Refs.\ \cite{Gri77,Dre81}.

If we wish to extract correctly from the color Coulomb potential the
$\beta$ function for the variational solutions of chapter
\ref{VarVac}, the form factors are to behave appropriately in the
ultraviolet. What exactly ``appropriately'' means will become clear in
the course of this chapter. For starters, consider the Coulomb potential
\begin{eqnarray}
  \label{VCpert}
  V_C^{(var)}(k)=\frac{g_B^2}{k^2}d^2(k)f(k)\; 
\end{eqnarray}
as the part of the energy in the variational state that depends on the
separation of the charges, see Eq.\
(\ref{VCmeson}). The self-energies, which only regulate infrared
divergences, are omitted. To lowest order in the bare gauge coupling $g_B$, $d(k)=f(k)=1$. The
$\cO(g_B^4)$ term of Eq.\ (\ref{VCpert}) can be found by expanding the DSEs for $d$
(\ref{dDSE}) and for $f$ (\ref{fDSE}) up to $\cO(g_B^2)$, using the
lowest-order expression $\omega(k)=k$,
\begin{eqnarray}
  \label{dundfPT}
  d(k)&=&1+g_B^2\frac{N_c}{2}\int \dbar^3\ell
\left(
1-(\hat{\ell}\cdot\hat{k})^2\right)
\frac{1}{\ell (\ell-k)^2}\\
  f(k)&=&1+g_B^2\frac{N_c}{2}\int \dbar^3\ell
\left(
1-(\hat{\ell}\cdot\hat{k})^2\right)
\frac{1}{\ell (\ell-k)^2}\; 
\end{eqnarray}
Note that $d(k)$ and $f(k)$ have the identical perturbative $\cO(g_B^2)$
expansion. Plugging these into Eq.\ (\ref{VCpert}), yields three times
the same $\cO(g_B^4)$ term,
\begin{eqnarray}
  \label{VCPT3}
  V_C^{(var)}(k) = \frac{g_B^2}{k^2}\left(1+3\cdot
    g_B^2\frac{N_c}{2}\int \dbar^3\ell
\left(
1-(\hat{\ell}\cdot\hat{k})^2\right)
\frac{1}{\ell (\ell-k)^2}\; \right)+\cO(g_B^6)\; ,
\end{eqnarray}
cf.\ the factor of $3$ in the expansion of the non-abelian Coulomb
operator (\ref{Fexpand}). Regularized by a UV cut-off $\Lambda$,
the integral in Eq.\ (\ref{VCPT3}) to $\cO(g_B^4)$ gives
\begin{eqnarray}
  \label{VCPTanti}
 V_C^{(var)}(k) = \frac{g_B^2}{k^2}\left(1+\frac{g_B^2}{(4\pi)^2}\frac{12 N_c}{3}\ln\frac{\Lambda^2}{k^2}\right)
\; ,
\end{eqnarray}
in agreement with Eq.\ (\ref{anti-screen}). Note, however, that only the
anti-screening contribution $V_C^{(+)}(k)$ is thus accounted for. Determining the
state $\ket{\Psi}$ by the variational principle in the absence of
color charges and deriving the color Coulomb potential $V_C(r)$ as a
perturbation around the vacuum, necessitates that the screening
contribution $V_C^{(-)}(k)$ is carried along in the full second-order perturbative
expression as in Eq.\ (\ref{PTdef}).

The renormalized coupling $g$ is defined so as to render the expression
for $V_C(k)$ finite, order by order in perturbation
theory. Decomposing an infinite quantity into a finite and an infinite
part always leaves an ambiguity that is controlled by the
renormalization scale $\mu$. Using 
\begin{eqnarray}
  \label{grendef}
  g_B(\Lambda)=Z_g(\Lambda,\mu)g(\mu)
\end{eqnarray}
as the definition of the renormalized, $\mu$-dependent $g(\mu)$, the
expression $V_C^{(var)}(k)$ in Eq.\ (\ref{VCPTanti}) turns out finite up to $\cO(g^4)$ if the
renormalization constant is chosen as\footnote{Swapping $g_B$ for $g$
  does not matter to the order considered.}
\begin{eqnarray}
  \label{Zg}
  Z_g(\Lambda,\mu)=
1-\frac{1}{2}\frac{g_B^2}{(4\pi)^2}\frac{12
  N_c}{3}\ln\frac{\Lambda^2}{\mu^2}\; .
\end{eqnarray}
% Plugging $Z_g$ into (\ref{VCPTanti}) yields
% \begin{eqnarray}
%   \label{VCPTantiren}
%  V_C^{(+)}(k) = \frac{g^2}{k^2}\left(1+\frac{g^2}{(4\pi)^2}\frac{12
%      N_c}{3}\ln\frac{\mu^2}{k^2}\right) + \cO(g^6)
% \; ,
% \end{eqnarray}
The $\beta$ function, as defined by Eq.\ (\ref{betafdef}), can now be
calculated from Eqs.\ (\ref{grendef}) and (\ref{Zg}) to lowest order,
\begin{eqnarray}
  \label{betaCoul}
  \beta(g)=-\frac{g^3}{(4\pi)^2}\frac{12 N_c}{3}\; .
\end{eqnarray}
This result does not agree with the canonical value of $\beta_0$, see
Eq.\ (\ref{beta0}), but it is clear why: only the contributions of
longitudinal gluons were considered to order $\cO(g^4)$, the
screening contribution from transversal gluons was omitted. For the
same reason, the perturbative renormalization group calculation of the color
Coulomb potential \cite{CucZwa01} relates $V_C(k)$ to the running coupling
$g^2(k)$ by $k^2 V_C(k)=x_0 g^2(k)$ with
$x_0=\frac{12}{11}$. In the next sections, we aim to use a
nonperturbative definition of the running coupling.

\section{Nonperturbative running coupling}
\label{secrunning}

Since the variational Coulomb gauge solution from
chapter \ref{VarVac} is nonperturbative, we seek here a
nonperturbative definition of the running coupling in order to match
$\alpha(M_Z)=0.1187$ to our solutions. In the Coulomb gauge, there are two
possibilities for its definition, either via the color Coulomb
potential, or by means of the ghost-gluon vertex \cite{FisZwa05}. The
nonrenormalization of the ghost-gluon vertex has proven above to be a useful
common property of both the Coulomb and Landau gauge. Here, it
is first discussed how the Landau gauge ghost-gluon vertex may define
a nonperturbative running coupling and afterwards we turn to the
Coulomb gauge.

\subsection*{Landau gauge running coupling from ghost-gluon vertex}

The local off-shell formulation of Landau gauge YM theory allows for
multiplicative renormalization of the fields by the introduction of
renormalization constants $Z(\Lambda,\mu)$. Some of these are related
to one another by the fact that $Z_g$, the renormalization constant of
the gauge coupling, can be defined by any of the interaction terms in
the YM Lagrangian, namely the ghost-gluon vertex, the quark-gluon
vertex, the three- and the four-gluon vertex. The equality of the
gauge coupling for all these interactions is expressed by the Slavnov--Taylor
identity
\begin{eqnarray}
  \label{STIfund}
  \frac{Z_1}{Z_3}=\frac{\tilde{Z}_1}{\tilde{Z}_3}
\end{eqnarray}
where $Z_1=Z_gZ_3^{3/2}$ is the renormalization constant of the three-gluon vertex,
$\tZ_1=Z_g\tZ_3Z_3^{1/2}$ the one of the ghost-gluon vertex, while $Z_3$ and
${\tilde{Z}_3}$ renormalize gluon and ghost fields,
\bee
A_B=Z_3^{1/2}A\; , \quad c_B=\tZ_3^{1/2}c\; , \quad
\oc_B=\tZ_3^{1/2}\oc\; .
\eeq
An unambiguous
definition of the $Z$'s requires renormalization
prescriptions. We define the theory at the scale $\mu$ by setting
\begin{eqnarray}
  \label{RSprop}
  D_A(k=\mu;\mu)=\frac{C_A}{k^2}\; ,\quad   D_G(k=\mu;\mu)=\frac{C_G}{k^2}\; .
\end{eqnarray}
The nonrenormalization of the ghost-gluon vertex in Landau gauge
infers that $\tZ_1$ is a finite number, independent of $\mu$.
% (in this sense a renormalization group invariant)
Therefore the quantity
\begin{eqnarray}
  \label{RGGGA}
  g_B^2(\Lambda)D_{A,B}(k;\Lambda)D_{G,B}^2(k;\Lambda)=\tZ_1^2g^2(\mu)D_A(k;\mu)D_G^2(k;\mu)
\end{eqnarray}
is finite, due to $\tZ_1<\infty$, and $\mu$-independent, since the
l.h.s.\ in $\mu$-independent. One may set $\tZ_1=1$ but
any other finite number will also do in an appropriate renormalization
prescription, recall section \ref{secGGA}. It is left arbitrary here.
In the perturbative momentum subtraction scheme, the values $C_A=C_G=1$ \emph{define} the
renormalized quantities on the r.h.s.\ of Eq.\ (\ref{RGGGA}). Nonperturbative Green functions might not allow for such a
simple subtraction scheme. We therefore generalize it
\cite{SmeHauAlk98} by setting
\begin{eqnarray}
  \label{nonptrs}
  C_AC_G^2=1\; .
\end{eqnarray}
Other schemes can be accomplished by performing a finite renormalization, see below. 

In Eq.\ (\ref{RGGGA}), one
may choose $\mu$ freely. Leaving $\mu$ once unspecified and
setting it once to $\mu=k$ leads with (\ref{RSprop}) to  a nonperturbative
definition of the running coupling \cite{SmeHauAlk98},
\begin{eqnarray}
  \label{runncoupdef}
  g^2(k)=k^6 g^2(\mu)D_A(k;\mu)D_G^2(k;\mu)\; .
\end{eqnarray}
While the factors on the r.h.s.\ are $\mu$-dependent, the running
coupling $g^2(k)$ on the l.h.s.\ is renormalization scale- and scheme
independent. Note that the finite value $\tZ_1$ drops out. Applying a finite renormalization such as
$A\to z_3^{1/2}A$, $c\to\tilde{z}_3^{1/2}c$ in order to discuss
the case $C_AC_G^2\neq 1$, 
%  and $g\to g$,\footnote{We did not scale $g$, $z_g=1$, otherwise the
% running coupling would naturally acquire a ultraviolet behavior that
% corresponds to the change
% $\beta_0\to z_g^2\beta_0$.}
the running coupling can be expressed by the transformed quantities,
\begin{eqnarray}
  \label{runncoupdeffinren}
  g^2(k)= \tZ_1^2\:k^6{g}^2(\mu)D_A(k;\mu)D_G^2(k;\mu)\; .
\end{eqnarray}
where we have set $\tZ_1^2z_3\tilde{z}_3^2\to\tZ_1^2$. Note that these simple scalings
are not renormalization group transformations in the usual sense since they do not depend on
the scale $\mu$. It will be shown
explicitly below that the (asymptotic) behavior of the running
coupling is ignorant to such scalings.

A different point of view is frequently found in the literature, see
e.g.\ Ref.\ \cite{AtkBlo98}. One may always absorb finite
transformations by a transformation in $g$ such that $\tZ_1^2=Z_g^2Z_3\tZ_3^2$ is left
unchanged (usually chosen to be unity). The running coupling (\ref{runncoupdef}) then
stays form-invariant. Here, we prefer to leave $g$ unchanged and allow
for a change in $\tZ_1$ and use Eq.\ (\ref{runncoupdeffinren}) as the
running coupling. 

\subsection*{Coulomb gauge running coupling from ghost-gluon vertex}

A running coupling in Coulomb gauge may be defined in analogy to the
Landau gauge from the ghost-gluon vertex. The multiplicative
renormalization, however, is not as well-understood in the Coulomb
gauge. From within a class of interpolating gauges where
multiplicative renormalization was discussed in Ref.\ \cite{BauZwa99}, the
Coulomb gauge limit may be taken, breaking the path for a definition
similar to Eq.\ (\ref{runncoupdeffinren}) in the Landau gauge
\cite{FisZwa05}. We adopt this definition here.

Using the notation $d(k)$ for the ghost form factor (\ref{ddef}) and $\omega(k)$
for the equal-time gluon propagator (\ref{Dhalf}), both renormalized
multiplicatively,
\begin{eqnarray}
  \label{Zdwdef}
  d_B(k;\Lambda)=\tZ_3(\Lambda,\mu)d(k;\mu)\; ,\quad
  \omega_B(k,\Lambda)=Z_3(\Lambda,\mu)\omega(k;\mu)\\
  d(k=\mu;\mu)=C_d\; , \quad \omega(k=\mu;\mu)=C_\omega k
\; ,
\end{eqnarray}
with $C_\omega^{-1}C_d^2=1$, we find from the ghost-gluon vertex
analogously to Eq.\ (\ref{RGGGA}),
\begin{eqnarray}
  \label{muindep2}
  g_B^2(\Lambda)d^2_B(k;\Lambda)\frac{1}{2}\omega^{-1}_B(k;\Lambda)=\tZ_1^2 g^2(\mu)d^2(k;\mu)\frac{1}{2}\omega^{-1}(k;\mu)\; .
\end{eqnarray}
Allowing for a change in the renormalization prescription,
$C_\omega^{-1}C_d^2\neq 1$, one may
either absorb it by a change in $g$, or (as above) redefine
$\tZ_1$. Accordingly to (\ref{runncoupdeffinren}), the latter choice leads directly to the running coupling 
\begin{eqnarray}
  \label{runncoupCoul}
  g^2(k)=\tZ_1^2\: k\, g^2(\mu)\:d^2(k,\mu)\:\omega^{-1}(k,\mu)
\end{eqnarray}
Note that $k$ is here $k=|\fk|$ and all interactions are
instantaneous.\footnote{For equal times, the tree-level gluon
  propagator is $D_A^0(k)=\frac{1}{2k}$, see Eq.\ (\ref{Dequaltime}). Therefore,
the definition (\ref{runncoupCoul}) is larger than in Landau gauge (\ref{runncoupdeffinren})
by a factor of $2$.} The running coupling (\ref{runncoupCoul}) was given in
Ref.\ \cite{FisZwa05} with a prefactor of $\frac{8}{3}$, chosen
as to recover the Landau gauge infrared limit from within the interpolating gauge. According to the
discussion above, this factor needs to be interpreted as $\tZ_1^2=\frac{8}{3}$.

\subsection*{Coulomb gauge running coupling from the color Coulomb potential}

Another renormalization group invariant that qualifies to define a
nonperturbative running coupling is the instantaneous color Coulomb
potential $V_C(k)$. It was shown in Ref.\ \cite{Zwa98} that the
product $g A_0$ is a renormalization group invariant and in Ref.\
\cite{CucZwa01} how the quantity $\lla g^2 A_0 A_0 \rra$ actually
relates to $V_C(k)$. We may accordingly define the running coupling in
Coulomb gauge also by
\begin{eqnarray}
  \label{couppot}
  g^2(k)=\frac{11}{12}\, g^2(\mu)k^2V_C(k;\mu)\; ,\quad k=|\fk| \; .
\end{eqnarray}
The factor of $\frac{11}{12}$ accounts for the fact that $V_C(k)$ only
comprises the anti-screening contribution and over-emphasizes this
effect. While in the ultraviolet the definition (\ref{couppot}) is
expected to agree with the definition (\ref{runncoupCoul}) of the running coupling from
the ghost-gluon vertex, the infrared behavior turns out to disagree
drastically. In view of the confining long-range part of $V_C(k)$, the
running coupling in Eq.\ (\ref{couppot}) diverges for $k\to 0$, a
clear instance of infrared slavery. On the
other hand, the definition (\ref{runncoupCoul}) will yield a running
coupling that freezes in the infrared, see below.

\section{Ultraviolet behavior of Green functions}
\label{propsinUV}

Given relations that express the running coupling in terms of the Green
functions nonperturbatively, one may extract information on these
Green functions in the UV domain, where the running
coupling is known from perturbation theory. Solving the
$\cO(g^3)$ expansion of the $\beta$ function (\ref{betafdef}) as a
differential equation for $g^2(k)$ yields for large $k^2=k_\mu k^\mu$
\begin{eqnarray}
  \label{coupasy}
  g^2(k)=\frac{1}{\beta_0\ln\frac{k^2}{\Lambda_{QCD}^2}}\; .
\end{eqnarray}
The ultraviolet behavior of the nonperturbative propagators has to be
such that the nonperturbative running couplings defined in section
\ref{secrunning} approach the function
(\ref{coupasy}). Further information can be extracted by solving the
ghost Dyson--Schwinger equation self-consistently with an ansatz for
the ultraviolet behavior. This
ansatz requires an educated guess. 

Despite asymptotic freedom, the Green functions do not
asymptotically become tree-level in the ultraviolet. Weinberg's
theorem states that there can be logarithmic corrections \cite{Wei60},
and an ansatz for the gluon and ghost propagators for $k\gg\mu$  might
therefore read
\begin{subequations}
\label{Weinberg}
\begin{eqnarray}
  D_A(k)&\sim&\frac{1}{k^2 \ln^{\gamma}(\frac{k^2}{\mu^2})}\\
  D_G(k)&\sim&\frac{1}{k^2 \ln^{\delta}(\frac{k^2}{\mu^2})}
\end{eqnarray}
\end{subequations}
The exponents of the logarithms are specified by the renormalization
group in relation to the anomalous dimensions
of the gluon and the ghost field \cite{GroWil73},
\begin{subequations}
  \label{anomdimdef}
\begin{align}
  \gamma_A(g)\, &=\, \frac{\del \ln Z_3}{\del \ln\mu^2}\, =\, \gamma_A^0g^2+\cO(g^4) \\
  \gamma_G(g)\, &=\, \frac{\del \ln \tZ_3}{\del \ln\mu^2}\, =\, \gamma_G^0g^2+\cO(g^4) 
\end{align}
\end{subequations}
In the Landau gauge, this relation is given by \cite{GroWil73}
\begin{subequations}
  \label{anovsexp}
\begin{align}
  \gamma\, &=\, \frac{\gamma_A^0}{\beta_0}\, =\, \frac{13}{22}\\
  \delta\, &=\, \frac{\gamma_G^0}{\beta_0}\, =\, \frac{9}{44}
\end{align}
\end{subequations}
Oftentimes, the logarithmic exponents $\gamma$ and $\delta$ in (\ref{Weinberg}) are
referred to as anomalous dimensions of the gluon and the ghost,
although strictly speaking they are only the lowest order coefficients
of the anomalous dimensions $\gamma_A$ and $\gamma_G$ defined in Eq.\
(\ref{anomdimdef}) and divided by $\beta_0$.

We now try to reproduce the well-established results (\ref{anovsexp}) from our
nonperturbative study of Dyson--Schwinger equations in Landau gauge. In
the same procedure, we may then go on to determine the UV behavior of
Coulomb gauge propagators.

\subsection*{Landau gauge anomalous dimensions}
Consider the ($d=4$) Landau gauge ghost DSE multiplicatively
renormalized in the scheme (\ref{RSprop}), where $\mu^2D_A(\mu)=C_A$,
$\mu^2D_G(\mu)=C_G$ and the ghost-gluon vertex is evaluated at some
momentum configuration to define a value of $\tZ_1$,
\begin{eqnarray}
  \label{ghostDSEmren}
     D_G^{-1}(k)=k^2\tZ_3-g^2N_c\tZ_1^2k^2\int \dbar^4\ell \: \left(1-(\hat{\ell}\cdot \hat{k})^2\right)D_A(\ell)D_G(\ell-k)\: .
\end{eqnarray}
A renormalization prescription that does not agree with (\ref{nonptrs}) is
accounted for by a change in $\tZ_1^2$. In order to study the ghost
DSE (\ref{ghostDSEmren}) in the asymptotic ultraviolet limit, the following
ans\"atze are made:
\begin{subequations}
  \label{UVpropLand}
\begin{eqnarray}
  D_A(k)&=&\frac{C_A}{k^2 \left(1+\frac{C_A}{A}\ln^{\gamma}(\frac{k^2}{\mu^2})\right)}\;\;\stackrel{k\to\infty}{\longrightarrow}\;\;\frac{A}{k^2\ln^{\gamma}(\frac{k^2}{\mu^2})}\\
  D_G(k)&=&\frac{C_G}{k^2 \left(1+\frac{C_G}{B}\ln^{\delta}(\frac{k^2}{\mu^2})\right)}\;\;\stackrel{k\to\infty}{\longrightarrow}\;\;\frac{B}{k^2\ln^{\delta}(\frac{k^2}{\mu^2})}
\end{eqnarray}
\end{subequations}
These satisfy the renormalization prescriptions (\ref{RSprop}) for
$k=\mu$ and behave like the proposed ultraviolet behavior (\ref{Weinberg}) for $k\to\infty$, with coefficients $A$ and
$B$. Rewrite the ghost DSE (\ref{ghostDSEmren}) by 
\begin{eqnarray}
  \label{ghostDSEUV}
  \frac{1}{k^2 D_G(k)}= \tZ_3-g^2N_c\tZ_1^2 AB \cI_G(k)\; ,
\end{eqnarray}
where the integral $\cI_G(k)$ can then be asymptotically evaluated for
$k\gg\mu$ with a dimensionally regularized UV divergence,\footnote{One can
  integrate for $\ell<k$ and $\ell>k$ and expand in $\ell/k$ and $k/\ell$,
  respectively. Only one term is dominant which turns out to be the one
  found by the ``angular approximation''. We restrict the calculation to $0<{1-\gamma-\delta}<1
$. Furthermore,
 the expansion of the incomplete gamma function \bee
 \int_\tau^\infty dx
 x^{z-1}\e^{-x}=\Gamma(z)-\sum_{n=0}^\infty\frac{(-1)^n\tau^{z+n}}{n!(z+n)} \nn\eeq is used.}
\begin{eqnarray}
  \label{DSEUVasy}
  \cI_G(k\to \infty)&=&\int
\dbar^{4-2\epsilon}\ell
\frac{1-\big(\hat{\ell}\cdot\hat{k}\big)^2}
{\ell^2(\ell-k)^2\ln^\gamma\left(\frac{\ell^2}{\mu^2}\right)\ln^\delta\left(\frac{(\ell-k)^2}{\mu^2}\right)}=\int_{\ell\geq k}\dbar^{4-2\epsilon}\ell
\frac{1-\big(\hat{\ell}\cdot\hat{k}\big)^2}
{\ell^4\ln^{\gamma+\delta}\left(\frac{\ell^2}{\mu^2}\right)}\nn\\
&=&\frac{2}{(2\pi)^3}\int_0^\pi d\eta\sin^4\eta(1+\cO(\epsilon))
\int_k^\infty\frac{d\ell}{\ell}\frac{1}{(\ell^2)^{\epsilon}\ln^{\gamma+\delta}\left(\frac{\ell^2}{\mu^2}\right)}\nn
\; ,\quad x:=\epsilon\ln\left(\frac{\ell^2}{\mu^2}\right)\\
&=&\frac{3}{32\pi^2}(1+\cO(\epsilon))\left(\mu^2\right)^{-\epsilon}\frac{1}{\epsilon^{1-\gamma-\delta}}\frac{1}{2}\int_{\epsilon\ln\left(\frac{k^2}{\mu^2}\right)}^\infty
dx x^{-\gamma-\delta}\e^{-x}\nn\; \\
&=&\frac{3}{64\pi^2}(1+\cO(\epsilon))\left(\frac{\Gamma({1-\gamma-\delta})}{\epsilon^{1-\gamma-\delta}}-\frac{1}{{1-\gamma-\delta}}\ln^{1-\gamma-\delta}\left(\frac{k^2}{\mu^2}\right)+\cO(\epsilon)\right)
\nn\\
&=&\frac{1}{(4\pi)^2}\frac{3\Gamma({1-\gamma-\delta})}{4\epsilon^{1-\gamma-\delta}}-\frac{1}{(4\pi)^2}\frac{3}{4(1-\gamma-\delta)}\ln^{1-\gamma-\delta}\left(\frac{k^2}{\mu^2}\right)\; .
\end{eqnarray}
Terms that vanish for
$\epsilon\to 0 $ were discarded. Matching this result to the l.h.s.\
of the ghost DSE (\ref{ghostDSEUV}) leads to a sum rule of the anomalous
dimensions,
\begin{eqnarray}
  \label{anomsumrule}
  \gamma+2\delta=1\; ,
\end{eqnarray}
as well as the relation
\begin{eqnarray}
  \label{UVLancoeff}
  g^2N_c\tZ_1^2AB^2\frac{1}{(4\pi)^2}\frac{3}{4\delta}=1\; .
\end{eqnarray}
The infinite part of the renormalization constant $\tZ_3$ is
identified by 
\begin{eqnarray}
  \label{tZ3}
  \tZ_3=\frac{\Gamma(\delta+1)}{B}\frac{1}{\epsilon^\delta}+\textrm{finite}
\end{eqnarray}
where setting the finite part to zero puts QCD into a nonperturbative
phase.\footnote{It was discussed in Ref.\ \cite{Zwa03a} that the
  horizon condition does not allow $\tZ_3\to 1$ for $g\to 0$ and hence
contradicts perturbation theory.}

We may now plug the asymptotic forms (\ref{UVpropLand}) into the
running coupling (\ref{runncoupdeffinren}) to obtain with Eq.\ (\ref{UVLancoeff})
\begin{eqnarray}
  \label{alphaUV}
  g^2(k\to\infty) = \tZ_1^2
  g^2\frac{AB^2}{\ln\left(\frac{k^2}{\mu^2}\right)} = \frac{1}{\frac{1}{(4\pi)^2}\frac{3N_c}{4\delta}\ln\left(\frac{k^2}{\mu^2}\right)}
\end{eqnarray}
Note that the finite value of $\tZ_1$ specific to
the renormalization scheme drops out.
The above UV limit expression for the nonperturbative $g^2(k)$ can be compared
to the perturbative expression (\ref{coupasy}) to yield
\begin{eqnarray}
  \label{deltares}
  \delta=\frac{1}{(4\pi)^2}\frac{3N_c}{4\beta_0}=\frac{9}{44}\; ,
\end{eqnarray}
in agreement with the perturbative result for $\delta$ in Eq.\
(\ref{anovsexp}). 
\subsection*{Coulomb gauge anomalous dimensions}

Since the above UV analysis of the nonperturbative ghost
Dyson--Schwinger equation successfully yielded the correct anomalous
dimensions for the Landau gauge, we use the same method to extract the
anomalous dimensions for the gluon and ghost fields in the Coulomb
gauge.

The multiplicatively renormalized ghost DSE reads in terms of $d(k)$ and $\omega(k)$,
\begin{eqnarray}
  \label{ghostDSEmultren}
  d^{-1}(k)=\tZ_3-g^2N_c\tZ_1^2\int \dbar^3 \ell\:
  \left(1-(\hat{\ell}\cdot
    \hat{k})^2\right)\frac{d(\ell-k)}{2\omega(\ell)(\ell-k)^2}\; ,
\end{eqnarray}
and the ans\"atze
\begin{eqnarray}
  \label{dwUV}
  \omega^{-1}(k)=\frac{C_\omega^{-1}}{k\left[1+\frac{C_\omega^{-1}}{A}\ln^\gamma\left(\frac{k^2}{\mu^2}\right)\right]}\; ,\quad  d(k)=\frac{C_d}{1+\frac{C_d}{B}\ln^\delta\left(\frac{k^2}{\mu^2}\right)}
\end{eqnarray}
give for $k=\mu$ the values $C_\omega^{-1}$ and $C_d$, resp., whereas the
$k\to\infty$ behavior is determined by the coefficients $A$ and $B$
and the exponents $\gamma$, $\delta$, similarly to the Landau
gauge. Setting the renormalization constant $\tZ_3$ to 
\begin{eqnarray}
  \label{tZ3Coul}
  \tZ_3=\frac{\Gamma(\delta+1)}{B}\frac{1}{\epsilon^\delta}\; ,
\end{eqnarray}
enforces the horizon condition (cf.\ Eq.\ \ref{tZ3}), and one may
calculate the integral in Eq.\ (\ref{dwUV}) for $k\to\infty$ by
\begin{eqnarray}
  \label{UVcalcCoul}
  B^{-1}\ln^{\delta}\left(\frac{k^2}{\mu^2}\right)&=&\tZ_3-g^2N_c\tZ_1^2\int_{\ell\geq k}\dbar^{3-2\epsilon}\ell\:  \left(1-(\hat{\ell}\cdot \hat{k})^2\right)\frac{AB}{2\ell^3\ln^{\gamma+\delta}\left(\frac{\ell^2}{\mu^2}\right)}\nn\\
&=&\tZ_3-g^2N_c\tZ_1^2AB\frac{1}{2}\frac{\frac{4}{3}}{(2\pi)^2}\frac{1}{2}\frac{1}{\epsilon^\delta}\int_{\epsilon\ln\left(\frac{k^2}{\mu^2}\right)}^\infty
dx\: x^{\delta-1}\e^{-x}\nn\\
&=&\tZ_3-g^2N_c\tZ_1^2AB\frac{1}{(4\pi)^2}\left(
\frac{4\Gamma(\delta)}{3}\frac{1}{\epsilon^\delta}-\frac{4}{3\delta}\ln^\delta\left(\frac{k^2}{\mu^2}\right)
\right)\nn\\
&=&g^2N_c\tZ_1^2AB\frac{1}{(4\pi)^2}\frac{4}{3\delta}\ln^\delta\left(\frac{k^2}{\mu^2}\right)
\end{eqnarray}
where we have matched the exponents by means of
the sum rule
\bee
\label{thelogsumrule}
\gamma+2\delta=1\; ,
\eeq
and the coefficients are found to to obey
\begin{eqnarray}
  \label{coeffCoulUV}
    g^2N_c\tZ_1^2AB^2\frac{1}{(4\pi)^2}\frac{4}{3\delta}=1\; .
\end{eqnarray}
With these results, the Coulomb gauge running coupling (\ref{runncoupCoul}) gives for the
ultraviolet limit
\begin{eqnarray}
  \label{alphaCoulUV}
    g^2(k\to\infty) = \tZ_1^2
  g^2\frac{AB^2}{\ln\left(\frac{k^2}{\mu^2}\right)} = \frac{1}{\frac{1}{(4\pi)^2}\frac{4N_c}{3\delta}\ln\left(\frac{k^2}{\mu^2}\right)}
\end{eqnarray} and in comparison to its known gauge invariant
behavior  (\ref{coupasy}) this yields
\begin{eqnarray}
  \label{deltaresCoul}
  \delta=\frac{1}{(4\pi)^2}\frac{4N_c}{3\beta_0}\; .
\end{eqnarray}
With the canonical value (\ref{beta0}) for $\beta_0$ and by virtue of the sum rule (\ref{thelogsumrule}), the logarithmic
exponents of ghost and gluon propagators can thus be found to yield
\begin{eqnarray}
  \label{gammadeltares}
  \gamma=\frac{3}{11}\; ,\quad \delta=\frac{4}{11}\; .
\end{eqnarray}
The lowest order anomalous dimensions (cf.\ Eq.\ \ref{anovsexp})
$\gamma_A^0=\beta_0\gamma=N_c$ and
$\gamma_G^0=\beta_0\delta=\frac{4N_c}{3}$ agree with the purely
perturbative calculation \cite{WatRei07a,privWeb,Zwa98}. Therefore, a full numerical
nonperturbative solution is expected to yield the ultraviolet
behavior specified by (\ref{dwUV}) with the logarithmic exponents
(\ref{gammadeltares}). 

Recalling the numerical
solution from chapter \ref{VarVac}, the values for $\gamma$ and
$\delta$, see Eq.\ (\ref{UVnumerics}), were
\begin{eqnarray}
  \label{gdnum}
  \gamma_{\textrm{num}}=0\; ,\quad \delta_{\textrm{num}}=\frac{1}{2}\; .
\end{eqnarray}
Note that the sum rule (\ref{thelogsumrule}) is satisfied by the above values.
In the calculation that led to (\ref{gammadeltares}), we required $\beta_0$ to be the canonical value. We
may turn the argument around, assume the numerical solutions (\ref{gdnum}) and calculate
$\beta_0$ from the UV behavior of the resulting running coupling,
cf.\ Eq.\ (\ref{deltaresCoul}). This procedure yields
\begin{eqnarray}
  \label{beta0num}
  \beta_0^{\textrm{num}}=
  \frac{1}{(4\pi)^2}\frac{4N_c}{3\delta_{\textrm{num}}}=\frac{1}{(4\pi)^2}\frac{8N_c}{3}\; .
\end{eqnarray}
The disagreement of $\beta_0^{\textrm{num}}$ with $\beta_0$ is small,
but in view of the exponential dependence of the QCD scale on
$\beta_0$, a slightly erroneous result for $\beta_0$ will make a
sensible assignment of $\Lambda_{QCD}$ hopeless. The failure to find
the correct $\beta_0$ is clearly due to the approximations made. The
Gaussian wave functionals used eliminate the gluon loop in the gap
equation since there exists no tree-level three-gluon vertex. From
purely perturbative calculations it is known that the gluon loop is
essential for the gluon self-energy to preserve gauge invariance and
produce the correct $\beta_0$. A nonperturbative calculation therefore
requires a truncation that involves the gluon loop as well. However,
DSE studies in Landau gauge showed that the dressing of the
three-gluon vertex in the 
gluon loop must be highly non-trivial \cite{SmeHauAlk98,AtkBlo98}, and to achieve results
with the canonical $\beta_0$ value a momentum dependent
renormalization constant $Z_1$ is proposed \cite{FisAlkRei02}. In the Hamiltonian
approach with a wave functional, the extraction of the right $\beta_0$ was
pursued from the Coulomb interaction using perturbative UV tails for
the form factors \cite{Bro98,BroKog99}. 
It is not clear at present how to manipulate the wave functional such that
the nonperturbative UV behavior (\ref{dwUV}) with correct values for $\delta$
and $\gamma$ is reproduced from analyzing the gap equation in the Coulomb gauge
Hamiltonian approach.

The agreement of the result (\ref{beta0num}) with Refs.\ \cite{SzcSwa02,FeuRei04} is coincidental. There, after
factoring out $g_B$ from the ghost form factor, the $\beta$ function was
defined as the (logarithmic) derivative of the tree-level term in the
ghost equation, being $g_B^{-1}(\Lambda)$. However, this term is actually $\tZ_3$, see Eq.\
(\ref{ghostDSEmultren}). What was proposed as the $\beta$ function was
actually
\bee
\bar{\beta}=\frac{\partial}{\partial\ln\mu}\ln \tZ_3=2\gamma_d
\eeq
i.e.\ twice the anomalous dimension of the ghost form factor. With the
specific value of $\delta=1/2$, these objects are indeed identical,
since $\delta=\frac{\gamma_d^0}{\beta_0}$. 

An alternative way to calculate the $\beta$ function from the
nonperturbative variational solutions in the Coulomb gauge is to
extract information from the heavy quark potential $V_C(k)$. Since the
nonperturbative running coupling can also be defined via $V_C(k)$, see
Eq.\ (\ref{couppot}), the
Coulomb form factor $f(k)$ must yield
\begin{eqnarray}
  \label{alphabyf}
  V_C(k)=\frac{d^2(k)f(k)}{k^2}\sim\frac{1}{\ln\frac{k^2}{\mu^2}}\; .
\end{eqnarray}
Making an ansatz for the Coulomb form factor $f(k)$ in the ultraviolet,
\begin{eqnarray}
  \label{fUVansatz}
  f(k)=\frac{C}{\ln^\varphi\frac{k^2}{\mu^2}}\; ,
\end{eqnarray}
therefore provides a sum rule for the logarithmic exponents $\delta$
and $\varphi$,
\begin{eqnarray}
  \label{dfsumrule}
  2\delta+\varphi=1\; .
\end{eqnarray}
Furthermore, the UV ansatz (\ref{fUVansatz}) for the form factor
$f(k)$ can be plugged into its DSE (\ref{fDSE}) to yield after UV
analysis\footnote{In the same manner as for the infrared analysis in
  section \ref{fsection}, the asymptotic evaluation of the DSE for
  $f(k)$ gives no additional condition on the (power or logarithmic)
  exponents. This is due to the fact the product
  $\frac{d^2(k)}{k^2\omega(k)}$ behaves like $\frac{1}{k}$ in the
  infrared and like $\frac{1}{k\ln k}$ in the ultraviolet, as a
  consequence of the ghost-gluon vertex' nonrenormalization.}
\bee
    g^2N_c\tZ_1^2AB^2\frac{1}{(4\pi)^2}\frac{4}{3\varphi}=1\; .
\eeq
By comparison of the above relation to the identity (\ref{coeffCoulUV}) from the UV
analysis of the ghost DSE, we infer that $\varphi=\delta$ and with the
sum rules (\ref{dfsumrule}) and (\ref{thelogsumrule}) this yields
\begin{eqnarray}
  \label{dfwathird}
  \delta=\varphi=\gamma=\frac{1}{3}\; .
\end{eqnarray}
One may now use Eq.\ (\ref{deltaresCoul}) from
the ghost-gluon vertex to calculate $\beta_0$ as a function of
$\delta$,
\begin{eqnarray}
  \label{beta0f}
    \beta_0^{\textrm{anti}}=
    \frac{1}{(4\pi)^2}\frac{4N_c}{3\delta_{\textrm{anti}}}=\frac{1}{(4\pi)^2}\frac{12N_c}{3}\; .
\end{eqnarray}
The occurrence of the anti-screening contribution $\frac{12N_c}{3}$ is
familiar from the consideration of the instantaneous potential, see section
\ref{antiscreen}. Apparently, the non-instantaneous part needs to be
taken into account, possibly in the manner proposed in Ref.\
\cite{CucZwa01}.

Apart from the technical difficulties to arrive at the correct value
for $\beta_0$ automatically, the above considerations showed that it
is possible to extract a certain ultraviolet behavior of the Coulomb
gauge Green functions, having \emph{imposed} the known value for
$\beta_0$. It may therefore be expected that, once approximations can
be relaxed, the ultraviolet behavior according to Eqs.\ (\ref{dwUV}) and (\ref{gammadeltares}), i.e.\
\begin{eqnarray}
  \label{UVres}
  \omega(k)\sim k\ln^{3/11}k\, ,\quad\, d(k)\sim \frac{1}{\ln^{4/11}k}\,  ,\quad
\end{eqnarray}
will be found numerically. Lattice calculations \cite{LanMoy04}
claimed a counter-intuitive UV power law $\omega\sim k^{3/2}$. The
data displayed in Fig.\ \ref{lattice} also indicate an UV
enhancement of $\omega(k)$. However, a stronger than linear rising of $\omega(k)$ might as well be an indication of the
logarithmic correction in (\ref{UVres}). The lattice calculations in
\cite{Qua+07} agree with such a statement. This should be investigated further.

\section{Infrared fixed point}
\label{IRfixedpoint}

\begin{figure}
\begin{center}
  \includegraphics[scale=1.3]{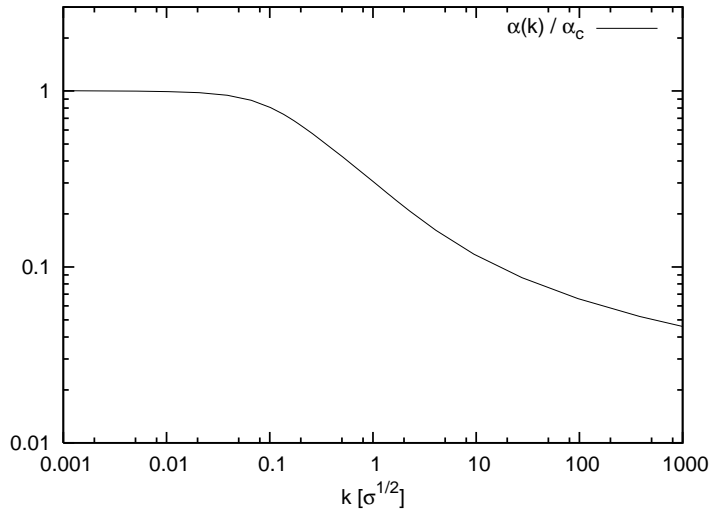}
\end{center}
\caption{The running coupling $\alpha(k)$ as defined by Eq.\
  (\ref{runncoupCoul}) with the variational solutions presented in
  chapter \ref{VarVac}.
\label{fig-alpha}}
\end{figure}

In this section we investigate the infrared behavior of the Coulomb
gauge running coupling as defined by Eq.\ (\ref{runncoupCoul}). Due to
the nonrenormalization property of the ghost-gluon vertex, the running
coupling attains a finite value in the infrared. The corresponding
$\beta$ function then has a fixed point in the infrared. This can be readily
seen by noting that the sum rule (\ref{sumrule}) for infrared exponents
yields
\begin{eqnarray}
  \label{nullexp}
1-2(2\alpha_G+1+\alpha_A)=0  
\end{eqnarray}
in the Coulomb gauge where $d=3$. Hence, with the infrared power laws (\ref{plawAnsatz})
with coefficients $2A$ for $\omega^{-1}(k)$ and $B$ for $d(k)$, we
find for the infrared limit of the (equal-time) running
coupling $\alpha(k)=\frac{g^2(k)}{4\pi}$ in Eq.\ (\ref{runncoupCoul})
\begin{eqnarray}
  \label{alphac}
   \alpha_c:=\lim_{k\to 0}\alpha(k)=\tZ_1^2g^2\frac{2AB^2}{4\pi}=\frac{1}{2\pi N_c I_G(\kappa)}\; ,
\end{eqnarray}
where we have made use of Eq.\ (\ref{coeffrule}) which relates the
infrared coefficients $A$ and $B$ to the function $I_G(\kappa)$
defined by Eq.\ (\ref{IGdef}).\footnote{A factor of $\tZ_1^2$ needs to
be reintroduced here, it was set to unity in the infrared analysis of
chapter \ref{ghostdom}.}
For the preferred solution with $\kappa=\frac{1}{2}$, we evaluate
$I_G(\kappa)$ to find
\begin{eqnarray}
  \label{alphac1}
  \left.\alpha_c\right|^{\kappa=\frac{1}{2}}_{d=3}=\frac{2\pi}{N_c}\; .
\end{eqnarray}

In Fig.\ \ref{fig-alpha}, the running coupling as given in Eq.\
(\ref{runncoupCoul}) is shown with the full numerical solutions for $d(k)$
and $\omega(k)$ given in chapter \ref{VarVac}. The numerical value of
$\alpha_c$ is in excellent agreement with Eq.\
(\ref{alphac1}).\footnote{In Ref.\ \cite{EppReiSch07}, the definition
  of the running coupling from Ref.\ \cite{FisZwa05} with a factor of
  $\frac{8}{3}$ in Eq.\ (\ref{runncoupCoul}) was used without
  adjustment of $\tZ_1$. Therefore, the value of $\alpha_c$ was found
  to be higher by that factor, i.e.\ $16\pi/(3N_c)$.} 

\begin{figure}
  \centering
  \inc[scale=1.4]{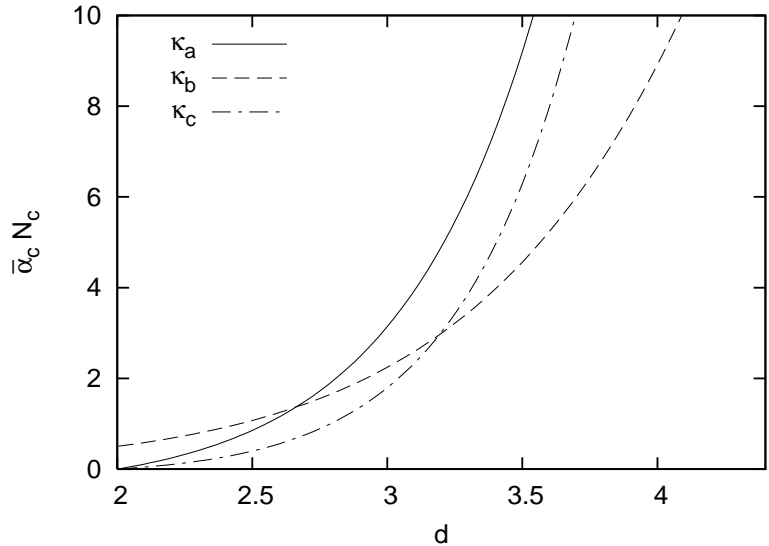}
  \caption{Infrared limit $\bar{\alpha}_c$ of the running coupling as function of $d$. The three solutions $\kappa_a$, $\kappa_b$ and $\kappa_c$ from Fig.\ \ref{kvond} are shown.}
  \label{figalphac}
\end{figure}

The other critical solutions for various spatial dimensions
$d$ and the respective solutions for $\kappa(d)$ yield different values
for the infrared limit of the running coupling. To study the
dependence on the dimension $d$, we use here the Landau gauge running
coupling $\bar{\alpha}(k):=\frac{g^2(k)}{4\pi}$ with $g^2(k)$ defined in Eq.\ (\ref{runncoupdeffinren}). The infrared
limit yields
\begin{eqnarray}
  \label{eq:alcL}
  \bar{\alpha}_c\, :=\, \lim_{k\to 0}\bar{\alpha}(k)=\frac{1}{4\pi
    N_cI_G(\kappa)}\; ,
\end{eqnarray}
cf.\  Eq.\ (\ref{alphac}). For the
three solution branches $\kappa_a(d)$, $\kappa_b(d)$ and $\kappa_c(d)$
shown in Fig.\ \ref{kvond}, the value of $\bar{\alpha}_c$ is shown in Fig.\
\ref{figalphac} as a function of the dimension $d$. It is seen that for large
$d$, $\bar{\alpha}_c(d)$ rises exponentially. For $d\to 2$, which
corresponds to the Landau gauge $1+1$ dimensional YM theory, the
solutions $\kappa_a(d)$ and $\kappa_c(d)$ yield $\bar{\alpha}_c(d=2)=0$
whereas for $\kappa_b(d)$ the infrared limit $\bar{\alpha}_c$ approaches a
finite value
\begin{eqnarray}
  \label{alphad2k02}
  \left.\bar{\alpha}_c\right|^{\kappa=\frac{1}{5}}_{d=2}\cdot N_c=\frac{2^{8/5}\Gamma^2(\frac{6}{5})\Gamma(\frac{13}{10})}{\sqrt{\pi}\:\Gamma(\frac{2}{5})\Gamma(\frac{4}{5})}N_c\approx 0.501N_c
\end{eqnarray}
The root $\kappa_b(d)$ does not show any qualitative change for
$d=2$ where all Landau gauge degrees of freedom are frozen. The solutions
$\kappa_a(d)$ and $\kappa_c(d)$, on the other hand, do show a
qualitative change at $d=2$. Granted
the approximations made, the latter solutions with  $\bar{\alpha}_c=0$ for
$d=2$ indicate what one may expect: the $1+1$ dimensional Landau gauge calls for a separate treatment.

\begin{figure}
  \centering
  \inc[scale=0.5]{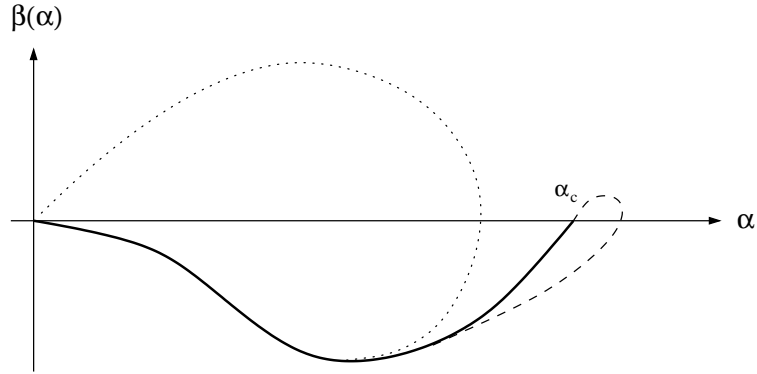}
  \caption{Sketch of the $\beta$ function. The solid line qualitatively
    describes the behavior of the running coupling as shown in Fig.\
    \ref{fig-alpha}. The dashed line shows the case where there is a small bump
    in the running coupling. With a dotted line, the $\beta$ function is
  shown for the subcritical solution.}
  \label{betafunc}
\end{figure}

Returning to the $d=3$ Coulomb gauge, it is reassuring to see that the running coupling in Fig.\ \ref{fig-alpha}
is a monotonic function. Otherwise the $\beta$ function would have
spurious zeros (a problem that occurs in the Landau gauge
\cite{FisAlk02}). The infrared finite value of the running coupling infers that the
$\beta$ function has an infrared fixed point for
$\frac{g^2}{4\pi}=\alpha_c$.\footnote{The infrared divergent running coupling (\ref{couppot}) from the heavy
quark potential would not exhibit this infrared fixed point.} In Fig.\
\ref{betafunc}, several cases of the $\beta$ function are sketched.
The subcritical solutions (section \ref{subcrit}), as opposed to the
critical solutions, do not have
a sum rule of infrared exponents since all form factors are infrared
finite. The running coupling $\alpha(k)$ should have an ultraviolet behavior
similar to the one shown in Fig.\ \ref{fig-alpha} since the form
factor are qualitatively the same in the UV. However, in the infrared one
finds
\begin{eqnarray}
  \label{alphasubcrit}
  \alpha(k\to 0)\sim k\; ,
\end{eqnarray}
that is the running coupling from the ghost-gluon vertex \emph{vanishes} in the infrared for
the subcritical solution. This behavior is unacceptable for a theory
that is strongly coupled in the infrared. In contrary to the $1+1$
dimensional theory, the $3+1$ dimensional case requires a
non-vanishing coupling strength for the infrared degrees of freedom. Moreover, if we hold onto
asymptotic freedom, the additional requirement (\ref{alphasubcrit}) yields a
multi-valued $\beta$ function. The interpretation of $\alpha(k)$ as a
running coupling is then questionable since the zeros of the $\beta$
function are not necessarily fixed points \cite{privSme}.\footnote{The
running coupling (\ref{couppot}) extracted from the Coulomb
potential would yield an infrared finite limit for the subcritical
solution and might serve as a more sensible definition in this case.} Aside from the discussion in section
\ref{subcrit}, this indicates that the subcritical solutions are unphysical. 

\subsection*{Vertex corrections}

There exists the possibility of nonperturbative corrections to the
running coupling. This might be interesting even for the extraction of
$\Lambda_{QCD}$ by comparison to experiment \cite{PDG04}. It was
suggested in Ref.\ \cite{CucMenMih04} to multiply the running coupling with
a contraction of the ghost-gluon vertex in a certain momentum
configuration. The infrared limit of the thus corrected running
coupling is denoted by
\begin{eqnarray}
  \label{alphaCucdef}
  \alpha_c^{(C)}=\lim_{k\to 0}\alpha(k) C^2\; .
\end{eqnarray}
The value $C$, defined by Eq.\ (\ref{Cdef2}), describes the dressing
of the vertex in the infrared limit where first the gluon momentum is
approached to zero and afterwards the ghost momentum is taken to be zero.
With the help of the result obtained in table \ref{Cresult} for
the number $C$, the corrections to $\alpha_c$ can be 
be specified. They are
shown in table \ref{alphavercorr} for the various solutions of
$\kappa(d)$ in question, in comparison to the uncorrected values $\alpha_c$.
Note that the solution $\kappa=\frac{1}{2}$ for $d=3$ is particular in
the sense that it is not corrected by the vertex contribution since
$C=1$.

\begin{table}[ht]
  \centering
  \begin{tabular}{c|c|c|c}
    $d$ & $\kappa$ & $\stackrel{{\scriptscriptstyle (-)}}{\alpha}_cN_c$ & $\stackrel{{\scriptscriptstyle (-)}}{\alpha}_c^{(C)}N_C$ \\
    \hline
     $3$ & $0.398$  & $4.488$ & $5.32$ \\
    $3$ & $\frac{1}{2}$ & $2\pi$ & $2\pi$ \\
    $4$ & $0.595$ & $8.915$ & $10.94$ 
  \end{tabular}
  \caption{Infrared finite values of the running coupling for the
    various infrared solutions, with and without vertex
    corrections. For $d=3$, we used the definition $\alpha_c$ in Eq.\
    (\ref{alphac}) whereas for $d=4$, $\bar{\alpha}_c$ from Eq.\
    (\ref{eq:alcL}) is used and reproduces the result in Ref.\
    \cite{LerSme02} without vertex correction.}
  \label{alphavercorr}
\end{table}

\bigskip

In concluding this chapter, we give a brief outlook on future
investigations with 
the nonperturbative running coupling in the Coulomb gauge. With an
improved ultraviolet behavior of the Green functions, see section
\ref{propsinUV}, it will be possible to adjust the scale in the
Hamiltonian approach to YM theory. This may be pursued by comparison
to experimental scattering amplitudes from deep inelastic
scattering. The nonperturbative feature of the variational solutions
makes it then possible to quantify other interesting objects numerically,
such as, e.g., the Coulomb string
tension, or correlators of the high-temperature phase of QCD.

%!!!!!!List of arguments for kappa=1/2,
%!!!!!!!asnswer the Qs of chapter 2!
\chapter{Inclusion of external charges}
\label{external}

In the Coulomb gauge, the Green functions were investigated in the
infrared limit in chapter \ref{ghostdom} and for the entire momentum
range with variational methods
in chapter \ref{VarVac}. These vacuum calculations include only
the interaction of gluons and ghosts, the coupling to the quark sector
is suppressed by setting $\rhoext^a(x)=0$. The translationally
invariant Green functions so obtained were discussed in the previous
chapters, and it was shown that the heavy quark potential is obtained
by a perturbation in $\rhoext^a(x)$. Although the crucial feature of
linear confinement is already given by the Green functions of the pure
glue theory, the back-reaction of the external charges on the wave
functional is lacking. One may expect that by extending the ansatz for
the wave functional in the presence of a quark-antiquark pair, for instance, the energy
can be lowered further by the formation of a flux tube connecting
these external charges. In this section, various approaches are
discussed by which the back-reaction of the external charges on the
gauge field sector can be studied.

\section{Gluons in a quasi-particle representation}
\label{sec:quasi}

The vacuum wave functional used in chapter \ref{VarVac} is of Gaussian
type. In analogy to the quantum harmonic oscillator where the Gaussian
ground state is annihilated by a linear combination of the coordinate
and momentum operators, the variational YM vacuum state $\ket{0}$ suggests the
introduction of an annihilation operator that yields
$a_k^a(x)\ket{0}=0$. Its hermitian conjugate $a_k^a(x)^\dagger$ may then be considered the
creation operator of a gluonic excitation at point $x$. By means of
these gluonic excitations, an orthonormal basis of
particle excitations can be constructed \cite{privRei}. In a normal order expansion of the
Yang--Mills Hamiltonian, the term $\omega a^\dagger a$ occurs so that
$\omega(k)$ represents part of the energy of a
gluonic excitation with momentum $k$. Since in the infrared, the
dispersion relation departs substantially from the behavior
$\omega(k)=k$ of a free particle, these particle excitations are
regarded as quasi-particles, incorporating nonperturbative effects.

Before we define the quasi-particle basis, recall that the choice
$\lambda=\frac{1}{2}$ for the wave functional in Eq.\ (\ref{Psi}),
\begin{eqnarray}
  \label{Psihalf}
  \Psi[A]=\braket{A}{0}
  =\cN\frac{1}{\sqrt{\cJ[A]}}\exp\left[-\frac{1}{2}\int
    d^3[xy]A_i^a(x)\omega(x,y)A_i^a(y)\right]\; ,
\end{eqnarray}
formally eliminates the Faddeev--Popov determinant $\cJ$ from
expectation values. The Coulomb gauge scalar product $\lla\:|\:\rra$ that comprises
$\cJ$ after gauge fixing can be related to a scalar product
$\lla\:|\:\rra_{\textrm{flat}}$ with a flat measure,
\begin{eqnarray}
  \label{flat}
  \bra{0}\cO[A,\Pi]\ket{0}&=& \int \cD A\,  \cJ[A]
  \Psi^*[A]\cO[A,\Pi]\Psi[A]\nn\\
  \quad &=& \int \cD A\,  \tP^*[A]\wtcO[A,\Pi]\tP[A]
  =: \bratvac\wtcO[A,\Pi]\kettvac_{\textrm{flat}}
\end{eqnarray}
if we identify
\begin{eqnarray}
  \label{tildedef}
  \tP[A]=\langle\,{A}\, |\, {\tilde{0}}\rangle=\sqrt{\cJ[A]}\Psi[A]\; ,\quad \wtcO[A,\Pi] =
  \sqrt{\cJ[A]}\cO[A,\Pi]\frac{1}{\sqrt{\cJ[A]}}\; .
\end{eqnarray}
With these manipulations, expectation values can be calculated
straightforwardly by means of
Wick's theorem. Note that the expectation values considered are not
taken in the auxiliary Gaussian state $\tP[A]$ but actually
in the state (\ref{Psihalf}). The Gaussian $\tP[A]$ is
suitable to define an annihilation operator
\begin{eqnarray}
  \label{adef}
  a_k^a(x)=\frac{1}{\sqrt{2}}\int d^3y\left(
\alpha^{-1}_{km}(x,y)A_m^a(y)+i\alpha_{km}(x,y)\Pi_m^a(y)
\right)
\end{eqnarray}
that yields\footnote{Strictly speaking, only the auxiliary vacuum $\kettvac$
  and not the true vacuum $\ket{0}$ is
  annihilated by the operator $a$. Employing the replacement
  $\ln\cJ=-\int A\chi A$ in the
  one-loop approximation, see Eq.\ (\ref{Jforchi}), a definition of
  the operator $a$ by means of $\Omega$ that annihilates $\ketvac$ could be pursued.}
\begin{eqnarray}
  \label{killvac}
  a_k^a(x)\kettvac=0\; .
\end{eqnarray}
Above, the function $\alpha_{ij}(x,y)$ is defined as the inverse
square root of the variational kernel $\omega$,
\begin{eqnarray}
  \label{alphaijdef}
  \alpha_{ij}(x,y)=t_{ij}(x)\alpha(x,y)\; ,\quad \int d^3y
  \alpha(x,y)\alpha(y,z)=\omega^{-1}(x,z)\; ,
\end{eqnarray}
and has the same dimension as the field operator $A$. Adopting the
canonical commutation relations\footnote{Here, we write the $\perp$
  explicitly in order to emphasize that all fields are
  transverse. The commutation relations (\ref{comrel}) that involve the
  field dependent operator $T_{ij}$ are not considered here.}
\begin{eqnarray}
  \label{cancomm}
  [A^{a\perp}_i(x),\Pi^{b\perp}_j(y)]=it^{ab}_{ij}(x,y)\; ,
\end{eqnarray}
the hermitian conjugate of Eq.\ (\ref{adef}) obeys
\begin{eqnarray}
  \label{aadagger}
  [a_i^a(x),a_j^{b}(y)^\dagger]=t_{ij}^{ab}(x,y)\; ,
\end{eqnarray}
and creates gluonic excitations by acting on the vacuum,
\begin{eqnarray}
  \label{excitdef}
  \ket{\{ x^{(k)}, a_k , i_k  \}}=\cN_n\prod_{k=1}^n
  {a_{i_k}^{a_k}(x^{(k)})}^\dagger\ket{\tilde 0}
\end{eqnarray}
which can be shown to be orthogonal. Normalization of the states
(\ref{excitdef}) by an appropriate
choice of $\cN_n$ may be issued, if needed, in a finite volume $V=\int
d^3x$. 

\section{Coherent states}

With the set of gluonic excitations formulated in the previous
section, we have a basis of the Hilbert space at our disposal in which a state can
be constructed taking the presence of external color charges into
account. This state, $\ket{Z}$, is here chosen as a coherent
superposition of the excited states (\ref{excitdef}), where a
localized function controls the excitation of gluons in between the external
color charges. This function, $Z_k^a(x)$, may be determined by the
variational principle, minimizing the expectation value of
$H'=H+H_{\textrm{ext}}$ with the external charges $\rhoext$ contained
in $H_{\textrm{ext}}$.

% \section{Harmonic oscillator in a magnetic field}

% In the following, the time-independent Schr\"odinger equation of
% quantum systems will be discussed. The quantum Hamiltonian $H[x,p]$ is given
% as a function of the coordinate operator $x$ and its conjugate
% momentum operator $p$, and
% is decomposed into $H=H_0+W$ where $H_0=a^\dagger\omega a+E_0$ is diagonal
% in the basis $\mathcal{B}$ spanned by $\{(n!)^{-1/2}(a^\dagger)^n\ketvac,\,
% n\in\bN\}$. The creation and annhiliation operators are functions of
% $x$ and $p$ and yield the commutation relation $[a,a^\dagger]=1$. We
% refer to the vacuum $\ketvac$ as the state that minimizes
% $\bravac H_0 \ketvac$ and satisfies the equation $a\ketvac =0$. It
% gives rise to an approximate vacuum energy $E_0=\bravac H\ketvac$. We
% define the coherent state $\ketZ$ as an arbitrary vector in the Hilbert
% space of $H_0$ and determine it by minimizing the expectation value
% $E_z = \braZ H\ketZ$ by the variational principle. In general, the
% coherent energy $E_z$ will be a lower bound to the vacuum
% energy, $E_z \leq E_0$.   

Coherent states were first introduced by Schr\"odinger \cite{Sch26} in 1926
and further established by Glauber \cite{Gla63b} in 1963 who laid the keystone
for developments in quantum optics and thus gained the 2005 physics
Nobel prize
 (along with Hall and H\"antsch). It will be instructive here to briefly discuss the simple
case of the three-dimensional harmonic oscillator and thus expose the main
features of the coherent state. The Hamiltonian reads
\bee
\label{3dharm}
H = \frac{{p}^2}{2m}+\frac{1}{2}m\omega^2{x}^2 % \nn \\
% &=&
% \label{harmYM}
% -\frac{1}{2m}\Delta_r+\frac{1}{2}m\omega^2{\fx}^2+\frac{1}{2mr^2}{\bf L}^2
% \nn\\
% &=&\omega(\ad_k a_k+\frac{3}{2})
=\omega\ad_k a_k+E_0\; ,
\eeq
where
\bee
\label{annidef}
a_k=\frac{1}{\sqrt{2}}\left(\frac{x_k}{\sigma}+i\sigma p_k\right)
\quad \textrm{and} \quad a_k^\dagger=\frac{1}{\sqrt{2}}\left(\frac{x_k}{\sigma}-i\sigma p_k\right)
\eeq
are linear combinations of $x_k$ and $p_k$ that diagonalize $H$ for $\sigma^{-1}=\sqrt{m\omega}$. Due to $[x_i,p_j]=i\delta_{ij}$, we
have $[a_i,\ad_j]=\delta_{ij}$ and the orthonormal states
$\ket{n}=\prod_k(n_k!)^{-1/2} (a_k^\dagger)^{n_k}\ketvac$ can be shown to yield the
equidistant spectrum $H\ket{n}=E_{n}\ket{n}$ with
$E_{n}= \omega(n_x+n_y+n_z+\frac{3}{2})$. The vacuum state
$\ketvac$ for ${n= {\bf 0}}$ has minimal energy $E_0 =\frac{3}{2}$ and is annihilated by
(\ref{annidef}), ${a}\ketvac = 0$. The solution to the latter
differential equation yields the vacuum wave function in coordinate
space,
\bee
\label{wavefkt}
\psi_0(x):=\braket{x}{0} =\frac{1}{\left(\pi
    \sigma^2\right)^{3/4}}\e^{-\frac{1}{2}x^2/\sigma^2}\; .
\eeq
%Some analogies of (\ref{3dharm}) to the Yang--Mills Hamiltonian in Coulomb gauge will be addressed
%below.
   
%\subsection*{Properties of the coherent state}
We now define the coherent state $\ket{z}$ by acting with a unitary
operator $U$ on the vacuum state $\ketvac$,
\bee
\label{zdef}
\ket{z}=U({z})\ketvac\; ,\quad   U(z)=\e^{z_k\ad_k-z_k^*a_k}\;  ,\quad
z_k\in\mathbb{C}\; .
\eeq
The coherent states so defined are normalized if $\ketvac$ is, but
they are not orthogonal. The continuous index $z$ makes the set of
coherent states over-complete in the given Hilbert space with a
countable basis. Further properties of the coherent state are listed below:\renewcommand{\labelenumi}{\alph{enumi})}
\begin{subequations}
\begin{enumerate}
\addtolength{\itemsep}{-0.9\baselineskip}
\item{eigenstate of annihilation operator}
  \begin{equation}
    \label{zprop1}
    a_k\ket{z}=z_k\ket{z}
  \end{equation}
\item{closure relation\footnote{The integral is taken over the complex
    planes of all components of $z$. By inserting the closure relation
    into $1=\braket{0}{0}$, the factor $\frac{1}{\pi^3}$ is cancelled
    by the integration of the polar angles in the complex planes.}}
  \begin{equation}
    \label{zprop2}
   1=\frac{1}{\pi^3}\int d^2z \ket{z}\bra{z}
  \end{equation}
\item{minimal uncertainty}
  \begin{eqnarray}
    \label{minunc}
    (\Delta x_k)(\Delta p_k) = \frac{1}{4}\quad \textrm{(no sum)}
  \end{eqnarray}
\item{Poissonian distribution}
  \begin{eqnarray}
    \label{Poissdistr}
    \left|\braket{n}{z}\right|^2=\prod_k\frac{|z_k|^2}{n_k!}\e^{-|z_k|^2}
  \end{eqnarray}
\item{shifted vacuum wave function}
  \label{shiftprop}
  \begin{eqnarray}
    \label{shiftstate}
    \braket{x}{z}\sim \big\langle x-\sqrt{2}\sigma\Re{z}\big | 0 \big\rangle
  \end{eqnarray}
\end{enumerate}
\end{subequations}
To see the property e) first note that the operator $U(z)$ in Eq.\ (\ref{zdef})
can be expressed by coordinate and momentum operators $x$ and $p$
using Eq.\ (\ref{annidef}),
\begin{eqnarray}
  \label{Ubyxp}
  U(z)=\e^{z_k\ad_k-z_k^*a_k}=\e^{iI_kx_k-iR_kp_k}\; ,
\end{eqnarray}
having abbreviated
\begin{eqnarray}
  \label{IandRdefqm}
  I_k:=\sqrt{2}\,\sigma^{-1}\Im(z_k)\; ,\quad
  R_k:=\sqrt{2}\,\sigma\Re(z_k)\; .
\end{eqnarray}
The well-known Campbell-Baker-Hausdorff formula
\bee
\label{CBH}
\ln\left(\e^X\e^Y\right)=X+Y+\frac{1}{2}[X,Y]+\frac{1}{12}[X,[X,Y]]-\frac{1}{12}[Y,[X,Y]]+\dots
\eeq
which infers $\e^{A+B}=\e^A\e^B\e^{-\frac{1}{2}[A,B]}$ for $[A,B]$ a
$c$-number, helps to rewrite the coherent state (\ref{zdef}) and
interpret it by a shifted vacuum wave function, 
\bee
\psi_z(x)&=&\braket{x}{z}=\e^{-\frac{1}{2}iI_kR_k}\e^{iI_kx_k}\e^{-R_k\partial^x_k}\psi_0(x)\nn\\
&=&\e^{-\frac{1}{2}iI_kR_k}\e^{iI_kx_k}\psi_0(x-R)\, .
\eeq
Here, we have used $\bra{x} ip_k
\ket{y}=\partial^x_k\delta^{(3)}(x-y)$ and assumed analyticity of $\psi_0(x)$.
That is, the wave function of the coherent state is, up to a
$x$-dependent phase, the vacuum wave function, evaluated at
$x-R$. This proves Eq.\ (\ref{shiftstate}). Equivalently, using
$\bra{p} ix_k \ket{q}=-\partial^p_k\delta^{(3)}(p-q)$, we find for the
momentum space wave function
\bee
{\psi}_z(p)=\braket{p}{z}=\e^{\frac{1}{2}iI_kR_k}\e^{-iR_kp_k}{\psi}_0(p-I)\; .
\eeq
Expectation values in the coherent state can thus be easily
related to vacuum expectation values, 
\bee
\bra{z}{\cal O}(x)\ket{z}&=&\int d^3x\:\psi_0^*(x-R){\cal O}(x)\psi_0(x-R)
\nn\\
&=&\int d^3x\:\psi_0^*(x){\cal O}(x+R)\psi_0(x)
\nn\\
&=&\bra{0}{\cal O}(x+R)\ket{0}
\eeq
For an operator $\cO(x,p)$ that depends on both $x$ and $p$, an
equivalent relation is derived. The most general form shall be
referred to as the \emph{shift rule}:
\bee
\label{shiftrule}
\bra{z}{\cal O}(x,p)\ket{z}&=&
\bra{0}\e^{-iI_kx_k}{\cal O}(x+R,p)\e^{iI_kx_k}\ket{0}\nn\\
&=&\bra{0}{\cal O}(x+R,p+I)\ket{0}
\eeq

The utility of the above identities can be presented in a simple
example. Consider the three-dimensional harmonic oscillator (\ref{3dharm}) and
construct coherent states as defined above, Eq.\ (\ref{zdef}). Now
think of the particles as charged and an external
electric field $\cE$ being switched on, $H\to H'=H-q\cE\cdot x$. A
prime in this chapter shall always indicate that external fields are
present. Let
$x=x_\parallel+x_\perp$ be split into a vector $x_\parallel$ parallel
to $\cE$ and one perpendicular to it (accordingly for $R$). The energy $E'(z)$ in the
coherent state is readily calculated using the shift rule (\ref{shiftrule}),
\bee
\label{Eznontr}
E'(z)&=&\bra{z}H'\ket{z} = E_0+\omega |z|^2-q\cE\cdot\bravac x+R
\ketvac\nn\\
&=&E_0+\frac{1}{2}\omega\sigma^2I^2+\frac{1}{2}\frac{\omega}{\sigma^2}R_\perp^2+\frac{1}{2}\frac{\omega}{\sigma^2}\left(R_\parallel-\frac{q\cE}{m\omega^2}\right)^2-\frac{q^2\cE^2}{2m\omega^2}\; .
\eeq
Minimizing $E'(z)$ clearly gives $I=R_\perp=0$ and $R_\parallel=\frac{q\cE}{m\omega^2}$. The wave
function $\psi_0(x)$ is therefore shifted to
$\psi_z(x)=\psi_0(x-\frac{q\cE}{m\omega^2})$. This happens to be the
exact vacuum wave function of $H'$, as seen directly by quadratic completion,
\bee
H'=\frac{p^2}{2m}+\frac{1}{2}m\omega^2x_\perp^2+\frac{1}{2}m\omega^2\left(x_\parallel -\frac{q\cE}{m\omega^2}\right)^2-\frac{q^2\cE^2}{2m\omega^2}
\eeq
The ground state energy (actually the entire spectrum) is shifted
to $E_0'=E_0-\frac{q^2\cE^2}{2m\omega^2}$.

In Yang--Mills theory, a less trivial situation is encountered. The
Hamiltonian $H$ is not diagonalized exactly, but by means of the
variational principle a quasi-particle basis can be set up. A
coherent state constructed in this basis is used to calculate the
expectation value of $H'$, and minimize the latter. We now discuss a
comparable quantum mechanic example. Let the anharmonic oscillator be
defined by
\begin{eqnarray}
  \label{Hanh}
  H_{\mathrm{anh.}}=\frac{{p}^2}{2m}+\frac{1}{2}m\omega^2{x}^2 +\lambda
  m^2\omega^3x^4\; ,\quad \lambda\in\R^+
\end{eqnarray}
may be a helpful quantum mechanic example. We also fail to diagonalize  this
Hamiltonian, but a Gaussian ansatz as
in Eq.\ (\ref{wavefkt})
with $\sigma$ as a variational parameter aids to find an approximate
ground state. Its width $\sigma$ can be shown to yield for 
\begin{eqnarray}
  \label{sigmares}
  \sigma^2=\frac{1}{30\lambda m
    \omega}\left(C^{1/3}+C^{-1/3}-1\right)\; ,\quad C=1350\lambda^2+30\sqrt{2025\lambda^4-3\lambda^2}-1
\end{eqnarray}
the minimal energy\footnote{A lowest-order expansion for small $\lambda$
  agrees with Ref.\ \cite{BenWu69} for the one-dimensional
  case. Particular care is required concerning the Riemann sheets when
taking the $\lambda\to 0$ limit.}
\begin{eqnarray}
  \label{E0}
  E_0=\frac{3}{4m\sigma^2}+\frac{3}{4}m\omega^2\sigma^2+\frac{15}{4}\lambda m^2\omega^3\sigma^4\; .
\end{eqnarray}

\begin{figure}
  \centering
  \inc[scale=1.0]{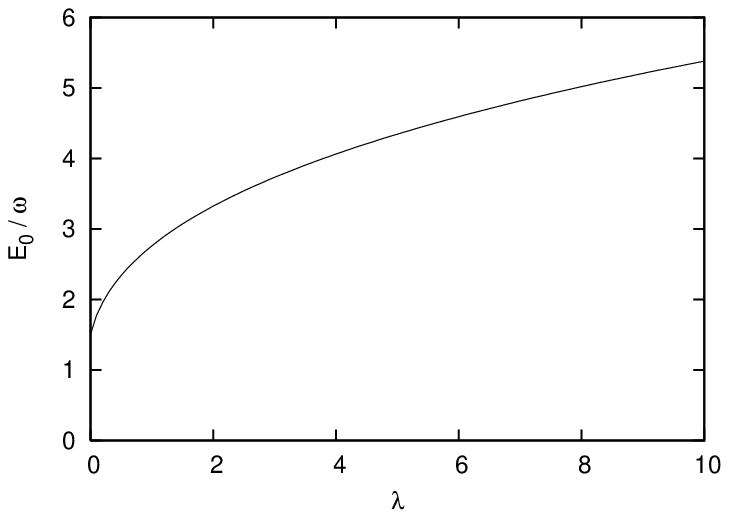}  \inc[scale=1.0]{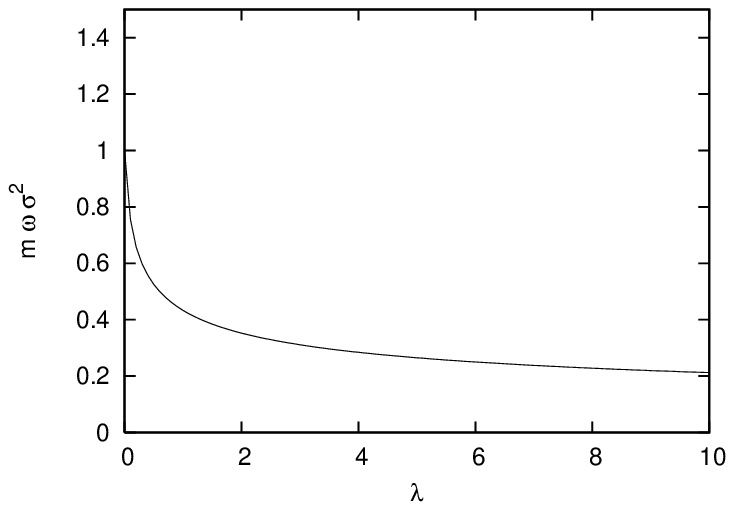}
  \caption{Variational solution for the ground state of the anharmonic oscillator. Left: Energy $E_0$ as a function of the parameter $\lambda$ of the quartic potential. Right: Width $\sigma$ of the Gaussian ground state, as a function of $\lambda$.}
  \label{anharmooszi}
\end{figure}

As can be seen in Fig.\ \ref{anharmooszi}, the Gaussian ground state wave
functional becomes narrower for increasing the coefficient $\lambda$
of the quartic potential, while the energy increases. Note that for
$\lambda\to 0$, the results for the harmonic oscillator are
recovered. For a comparison to field theory, we discuss the
variational solution to the $\phi^4$ theory in the Hamiltonian approach in appendix \ref{Appp4}.

Similarly to the variational approach to YM theory, we have found an
approximate Gaussian ground state and can create quanta by the
creation operator in Eq.\ (\ref{annidef}) with the Gaussian width $\sigma$ given by Eq.\ (\ref{sigmares}).
Now switch on an external electric field,
\begin{eqnarray}
  \label{Hstr}
  H'_{\mathrm{anh.}}=H_{\mathrm{anh.}}-q\cE \cdot x
\end{eqnarray}
and calculate the energy in the coherent state. The result gives the
ground state energy $E_0$, see Eq.\ (\ref{E0}), and a correction
dependent on $z$,
\begin{eqnarray}
  \label{Estrz}
  E'(z)&=&\bra{z}H'_{\mathrm{anh.}}\ket{z}\nn\\
&=&E_0+\frac{I^2}{2m}+ \lambda m^2\omega^3\sigma^4R^4+
  \frac{1}{2}m\omega^2(1+ 18 \lambda m\omega\sigma^2)R^2
  -q \cE\cdot R
\end{eqnarray}
For $z=0$, the energy $E'(z)$ is
equal to $E_0$, as expected. Minimizing $E'(z)$ w.r.t.\ $z$ yields
some value $z^{\textrm{min}}$. The result will depend on the parameter $\lambda$ of
the quartic potential and the electric force $F=-q\cE$. Without giving
the result explicitly, we can see from Eq.\ (\ref{Estrz}) that for all
values of $\lambda>0$ and $F\neq 0$ there exists a $z^{\textrm{min}}=R_\parallel^{\textrm{min}}\neq 0$ such that
$E'(R_\parallel^{\textrm{min}})<E_0$, i.e.\ the energy is lowered in the coherent state.

We now turn back to the original task, writing down a coherent state
in the quasi-particle basis of Yang--Mills theory. The over-complete set of 
states, here given by Eq.\ (\ref{excitdef}), incorporates already the
nontrivial effects of the Hamiltonian $H$ (\ref{CLHam}) without external
charges, similar to the treatment of the quartic quantum mechanic
potential above. Using these states, the coherent state for
Yang--Mills theory is introduced here by
\begin{eqnarray}
  \label{ketZdef}
  \kettZ=\tU\kettvac
\end{eqnarray}
with
\begin{eqnarray}
  \label{Ucohdef}
  \tU=\exp\left\{\int d^3x
    \left(Z_k^a(x)a_k^a(x)^\dagger-Z_k^a(x)^*a_k^a(x)\right)\right\}\; ,
\end{eqnarray}
where the field $Z_k^a(x)$ controls the excitations of gluons. By
means of Eq.\ (\ref{adef}), one can rewrite $U$ in terms of field operators,
\begin{eqnarray}
  \label{UcohbyIR}
  \tU=\exp\left\{\int d^3x
    \left(iI_k^a(x)A_k^a(x)-iR_k^a(x)\Pi_k^a(x)\right)\right\}\; ,
\end{eqnarray}
where the abbreviations
\begin{subequations}
  \label{IandRdef}
\begin{eqnarray}
  R_j^a(x)&=&\sqrt{2}\int d^3x'
\alpha_{jk}(x,x')\Re\left(Z^a_k(x')\right)
\\
I_j^a(x)&=&\sqrt{2}\int d^3x'
\alpha^{-1}_{jk}(x,x')\Im\left(Z^a_k(x')\right)
\end{eqnarray}
\end{subequations}
were used. The
expectation value of the Hamiltonian with charges,
$H'=H+H_{\textrm{ext}}$, in the state (\ref{ketZdef}) will be calculated
in the next section. Optimistically, we may hope that in analogy to
the quantum mechanical example, a nontrivial solution for a coherent
state exists that lowers the energy.

We provide here the definition of the coherent state energy,
\begin{eqnarray}
  \label{cohEdef}
  E[Z]:=\bratZ\tH\kettZ_{\textrm{flat}}=\int \cD A\,
  \left(\tU\tP\right)^*\tH\left(\tU\tP\right)\; .
\end{eqnarray}
It is important to clarify which state it refers to as a Coulomb gauge
expectation value of $H$ (not $\tH$). Carefully treating
operator ordering, the energy (\ref{cohEdef}) can be rewritten as a Coulomb gauge
expectation value,
\begin{eqnarray}
  \label{EbyZ}
  E[Z]&=&\int \cD A\, \tP^*\tU^\dagger\tH\tU\tP\nn\\
  &=&\int \cD A\, \left(\sqrt{\cJ}\Psi^*\right)
  \left(\frac{1}{\sqrt{\cJ}}U^\dagger\sqrt{\cJ}\right)
  \left(\sqrt{\cJ}H\frac{1}{\sqrt{\cJ}}\right)
  \left(\sqrt{\cJ}U\frac{1}{\sqrt{\cJ}}\right)
  \left(\sqrt{\cJ}\Psi\right)\nn\\
  &=&\int \cD A\,  \cJ
  \Psi^*\left(\frac{1}{\cJ}U^\dagger{\cJ}\right)HU\Psi\nn\\
  &=&\int \cD A\,  \cJ \left(U\Psi\right)^*H\left(U\Psi\right)\nn\\
  &\equiv&\bra{U\Psi}H\ket{U\Psi}
\end{eqnarray}
It is seen that by calculating $E[Z]$ via the auxiliary expectation
value (\ref{cohEdef}) with a flat measure, the Coulomb gauge expectation value of
the Hamiltonian $H$ is taken in the state
\begin{eqnarray}
  \label{Znotilde}
  \ket{Z}=U\ket{0}\; ,
\end{eqnarray}
cf.\ Eq.\ (\ref{ketZdef}). The corresponding wave functional is obtained by acting with
$U$ on the vacuum wave functional
$\Psi[A]$. With $\tU$ given by Eq.\ (\ref{Ucohdef}), a
Campbell-Baker-Hausdorff formula is needed to find
$U=\cJ^{-1/2}\tU\cJ^{1/2}$. From Eq.\ (\ref{CBH}), it can be derived
that for $[X,[X,Y]]$ and $[Y,[X,Y]]$ $c$-numbers,
\begin{eqnarray}
  \label{CBH2}
  \e^{Y}\e^X\e^{-Y}=\e^{X-[X,Y]-\frac{1}{2}[Y,[X,Y]]}\; .
\end{eqnarray}
Using this relation, and the one-loop Gaussian approximation for the
Faddeev--Popov determinant (\ref{Jforchi}), we find after some
algebra
\begin{eqnarray}
  \label{UbytU}
  U=\e^{-\frac{1}{2}\int R\chi R+\int A\chi R}\tU\; .
\end{eqnarray}

\section{Energy of Yang--Mills theory in the coherent state}
\label{energycoh}

In this section, the calculation of the energy $E'[Z]$ in the coherent state
is presented. First of all, let us clarify how
this energy is defined. The coherent state (\ref{Znotilde}) is used to calculate the
expectation value of the Yang--Mills Hamiltonian (\ref{CLHam}) in the presence of
external charges $\rhoext^a(x)$. Without external charges, the
Hamiltonian reads
\begin{eqnarray}
  \label{Hnomatter}
  H=H_k+H_p+H_C^{(0)}
\end{eqnarray}
with the definitions of the kinetic part,
\begin{eqnarray}
  \label{Hkdef}
  H_k=\frac{1}{2}\int d^d x\, \frac{1}{\cJ[A]}\,\Pi_k^a(x)\,\cJ[A]\,\Pi_k^a(x)
\end{eqnarray}
the magnetic potential part,
\begin{eqnarray}
  \label{Hpdef}
  H_p=\frac{1}{4}\int d^dx \, F_{ij}^a(x)F_{ij}^a(x)
\end{eqnarray}
and the Coulomb part $H_C$ of the Hamiltonian which involves a contribution
coming from dynamical gluonic charges,
\begin{eqnarray}
  \label{HC0def}
  H_C^{(0)}=\frac{g^2}{2}\int d^d[xy]\,\frac{1}{\cJ[A]}\,\rhodyn^a(x)\,\cJ[A]\, F^{ab}(x,y)\,\rhodyn^b(y)\; .
\end{eqnarray}
In the presence of external charges, the total charge distribution
$\rho^a(x)=\rhodyn^a(x)+\rhoext^a(x)$ adds extra terms to the Coulomb
part $H_C$, so that the Hamiltonian under consideration reads
\begin{eqnarray}
  \label{Hwm}
  H'=H+H_C^{(1)}+H_C^{(2)}
\end{eqnarray}
with 
\begin{eqnarray}
  \label{HC1def}
  H_C^{(1)}=\frac{g^2}{2}\int d^d[xy]\left(
\frac{1}{\cJ[A]}\rhodyn^a(x)\cJ[A]\: F^{ab}(x,y)\rhoext^b(y)
+\rhoext^a(x) F^{ab}(x,y)\rhodyn^b(y)
\right)
\end{eqnarray}
and
\begin{eqnarray}
  \label{HC2def}
  H_C^{(2)}=\frac{g^2}{2}\int d^d[xy]\rhoext^a(x)
  F^{ab}(x,y)\rhoext^b(y)\; .
\end{eqnarray}
The energy in the presence of external charges in the coherent state
$\ket{Z}$ is defined by
\begin{eqnarray}
  \label{E'def}
  E'[Z]=\bra{Z}H'\ket{Z}\; ,
\end{eqnarray}
where this expectation value involves after Coulomb gauge fixing the
Faddeev--Popov determinant $\cJ$. For the calculation of $E'[Z]$, it
will be extremely helpful to employ the shift rule as it was written
down in Eq.\ (\ref{shiftrule}) for the quantum mechanic
oscillator. Here, we can find the equivalent ``shifting effect'' of
the operator $\tU$, given in Eq.\ (\ref{Ucohdef}), on the operator
$\tH'$ (the use of the tilde was explained in Eq.\ (\ref{tildedef})),
\begin{eqnarray}
  \label{E'flat}
  E'[Z]&=&\bratZ \tH'[A,\Pi]\kettZ_\flach\nn\\
  &=&\bratvac \tH'[A+R,\Pi+I]\kettvac_\flach
\end{eqnarray}
In the calculations below, we always use the transformations
(\ref{tildedef}), ending up with the flat scalar product
$\lla\:|\:\rra_{\textrm{flat}}$. For brevity, the notation
\begin{eqnarray}
  \label{flatbrief}
  \langle \wtcO \rangle_\omega = \bratvac \wtcO \kettvac_\flach\; ,\quad  \langle \wtcO \rangle_Z = \bratZ \wtcO \kettZ_\flach
\end{eqnarray}
is used. Furthermore, recall the matrix notation in coordinate space
from chapter \ref{ch:YMconstr}. It will sometimes be convenient in this
chapter as well. The spatial dimension $d$ is left unspecified.

The energy $E'[Z]$ can be expressed by the results from chapter
\ref{VarVac} if we set $Z=0$,
\begin{eqnarray}
  \label{E'Z0}
  E'[0]=E+V_C\; .
\end{eqnarray}
Here, $E$ is the vacuum energy displayed below Eq.\ (\ref{Hvev}) and
$V_C$ is the Coulomb potential (\ref{potential}) in the vacuum
state. In Eq.\ (\ref{E'Z0}), there are no contributions coming from
$H_C^{(1)}$,
\begin{eqnarray}
  \label{nomixedterms}
  E_C^{(1)}[0]&=&\langle \tH_C^{(1)} \rangle_\omega\nn\\
&=&\frac{1}{2}\langle \trhodyn^{\,\dagger} F\, \rhoext + \rhoext F \,\trhodyn \rangle_\omega\nn\\
&=&\frac{g^2}{2}\lla \big( i\hat AQ\big)^* F \rhoext + \rhoext F
\big( i\hat AQ\big) \rra_\omega\nn\\
&=&0
\end{eqnarray}
We will see that in the coherent state the mixed term $H_C^{(1)}$
does have a contribution to the energy.
Let us emphasize that it is
instructive to stick to the coordinate space representation of the
Green functions. The
transformation to momentum space is helpful in the vacuum where thus
an overall infinite factor $(2\pi)^d\delta^d(0)$ can be extracted from the
energy, due to translational invariance. The coherent state, on the
other hand, accounts for localized external color charges and translational
invariance is lost. Therefore, an overall volume factor cannot be
expected and the Fourier transformation provides no simplification whatsoever.

Let us turn to the calculation of the kinetic energy in the coherent
state. The kinetic part $\tH_k$ in Eq.\ (\ref{Hkdef}) comprises the operator
\begin{eqnarray}
  \label{tPidef}
  \tPi_k^a(x)[A,\Pi]=\Pi_k^a(x)-\frac{1}{2i}\frac{\delta\ln\cJ[A]}{\delta A_k^a(x)}
\end{eqnarray}
that may be understood as a function of $A$ and $\Pi$. Applying the
shift rule (\ref{E'flat}), we obtain
\bee
\label{Ekshift}
E_k[Z]&=&\lla H_k[A,{\tPi}[A,\Pi]] \rra_Z = \lla
H_k[A+R,{\tPi}[A+R,\Pi+I]] \rra_\omega \nn\\
&=&\frac{1}{2}\int d^dx\int\cD
A\left|{\tPi}^a_i(x)[A+R,\Pi+I]\tP[A]\right|^2\; .
\eeq
With the explicit form (\ref{tildedef}) of the Gaussian vacuum wave functional
$\tP[A]=\langle A\kettvac$, we use
\bee
\label{defQ}
\tPi_k[A+R,\Pi]\tP[A]=iQ_k[A;R]\tP[A]
\eeq
 to define
\bee
\label{resQ}
Q_k^a(x)[A;R]=\int d^dx'\omega_{kj}^{ab}(x,x')A_j^b(x')+\frac{1}{2}\frac{\delta\ln\cJ[A+R]}{\delta
  A^a_k(x)}
\eeq
with the obvious notation
$\omega_{ij}^{ab}(x,y)=\delta^{ab}t_{ij}(x)\omega(x,y)$. The kinetic
energy in the coherent state thus yields schematically
\begin{eqnarray}
  \label{EkZsch}
  E_k[Z]=\frac{1}{2}\Tr\left(\omega\lla AA\rra\omega+\omega\lla A\frac{\delta\ln\cJ}{\delta
      A}\rra+\frac{1}{4}\lla\frac{\delta\ln\cJ}{\delta
      A}\frac{\delta\ln\cJ}{\delta A}\rra\right)+\frac{1}{2}\int d^dx I^2(x)
\end{eqnarray}
which we calculate, following Ref.\ \cite{FeuRei04}, with a quadratic
completion. To this end, define the curvature $\chi_{{}_Z}$ by
\begin{eqnarray}
  \label{chiZdef}
  ({\chi_{{}_Z}})_{ij}^{ab}(x,y)&:=&-\frac{1}{2}\lla
  \frac{\delta^2\cJ[A+R]}{\delta A_i^a(x)\delta A_j^b(y)}
  \rra_\omega\nn\\
  &=&+\frac{1}{2}\Tr\lla G[A+R] \Gamma_i^{0,a}(x) G[A+R]
  \Gamma_j^{0,b}(y) \rra_\omega 
\end{eqnarray}
With the definition of the ghost propagator $D_G^Z$ with a coherent
background field,
\begin{eqnarray}
  \label{DGZdef}
  D_G^{Z\, ab}(x,y)=\lla G[A]\rra_Z^{ab}(x,y)\; , 
\end{eqnarray}
the curvature (\ref{chiZdef}) can be expressed by
\begin{eqnarray}
  \label{chiZ}
  ({\chi_{{}_{Z}}})_{ij}^{ab}(x,y) = \frac{1}{2}\Tr \, D_G^Z\,
  \Gamma_i^{Z,a}(x) \, D_G^Z  \, \Gamma_j^{0,b}(y)\; .
\end{eqnarray}
This implicitly establishes a definition of the proper ghost-gluon vertex
$\Gamma_i^{Z,a}(x)$ in the coherent state; the latter will in actual
calculations be rendered tree-level, as before. Now, the second term
in Eq.\ (\ref{EkZsch}) can be written using a partial integration as
\begin{eqnarray}
  \label{AdlnJ}
  \lla A_i^a(x) \frac{\delta\ln\cJ[A+R]}{\delta A_j^b(y)} \rra_\omega
  &=& \frac{1}{2}\int d^dz (\omega^{-1})_{ik}^{ac}(x,z)\lla \frac{\delta^2\cJ[A+R]}{\delta A_k^c(z)\delta A_j^b(y)}
  \rra_\omega\nn\\
  &=& -\int d^dz (\omega^{-1})_{ik}^{ac}(x,z) (\chi_{{}_{Z}})_{kj}^{cb}(z,y)
\end{eqnarray}
In the one-loop approximation, one can realize by expansion of the
Faddeev--Popov operators that the third term in Eq.\ (\ref{EkZsch}),
$\lla \frac{\delta\ln\cJ}{\delta A}\frac{\delta\ln\cJ}{\delta A}\rra$,
is simply obtained by writing the first two terms as a modulo
square. With $Q=Q[A;R]$ as in Eq.\ (\ref{resQ}), one thus finds
\begin{eqnarray}
  \label{QQ}
  \lla Q^a_i(x) Q^b_j(y)\rra_\omega = \frac{1}{2}\int
  d^d[uv][\omega-\chi_{{}_Z}]^{ac}_{ik}(x,u)\, \omega^{-1}(u,v)[\omega-\chi_{{}_Z}]^{cb}_{kj}(v,y)
\end{eqnarray}
and is ready to write down the result for the kinetic energy,
\begin{eqnarray}
  \label{EkZres}
  E_k[Z]=\frac{1}{4}\int d^d[xyz][\omega-\chi_{{}_Z}]^{ab}_{ij}(x,y)\,
  \omega^{-1}(y,z)[\omega-\chi_{{}_Z}]^{ba}_{ji}(z,x)+ \frac{1}{2}\int
  d^dx I^a_k(x) I^a_k(x)\; .
\end{eqnarray}
By setting the coherent field to zero, $Z=0$, the vacuum result (\ref{Ek}) is recovered.

We now proceed to the Coulomb part $H_C$ of the Hamiltonian where
Eq.\ (\ref{HC0def}) amounts to the Coulomb interaction of gluonic charges,
\begin{eqnarray}
  \label{trhodyn}
  \trhodyn^{\, a}(x)[A,\Pi]=\hat A_k^{ab}(x)\tPi_k^b(x)[A,\Pi]
\end{eqnarray}
Taking the expectation value of such operators, relation (\ref{defQ})
can be used. Noting that $E_C^{(0)}[Z]=\big\langle
\tH_C^{(0)}[A,{\Pi}] \big\rangle_Z$ is a modulo square, the imaginary terms can
be explicitly eliminated to find
\bee
\label{EC0uno}
E_C^{(0)}[Z]&=&\frac{g^2}{2}\int d^d[xy]\lla (\hat A + \hat
R)^{ab}_k(x)I^b_k(x)F^{ac}(x,y)[A+R] (\hat A + \hat R)^{cd}_j(y)I^d_j(y)
\rra_\omega \nn \\
&&\hspace{-1cm}+\,\frac{g^2}{2}\int d^d[xy]\lla (\hat A + \hat
R)^{ab}_k(x)Q^b_k(x)F^{ac}(x,y)[A+R] (\hat A + \hat R)^{cd}_j(y)Q^d_j(y)
\rra_\omega
\eeq
with $Q=Q[A;R]$ as given in Eq.\ (\ref{resQ}).
To $2$-loop order in the
energy, there are no contractions with the operator $F$.
On the assumption that $D_G^Z$ is symmetric in color and coordinate
space separately (just like $D_G$), we have 
\begin{eqnarray}
  \label{Qexp0}
  \lla Q_k(x)[A;R]
  \rra_\omega=\frac{1}{2}\mathrm{Tr}\:D_G^Z\Gamma_k^{0}(x)= 0\; .
\end{eqnarray}
The non-vanishing contractions in Eq.\ (\ref{EC0uno}) then are
\bee
%\contraction{E_C^{(0)}[Z]=\frac{g^2}{2}\lk\lla}{\hat A}{Q F }{\hat A Q \big\rangle_\omega \rk}
\left\{ \big\langle\wick{12}{<1A <2Q F >1A >2Q} \big\rangle\, ,\quad \big\langle\wick{12}{<1A <2Q F >2A >1Q} \big\rangle \, ,\quad RI
\big\langle F\big\rangle RI \, ,\quad \big\langle\wick{1}{<1A I F >1A I} \big\rangle\, ,\quad \big\langle\wick{1}{R <2Q F R >2Q}
\big\rangle\right\}\nn\; .
\eeq

A careful treatment of all indices gives\footnote{The vacuum result
  (\ref{EC}) can be recovered for $Z=0$ by using $\mathrm{tr}\lk \hat T^a
  \hat T^a\rk=-N_c(N_c^2-1)$.}
\bee
E_C^{(0)}[Z]&=&-\frac{g^2}{8}\int d^d[xy] (\hat T^g)^{bb'}
F_Z^{b'd'}(x,y)(\hat T^h)^{d'd}\nn\\
&&\left\{\lk\omega^{-1}\rk^{gh}_{ij}(x,y)\int d^d[uv][\omega-\chi_{{}_Z}]^{be}_{ik}(x,u)\omega^{-1}(u,v)[\omega-\chi_{{}_Z}]^{ed}_{kj}(v,y)\right.
\nn\\
&& + \lk t_{ij}^{gd}(x)\delta^d(x,y)-\int
d^dw\,\omega^{-1}(x,w)(\chi_{{}_Z})^{gd}_{ij}(w,y)\rk\nn\\&&\quad\left. \lk t_{ij}^{bh}(x)\delta^d(x,y)-\int
d^dw\,\omega^{-1}(x,w)(\chi_{{}_Z})^{bh}_{ij}(w,y)\rk\right\} \\
&& + \frac{g^2}{2}\int d^d[xy] \hat
R^{ab}_k(x)I^b_k(x)F_Z^{ac}(x,y)  \hat R^{cd}_j(y)I^d_j(y)\nn\\
&& -\frac{g^2}{4}\int d^d[xy]\omega_{ij}^{-1}(x,y)\mathrm{tr}\lk\hat
  I_i(x)F_Z(x,y)\hat I_j(y)\rk\nn\\
\label{resEC0}
&& -\frac{g^2}{4}\int d^d[xyuv][\omega-\chi_{{}_Z}]^{ac}_{ik}(x,u)\omega^{-1}(u,v)[\omega-\chi_{{}_Z}]^{cb}_{kj}(v,y)\hat
R_i^{ad}(x)F_Z^{de}(x,y)\hat R_j^{eb}(y)\nn
\eeq
where we assigned
\begin{eqnarray}
  \label{FZdef}
  F_Z^{ab}(x,y):=\big\langle F^{ab}(x,y)[A]\big\rangle_Z \; .
\end{eqnarray}
The expectation value of the dynamical charge $\rhodyn=-\hat{{\bf
    A}}\cdot {\bf \Pi}$ in the coherent state is in view of Eq.\ (\ref{Qexp0})
\begin{eqnarray}
  \label{rhodynexp}
  \lla \trhodyn^{\, a}(x)[A,\Pi]\rra_Z=-\hat R^{ab}_k(x)I^b_k(x)\; .
\end{eqnarray}
Hence, the mixed terms $\:\sim\rhodyn F \rhoext$ in the Coulomb energy
yield without contractions of the Coulomb operator $F$
\begin{eqnarray}
  \label{EC1}E_C^{(1)}[Z]=-g^2\int d^d[xy]\hat R^{ab}_k(x)I^b_k(x)
F_Z^{ac}(x,y)\rho_{\mathrm{ext}}^c(y)\; .
\end{eqnarray}
The Coulomb energy of solely external charges obviously yields
\bee
\label{EC2}
E_C^{(2)}[Z]=\frac{g^2}{2}\int d^d[xy]\rho_{\mathrm{ext}}^a(x)
F_Z^{ab}(x,y)\rho_{\mathrm{ext}}^b(y)\; .
\eeq

The magnetic potential energy in the coherent state is yet to be calculated. With the
Hamiltonian given by Eq.\ (\ref{Hpdef}), we find that the vacuum energy
$E_p[0]$ as given in Eq.\ (\ref{Ep}) is found in addition to terms
involving the coherent field,
\begin{eqnarray}
  \label{resEp}
  E_p[Z]&=&E_p[0]\nn\\&&+\,\frac{1}{4}\, g^2\int d^dx\,
f^{eab}f^{ecd}R^a_i(x)R^b_j(x)R^c_i(x)R^d_j(x)\nn\\
&&
+\, g\int d^dx\,f^{abc}\lk\partial_iR^a_j(x)\rk R^b_i(x)R^c_j(x)\nn\\&&
+\, \frac{1}{2}\int d^dx\, {R}^a_i(x)\lk -\partial^2\rk R_i^a(x)\nn\\&&
+\, \frac{1}{4}\, g^2N_c\frac{(d-1)^2}{d}\omega^{-1}(0,0)\int d^dx \lk R_k^a(x)\rk^2
\end{eqnarray}
All but the last term, call it $E_p^{\textrm{div}}[Z]$, are quite trivially obtained using the shift
rule (\ref{flatbrief}). Since it is divergent, let us look
at it more carefully. It comes from the quartic part of $H_p$ after
applying the shift rule and contracting two gauge fields $A$, leaving
two coherent fields $R$,
\bee
 E_p^{\textrm{div}}[Z]&=&\frac{1}{4}\, g^2\int d^dx\left\{\big\langle A_i(x)\hat R_j(x)\hat
R_i(x)A_j(x)\big\rangle_\omega+ \big\langle R_i(x)\hat A_j(x)\hat
A_i(x)R_j(x)\big\rangle_\omega\right.\nn\\
&& +\, \big\langle A_i(x)\hat R_j(x)\hat A_i(x)R_j(x)\big\rangle_\omega + \big\langle R_i(x)\hat A_j(x)\hat
 R_i(x)A_j(x)\big\rangle_\omega\nn\\
&& +\,  \big\langle A_i(x)\hat A_j(x)\hat
R_i(x)R_j(x)\big\rangle_\omega \left. + \, \big\langle R_i(x)\hat R_j(x)\hat A_i(x)A_j(x)\big\rangle_\omega\right\}
\eeq
Using the symmetry of the vacuum gluon propagator, we get
\bee
 E_p^{\textrm{div}}[Z]&=&\frac{1}{4}\, g^2\int d^dx\,\left\{2\, \big\langle A_i(x)\hat R_j(x)\hat
R_i(x)A_j(x)\big\rangle_\omega-2\, \big\langle A_i(x)\hat R_j(x)\hat
R_j(x)A_j(x)\big\rangle_\omega\right\}\nn\\
&=&\frac{1}{4}\, g^2\int d^dx\,\Big\{\left.\lk
    t_{ij}(x)\omega^{-1}(x,y)\rk\right|_{y=x}\mathrm{tr}\hat
  R_j(x)\hat R_i(x)
\nn\\&& \qquad\qquad\qquad  
- \left.\lk
    t_{ii}(x)\omega^{-1}(x,y)\rk\right|_{y=x}\mathrm{tr}\hat
  R_j(x)\hat R_j(x) \Big\}\; .
\eeq
Using a Fourier transform for $\omega^{-1}(x,y)$ and $\mathrm{tr}\big( \hat
T^a \hat T^b\big)=-N_c\,\delta^{ab}$, we recover the proposed expression in
the last line of Eq.\ (\ref{resEp}):
\bee
\label{Epdivres}
 E_p^{\textrm{div}}[Z]&=&\frac{1}{4}g^2\mathrm{tr}\lk{\hat T^a
  \hat T^b}\rk\nn\\&&\hspace{-1cm}\int d^dx\left\{\lk\int \dbar^dp
\,  t_{ij}(p)\omega^{-1}(p)\rk R^a_j(x)\hat R^b_i(x)-
  \lk\int \dbar^dp\, t_{ii}(p)\omega^{-1}(p)\rk
  R^a_j(x)\hat R^b_j(x) \right\}\nn\\
&=&\frac{1}{4}g^2N_c\int \dbar^dp\, \omega^{-1}(p)\int
d^dx\:R_i^a(x)\lk\delta_{ij}(d-1)-\delta_{ij}\frac{d-1}{d}\rk
R_i^a(x)\nn\\        
&=&\frac{1}{4}g^2N_c\frac{(d-1)^2}{d}\omega^{-1}(0,0)\int d^dx \lk
R_k^a(x)\rk^2\; ,
\eeq
where the identity
\bee
\int d^dp\, f(p^2)\, t_{ij}(p)=\delta_{ij}\frac{d-1}{d}\int
d^dp\,f(p^2)\, %,\quad \textrm{if}\:\: I_{ij}<\infty
\eeq
has been employed. We note here, however, that $\omega^{-1}(0,0)=\int
\dbar^{\, d}p\,\omega^{-1}(p)$ is not necessarily a well-defined quantity.
Actually, if we consider that the gluon will essentially behave like a free
particle for high momentum, $\tilde\omega (p)\to p$, there is an
ultraviolet divergence of degree $d-1$, i.e.\ quadratic for $d=3$. We
therefore have an undefined expression in the coherent state magnetic
potential (\ref{resEp}) which calls for renormalization.

%\vspace{2cm}
%\vfill
\subsection*{Renormalization}

The renormalization of the vacuum Green functions as discussed in
chapter \ref{VarVac}, suffices to render the shift $\Delta E'[Z]:=
E'[Z]-E'[0]$ of the coherent state energy from the vacuum energy finite, as will be shown now. The ultraviolet divergence in the
magnetic potential energy (\ref{resEp}) comes as no surprise, for the magnetic potential
operator (\ref{Hpdef}) is a
highly local object. We now try to single out the divergent object
from the magnetic energy (\ref{resEp}). To this end, we first realize that it is
contained in
the quadratic term in the (terminating) Taylor series of $E_p[Z]$ about $Z=0$,
\bee
\label{Epexpand}
E_p[Z]&=&\lla H_p[A+R]\rra_\omega \nn\\
&=&\lla H_p[A]\rra_\omega+\lla \frac{\delta^{}H_p[A]}{\delta A_i}
\rra_\omega R_i + \frac{1}{2!}\lla
\frac{\delta^{2}H_p[A]}{\delta A_i A_j}
\rra_\omega R_i R_j\nn\\
&&+\frac{1}{3!}\lla
\frac{\delta^{3}H_p[A]}{\delta A_i A_jA_k}
\rra_\omega R_i R_j R_k+\frac{1}{4!}\lla
\frac{\delta^{4}H_p[A]}{\delta A_i A_jA_kA_m}
\rra_\omega R_i R_j R_kR_m
\eeq
Such a term with a second derivative inside a Gaussian
expectation value  can be rewritten as, see Eq.\ (\ref{del2O}) in the appendix,
\begin{eqnarray}
  \label{divbyEp}
  E_p^{(R^2)}[Z]&=&\frac{1}{2}\int d^d[xy]R_i^a(x)
  \bigg\langle \frac{\delta^2H_p}{\delta A_i^a(x)\delta A_j^b(y)}\bigg\rangle_\omega R_j^b(y)\nn\\ &=& -2\int
  d^d[xyuv]R_i^a(x)\omega_{im}^{ac}(x,u)\frac{\delta\lla  H_p
    \rra_\omega}{\delta\omega_{mn}^{cd}(u,v)}\omega_{nj}^{db}(v,y)R_j^b(y)
\end{eqnarray}
Note that $ E_p^{(R^2)}[Z]$ is not given by only $
E_p^{\textrm{div}}[Z]$ in Eq.\ (\ref{Epdivres}) but also includes the other term in
Eq.\ (\ref{resEp}) that is quadratic in $R$.

In view of the gap equation, we know that $\omega$ was chosen to
minimize the vacuum energy,
$\delta\langle\tH\rangle_\omega/\delta\omega = 0$. Therefore, the
divergent expression $\delta\langle H_p\rangle_\omega/\delta\omega$ in
Eq.\ (\ref{divbyEp}) can be expressed in terms of the kinetic vacuum energy
$\langle\tH_k\rangle_\omega$ and the Coulomb vacuum energy
$\langle\tH_C^{(0)}\rangle_\omega$,
\begin{eqnarray}
  \label{divbyEkC}
  E_p^{(R^2)}[Z]=+2\int
  d^d[xyuv]R_i^a(x)\omega_{im}^{ac}(x,u)\frac{\delta\big\langle  H_k + H_C^{(0)}
    \big\rangle_\omega}{\delta\omega_{mn}^{cd}(u,v)}\, \omega_{nj}^{db}(v,y)R_j^b(y)\; .
\end{eqnarray}
The above expression is now finite if we replace the bare form factors
by the renormalized ones.
%  Thus, the renormalization in the vacuum is
% sufficient to make the change in the energy $\Delta E' = E'[Z]-E'[0]$ finite.

To arrive at the finite expression for $E_p^{(R^2)}[Z]$, 
one needs to evaluate% \footnote{It is important to realize that
%   $\int d^d[xy]\frac{\delta
%     E_p[0]}{\delta\omega(x,y)}R^a_i(x)R^a_i(y)\neq
%     \int d^d[xy]\frac{\delta
%     E_p[0]}{\delta\omega^{ab}_{ij}(x,y)}R^a_i(x)R^b_j(y)$. Thus, we
%   cannot use the results in Ref.\ \cite{FeuRei04}.}
\begin{subequations}
\label{auxrel}
\bee
\frac{\delta \lla
  A_m^c(u)A_n^d(v)\rraw}{\delta\omega_{ij}^{ab}(x,y)} &=&
 -\frac{1}{2}(\omega^{-1})_{im}^{ac}(x,u)(\omega^{-1})_{nj}^{db}(v,y)
\\
\frac{\delta \lla
  A_m^c(u)Q_n^d(v)\rraw}{\delta\omega_{ij}^{ab}(x,y)} &=&
\frac{1}{2}(\omega^{-1})_{im}^{ac}(x,u)\int d^dz
(\omega^{-1})_{jk}^{bg}(y,z)\chi_{kn}^{gd}(z,v)\\
\frac{\delta \lla
  Q_m^c(u)Q_n^d(v)\rraw}{\delta\omega_{ij}^{ab}(x,y)} &=&
\frac{1}{2}t_{im}^{ac}(x,u)t_{jn}^{bd}(y,v)  \nn\\&&\hspace{-3cm}-
\frac{1}{2}\int d^dx'(\omega^{-1})_{ii'}^{aa'}(x,x')\chi_{i'm}^{a'c}(x',u)
\int d^dy'(\omega^{-1})_{jj'}^{bb'}(y,y')\chi_{j'n}^{b'd}(y',v)
\eeq
\end{subequations}
The implicit $\omega$ dependence of the form factors $\chi(k)$ and of
the vacuum Coulomb propagator,
\begin{eqnarray}
  \label{Fomegadef}
  F_\omega^{ab}(x,y):= \lla F[A] \rraw^{ab}(x,y)\; ,
\end{eqnarray}
is omitted in the computation of Eq.\ (\ref{divbyEkC}); it would bring about two-loop terms in the equations of motion.
With the relations (\ref{auxrel}), Eq.\ (\ref{divbyEkC}) can be shown to give
\begin{eqnarray}
\label{infsubst}
  E_p^{(R^2)}[Z] &=&\frac{1}{2}\int d^d[xyz]R^a_i(x)\lk
\omega(x,y)\omega(y,z)-\chi(x,y)\chi(y,z)\rk R^a_i(z)\nn\\
&&\hspace{-1cm} +\;\frac{g^2}{4}N_c \int d^d[xy] F_\omega(x,y)\nn\\
&&\hspace{-.7cm}\times\left\{\;-\int d^d[uv]R^a_i(x)[\omega(x,u)-\chi(x,u)]
t_{ij}(u)\omega^{-1}(u,v)[\omega(v,y)-\chi(v,y)] R^a_j(y)\right.\nn\\
&&\hspace{-.1cm}\;\;\, +\lk t_{ij}(x)\omega^{-1}(x,y)\rk
\int d^d[uv]R^a_i(u)[\omega(u,x)\omega(y,v)-\chi(u,x)\chi(y,v)] R^a_j(v)\nn\\
&&\hspace{-.1cm}\;\; \left. -2\int d^du\: R^a_i(x)\chi(y,u)R^a_j(u)t_{ji}(y)\lk\delta^d(y,x)-\int
d^d v\,\omega^{-1}(y,v)\chi(v,x)\rk\right\}\nn\\
\end{eqnarray}
The magnetic potential energy in the coherent state thus finally
yields
\begin{eqnarray}
  \label{Epfin}
  E_p[Z]-E_p[0]&=&\frac{1}{4}g^2\int d^dx
f^{eab}f^{ecd}R^a_i(x)R^b_j(x)R^c_i(x)R^d_j(x)\nn\\
&&
+g\int d^dxf^{abc}\lk\partial_iR^a_j(x)\rk R^b_i(x)R^c_j(x) + E_p^{(R^2)}[Z]
\end{eqnarray}
where $E_p[0]$ is given by Eq.\ (\ref{Ep}) and $E_p^{(R^2)}[Z]$ by
Eq.\ (\ref{infsubst}). The shift $\Delta E_p=E_p[Z]-E_p[0]$ is indeed
finite for the variational solution of the coherent field (any
divergence will be suppressed, for $Z=0$ is an option).

\subsection*{Solution for the imaginary part of the coherent field}

The energy $E'[Z]$ in the coherent state sums up the magnetic energy
(\ref{Epfin}), the kinetic energy (\ref{EkZres}), and the Coulomb
energy contributions (\ref{resEC0}), (\ref{EC1}) and (\ref{EC2}). The
real and imaginary parts of the coherent field $Z_k^a(x)$ are to be
determined by the variational principle,
\begin{eqnarray}
  \label{RitzforZ}
  \frac{\delta E'[Z]}{\delta R_k^a(x)}= 0\; ,\quad \frac{\delta E'[Z]}{\delta
    I_k^a(x)}= 0\; .
\end{eqnarray}
Since the Yang--Mills
Hamiltonian has cubic and quartic terms in the fields, we cannot
algebraically solve these equations. However, the energy expression is
quadratic in the imaginary part of $Z$. The relevant terms can be
written in matrix notation as
\bee
\label{quadI}
E^{(I)}=\frac{1}{2}IMI-bI\; .
\eeq
This defines the matrix $M$ as well as $b$ by
\bee
M_{ij}^{ab}(x,y)&=& t_{ij}^{ab}(x,y) \nn\\
\label{defM}
&& - g^2 \hat R^{ad}_i(x)F_Z^{dc}(x,y)\hat
R^{cb}_j(y) 
-\frac{g^2}{2}\big(\hat T^a\big)^{cd}\omega_{ij}^{-1}(x,y) {F_Z}^{de}(x,y)
\big( \hat T^b \big)^{ec} \\
b_k^a(x)&=& g^2\int d^dy\,\rhoext^c(y)F_Z^{cb}(y,x)\hat R_k^{ba}(x)
\eeq
The quadratic form (\ref{quadI}) attains its minimal value
$E^{(I)}=-\frac{1}{2}b A^{-1}b$ for
\bee
\label{minI}
I_k^a(x)=\int d^dy\,(M^{-1})_{kj}^{ab}(x,y)b_j^b(y)\; .
\eeq
In order to write down the coherent state energy $E'[Z]$ minimal for the
imaginary part of the coherent field, replace the fields $I_k^a(x)$
by the expression (\ref{minI}). This yields the complete energy
expression
\bee
\label{completeE}
E'[Z]&=&\frac{1}{4}\int
d^d[xyz][\omega-\chi_{{}_Z}]^{ab}_{ij}(x,y)\omega^{-1}(y,z)[\omega-\chi_{{}_Z}]^{ba}_{ji}(z,x)\nn\\
&&
-\frac{g^2}{8}\int d^d[xy] (\hat T^g)^{bb'}
F_Z^{b'd'}(x,y)(\hat T^h)^{d'd}\nn\\
&&\left\{\lk\omega^{-1}\rk^{gh}_{ij}(x,y)\int d^d[uv][\omega-\chi_{{}_Z}]^{be}_{ik}(x,u)\omega^{-1}(u,v)[\omega-\chi_{{}_Z}]^{ed}_{kj}(v,y)\right.
\nn\\
&& + \lk t_{ij}^{gd}(x)\delta^d(x,y)-\int
d^dw\,\omega^{-1}(x,w)(\chi_{{}_Z})^{gd}_{ij}(w,y)\rk\nn\\&&\quad\left. \lk t_{ij}^{bh}(x)\delta^d(x,y)-\int
d^dw\,\omega^{-1}(x,w)(\chi_{{}_Z})^{bh}_{ij}(w,y)\rk\right\} \nn\\
&& -\:\frac{g^2}{4}\int d^d[xyuv][\omega-\chi_{{}_Z}]^{ac}_{ik}(x,u)\omega^{-1}(u,v)[\omega-\chi_{{}_Z}]^{cb}_{kj}(v,y)\lk\hat
R_i(x)F_Z(x,y)\hat R_j(y)\rk^{ab}\nn\\
&&+\:\frac{g^2}{2}\int d^d[xy]\rho_{\mathrm{ext}}^a(x)
F_Z^{ab}(x,y)\rho_{\mathrm{ext}}^b(y)\nn\\
&& +\:g^4\int d^d[xyuv]\rho_{\mathrm{ext}}^a(x)F_Z^{ab}(x,u)\hat R^{bc}_i(u)\lk
M^{-1} \rk^{cc'}_{ij}(u,v)\hat R^{c'd}_j(v)F_Z^{de}(v,y)\rho_{\mathrm{ext}}^e(y)\nn\\
&&+\: E_p[0]\nn\\
&&+\frac{1}{4}\,g^2\int d^dx
f^{eab}f^{ecd}R^a_i(x)R^b_j(x)R^c_i(x)R^d_j(x)\nn\\
&&+\:g\int d^dx\, f^{abc}\lk\partial_iR^a_j(x)\rk R^b_i(x)R^c_j(x)\nn\\
&&+\frac{1}{2}\int d^d[xyz]R^a_i(x)\lk
\omega(x,y)\omega(y,z)-\chi(x,y)\chi(y,z)\rk R^a_i(z)\nn\\
&&\hspace{-0cm} +\;\frac{g^2}{4}N_c \int d^d[xy] F_\omega(x,y)\nn\\
&&\hspace{-0cm}\quad\times\left\{\;-\int d^d[uv]R^a_i(x)[\omega(x,u)-\chi(x,u)]
t_{ij}(u)\omega^{-1}(u,v)[\omega(v,y)-\chi(v,y)] R^a_j(y)\right.\nn\\
&&\hspace{-0cm}\quad\;\;\, +\lk t_{ij}(x)\omega^{-1}(x,y)\rk
\int d^d[uv]R^a_i(u)[\omega(u,x)\omega(y,v)-\chi(u,x)\chi(y,v)] R^a_j(v)\nn\\
&&\hspace{-0cm}\quad\;\; \left. -2\int d^du\: R^a_i(x)\chi(y,u)R^a_j(u)t_{ji}(y)\lk\delta^d(y,x)-\int
d^d v\,\omega^{-1}(y,v)\chi(v,x)\rk\right\}\nn\\
\eeq
with the abbreviation $\lk M^{-1} \rk^{bc}_{ij}(u,v)$ given as the inverse
of the matrix $M$ in Eq.\ (\ref{defM}).

\subsection*{Estimate for translationally invariant form factors}
The expression (\ref{completeE}) for the energy in the coherent state is quite
large and it might be useful to make crude approximations to get an
idea about its main characteristics. In Eq.\ (\ref{completeE}), the kernel $\omega(x,y)$ is fixed by
the variational calculation in the absence of external charges, see
chapter \ref{VarVac}. The other Green functions, $(D_G^Z)^{ab}(x,y)$,
$F_Z^{ab}(x,y)$ and $(\chi_{{}_Z})_{ij}^{ab}(x,y)$, are dependent on the
location of the charges and thus have lost their translational
invariance. Moreover, neither the Lorentz nor the color structure can
be expected to be trivial. Therefore, a minimization of the energy
$E'[Z]$ is a very costly calculation, not even vaguely comparable to
the challenge in solving the integral equations in the vacuum.

A possible simplification is achieved by rendering the Green functions
insensitive to the presence of quarks. Setting $D_G^Z=D_G$,
$\chi_{{}_Z}=\chi$ and $F_Z=F_\omega$, we can simplify the expression (\ref{completeE}) for
$E'[Z]$ quite significantly,
\begin{eqnarray}
  \label{E'trinv}
  E'[Z]&=&E_k[0]+E_C^{(0)}[0]+E_C^{(2)}[0]+E_p[R]\nn\\
  &&\hspace{-1.5cm}+\, N_c\frac{g^2}{4}\int
  d^d[xyuv]F_\omega(x,y)R_i^a(x)[\omega-\chi](x,u)\omega_{ij}^{-1}(u,v)[\omega-\chi](v,y)R_j^a(y)\nn\\
  &&\hspace{-1.5cm}+\, g^4\int d^d[xyuv]\rhoext^a(x)F_\omega(x,u)\hat
  R_i^{ab}(u)(M^{-1})_{ij}^{bc}(u,v)\hat
  R_j^{cd}(v)F_\omega(v,y)\rhoext^d(y)\; .
\end{eqnarray}
The imaginary part of $Z$ is already chosen so that $E'[Z]$ is
minimal, this gives the last term quadratic in $F_\omega$. The term in
the second line of Eq.\ (\ref{E'trinv}) is exactly cancelled by the
term in $E_p[R]$ that came from the expectation value $\langle\hat
RQF\hat R Q\rangle_\omega$, so that the energy shift $\Delta E' = E'[Z] -
E'[0]$ yields
\begin{eqnarray}
  \label{deltaE'tr}
  \Delta E'[Z] &=&\frac{1}{4}\,g^2\int d^dx
f^{eab}f^{ecd}R^a_i(x)R^b_j(x)R^c_i(x)R^d_j(x)\nn\\
&&+\:g\int d^dxf^{abc}\lk\partial_iR^a_j(x)\rk R^b_i(x)R^c_j(x)\nn\\
&&+\frac{1}{2}\int d^d[xyz]R^a_i(x)\lk
\omega(x,y)\omega(y,z)-\chi(x,y)\chi(y,z)\rk R^a_i(z)\nn\\
&&+\;\frac{g^2}{4}N_c \int d^d[xyuv] F_\omega(x,y)\omega_{ij}^{-1}(x,y)
R^a_i(u)[\omega(u,x)\omega(y,v)-\chi(u,x)\chi(y,v)]
R^a_j(v)\nn\\
&&-\;\frac{g^2}{2}N_c \int d^d[xyuv] F_\omega(x,y) R^a_i(x)\chi(y,u)R^a_j(u)t_{ji}(y)\nn\\&&\qquad\qquad\qquad\lk\delta^d(y,x)-\int
d^d v\,\omega^{-1}(y,v)\chi(v,x)\rk\nn\\
  && + g^4\int d^d[xyuv]\rhoext^a(x)F_\omega(x,u)\hat
  R_i^{ab}(u)(M^{-1})_{ij}^{bc}(u,v)\hat
  R_j^{cd}(v)F_\omega(v,y)\rhoext^d(y)\; .
\end{eqnarray}
The prominent terms are the very first and the very last ones,
resembling the energy of the anharmonic oscillator (\ref{Estrz}) in an
external electric field.\footnote{Studying the infrared limit, the
 third, fourth and fifth terms can be neglected since $\omega(k)=\chi(k)$ for $k\to 0$.} Approximating $M\approx\id$,
\begin{eqnarray}
  \label{DEM1}
  \Delta E'[Z] &=& -\frac{g^2}{4} \int d^dx\, R_j^a(x)\hat
  R_k^{ab}(x)\hat R_k^{bc}(x) R_j^c(x) \nn\\
&& + g^4\int d^d[xyu]\rhoext^a(x)F_\omega(x,u)\hat
  R_k^{ab}(u)\hat
  R_k^{bc}(u)F_\omega(u,y)\rhoext^c(y)\; ,
\end{eqnarray}
and furthermore replacing the matrix $\hat R_k\hat R_k$ by its mean
diagonal elements,
\begin{eqnarray}
  \label{RRappr}
  \hat R_k(x)\hat R_k(x) \approx \id \frac{\tr \hat R_k(x)\hat R_k(x)}{\tr
    \id} = -\id \frac{N_c}{N_c^2-1}\phi(x)\; ,\quad
  \phi(x):=R_k^a(x)R_k^a(x)\; ,
\end{eqnarray}
we get
\begin{eqnarray}
  \label{DEsimple}
  \Delta E'[Z] = +\frac{g^2N_c}{4(N_c^2-1)}\int d^d x\,\phi^2(x) -
  \frac{ g^4N_c}{N_c^2-1} \int d^dx\,\phi(x)\left(\int d^dy\,
    F_\omega(x,y)\rhoext^a(y)\right)^2\; .
\end{eqnarray}
The expression (\ref{DEsimple}) attains for $\phi(x) = 2 g^2
\big(\int d^dy\,
    F_\omega(x,y)\rhoext^a(y)\big)^2 $ its
minimal value
\begin{eqnarray}
  \label{DE'min}
  \Delta E'[Z] = - \frac{g^6 N_c}{N_c^2-1}\int d^d x\left(\left(\int
      d^dy\,  F_\omega(x,y)\rhoext^a(y)\right)^2\right)^2\; .
\end{eqnarray}
By dimensional analysis
we find in the infrared limit of quark separation $r$, noting that
$F_\omega\sim \sigma_{{}_C} r^{4\kappa+2-d}$ with Coulomb string
tension $\sigma_{{}_C}$, we get in general dimensions $ \Delta E'[Z]\,\sim\, - \sigma_{{}_C}^4
  r^{16\kappa+11-4d}$ and hence for $d=3$ and $\kappa=\frac{1}{2}$
\begin{eqnarray}
  \label{WeirdPot}
\Delta E'[Z]\,\sim\,-\sigma_{{}_C}^4 r^7\; ,
\end{eqnarray}
a completely senseless result. First of all, confinement is lost since
Eq.\ (\ref{WeirdPot}) overwhelms the confining potential for large
$r$. But even worse, the energy $E'[Z]$ is negative which
contradicts the positive definiteness of the YM Hamiltonian. 

What has been done wrong? With various ans\"atze, such as for $\mathit{SU(2)}$
\begin{eqnarray}
  \label{WuYang}
  R_k^a(x)=f^{akm}\del_m\varphi(x)\; ,
\end{eqnarray}
the energy in the form (\ref{E'trinv}) was minimized and the erroneous
large $r$ behavior (\ref{WeirdPot}) was confirmed.  Therefore, the
approximations done after Eq.\ (\ref{E'trinv}) cannot be blamed, the
awkward result (\ref{WeirdPot}) must come from the use of vacuum Green
functions. In a sensible approach, the Green functions must take into account
the presence of external charges.

The proper way to go about this has to involve solving a set of
Dyson--Schwinger equations without translational invariance. One of
these equations follows from the variational principle, \mbox{$\delta
E'[Z]/\delta Z_k^a(x)=0$}. The ghost propagator $D_G^Z$ can be
determined by the DSE
\begin{eqnarray}
  \label{DGZDSE}
  (D_G^Z)^{-1} = G^{-1}[R] - \Sigma_Z
\end{eqnarray}
where the self-energy $\Sigma_Z$ is obtained by using the propagator
$D_G^Z$ in the definition of $\Sigma$ in Eq.\ (\ref{Sigma}). A
relation for the inverse ghost propagator, such as Eq.\
(\ref{DGZDSE}), might not be so useful. Instead, the equation
\begin{eqnarray}
  \label{DGZstraight}
  D_G^Z=D_G + \int d^dx\,\lla G[A]\,\Gamma_k^{0,a}(x)R_k^a(x)\, G[A+R]\rraw
\end{eqnarray}
seems more applicable for solving the set of integral equations.

It was not attempted to solve the DSEs in the variational coherent
state. Immense computer power would be necessary and such a project is
at best problematical. In the upcoming section, the effect of the
coherent field on the confining property of the heavy quark potential
will be discussed with alternative methods.

\section{The gluon chain}
\label{gluonchain}

Having learned from first approximations that the Green function need
to be sensitive to the presence of external charges in order to arrive
at a sensible result, we try to make estimates for such Green
functions in this section. The quest for the correct wave functional
has been long-standing. After Dirac's description of the abelian case
\cite{Dir55}, the generalization to YM theories proved to be
difficult \cite{HaaJoh97}. There have been various attempts of formulating flux
tube models (see e.g.\ Ref.\ \cite{IsgPat84}), calculations in the
Hamiltonian approach to QCD \cite{SzcKru06}, and also lattice
calculations \cite{Hei+07} that elaborate on the subject of the ground state in the
presence of heavy quarks. One of the most intuitive notions might be
the gluon chain model \cite{Tho79,GreTho02,Gre02}, and we will come
back to it below.

An advantage of the variational approach is that any ansatz for the
wave functional will do. A too wild guess will be rejected if it 
leads to a rise in the energy. If the ansatz is somewhat sensible,
however, and a decrease in the energy can be detected, this wave
functional is to be favored over the vacuum wave functional. A
description of the back reaction of the presence of external charges
on the vacuum wave functional is then found. With this reasoning, we here
try to find a coherent field $Z_k^a(x)$ that leads to a lowering of the
infrared Coulomb propagator. At very high quark separation, the
Coulomb energy $E_C^{(2)}[Z]$ can be expected to give a dominant
contribution. We will now focus on the calculation of the respective
expectation value in the coherent field.

Note that the Coulomb propagator $F_Z$ corresponds to the vacuum
counterpart $F_\omega$ with the coherent field $R$ as a background field,
\begin{eqnarray}
  \label{FZexp}
  F_Z=\lla F[A+R] \rraw = \int \cD A\,  |\tP[A]|^2 \exp\left[\int d^3x\, 
    R_k^a(x)\frac{\delta}{\delta A_k^a(x)}\right] F[A]\; .
\end{eqnarray}
We find a series in the coherent field where each term is treated
using the identity
\begin{eqnarray}
  \label{delG}
  \frac{\delta G[A]}{\delta A_k^a(x)} = G[A]\Gamma_k^{0,a}(x)G[A]\; .
\end{eqnarray}
Acting on the Coulomb operator $F=GG_0^{-1}G$, the variational
derivative generates a sum of terms,
\begin{equation}
\label{delF}
  \frac{\delta F[A]}{\delta A_k^a(x)} = F[A]\Gamma_k^{0,a}(x)G[A]+
  G[A]\Gamma_k^{0,a}(x)F[A]\; .
\end{equation}
One can check that the combinatorial factors cancel the $1/n!$
from the expansion of the exponential function in Eq.\ (\ref{FZexp}),
and one obtains\footnote{A proof is given by showing
  $(R_k\frac{\delta}{\delta A_k})^nG=n!(GM)^nG$ by induction, which is
  trivial, and plugging it into
  $G[A+R]=\sum_n\frac{1}{n!}(R_k\frac{\delta}{\delta A_k})^nG[A]$
  which leads with $F=G\, G_0^{-1}\, G$ directly to the second line in Eq.\ (\ref{Fseries}).}
\begin{eqnarray}
  \label{Fseries}
  F_Z&=&\lla F +  FMG + GMF + FMGMG + GMFMG + GMGMF + \dots \rraw \nn\\
  &=&\lla (\id + GM + GMGM +\dots)\, F\, (\id + MG + MGMG + \dots) \rraw \nn\\
  &\approx&\sum_m\sum_n(D_GM)^m F_\omega (MD_G)^n 
\end{eqnarray}
with the abbreviation
\begin{eqnarray}
  \label{MvonRdef}
  M^{ab}(x,y)=\int d^3z R_k^c(z)\big (\Gamma_k^{0,c}(z)\big )^{ab}(x,y)\; ,
\end{eqnarray}
where the tree-level ghost-gluon vertex $\Gamma_k^{0,a}(x)$ was
defined in Eq.\ (\ref{GGAdef}). In the last line of Eq.\
(\ref{Fseries}), the approximation of factorizing the Coulomb
expectation value was employed (for a discussion, see section
\ref{fsection}). We have thus expressed the Coulomb propagator $F_Z$
in the coherent field by vacuum expectation values, i.e.\ the vacuum
ghost propagator $D_G$ and the vacuum Coulomb propagator $F_\omega$ (\ref{Fomegadef}),
and the coherent field itself via $M$ in Eq.\ (\ref{MvonRdef}).

\begin{figure}
  \centering
  \ing[scale=0.88,bb= 100 490 595 700,clip=]{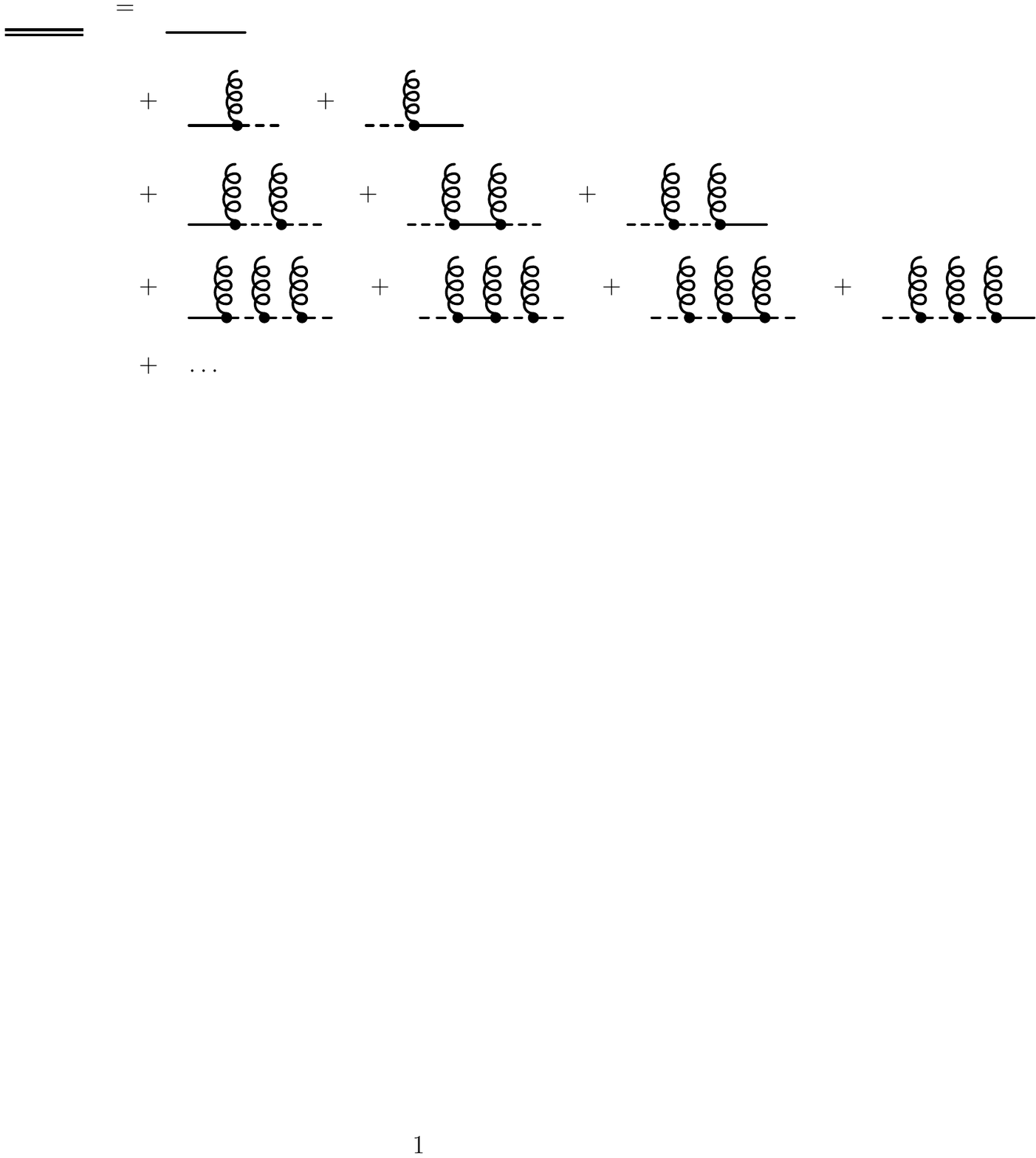}
  \caption{The Coulomb propagator in the coherent field, denoted by
    the double line on the l.h.s., and its expansion
    (\ref{Fseries}). The solid lines represent vacuum Coulomb
    propagators, the dashed lines vacuum ghost propagators, the dots are tree-level ghost-gluon vertices, and the wiggly lines indicate that gluons are excited by the coherent field.}
  \label{chainpic}
\end{figure}

The Coulomb potential $V_C^Z(r)$ in the coherent field is defined for
a given charge distribution $\rhoext^a(x)$ by
\begin{eqnarray}
  \label{VCZdef}
  V_C^Z(r):= \frac{g^2}{2}\int
  d^3[xy]\rhoext^a(x)F_Z^{ab}(x,y)\rhoext^b(y)\; ,
\end{eqnarray}
cf.\ the vacuum Coulomb potential $V_C$ in Eq.\
(\ref{potential}). With two point-like heavy charges located at the
origin and at $r$, 
\begin{equation}
  \label{rhostr}
  \rhoext^a(x)=\delta^{a3}\left(\delta^{3}(x)-\delta^{3}(x-r)\right)\; ,
\end{equation}
one may think of the expectation value in Eq.\ (\ref{VCZdef}) as shown
in Fig.\ \ref{chainpic}. As the charges are separated by a large
distance $r$, it can be energetically more favorable to excite a
gluon in between the charges. This notion basically corresponds to the
gluon chain model proposed in Refs.\ \cite{Tho79,GreTho02,Gre02}. These gluons possibly screen the
potential and thus account for a lowering of the Coulomb string
tension $\sigma_C$ to the string tension $\sigma_W$ from the gauge
invariant Wilson loop. Moreover, with excited gluons along the
flux tube, the thickness of the string (lacking for the one-gluon exchange) may be generated. The coherent state is suitable to check
whether this mechanism is realized in the framework of the variational
approach. 

The guiding properties for choosing an ansatz for the coherent field
are {\it transversality} and {\it symmetry}. Firstly, let us exploit
transversality. The coherent field $Z_k^a(x)$, here the real
part $R_k^a(x)$, is always projected transversally, i.e.\ it occurs only
in the form $t_{jk}(x)R_k^a(x)$. Without loss of generality, we can
therefore set
\begin{eqnarray}
  \label{hdef}
  R_k^a(x)=\epsilon_{kij}\del_i h_j^a(x)
\end{eqnarray}
and determine the field $h_k^a(x)$ instead. 

Secondly, we can assume that the symmetry of the charge distribution
is mirrored in the field $R_k^a(x)$. While in the absence of charges,
the Green functions are translationally invariant, a charge
distribution $\rhoext^a(x)$ will remove this invariance, yet leaving
us with some weaker symmetry. Assume for instance, one of the
charges is located at the origin of our coordinate system, and the
second charge is smeared on the surface of a concentric sphere with
radius~$r$,\footnote{The bold type notation for vectors is
  reinstated for the remainder of this chapter.}
\begin{eqnarray}
  \label{spherepic}
  \rhoext^a(\fx)=\delta^{a3}\delta^3(\fx)-\delta^{a3}\frac{1}{4\pi \fx^2}\,\delta(|\fx|-r)
\end{eqnarray}
The exchange energy is the same as for the charge distribution Eq.\ 
(\ref{rhostr}), and  the choice (\ref{spherepic}) may
therefore equally be used for the purpose of these investigations. We
refer to the setting (\ref{spherepic}) as the {\it sphere picture} and
Eq.\ (\ref{rhostr}) as the {\it string picture}. For
the charge contribution of the sphere picture (\ref{spherepic}), the
coherent field $Z_k^a(x)$ can be expected to be point symmetric,
\begin{eqnarray}
  \label{Zisotrop}
  {\bf Z}^a(D\fx)=D{\bf Z}^a(\fx)\; ,
\end{eqnarray}
that is, for any rotation $D\in\mathit{SO(3)}$ about an axis that
contains the origin $\fx=0$, it behaves according to Eq.\
(\ref{Zisotrop}). 
Expanding ${\bf Z}^a(\fx)$ in spherical coordinates, this
means that the coefficients merely depend on the distance $|\fx|$ from
the origin,
% \begin{eqnarray}
%   \label{Zexp}
%   {\bf Z}^a(\fx)=Z_r(r){\bf e}_r(\fx) + Z_\theta(r){\bf e}_\theta(\fx)+
%   Z_\varphi(r){\bf e}_\varphi(\fx)
% \end{eqnarray}
The field ${\bf h}^a(x)$ defined in Eq.\ (\ref{hdef}) inherits this
property from $Z_k^a(x)$ via Eq.\ (\ref{IandRdef})
% \begin{eqnarray}
%   \label{hisotrop}
%   {\bf h}^a(D\fx)=D{\bf h}^a(\fx)\; ,
% \end{eqnarray}
and suggests an expansion into spherical coordinates.
% \begin{eqnarray}
%   \label{hexp}
%   {\bf h}^a(\fx)=h_r(r){\bf e}_r(\fx) + h_\theta(r){\bf e}_\theta(\fx)+
%   h_\varphi(r){\bf e}_\varphi(\fx)
% \end{eqnarray}
On the other hand, if we use the string picture, see Eq.\ (\ref{rhostr}),
only a cylindrical symmetry may be expected, i.e.\ the rotations $D$
are restricted to those about the symmetry axis along ${\bf r}_0$. It is then
useful to decompose ${\bf h}^a(\fx)$ in the cylindrical basis $\mathfrak{B}_{\fx}=\{{\bf e}_\rho(\fx),{\bf e}_\varphi(\fx),{\bf e}_z\}$ with
${\bf e}_z\parallel{\bf r}_0$,
\begin{eqnarray}
  \label{hstring}
  {\bf h}^a(\fx)=h_\rho^a(\rho,z){\bf e}_\rho(\fx) +
  h_\varphi^a(\rho,z){\bf e}_\varphi(\fx)+
  h_z^a(\rho,z){\bf e}_z\; .
\end{eqnarray}

Now consider the gluon chain integral (\ref{Fseries}).\footnote{A
  transformation to momentum space would be, on the one hand, convenient
  since the infrared asymptotic vacuum Green functions from chapter
  \ref{ghostdom} can be used. However, a subtlety is accompanied by
  the Fourier transformation. The real part of the coherent field in
  coordinate space,
  $\frac{1}{2}(Z(x)+Z(x)^*)$, transfers to $\frac{1}{2}(Z(k)+Z(-k)^*)$
and only by knowledge of the parity properties of $Z(x)$ can we tell
whether the above momentum space quantity is real, purely imaginary,
or complex.} Plugging in the definition of the field $h_k^a(x)$ into
the matrices $M$, any integral involving $h_k^a(x)$ is of the
form\footnote{Recall that a caret denotes an object in the adjoint
  representation. Here: $\hat h_j(\fx)=\hat T^a h^a_j(\fx)$.}
\begin{eqnarray}
  \label{GhG}
[D_GMD_G](\fz,\fy)= g \epsilon_{ijk} \int d^3x\, \del_i^{\fz} D_G(\fz,\fx)\hat h_j(\fx){\del}^{\fx}_k D_G(\fx,\fy) = \;\ing{eps/onelink}
\end{eqnarray}
The above object corresponds to one ''chain link'' and the entire
gluon chain is a convolution of chain links we intend to re-sum.

We may choose the first integrations as the one where the charge
sitting at the origin is attached to the ghost line. This corresponds to
setting $\fz=0$ in Eq.\ (\ref{GhG}). The distance $r$ between the
charges being considered very large, the infrared asymptotic forms of
the vacuum ghost propagator may be used. From the momentum space
expression (see table \ref{tab:sol}), we find the dimensionally regularized
($d=3+2\epsilon$) coordinate space ghost propagator
\begin{eqnarray}
  \label{DGcoo}
  D_G(\fx,\fy)=\int \dbar^dk \frac{B}{({\bf k}^2)
^{1+\kappa}}\e^{i{\bf k}\cdot (\fx-\fy)} =
\frac{B}{(4\pi)^{d/2}}\frac{\Gamma\big(\frac{d}{2}-(1+\kappa)\big
  )}{\Gamma(1+\kappa)}\bigg(\frac{{(\fx-\fy)}^2}{4}\bigg)^{(1+\kappa)-\frac{d}{2}}\; ,
\end{eqnarray}
and taking the derivative as in Eq.\ (\ref{GhG}) gives for
$\kappa=\frac{1}{2}$ and $d=3$,
\begin{eqnarray}
  \label{delD_G}
  \del_k^{\fx}D_G(\fx,\fy) =
  \frac{B}{4\pi^2}\left.\del_k^{\fx}\Gamma(\epsilon)\e^{-\epsilon\ln\frac{(\fx-\fy)^2}{4}}\right|_{\epsilon\to 0}=\, -\frac{B}{2\pi^2}\frac{x_k-y_k}{(\fx-\fy)^2} \; .
\end{eqnarray}
Plugging this into the chain link (\ref{GhG}), we get
\begin{eqnarray}
  \label{GhGe}
[D_GMD_G]({\bf 0},\fy)=\, -g \frac{B^2}{4\pi}\int d^3x \,\frac{{\hat{\bf h}(\fx)}\cdot (\fx\times\fy)}{\fx^2(\fx-\fy)^2}
\end{eqnarray}
The vector ${\bf h}(\fx)$ is expanded in the cylindrical basis
$\mathfrak{B}_{\fx}$, see Eq.\ (\ref{hstring}), and we expand
$\fx = \rho_x\,{\bf e}_\rho(\fx)+z_x\,{\bf e}_z$ and ${\bf y} =
\rho_y\cos\varphi\,{\bf e}_\rho(\fx)-\rho_y\sin\varphi\,{\bf
  e}_\varphi(\fx)+z_y\,{\bf e}_z$ accordingly. The outer product in Eq.\
(\ref{GhGe}) then reads
\begin{eqnarray}
  \label{xxy}
  {\bf x}\times \fy =
\left(
  \begin{array}[c]{c}
    \rho_yz_x\sin\varphi \\ \rho_yz_x\cos\varphi - \rho_xz_y\\ - \rho_x\rho_y\sin\varphi
  \end{array}
\right)_{\mathfrak{B}_{\fx}}\; .
\end{eqnarray}
The $\varphi$-integration eliminates both the $\rho$- and the
$z$-component of ${\bf h}^a(\fx)$ in Eq.\ (\ref{GhGe}). To prove this, note that
both $\hat h_k=\hat h_k(\rho,z)\;\forall k$ and the product $\fx\cdot\fy=\rho_x\rho_y\cos\varphi+z_xz_y$ that
occurs in the denominator are even functions in
$\varphi$. The components $\hat h_\rho$ and $\hat h_\varphi$ contribute
with the outer product (\ref{xxy}) a factor of $\sin\varphi$, odd in
$\varphi$, and hence must vanish. We are left with
$\hat h_\varphi(\rho,z)$, the only function that will contribute to the
corrections to the Coulomb propagator.

In order to estimate the effect of the infrared ($|\fx|\to\infty$)
behavior of ${\bf h}^a(\fx)$, we switch to the sphere picture. This is
achieved by letting $\fy\parallel {\bf e}_z$ which infers $\rho_y=0$,
$\rho_y\equiv y$, and $\cos\varphi = 1$. On the assumption that a
non-vanishing $\hat h_\varphi$ component can be realized in the sphere
picture, use the ansatz 
\begin{eqnarray}
  \label{hansatz}
  \hat h_\varphi(\fx)\sim \frac{1}{|\fx|^{\alpha_h}}\; .
\end{eqnarray}

By virtue of the homogeneity of the functions in the chain link
integral (\ref{GhGe}), one can show that for $\alpha_h<1$, the infrared
strength of the Coulomb propagation is weakened by each chain link in
the series in Fig.\ \ref{chainpic}. This means that for
$r\to\infty$, the only linearly rising term would be the very first
one of the expansion, the vacuum Coulomb propagator $F_\omega$. On the
other hand, if $\alpha_h>1$, the infrared strength is enhanced by each
chain link integral, meaning that for some minimal number of chain
links, the integral would be divergent. Such a choice will be avoided
by the variational principle. If (and only if) $\alpha_h=1$, each
chain link produces a constant factor and all terms of the expansion
contribute to a linearly rising potential. The numerical value of this
factor can be computed by means of the two-point integral (\ref{Xres}) in
the appendix which applies to Eq.\ (\ref{GhGe}) by
\begin{eqnarray}
  \label{2ptvecprod}
  \int d^3x\, \frac{|\fx\times\fy|}{|\fx|^3(\fx-\fy)^2} = 2\pi^2\; .
\end{eqnarray}

The string
picture will require a more involved calculation but the statement
aimed at here is only on a qualitative level. It shall only be noted
that with the choice $\alpha_h=1$, the chain link integrals
in\footnote{The coordinate space ghost propagator can only be defined
  as a distribution. Therefore, we act with a derivative from the
  right on it which is always present in the gluon chain (see Fig.\ \ref{chainpic}).} 
\begin{eqnarray}
  \label{Ldef}
  [D_GMD_G\overleftarrow{\del_k}](0,\fy)=\hat L D_G(0,\fy)\overleftarrow{\del_k}
\end{eqnarray}
yield a calculable constant $\hat L$, here not given
explicitly. It is only important to remark that $\hat L$ is a matrix
made of real numbers.\footnote{It is not as straightforward to keep
  track of the complex phase in momentum space.} Integrating the gluon
chain (\ref{Fseries}) ``from left to right'', every
chain link produces a factor of $\hat L$ in the infrared
limit, 
\begin{eqnarray}
  \label{VCZres}
  V_C^Z(r\to\infty)&& \nn\\&&\hspace{-2cm}=\frac{g^2}{2}\int
  d^3[xy]\rhoext^a(\fx)\:
\left(
\id+\hat L+\hat L\hat L+\dots
\right)^{ab}
F_\omega^{bc}(\fx,\fy)\:
\left(
\id+\hat L+\hat L\hat L+\dots
\right)^{cd}
  \rhoext^d(\fy)\;\nn\\
\end{eqnarray}
The excitation of a single gluon at first sight seems to lower the Coulomb
interaction if $\hat L<0$ (which depends on the self-consistent
solution for $Z_k^a(\fx)$).

So far, the color structure has been suppressed. Recall that $\hat L =
L^a\hat T^a$ is a color matrix in the adjoint representation. The
Coulomb energy $V_C^Z(r)$ in Eq.\ (\ref{VCZres}) is sensitive only to the diagonal element
$F_Z^{33}$. The diagonal elements vanish for $\hat L$. Therefore, the excitation of
a single gluon does not contribute to the Coulomb energy. The gluons
have to come in pairs. If we turn now to $\mathit{SU(2)}$, two chain
  links are seen to produce a factor
\begin{eqnarray}
  \label{twoneg}
\ing[scale=0.9,bb= 140 672 460 730,clip=]{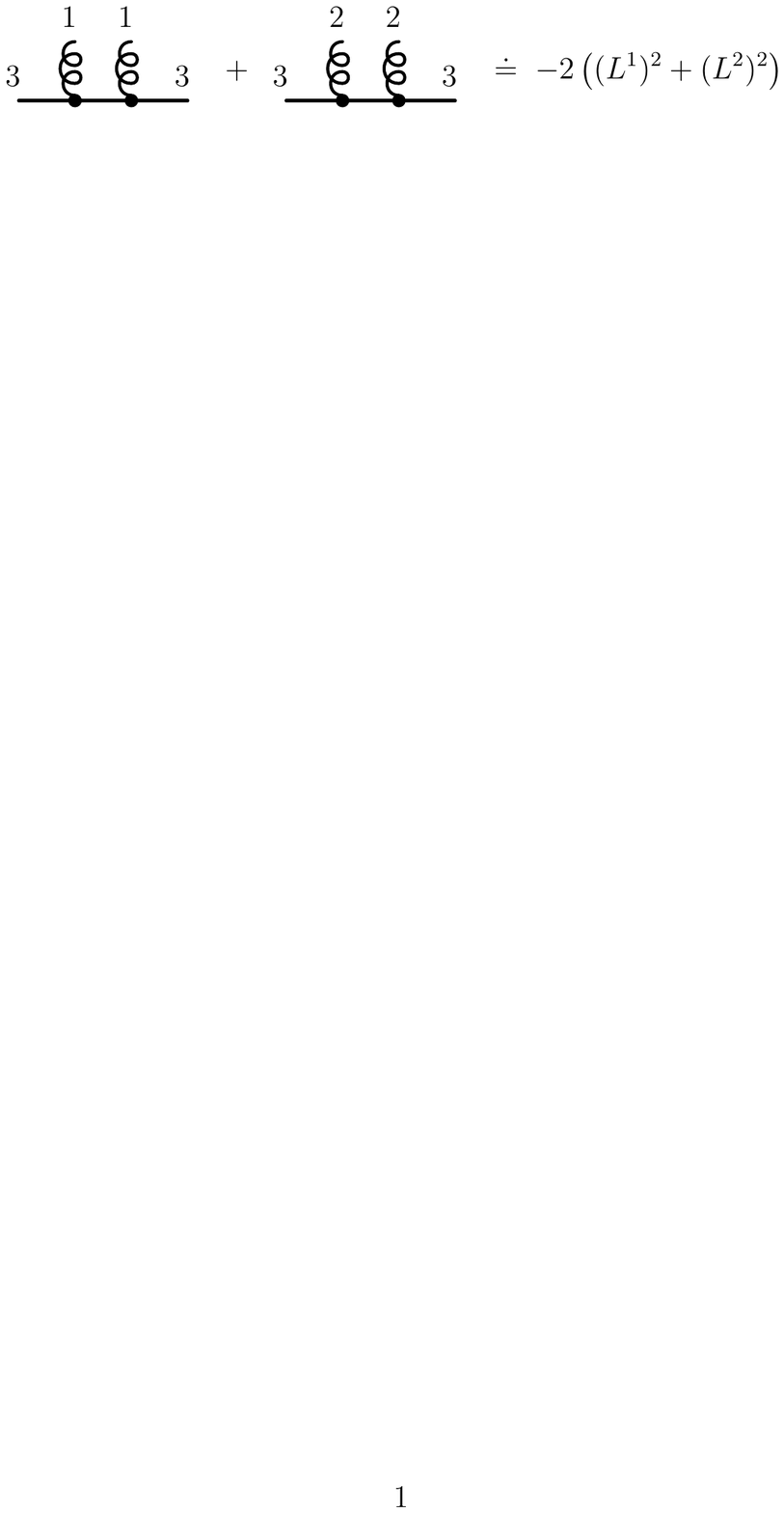}
\nn
\end{eqnarray}
which is manifestly negative. These contributions therefore lower the
Coulomb interaction, due to the color structure. We will now proceed
to re-sum the entire series (\ref{VCZres}) for $\mathit{SU(2)}$.

% Actually, we will now proceed to calculate odd
% and even numbers of gluon excitation, meaning odd and even powers of the matrices $\hat L$, in order to resum the entire
% series (\ref{VCZres}). It will become evident that only even numbers of gluons
% give a contribution.

There are two geometric series in $\hat L$ occurring in the gluon chain
(\ref{VCZres}). The value of the series can be computed by noting that
with ${\bf L}^2=\sum_a L^aL^a$,
\begin{eqnarray}
  \label{Lqua}
  \big (\hat L\hat L\big )^{ab} = L^aL^b -  \delta^{ab}{\bf L}^2\; ,
\end{eqnarray}
which leads to the following result for odd and even powers of $\hat L$,
\begin{eqnarray}
  \label{Loddeven}
  \big(\hat L\big)^{2n+1} = \left(-{\bf L}^2\right)^n\hat L\; ,\quad
  \big(\hat L\big)^{2n+2} = \left(-{\bf L}^2\right)^n\hat L\hat L\; .
\end{eqnarray}
Hence, a geometric series in the matrix $\hat L$ can be expressed by a
geometric series in the number $-{\bf L}^2$,
\begin{eqnarray}
  \label{cdef}
  C&=&1+\hat L+\hat L\hat L+\dots = \id + \sum_{n=0}^{\infty}\big(\hat
  L\big)^{2n+1} + \sum_{n=0}^{\infty}\big(\hat
  L\big)^{2n+2}\nn\\
&=& \id + \frac{1}{1+{\bf L}^2}\hat L + \frac{1}{1+{\bf L}^2}\hat
L\hat L\; .
\end{eqnarray}
The series converges if 
\begin{eqnarray}
  \label{geomconv}
  {\bf L}^2\, <\, 1\; ,
\end{eqnarray}
and the elements of $C$ are found in view of Eq.\
(\ref{Lqua}),\footnote{This matrix is the inverse to $\id-\hat L$, as
  one may expect.}
\begin{eqnarray}
  \label{celem}
  C^{ab} = \frac{1}{1+{\bf L}^2}\left(\delta^{ab}+L^aL^b+\hat
    L^{ab}\right)\; .
\end{eqnarray}
The Coulomb potential in the coherent field $V_C^Z(r)$, see Eq.\
(\ref{VCZres}), depends on the matrix element $(CC)^{33}$. Defining a
string tension $\sigma_Z$ by the infrared limit of
the Coulomb potential $V_C^Z(r)$ in the coherent field,
\begin{eqnarray}
  \label{sigmaZdef}
  V_C^Z(r\to\infty)=(CC)^{33}\sigma_C r=:\sigma_Z r\; ,
\end{eqnarray}
we get from Eq.\ (\ref{celem}) that\footnote{This result is sensitive to the
  approximations made above. At this stage, only qualitative
  statements can be made.}
\begin{eqnarray}
  \label{sigmaZ}
  \sigma_Z = \frac{1-{\bf L}^2+\left(3+{\bf
        L}^2\right)(L^3)^2}{\left(1+{\bf L}^2\right)^2}\sigma_C\; .
\end{eqnarray}
The variables ${\bf L}^2$ and $L^3$ are entangled by the convergence
condition (\ref{geomconv}). With spherical coordinates in color space,
$\xi:=|{\bf L}|$ and $L^3=:|{\bf L}|\cos\eta$, Eq.\ (\ref{sigmaZ}) can be
expressed in terms of independent variables $0<\xi<1$ and $0<\eta<\pi$,
\begin{eqnarray}
  \label{sigmasphcoo}
  \sigma_Z =
  \left(1-\sin^2\eta\:\frac{(3+\xi^2)\xi^2}{(1+\xi^2)^2}\right)\,\sigma_C\; .
\end{eqnarray}
It is thus seen that 
\begin{eqnarray}
  \label{simgaZrange}
  0\, <\,\sigma_Z\, <\sigma_C\; ,
\end{eqnarray}
i.e.\ the Coulomb string tension is indeed lowered by the excitation
of gluons. This very notion was proposed in the gluon chain model and
it follows here from an ansatz for the wave functional of
$\mathit{SU(2)}$ YM theory with heavy quarks. The upper bound,
$\sigma_Z = \sigma_C$ is reached only if $\eta = 0$, i.e.\ $|{\bf L}| = L^3$.

Depending on the coherent field $Z_k^a(\fx)$, the string
tension $\sigma_Z$ could in principle be lowered just above zero. 
In this case, ${\bf L}$ has no component along the color vector
${\boldsymbol{\rho}}_{\mathrm{ext}}(\fx)$, i.e.\ $\eta = \frac{\pi}{2}$, and its length is
maximal, i.e.\ $|{\bf L}| \nearrow 1$. Thus, we find $\sigma_Z\searrow
0$.
However, one
may expect that this leads to a rise in the magnetic potential
energy. Adding more and more gluons into the Coulomb interaction must
increase this part of the energy since it is ultralocal in the
coherent field, see Eq.\ (\ref{Epfin}). At a given
mean number of excited gluons (the peak of the Poisson distribution), a
saturation of the energy must occur. Realizing this effect
computationally may be difficult, though. If the magnetic potential
energy indeed saturated the number of gluons for large separations
$r$, it would have to also rise linearly in $r$, otherwise it
would be negligible in the asymptotic infrared. 

\section{Squeezed states}
\label{squeeze}

Although the estimate of the Coulomb propagator in the coherent
state in the previous section indicated a lowering of the string
tension, it is not clear whether a full self-consistent solution will
exhibit the same feature. A possible alternative description of the
back reaction of the gluonic sector on the external charges may be
motivated, again, from the quantum mechanic harmonic oscillator. In
three dimensions, the charged oscillator can be studied in an external
uniform magnetic field ${\bf B}=B{\bf e}_\parallel$ which couples to the angular momentum. To this
end, the Hamiltonian (\ref{3dharm}) is supplemented by the substitution
\begin{eqnarray}
  \label{minsubst}
  p_k\rarr p_k-qA_k\; ,\quad {\bf A} =-\frac{1}{2}\fx\times{\bf B}\; .
\end{eqnarray}
The new Hamiltonian $H'=H+H_P+H_D$ contains the Larmor frequency
$\omega_L=-\frac{qB}{2m}$ linearly in the paramagnetic term
\begin{eqnarray}
  \label{HPsq}
  H_P=\omega_L L_\parallel\; ,
\end{eqnarray}
and quadratically in the diamagnetic term
\begin{eqnarray}
  \label{HDsq}
  H_D = \frac{1}{2}m\omega_L^2\fx_{\perp}^2\; .
\end{eqnarray}
The coupling of the angular momentum operator ${\bf L}=\fx\times{\bf
p}$ to the external magnetic field ${\bf B}$ in $H_P$ (\ref{HPsq}) is
reminiscent of the coupling of the dynamical charge $\rhodyn^a(x)$ of
YM theory to the external charge $\rhoext^a(x)$ in the Coulomb
interaction $\rhodyn F \rhoext$. This similarity is acknowledged by
noting that $\rhodyn^a = -\hat{\bf A}^{ab}\cdot{\bf\Pi}^b$ is an outer product of coordinate and
momentum operators, albeit in color space, just like the angular momentum ${\bf L}$. For
the calculation of the energy in the coherent state, this has
important consequences, as will be seen.

Before we focus on the coherent state, let us give the exact ground
state solution of the rather simple Hamiltonian $H'$. The primed
quantities always include the effect of the external field. Thus, we can interpret
\begin{eqnarray}
  \label{H+HD}
  H+H_D = \frac{1}{2}\omega\left(\sigma^2{\bf
      p}_\parallel^2+\sigma^{-2}\fx_\parallel^2\right) + \frac{1}{2}\omega'\left(\sigma'^2{\bf
      p}_\perp^2+\sigma'^{-2}\fx_\perp^2\right)
\end{eqnarray}
as one harmonic oscillator of frequency $\omega$ (and width $\sigma^2 =
\frac{1}{m\omega}$) in the ${\bf e}_\parallel$-direction, and two
oscillators perpendicular to it with frequency $\omega' =
({\omega^2+\omega^2_L})^{1/2}$ (and width
$\sigma'^2=\frac{1}{m\omega'}$).
Using the annihilation operators
\begin{eqnarray}
  \label{asqdef}
  a_\parallel = \frac{1}{\sqrt{2}}\left(\sigma^{-1}x_\parallel+i\sigma
    p_\parallel\right)\; ,\quad
  a_k = \frac{1}{\sqrt{2}}\left(\sigma'^{-1}x_k+i\sigma' p_k\right)\;
  ,\quad k=x,y\; ,
\end{eqnarray}
where ${\bf a}_\perp=a_x{\bf e}_x+a_y{\bf e}_y$, the Hamiltonian $H'$
can be re-expressed, but the paramagnetic part $H_P$ is still not diagonal. By
means of a Bogoliubov transformation,
\begin{eqnarray}
  \label{Bogo}
  a_\pm = \frac{1}{\sqrt{2}}\left(a_x\pm ia_y\right)\; ,
\end{eqnarray}
the complete Hamiltonian is diagonal,
\begin{eqnarray}
  \label{H'sq}
  H'=\omega \left(a_z^\dagger a_z +
    \frac{1}{2}\right)+\omega'\left(a_+^\dagger a_+ + a_-^\dagger a_-
    + 1\right) - \omega_L\left(a_+^\dagger a_+ - a_-^\dagger a_-
    + 1\right)\; ,
\end{eqnarray}
in the orthogonal basis of the states
$(a_+^\dagger)^{n_+}(a_-^\dagger)^{n_-}(a_z^\dagger)^{n_z}\ketvac$.
The exact ground state of $H'$ therefore has the energy
\begin{eqnarray}
  \label{E'sq}
  E'_0 = \frac{1}{2}\omega + \sqrt{\omega^2+\omega_L^2}\, >
  \,\frac{3}{2}\omega = E_0\; ,
\end{eqnarray}
i.e.\ it is higher than without the external field ($\omega_L=0$). The
interesting part is that the exact ground state $\psi_0'(\fx) =
\psi_{0,\parallel}'(\fx)\psi_{0,\perp}'(\fx)$ is Gaussian,
\begin{eqnarray}
  \label{psi'sq}
  \psi_{0,\parallel}'(\fx)&=&\frac{1}{\left(\pi
    \sigma^2\right)^{1/4}}\e^{-\frac{1}{2}\fx_\parallel^2/\sigma^2}\; ,\\
  \psi_{0,\perp}'(\fx)&=&\frac{1}{\left(\pi
    \sigma'^2\right)^{1/2}}\e^{-\frac{1}{2}\fx_\perp^2/\sigma'^2}\; ,
\end{eqnarray}
but the width $\sigma'$ in the directions perpendicular to the external field
${\bf B}$ is changed, as compared to the width $\sigma$ of the
Gaussian in the direction parallel to ${\bf B}$. By switching on the
external field ${\bf B}$, the ground state wave function is squeezed
by a factor smaller than $1$,
\begin{eqnarray}
  \label{squeezefactor}
  \sigma' =
  \sqrt[4]{\frac{1}{1+\frac{\omega_L^2}{\omega^2}}}\:\sigma\; ,
\end{eqnarray}
in perpendicular directions.

The effect of squeezing cannot be expected from a coherent
state which merely shifts the vacuum wave
functional. We now construct a coherent state 
in the eigenbasis of $H$, and determine the parameters $z_k$ by
minimizing $E'({\bf z}):=\bra{{\bf z}}H'\ket{{\bf z}}$. This is equivalent to the
approach in YM theory, see section \ref{energycoh}. The minimal value
$E'({\bf z})$
can then be compared to $E'_0$ in Eq.\ (\ref{E'sq}). 

Define the coherent state by 
\begin{eqnarray}
  \label{Zsqdef}
  \ket{{\bf z}} = \e^{{\bf z}\cdot {\bf a}^\dagger - {\bf z}^*\cdot{\bf a}}\ketvac\; ,\quad
    {\bf a}=\frac{1}{\sqrt{2}}\left(\sigma^{-1}\fx+i\sigma {\bf p}\right)
\end{eqnarray}
and calculate the coherent state energy $E'({\bf z})=E({\bf
  z})+E_P({\bf z})+E_D({\bf z})$,
\begin{subequations}
 \label{Esqres}
\begin{eqnarray}
   E({\bf z}) &=& \omega\left((\Im {\bf z})^2+(\Re {\bf z})^2 + \frac{3}{2}\right)\\
  E_P({\bf z}) &=& 2\omega_L\left(\Re {\bf z}_\perp\times\Im {\bf
      z}_\perp\right)\cdot {\bf e}_\parallel\\
\label{Ediam}
  E_D({\bf z}) &=&
  \frac{1}{2}m\omega_L^2\left(\sigma^2+2\sigma^2(\Re {\bf z}_\perp)^2\right)
\end{eqnarray}
\end{subequations}
Note that the coupling of the angular momentum to the magnetic field
brings about a term $E_P(z)$ linear in $\Im {\bf z}_\perp$. This also
occurs in the energy expression (\ref{completeE}) of YM theory where the coupling of
$\rhodyn$ to $\rhoext$ is responsible for such a term. By quadratic
completion, it is easily found that for $\Im {\bf
  z}=\frac{\omega_L}{\omega}{\Re {\bf z}}\times{\bf e}_\parallel$ the
energy $E'({\bf z})$ is minimal. In this case,
\begin{eqnarray}
  \label{E'Imin}
  E'({\bf z})& =& E_0+\omega\left(1-\frac{\omega_L^2}{\omega^2}\right)(\Re {\bf
    z}_\perp)^2+\omega(\Re {\bf z}_\parallel)^2 + E_D(z)\nn\\
&=& \frac{3}{2}\omega + \frac{1}{2}\frac{\omega_L^2}{\omega} +
\omega (\Re {\bf z})^2\; .
\end{eqnarray}
Minimizing this expression w.r.t.\ $\Re{\bf z}$, one finds that $\Re {\bf z}=0$ and thus $\Im
{\bf z}=0$. Therefore, the trivial solution $\ket{{\bf z}}=\ket{\bf 0}$ is
energetically most favorable. In the first line of Eq.\ (\ref{E'Imin})
one might hope for a non-trivial solution due to the minus sign. Such
a situation occurs also in YM theory. However, this negative term is
exactly cancelled by the diamagnetic energy (\ref{Ediam}) with
$\sigma^2=\frac{1}{m\omega}$. Since the coherent state with ${\bf Z=0}$
still gives a higher energy than the exact ground state energy (\ref{E'sq}), 
\begin{eqnarray}
  \label{whoisbest}
  E'({\bf 0}) - E'_0 =
  \omega\left(1+\frac{1}{2}\frac{\omega_L^2}{\omega^2} -
    \sqrt{1+\frac{\omega_L^2}{\omega^2}}\right)\, > 0\; ,
\end{eqnarray}
it must be inferred that the coherent state cannot
mimic the squeezing of a Gaussian wave function.

Of course, the comparison of the coherent state to the quantum mechanical harmonic
oscillator and the Yang--Mills theory is by no means supposed to
indicate a complete analogy. For instance, quartic and cubic terms may
allow for a nontrivial shifted ground state---the coherent state---to be
energetically more favorable. Also, the coupling of the ''outer
product'' $\rhodyn$ to the external field in YM theory is mediated by
the Coulomb operator $F[A]$ which is dependent on the field operator
$A$ itself. Such a situation is not encountered for the harmonic
oscillator and we have seen in the previous section that the
expectation value of $F[A]$ in the coherent state may alone provide
part of the desired effect. Despite these differences of the
oscillator coupled to a magnetic field on the one hand and the gluonic vacuum with
external charges on the other, it might be a useful idea to consider
squeezed states as an alternative to coherent states. This is a
possible future investigation. An introduction to squeezed states
can be found in the literature on quantum optics \cite{Gerry}.

\addcontentsline{toc}{chapter}{Summary and outlook}
\fancyhead[LE]{\scshape Summary and outlook}        % chapter to left
\fancyhead[RO]{\scshape Summary and outlook}       % section to right
\chapter*{Summary and outlook}
%\addcontentsline{toc}{chapter}{Conclusions and outlook}

The gauge principle, which is the main ingredient in the very
definition of the gauge field sector, recurs in all aspects of
Yang--Mills theory. Starting with the quantization procedure, discussed in
chapter \ref{ch:YMconstr}, gauge invariance requires a careful
treatment which is particularly difficult for nonabelian gauge
groups. The generator of (time-independent) gauge transformations is
the Gauss law operator. In the canonical quantization approach, it is
not possible to promote the Gauss law to an operator identity. We have
shown that by means of fixing the gauge completely on the classical
level and imposing quantization conditions within the Dirac bracket
formalism, the Gauss law does hold on the quantum level. The
gauge-fixed Hamiltonian operator thus derived determines an
unambiguous time evolution of the quantum system. It agrees with the
Christ-Lee Hamiltonian \cite{ChrLee80} where a projection on gauge
invariant states is employed. This agreement is notably reassuring in
the sense that there is no ambiguity in the choice of the quantization
procedure. For Yang--Mills theory in the temporal Coulomb gauge, it
makes no difference whether Dirac quantization ({\it first quantize, then
constrain}) or constrained quantization ({\it first constrain, then
quantize}) is used. In the light of the gravitational force, the
method of quantization is to date a controversial issue.

In the temporal Coulomb gauge-fixed configuration space of Yang--Mills
theory, the uniqueness of the gauge fixing condition demands the
restriction to a compact region, the fundamental modular region. In
the Gribov--Zwanziger scenario, this infers the enhancement of
boundary effects in that region. We have investigated in chapter
\ref{ghostdom} the consequences for the Green functions in the temporal
Coulomb gauge in $d$ spatial dimensions of $(d+1)$-dimensional
Minkowski spacetime. The Landau gauge results for
$(d+1)$-dimensional Euclidean spacetime can be obtained by shifting $d\to d+1$. In
the temporal Coulomb gauge, the stochastic vacuum ($\Psi[A]=1$) is
sufficient to account for the infrared asymptotics of the theory.
The horizon condition, which enhances the ghost propagator in the
infrared, is along with the nonrenormalization of the ghost-gluon
vertex the most important feature for the infrared behavior of
propagators in the temporal Coulomb or the Landau gauge. Using power
law ans\"atze for these propagators, the set of Dyson--Schwinger
equations was solved with a tree-level ghost-gluon vertex. A single
infrared exponent $\kappa$ can be extracted to specify the infrared
power laws of both the ghost and gluon propagators, for a given dimension
$d$. In the $3+1$ dimensional Coulomb gauge, the result
$\kappa=\frac{1}{2}$ is shown to yield a heavy quark potential
which rises linearly. Quark confinement is thus, qualitatively speaking, a
consequence of the horizon condition. It was argued that this
result, is not necessarily unique. First of all, there is one further
solution for $d=3$, $\kappa\approx 0.398$, which leads to a Coulomb potential $V_C(r)$ that
rises less than linearly \cite{FeuRei04}. With the Coulomb potential
being an upper bound to the gauge invariant quark potential $V_W(r)$, the
latter solution would indicate that $V_W(r)$ also rises less than
linearly, contradicting the lattice results. In this sense, it is more
likely that the solution $\kappa=\frac{1}{2}$ is realized. Moreover,
the off-shell gauge condition introduces an additional parameter
$\zeta$ into the infrared analysis, on which the solution for $\kappa$ generally depends, due to the
approximation of the ghost-gluon vertex. It was found that all
solutions but the one for $d=3$ and $\kappa=\frac{1}{2}$ depend on the
value of $\zeta$. This infers that the latter solution is less sensitive
to the approximations made. Its corresponding continuous branch of
solutions $\kappa_a(d)=\frac{d}{2}-1$ as a function of the dimension $d$ 
exists for all $d$, even for very high dimensions
where the other solutions cease to exist (see Fig.\
\ref{highd}). The infrared power laws were furthermore extended by an
ansatz that incorporates powers of logarithms in the infrared. It was
found that values for these exponents exist for which the set of
integral equations is also solved in the asymptotic infrared. The
power law exponents, however, remain unchanged. Hence, there is a
degeneracy in the asymptotic power law solutions, also for $d=3$ and
$\kappa=\frac{1}{2}$. Qualitatively, the logarithmic corrections do
not change the power law behavior and thus also not the confining property of
the Coulomb potential.

A full solution for the propagators of Yang--Mills theory in the
temporal Coulomb gauge is expected to show both, the infrared behavior that
infers confinement and the ultraviolet behavior that agrees with
perturbation theory. The variational solution to the Yang--Mills \Sch
equation, discussed in chapter \ref{VarVac}, provides such a solution approximatively. The Gaussian types of
wave functionals were shown to reproduce exactly the infrared behavior ($\kappa=\frac{1}{2}$)
anticipated in the previous chapter. Thus, numerical results
with a linearly confining potential were obtained
\cite{EppReiSch07}. While the propagators are insensitive to the
details of the wave functional, it was found that the three-gluon
vertex is extremely dependent on it. For one wave functional
considered, it is equally zero, whereas for another it shows a strong
infrared enhancement. However, in the Dyson--Schwinger equations of the
propagators the dressing of the three-gluon vertex was found to have
no effect in the infrared \cite{SchLedRei06}. The ghost-gluon vertex
was studied in the infrared gluon limit from its corresponding
Dyson--Schwinger equation. It was shown that generally
the infrared limits of the vertex' momenta are not interchangeable,
except for the solution $d=3$ and $\kappa=\frac{1}{2}$. Again, the
regularity of the vertex indicates that this solution may be favored. 
Chapter \ref{VarVac} is furthermore concerned with the assessment of
the approximation of the Coulomb form factor. It was shown that
the horizon condition needs to be relaxed in order to arrive at a
solution that satisfies the Coulomb form factor DSE. The infrared
analysis of the ghost DSE and the Coulomb form factor DSE proves that
a simultaneous solution with infrared divergent form factors is impossible.
The so-called ''subcritical solution'' \cite{Epp+07} where the form factors are
infrared finite solves all DSEs but does not provide the expected
infrared behavior of the quark potential. If a critical solution
exists, then the Coulomb form factor DSE, as it stands, will be
violated. Higher-order effects must then be included. An approach that
includes higher-order effects may be to calculate the gap equation in
the form of Eq.\ (\ref{gapAP}) up to two loops (avoiding a three-loop
expression for the energy density). This may be tedious but possible.

The ultraviolet tails of the variational solutions for the Coulomb
gauge Green functions have the correct power law behavior but the
anomalous dimensions, related to the powers of logarithms, turn out
incorrect. This might be an effect of the simple Gaussian shape of the
wave functional. In chapter \ref{UVchap}, the ultraviolet gluon energy
$\omega(k)\sim k\ln^{3/11}k$ and ghost form factor $d(k)\sim
1/\ln^{4/11}k$ were motivated. This followed from a nonperturbative
definition of the running coupling by the ghost-gluon vertex. In a
calculation carried out in both Landau and Coulomb gauge, the ghost
DSE was analyzed in the asymptotic ultraviolet using appropriate
ans\"atze. From the requirement that the nonperturbative running coupling
coincide with the ultraviolet limit of perturbation theory, it can be
deduced what the anomalous dimensions are to leading order. The way to
reproduce these anomalous dimensions must be to extend the Gaussian
wave functional. At the very least, the gluon loop must be included. A
non-trivial dressing of the three-gluon vertex, similar to Landau
gauge studies, is most probably necessary. This is a possible future investigation.

Moreover, chapter \ref{UVchap} dealt with the infrared limit of the
running coupling which can be calculated  analytically. In all gauges that interpolate between the
Landau and the Coulomb gauge, this value is the same, but changes discontinuously in
the Coulomb gauge limit \cite{SchLedRei06}. Vertex corrections to the
infrared limit of the running coupling were also
calculated, showing that the only solution that has no
vertex correction is the one where $d=3$ and $\kappa=\frac{1}{2}$.

The final chapter \ref{external} uses a quasi-particle representation
of gluonic excitations to incorporate the back reaction of the presence
of external charges on the gluon sector. To this end, coherent states
are motivated from quantum mechanical examples. The calculation of the
Yang--Mills energy in the coherent state produces a vast expression
that depends on a localized background field. It was possible to remove
divergences from this expression by usage of the gap equation. Even
after the explicit variational solution for the imaginary part of the
background field, the energy expression requires some approximation to
be processed further. A first approximation, rendering the Green
function insensitive to the presence of external charges, yielded
senseless results. Subsequently, the Coulomb potential was estimated
in the asymptotic infrared by an ad hoc ansatz for the background
field. It was possible to expand, calculate, and resum this expectation
value and indeed show that the string tension can be lowered by such an
ansatz. The determination of the exact factor by which the string
tension is lowered, requires further investigation. Another quantum
mechanical example was put forward to motivate that squeezed states
may serve equally well for the incorporation of external charges.

\fancyhead[LE]{\scshape\leftmark}        % chapter to left
\fancyhead[RO]{\scshape\rightmark}       % section to right
\appendix
\chapter{Conventions and notation}
\label{AppA}
\section{Units, metric and group conventions}
Throughout this thesis, natural units are used,
\begin{eqnarray}
  \label{unitsdef}
  \hbar = c = 1\; .
\end{eqnarray}
Since $\hbar c \approx 200\, \mathrm{MeV\, fm}^{-1}$, Eq.\
(\ref{unitsdef}) infers that $1\,$fm
corresponds to $200\, \mathrm{MeV}$. The Minkowski metric is here defined by the
metric tensor
\begin{eqnarray}
  \label{metrdef}
  \left( g_{\mu\nu} \right) = \left(
    \begin{array}{cccc}
      1 & 0 & 0 & 0 \\
      0 &-1 & 0 & 0 \\
      0 & 0 &-1 & 0 \\
      0 & 0 & 0 &-1
    \end{array}
    \right)
\end{eqnarray}
Antihermitian generators of group transformation in $SU(N_c)$ are normalized as
follows. For the fundamental representation, we choose
\begin{eqnarray}
  \label{trfun}
  \tr \left(T^aT^b\right) =-\frac{1}{2}\delta^{ab}
\end{eqnarray}
and for the adjoint representation
\begin{eqnarray}
  \label{tradj}
  \tr \left(\hat T^a\hat T^b\right)= -f^{acd}f^{bcd} = -N_c\delta^{ab}\; .
\end{eqnarray}
It follows that
\begin{eqnarray}
  \label{3Ts}
  \tr \left(\hat T^a\hat T^b \hat T^c \right)= -f^{ade}f^{beg}f^{cgd} = -\frac{N_c}{2}f^{abc}\; .
\end{eqnarray}

\section{Notation% for integration, functional traces \\   and Fourier
                 % transformations
}
In momentum space integration, the phase space factor $2\pi$ is
absorbed by the definition 
\begin{eqnarray}
  \label{dbardef}
  \dbar^dk:=\frac{d^dk}{(2\pi)^d}\; ,
\end{eqnarray}
in analogy to $\hbar$. In the case of multiple integrations, the
abbreviation
\begin{eqnarray}
  \label{intmult}
  d[xyz]=dx\, dy\, dz
\end{eqnarray}
is frequently used.

Matrices in coordinate and intrinsic space, for instance $G$, have the
components $G^{ab}(x,y)$. The functional trace ``$\Tr$'' is then
defined by
\begin{eqnarray}
  \label{Trdef}
  \Tr G = \int d^d[xy]\delta^d(x,y)\delta^{ab}G^{ab}(x,y)
\end{eqnarray}
Functional determinants can be related to the functional traces
(\ref{Trdef}) by the identity $\Det G = \exp\Tr\ln G$.

The Fourier transformation of a function ${\cal O}(x)$, operating
on spacetime, defines by
\be
\label{FourDef}
\cO(x)=\int\dbar^dp\:\tcO(p)\e^{ip\cdot x}\: 
\ee
the function $\tcO(p)$ operating on momentum space. Although these are
different functions, the tilde is often omitted. For a function of
two variables, we may define a Fourier transform by
\be
\cO(x,y)=\int\dbar^d[pq]\:\e^{ip\cdot x}\tcO(p,q)\e^{-iq\cdot y}\; .
\ee
In translationally invariant systems, $\cO(x,y)=\cO(x-y)$, and
\be
\tcO(p,q)&=&\int d^d[xy]\:\e^{-ip\cdot x}\cO(x,y)\e^{iq\cdot y}\nn \\
&=&\int d^dy\:\e^{i(q-p)\cdot y}\left\{\int d^dx
\e^{-ip(x-y)}\cO(x-y)\right\}\nn\\ &=& (2\pi)^d\delta^d(p-q)\tcO(p)\: ,
\ee
such that the propagator-like object $\cO(x-y)$ is directly related to
$\tcO(p)$ via Eq.\ (\ref{FourDef}). Whereas in coordinate space, a
matrix in coordinate and intrinsic space is denoted by a single
symbol, for instance $G$, the Fourier transformed object is usually a
function of a single momentum scale and has no color structure, cf.\
Eq.\ (\ref{DGdef}).

Higher $n$-point functions are defined equivalently. A vertex function
with translational invariance, $\cO(x,y,z)=\cO(x-y,x-z)$, leads to
\be
\tcO(k,q,p)&=&\int d^d[xyz]\:\e^{-ik\cdot x}\cO(x,y,z)\e^{-iq\cdot
  y}\e^{-ip\cdot z}\nn\\
&=& \int d^dz\:\e^{-i(p+q+k)\cdot z}\left\{\int
d^d[xy]\:\e^{-ik\cdot(x-z)}\cO(x-y,x-z)\e^{-iq\cdot(y-z)}\right\}\nn\\
&=& (2\pi)^d\delta^d(p+q+k)\tcO(k,q)
\ee
and momentum is conserved by the delta function. Choosing the momentum
routing sets the signs in the exponent of the definition of
the Fourier transform.

Functional derivatives can be shown to yield
\be
\frac{\delta}{\delta \cO(x)}=\int\dbar^d p\, \e^{-ip\cdot
  x}\frac{\delta}{\delta \tcO(p)}\; .
\ee
Note that the sign in the exponent is opposite of that in Eq.\
(\ref{FourDef}).

\chapter{Two-point integrals}
\label{app:2ptint}
Here we sketch the evaluation of nonperturbative Euclidean loop integrals in asymptotic limits. According to the discussion on page \pageref{page:asymp}, we
can replace the integrands by their asymptotic forms and encounter the two-point integrals
\begin{equation}
 \Xi_m(\alpha,\beta):=\int \dbar^d\ell\:\frac{(\ell\cdot
   k)^m}{(\ell^{2})^{\alpha}((\ell-k)^{2})^{\beta}}\; ,\quad \alpha
 ,\beta\in\mathbb{R}\, ,\,\, m\in\mathbb{N}\; .
\label{eq:2ptint}
\end{equation}
These can be shown to be homogeneous functions of the momentum $k$. By a scaling of
the integration variable, $\ell\rightarrow\lambda\ell$, one readily finds that since the two-point integral can only depend on the scale, it should obey $\Xi_m(\alpha,\beta) \sim (k^2)^\kappa$ and the exponent of
the power law can be determined to be $\kappa=d/2-\alpha-\beta+m$. After applying the usual trick of introducing Feynman parameters,
\begin{equation}
\label{Feynmantr}
\frac{1}{C_{1}^{\alpha}C_{2}^{\beta}}=\int_{0}^{1}dx\int_0^1dy\:\delta(x+y-1)\:\frac{x^{\alpha-1}y^{\beta-1}}{(xC_{1}+yC_{2})^{\alpha+\beta}}\frac{1}{B(\alpha,\beta)}
\: ,
\end{equation}
where $B(\alpha,\beta)$ is the Euler beta function, we can shift the
integration variable, $\ell\rarr\ell-yk$. The integrand then depends
on $\ell^2$ only and we can integrate it out. For $m=0,1,2,3$ we need
the following standard integrals \cite{Peskin}:
\bee
\int\frac{\dbar^{d}\ell}{(\ell^{2}+\Delta)^{n}}&=&\frac{1}{(4\pi)^{d/2}}\frac{\Gamma(n-d/2)}{\Gamma(n)}\left(\frac{1}{\Delta}\right)^{n-d/2} \\
\int\frac{\dbar^{d}\ell\:\ell_i\ell_j}{(\ell^{2}+\Delta)^{n}}&=&\frac{1}{2}\:\delta_{ij}\frac{1}{(4\pi)^{d/2}}\frac{\Gamma(n-d/2-1)}{\Gamma(n)}\left(\frac{1}{\Delta}\right)^{n-d/2-1}
\eeq
Integrals with an odd number of vectors $\ell$ in the numerator vanish
by symmetry. For our purposes, we have $\Delta=xyk^2$. The two-point integrals can be straightforwardly computed using the identity
\bee
\int_0^1dx\int_0^1dy\:\delta(x+y-1)x^{\alpha-1}y^{\beta-1}=B(\alpha,\beta) = \frac{\Gamma(\alpha)\Gamma(\beta)}{\Gamma(\alpha+\beta)}
\eeq
for the Euler beta function. One then finds the results
\begin{subequations}
\label{2ptints}
\begin{eqnarray}
\label{eq:2pt}
\Xi_0(\alpha,\beta)&=& \frac{1}{(4\pi)^{d/2}}\frac{\Gamma(d/2-\alpha)\Gamma(d/2-\beta)\Gamma(\alpha+\beta-d/2)}{\Gamma(\alpha)\Gamma(\beta)\Gamma(d-\alpha-\beta)}(k^{2})^{d/2-\alpha-\beta}\\
\label{eq:2ptNum1}
\Xi_1(\alpha,\beta) &=& \frac{1}{(4\pi)^{d/2}}\frac{\Gamma(d/2-\alpha+1)\Gamma(d/2-\beta)\Gamma(\alpha+\beta-d/2)}{\Gamma(\alpha)\Gamma(\beta)\Gamma(d-\alpha-\beta+1)}(k^{2})^{d/2-\alpha-\beta+1}\\
\label{eq:2ptNum2}
\Xi_2(\alpha,\beta)\nonumber &=&
 \frac{1}{(4\pi)^{d/2}}\frac{\Gamma(d/2-\alpha+2)\Gamma(d/2-\beta)\Gamma(\alpha+\beta-d/2)}{\Gamma(\alpha)\Gamma(\beta)\Gamma(d-\alpha-\beta+2)}(k^{2})^{d/2-\alpha-\beta+2} \\
&& \hspace{-1.9cm}+ \frac{1}{2} \frac{1}{(4\pi)^{d/2}}\frac{\Gamma(d/2-\alpha+1)\Gamma(d/2-\beta+1)\Gamma(\alpha+\beta-d/2-1)}{\Gamma(\alpha)\Gamma(\beta)\Gamma(d-\alpha-\beta+2)}(k^{2})^{d/2-\alpha-\beta+2}\\
\Xi_3(\alpha,\beta) \nonumber &=&
 \frac{1}{(4\pi)^{d/2}}\frac{\Gamma(d/2-\alpha+3)\Gamma(d/2-\beta)\Gamma(\alpha+\beta-d/2)}{\Gamma(\alpha)\Gamma(\beta)\Gamma(d-\alpha-\beta+3)}(k^{2})^{d/2-\alpha-\beta+3}\\ 
&& \hspace{-1.9cm}+ \frac{3}{2}\frac{1}{(4\pi)^{d/2}}\frac{\Gamma(d/2-\alpha+2)\Gamma(d/2-\beta+1)\Gamma(\alpha+\beta-d/2-1)}{\Gamma(\alpha)\Gamma(\beta)\Gamma(d-\alpha-\beta+2)}(k^{2})^{d/2-\alpha-\beta+3}
\label{eq:2ptNum3}
\end{eqnarray}
\end{subequations}

The above formulae are valid for those values of $\alpha$, $\beta$, $m$ only for which the integrals converge. At $\ell=k$, a pole is integrable as long as $\beta<d/2$. On the other hand, the infrared convergence at $\ell=0$ depends on $m$:
\bee
\label{IRconv}
\alpha < \left\{%\right.
\begin{array}{cc}
d/2+ m/2 & \quad \textrm{for even}\,\, m \\
d/2+m/2+1/2 & \quad \textrm{for odd}\,\, m 
\end{array}\quad .
\right.
\eeq

One can relate this inequality to the requirement that the arguments
of the first gamma functions in the numerators of Eqs.\ (\ref{2ptints}) be positive. The ultraviolet convergence is contained in the third gamma function of the numerators. For convergence, the relation
\bee
\label{UVconv}
\alpha+\beta > \left\{%\right.
\begin{array}{cc}
d/2+ m/2 & \quad \textrm{for even}\,\, m \\
d/2+m/2-1/2 & \quad \textrm{for odd}\,\, m 
\end{array}
\right.
\eeq
has to be satisfied. Obviously, odd values of $m$, compared to even
values, work ``in favor'' of convergence both in the infrared and in
the ultraviolet, due to the angular integration. In sums of UV
divergent two-point integrals, the angular integration eliminates the
divergence in some cases. For instance, the Brown-Pennington
projection of the gluon propagator in Landau gauge \cite{BroPen88a},
\begin{eqnarray}
  \label{BPgluon}
  R_{\mu\nu}^{\zeta=d}(k)D_{\mu\nu}(k)\sim\ln\Lambda
\end{eqnarray}
eliminates the unphysical quadratic divergences proportional to the
metric tensor, by virtue of $R_{\mu\nu}^d(k)$ in Eq.\
(\ref{Rtens}). We demonstrate that this elimination can be due to the
angular integration for the Brown-Pennington
projection of the off-shell ghost loop (\ref{gluonDSEoffij}) for $d=3$ and the
ultraviolet ghost propagator (\ref{UVd}),
\begin{eqnarray}
  \label{chiBP}
  \chi^{\zeta}(k) &=& g^2\frac{N_c}{4}\int\dbar^3\ell\left[1-\zeta(\hat{\ell}\cdot\hat{k})^2+(\zeta-1)\frac{\ell\cdot
    k}{\ell^2}\right]\frac{d(\ell)d(\ell-k)}{(\ell-k)^2}\nn\\
&=&\frac{g^2N_c}{16\pi^2}\int^\Lambda d\ell
\frac{\ell^2}{\sqrt{\ln(\ell^2)}}\int_{-1}^1 dx\left[1-\zeta
  x^2+(\zeta-1)\frac{k x}{\ell}\right]\nn\\&&\qquad\qquad\frac{1}{(\ell^2+k^2-2\ell k
  x)\sqrt{\ln(\ell^2+k^2-2\ell k x)}}\nn\\
&=&\frac{g^2N_c}{16\pi^2}\int^\Lambda d\ell
\frac{1}{\ln(\ell^2)}\left(\frac{2}{3}(3-\zeta)+\cO(\frac{k^2}{\ell^2})\right)\nn\\
&=&(3-\zeta)\:\cO\left(\frac{\Lambda}{\ln(\Lambda^2)}\right)+\mathrm{finite}
\end{eqnarray}
For $\zeta=d=3$, the linear divergence of the ghost loop vanishes.

Another cancellation of UV divergences is encountered when
implementing the horizon condition in the ghost DSE
(\ref{ghostDSEren}). The proposition claimed in section \ref{rechnung}
is that although the ghost self-energy $\Sigma$ is UV divergent, the
naive usage of the formulae (\ref{2ptints}) yields the correct result
for the finite difference $d^{-1}(k)-d^{-1}(0)$. The proof is here
restricted to two-point integrals $I(k)=\Xi_0(\alpha,\beta)$ with
$n:=\alpha+\beta$ such that $I(k)$ does not converge in the UV. Let
$\frac{d}{2}>n>\frac{d}{2}-1$. With the Feynman parameter trick
(\ref{Feynmantr}), we get
\begin{eqnarray}
  \label{IbyJ}
  I(k)&=&
  \frac{1}{B(\alpha,\beta)}\int_0^1dx\int_0^1dy\:\delta(x+y-1)x^{\alpha-1}y^{\beta-1}\int \dbar^d\ell\frac{1}{(x\ell-y(\ell-k))^n}\nn\\
&=& \frac{\Omega_d}{(2\pi)^dB(\alpha,\beta)}\int_0^1dx\int_0^1dy\:\delta(x+y-1)x^{\alpha-1}y^{\beta-1}J(x,y;k)
\end{eqnarray}
where $\Omega_d=2\pi^{d/2}/\Gamma(d/2)$ and the function
\begin{eqnarray}
  \label{JUVdef}
  J(x,y;k)=\frac{1}{2}\int_0^{\Lambda^2}d\ell^2\frac{(\ell^2)^{\frac{d}{2}-1}}{(\ell^2+\Delta)^n}
\end{eqnarray}
with $\Delta=xyk^2$ is regulated by the UV cut-off $\Lambda$. Employ
the mean-value theorem,
\begin{eqnarray}
  \label{meanvalue}
  \frac{1}{(\ell^2+\Delta)^n}=\frac{1}{(\ell^2)^n}-\frac{n\Delta}{(\ell^2+\xi\Delta)^{n+1}}\; ,\quad 0<\xi<1\; ,
\end{eqnarray}
to realize that $J$ can be written as $J=J_{\textrm{reg}}+J_{\infty}$
with
\begin{eqnarray}
  \label{Js}
  J_{\infty}&=&\frac{1}{2}\frac{1}{\frac{d}{2}-n}(\Lambda^2)^{\frac{d}{2}-n}\stackrel{\Lambda\to\infty}{\longrightarrow}\infty\\
  J_{\textrm{reg}}(k)&=&-\frac{n\Delta}{2}\int_0^{\Lambda^2}d\ell^2\frac{(\ell^2)^{\frac{d}{2}-1}}{(\ell^2+\xi\Delta)^{n+1}}\stackrel{\Lambda\to\infty}{<}\infty
\end{eqnarray}
for the values $\frac{d}{2}>n>\frac{d}{2}-1$ of the exponent $n=\alpha+\beta$.
In order to determine $J_{\textrm{reg}}=J-J_{\infty}$ explicitly, write
\begin{eqnarray}
  \label{Jreg}
  J_{\textrm{reg}}(k)&=&\frac{1}{2}\int_0^{\Lambda^2}d\ell^2(\ell^2)^{\frac{d}{2}-1}\left[\frac{1}{(\ell^2+\Delta)^n}-\frac{1}{(\ell^2)^n}\right]\nn\\
&=&\frac{1}{2}\int_0^{\Lambda^2}d\ell^2(\ell^2)^{\frac{d}{2}-1}\frac{1}{\Gamma(n)}\int_0^\infty
dt \, t^{n-1}\left[\e^{-(\ell^2+\Delta)t}-\e^{-\ell^2 t}\right]
\end{eqnarray}
With help of the identity
\begin{eqnarray}
  \label{auxid}
  \e^{-at}-\e^{-bt}=\e^{-bt}(b-a)t\int_0^1ds\e^{-(a-b)ts}
\end{eqnarray}
we can now compute $J_{\textrm{reg}}$:
\begin{eqnarray}
  \label{Jregres}
   J_{\textrm{reg}}(k)&=&
\frac{1}{2} \int_0^{\Lambda^2}d\ell^2(\ell^2)^{\frac{d}{2}-1}\frac{1}{\Gamma(n)}\int_0^\infty
dt \, t^{n-1}\e^{-\ell^2 t}\int_0^1ds\e^{-st\Delta}(-t\Delta)\nn\\
&=&-\frac{1}{2}\frac{\Delta}{\Gamma(n)}\,\Gamma\big(\frac{d}{2}\big)\int_0^1ds\int_0^\infty
dt\: t^{n-\frac{d}{2}}\e^{-st\Delta}\; ,\quad n+1>\frac{d}{2}\nn\\
&=&-\frac{1}{2}\Delta\frac{\Gamma(\frac{d}{2})\Gamma(n-\frac{d}{2}+1)}{\Gamma(n)}\int_0^1ds\frac{1}{(s\Delta)^{n-\frac{d}{2}+1}}\;  ,\quad n<\frac{d}{2}\nn\\
&=&+\frac{1}{2}\Delta^{\frac{d}{2}-n}\frac{\Gamma(\frac{d}{2})\Gamma(n-\frac{d}{2})}{\Gamma(n)}
\end{eqnarray}
Plugging this result back into the definition of $I(k)$ in Eq.\
(\ref{IbyJ}), one finds
\begin{eqnarray}
  \label{IUVres}
  I(k)&=&I_\infty+I_{\textrm{reg}}(k)\\
  I_\infty&=&\frac{1}{(4\pi)^{\frac{d}{2}}\Gamma(\frac{d}{2})(\frac{d}{2}-n)}(\Lambda^2)^{\frac{d}{2}-n}\\
  I_{\textrm{reg}}(k)&=&\frac{2\pi^{\frac{d}{2}}}{(2\pi)^d\Gamma(\frac{d}{2})}\frac{\Gamma(\frac{d}{2})\Gamma(n-\frac{d}{2})}{B(\alpha,\beta)\Gamma(n)}\frac{1}{2}B\big(\frac{d}{2}-\alpha,\frac{d}{2}-\beta\big)(k^2)^{\frac{d}{2}-n}\nn\\
&=&\frac{1}{(4\pi)^{d/2}}\frac{\Gamma(d/2-\alpha)\Gamma(d/2-\beta)\Gamma(\alpha+\beta-d/2)}{\Gamma(\alpha)\Gamma(\beta)\Gamma(d-\alpha-\beta)}(k^{2})^{d/2-\alpha-\beta}\; .
\end{eqnarray}
The regular part of $I(k)$ would have been found correctly by direct
usage of Eq.\ (\ref{eq:2pt}), despite the UV divergence.

Finally, another result of a two-point sort of integral is provided that
comes into play in the gluon chain in chapter \ref{gluonchain}. It reads
\begin{eqnarray}
  \label{Xdef}
  X(\alpha,\beta):=\int\dbar^d\ell\:\frac{|\ell\times
    k|}{(\ell^{2})^{\alpha}((\ell-k)^{2})^{\beta}}\; ,
\end{eqnarray}
where in $d$ dimensions $|\ell\times k|$ can be understood  as the
projection of $\ell$ to the hypersurface perpendicular to $k$. The
Feynman trick (\ref{Feynmantr}) yields after the shift $\ell\to\ell-yk$ (which does
not affect the numerator in Eq.\ (\ref{Xdef}))
\begin{eqnarray}
  \label{XFeyn}
   X(\alpha,\beta)=\frac{1}{B(\alpha,\beta)}\int_{0}^{1}dx\int_0^1dy\:\delta(x+y-1)\:x^{\alpha-1}y^{\beta-1}\int\dbar^d\ell\:\frac{|\ell\times
    k|}{(\ell^2+\Delta)^{\alpha+\beta}}\; .
\end{eqnarray}
If $\theta$ is the enclosed angle of $\ell$ and $k$, $|\ell\times k| =
\ell k \sin\theta$, its integration gives
\begin{eqnarray}
  \label{1stangle}
  \int_0^\pi d\theta\,\sin^{d-1}\theta = B\big(\frac{d}{2},\frac{1}{2}\big)\; ,
\end{eqnarray}
whereas the remnant angular integrals simple yield surface of the unit
sphere with $d\to d-1$,
\begin{eqnarray}
  \label{Omegad-1}
  \Omega_{d-1}=\frac{2\sqrt{\pi}^{d-1}}{\Gamma(\frac{d}{2}-\frac{1}{2})}\; .
\end{eqnarray}
Thus, Eq.\ (\ref{XFeyn}) is calculated by
\begin{eqnarray}
  \label{Xres}
   X(\alpha,\beta) &=&
   \frac{1}{(2\pi)^d}
\frac{B(\frac{d}{2},\frac{1}{2})\Omega_{d-1}}{B(\alpha,\beta)}
\int_0^1dy\:\delta(x+y-1)\: x^{\alpha-1}y^{\beta-1}
\frac{1}{2}\int_0^\infty d\ell^2
\frac{k \ell^{\frac{d}{2}-\frac{1}{2}}}
{(\ell^2+\Delta)^{\alpha+\beta}}\nn\\
&=& \frac{1}{(4\pi)^{d/2}} 
\frac{
\Gamma(\frac{d}{2}+\frac{1}{2}-\alpha)\Gamma(\frac{d}{2}+\frac{1}{2}-\beta)\Gamma(\frac{d}{2})\Gamma(\alpha+\beta-\frac{d}{2}-\frac{1}{2})
}{
\Gamma(\alpha)\Gamma(\beta)\Gamma(\frac{d}{2}-\frac{1}{2})\Gamma(d+1-\alpha-\beta)
}
(k^2)^{\frac{d}{2}+1-\alpha-\beta}\nn\\
\end{eqnarray}

\chapter{Infrared limit of the ghost-gluon vertex}
\label{AppC}

The rather technical project of determining the infrared limit of the
ghost-gluon vertex to yield the solution shown in Table \ref{Cresult}, is
presented here. We use the truncated vertex DSE as shown in Fig.\
\ref{fig:GGADSE}. To illustrate how to keep track of color group and
vertex factors, first note that the triangle diagrams comprise
a trace of generators, $\tr\left(\hat T^a \hat T^b \hat T^c\right)$,
as well as three vertex factors of $ig$, i.e.\
\begin{equation}
  \label{vertexfac}
  \inc[bb = 100 600 500 710, clip=]{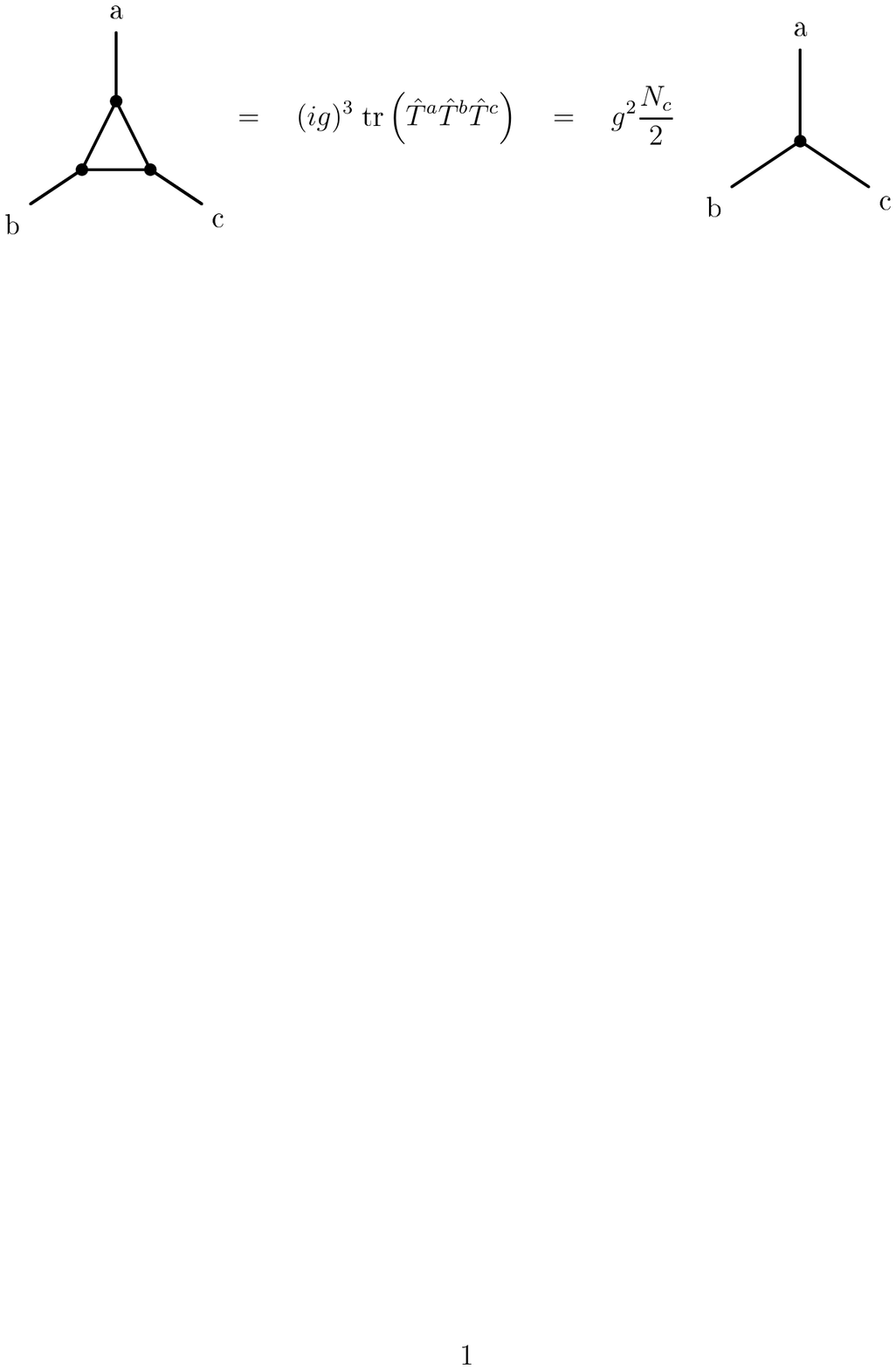}
\end{equation}
due to Eq.\ (\ref{3Ts}). We now consider the infrared gluon limit of
the first loop diagram (\ref{GGADSE1stk0}). With the identity
\begin{equation}
  \label{auxid1}
\ell_{\alpha\beta}(q)t_{\alpha\beta}(\ell-q)=\frac{\ell^2}{(\ell-q)^2}-\frac{(\ell\cdot q)^2}{q^2(\ell-q)^2}  
\end{equation}
we can express this diagram in the infrared limit of the remnant ghost
momentum $q$ by two-point integrals,
\bee
\label{GGAk0res}
\lim_{q\rarr 0}\Gamma^{(GGA)}_\mu(0;q,q)&=&\Gamma_\mu^{(0)}(q)\:Cg^2\frac{N_c}{2}AB^2\left(
\Xi_1(1+2\kappa,\frac{d}{2}-2\kappa)-\Xi_3(2+2\kappa,\frac{d}{2}-2\kappa)/q^2
\right)\nn\\
&=&\Gamma_\mu^{(0)}(q)\:C\:\frac{1}{2}\:\frac{I_1}{I_A}\; ,
\eeq
where 
\bee
I_1=\frac{1}{(4\pi)^{d/2}}\:
\frac{d-1}{d\,\left( 1 + 2\,\kappa  \right)\Gamma(d/2)}\: .
\eeq
and we have used Eq.\ (\ref{coeffrule}). The factor of $C$ appears in
Eq.\ (\ref{GGAk0res}) since the infrared limit is taken as in Eq.\
(\ref{Cdef2}). A factor of $\frac{1}{2}$ stemming from the color trace
(\ref{vertexfac}) is left explicit. Considering the second loop
integral of the ghost-gluon vertex DSE, the result (\ref{ZZZk0}) for the three-gluon vertex in the infrared
limit is plugged into the expression (\ref{GGA1loopGAA}) to find that 
\bee
\label{GAAIRkres}
\lim_{q\rarr 0}\Gamma^{(GAA)}_\mu(0;q,q)&=&(-2)ig^5A^2B^4I_3\, q_\mu\frac{N_c^2}{4}\int \dbar^d\ell\:\ell\cdot q\left(1-(\hat p\cdot\hat\ell)^2\right)\nn\\&&\hspace{4cm}D_Z^2(\ell)D_G(\ell-p)(\ell^2)^{-(\alpha_G-\alpha_Z)} \nn\\
&=& (-2)igq_\mu\:\frac{1}{4}\:\frac{I_3}{I_A^2}\left(\Xi_1(d/2-\kappa,1+\kappa)-\Xi_3(d/2-\kappa+1,1+\kappa)/q^2
\right)\nn\\
&=& \Gamma_\mu^{(0)}(q)\:(-2)\frac{1}{4}\:\frac{I_2I_3}{I_A^2}
\eeq
where $I_3$ was defined in Eq.\ (\ref{I3def}) and
\bee
I_2=\frac{1}{(4\pi)^{d/2}}\:\frac{d-1}{(d-2\kappa)\Gamma(d/2+1)}\; .
\eeq
The factor of $(-2)$ in  Eq.\ (\ref{GAAIRkres}) is the prefactor of the ghost triangle in the
three-gluon vertex DSE, see Fig.\ \ref{DSE diag}.
Altogether, the infrared gluon limit of the ghost-gluon vertex, defined by Eq.\ (\ref{Cdef2}) yields,
\bee
\label{DSEbyC}
C&=&1+\frac{1}{2}C\frac{I_1}{I_A}-\frac{1}{2}\frac{I_2I_3}{I_A^2}
\eeq
and one can find $C$ to obey
\bee
\label{constCfin}
C&=&2\left\{ 4^{\kappa }\left(d-1 \right) d\left( \Gamma(1 + \frac{d}{2}) - \kappa\: \Gamma(\frac{d}{2}) \right) 
     \Gamma(d - 3\kappa )\Gamma(\frac{d}{2} - \kappa ){\Gamma(-\kappa )}^2\Gamma(\frac{1}{2} + \kappa ) \Gamma(1 + 2\kappa )\right.\nn\\
&&\left.\left.\Gamma(2 + 2\kappa ) - 
    \sqrt{\pi }\:\Gamma(\frac{d}{2} - 2\kappa )^3\Gamma(1 + \frac{d}{2} + \kappa )^2\Gamma(2 - \frac{d}{2} + 3\kappa )\right\}\right/\nn\\
&&\bigg\{\left(d-1 \right) \,\left( d - 2\,\kappa  \right) \,\Gamma(d - 3\,\kappa )\,\Gamma(-\kappa )\,\Gamma(1 + 2\,\kappa )^2\nn\\
&&  \left( 4^{1 + \kappa }\Gamma(1 + \frac{d}{2})\Gamma(\frac{d}{2} - \kappa )\Gamma(-\kappa )
     \Gamma(\frac{3}{2} + \kappa ) + {\sqrt{\pi }}\Gamma(\frac{d}{2} - 2\kappa )\Gamma(1 + \frac{d}{2} + \kappa ) \right)\bigg\}
\eeq

As can be checked, this leads to the numerical values of $C$ given by
table \ref{Cresult} for the various solutions of $\kappa$. A
cancellation of the two loop diagrams, see Eq.\ (\ref{cancel}), must yield $C=1$ in
Eq.\ (\ref{DSEbyC}). In this case
\begin{eqnarray}
  \label{Cbitte1}
  1=\frac{I_1I_G}{I_2I_3}=\frac{(d-1)(d-2\kappa)\Gamma(d-3\kappa)\Gamma(1-\kappa)\Gamma(2\kappa)\Gamma(2\kappa+2)}{\Gamma^2(\frac{d}{2}-2\kappa)\Gamma(\frac{d}{2}+\kappa+1)\Gamma(2+3\kappa-\frac{d}{2})}
\end{eqnarray}
and among the solution $\kappa_a(d)$ and $\kappa_b(d)$ that solve
$I_G=I_A$, see Fig.\ \ref{kvond}, Eq.\ \ref{Cbitte1} uniquely yields
$\kappa=\frac{1}{2}$ with $d=3$.

One might object that the Bose symmetry of the ghost triangle
contribution to the three-gluon vertex is broken, for only the upper
vertex is dressed while the others are at tree-level (see Fig.\
\ref{DSE diag}). In the derivation of the DSE, the channel to start
the calculation in is an arbitrary choice. Therefore, one might as
well find the upper vertex dressed. If we repeat the calculation, the
DSE for the ghost-gluon vertex now reads
\begin{eqnarray}
  C&=&1+\frac{1}{2}C\frac{I_1}{I_A}-\frac{1}{2}C\frac{I_2I_3}{I_A^2}\; ,
\end{eqnarray}
cf.\ Eq.\ (\ref{DSEbyC}), and the solution can be found to be
\begin{eqnarray}
  C&=&4(d-1)(2 \kappa +1)^2 \Gamma^2(\frac{d}{2}+1)\Gamma(d-3\kappa)\Gamma(\frac{d}{2}-\kappa+1)\Gamma^2(-\kappa)\Gamma^3(2\kappa+1)
/
\nn\\
&&\left\{
d(2\kappa
+1)\Gamma(\frac{d}{2})\Gamma(\kappa+1)\Gamma^2(\frac{d}{2}+\kappa
+1)\Gamma(3\kappa+2-\frac{d}{2})\Gamma^3(\frac{d}{2}-2\kappa)\nn\right.\\
&&+(d-1)(d-2\kappa)\Gamma(\frac{d}{2}+1)\Gamma(d-3\kappa)\Gamma(-\kappa
  )\Gamma(2\kappa+1)\Gamma(2\kappa +2)\nn\\
&&\left. \left(\Gamma(\frac{d}{2}-2 \kappa)\Gamma(\kappa +1)\Gamma(\frac{d}{2}+\kappa +1)+2\Gamma(\frac{d}{2}+1)\Gamma(\frac{d}{2}-\kappa)\Gamma(-\kappa)\Gamma(2\kappa+2)\right)\right\}\nn\\
\end{eqnarray}
Again, the limit $d\to 3$ and $\kappa\to \frac{1}{2}$ yields the
result $C=1$. Apparently, it makes no difference which one of the
vertices is at tree-level.

\chapter{Hamiltonian approach to \texorpdfstring{$\phi^4$}{phi**4} theory}
\label{Appp4}

The $\phi^4$ theory is a popular toy model to illustrate some aspect
of a physical field theory in a simpler context. Unfortunately, the
$\phi^4$ theory in $d=4$ dimensions suffers from the {\it triviality
  problem} which states that after renormalization, the coupling
parameter is equally zero and a free field theory is considered \cite{Fro82}. This
problem can be circumvented by choosing a negative coupling parameter
which is at the prize of losing the boundedness of the spectrum. In
this appendix, we will use the $\phi^4$ theory in order to show how
the variational principle can be applied, in accordance to chapter
\ref{VarVac}.

With the classical Lagrangian density
\begin{eqnarray}
  \label{Lp4}
  \cL = \frac{1}{2}(\del_\mu\phi)\del^\mu\phi - \frac{1}{2}m^2\phi^2 - \frac{\lambda}{4!}\phi^4
\end{eqnarray}
and the conjugate momenta
\begin{eqnarray}
  \label{pip4}
  \pi=\frac{\delta\cL}{\delta\del_0\phi}=\del^0\phi
\end{eqnarray}
we can define the classical Hamiltonian
 \begin{eqnarray}
   \label{Hp4}
   H = \int d^3x
\left(
 \frac{1}{2}\pi^2 +
 \frac{1}{2}\left(\nabla\phi\right)^2+\frac{1}{2}m^2\phi^2+\frac{\lambda}{4!}\phi^4\right)\; .
 \end{eqnarray}
The equal-time canonical quantization relations,
$[\phi(x,t),\pi(y,t)] = i\delta^3(x,y)$, lead to the representation
\begin{eqnarray}
  \label{picoo}
  \pi(x)=\frac{\delta}{i\delta\phi(x)}
\end{eqnarray}
of the momentum operator, where $x$ is a $3$-vector and the time is
fixed in the Schr\"odinger picture. Thus, the quantum theory is
defined by the Hamiltonian (\ref{Hp4}) in terms of the field operators.

The time-independent Schr\"odinger equation $H\ket{\psi}=E\ket{\psi}$,
along with the semi-definiteness of $H$ for $\lambda>0$ infers that the vacuum
wave functional $\psi[\phi]=\braket{\phi}{\psi}$ can be determined by
a variational principle
\begin{eqnarray}
  \label{varprinc}
 E_0= \bra{\psi}\,H\,\ket{\psi}\longrightarrow \,\textrm{min.}\; 
\end{eqnarray}
As an ansatz, we choose a Gaussian wave functional
\begin{eqnarray}
  \label{Gaussp4def}
  \psi[\phi] = \cN\exp\left(-\frac{1}{2}\int d^3x\int d^3y\,
    \phi(x)\omega(x,y)\phi(y)\right) \equiv \cN\exp\left(-\frac{1}{2}\phi_x\omega_{xy}\phi_y\right)
\end{eqnarray}
where the short-hand notation keeps the coordinate dependence as an index
and the integration is implicit by means of the sum convention. The
kernel $\omega$ is determined by the variational principle~(\ref{varprinc}).

Expectation values are computed by
\begin{eqnarray}
  \label{vevdef}
  \lla \cO[\phi]\rra &=& \bra{\phi}\,\cO[\phi]\,\ket{\phi} = \int
  \cD\phi\, \, \psi^*[\phi]\cO[\phi]\psi[\phi]=  \left.\cO\left[\frac{\delta}{\delta\phi}\right]Z[j]\right|_{j=0}
\end{eqnarray}
where the generating functional $Z[j]$ is defined as
\begin{eqnarray}
  \label{Zp4def}
  Z[j]&=& \cN^2\int
\cD\phi\, \exp\left(-\phi_x\omega_{xy}\phi_y+j_x\phi_y\right)\nn\\
&=& \left\{ \cN^2\int
  \cD\phi\, '\exp\left(-\phi'_x\omega_{xy}\phi'_y\right)\right\}\cdot
\exp\left(\frac{1}{4}j_x\omega^{-1}_{xy}j_y\right)\nn\\
&=&\exp\left(\frac{1}{4}j_x\omega^{-1}_{xy}j_y\right)
\end{eqnarray}
Above, after a quadratic completion, the expression in the curly
brackets is $\lla 1\rra=1$ by virtue of the normalization of the wave
functional via $\cN$.

The vacuum energy $E_0$ is now calculated term by term, as a
functional of $\omega$. First of all, the mass term yields
\begin{eqnarray}
  \label{massp4}
  \frac{m^2}{2}\lla\phi_x\phi_y\rra& =&
  \frac{m^2}{2}\frac{\delta}{\delta j_x}\frac{\delta}{\delta
    j_x}\left.\e^{\frac{1}{4}j_u\omega^{-1}_{uv}j_v}\right|_{j=0}\nn\\
  &=& \frac{m^2}{2}\frac{\delta}{\delta
    j_x}\frac{1}{2}\omega^{-1}_{xx'}j_{x'}\left.\e^{\frac{1}{4}j_u\omega^{-1}_{uv}j_v}\right|_{j=0}\nn\\
&=&\frac{m^2}{4}\omega^{-1}_{xx}
\end{eqnarray}
The above term is basically the trace of the propagator
\begin{eqnarray}
  \label{propp4}
  \lla\phi(x)\phi(y)\rra=\frac{1}{2}\omega^{-1}(x,y)\; .
\end{eqnarray}
In order to understand better the meaning of the quantity
$\omega^{-1}_{xx}$, use the explicit notation and relate it to its
Fourier transform,
\begin{eqnarray}
  \label{wp4Four}
  \omega^{-1}_{xx} &=& \int d^3[xy]\omega^{-1}(x,y)\delta^3(x,y) = \int
  d^3[xy]\int\dbar^3k\,\omega^{-1}(k)\e^{ik\cdot(x-y)}\delta^3(x,y)\nn\\
&=&V\int\dbar^3k\,\omega^{-1}(k)
\end{eqnarray}
where $V=\int d^3x$ is the volume of coordinate space. The first variational
derivative in the calculation (\ref{massp4}) brings down a source term
$j_{x'}$ from the exponential so that the second variational
derivative acts not only on the exponential but also on the latter
source term $j_{x'}$. Setting sources to zero at the end of the
calculation generally eliminates some terms. Some combinatorics is
necessary to deal with the quartic term in the energy, we just write
down the result:
\begin{eqnarray}
  \label{quartp4}
  \frac{\lambda}{4!}\int
  d^3x\lla(\phi(x))^4\rra=\frac{\lambda}{8}V\int \dbar^3\ell\,\omega^{-1}(\ell)\omega^{-1}(\ell-k)
\end{eqnarray}
The kinetic energy is found to give
\begin{eqnarray}
  \label{Ekp4}
  \frac{1}{2}\lla\pi_x\pi_x\rra &=& \frac{1}{2}\int \cD\phi\, 
  \left|\frac{\delta}{i\delta\phi_x}\psi[\phi]\right|^2=\frac{1}{2}\omega_{xy}\omega_{xz}\lla
  \phi_y\phi_z\rra=\frac{1}{4}\omega_{xx}\nn\\
&=&\frac{1}{4}V\int \dbar^3k\,\omega(k)
\end{eqnarray}
where the propagator (\ref{propp4}) was used.

Furthermore, the spatial derivative terms of the Hamiltonian (\ref{Hp4}) yield explicitly
\begin{eqnarray}
  \label{Ekspp4}
  \frac{1}{2}\int d^3x\lla\phi(x)(-\del^2_x)\phi(x)\rra &=&\int
  d^3[xy]\delta^3(x,y)(-\del^2_x)\lla\phi(y)\phi(x)\rra\nn\\
  &=& \int
  d^3[xy]\delta^3(x,y)(-\del^2_x)\int\dbar^3k\,\frac{1}{2}\omega^{-1}(k)\e^{ik\cdot(x-y)}\nn\\
  &=& \frac{1}{2}\, V \int \dbar^3k\,\omega^{-1}(k)k^2
\end{eqnarray}

Summing up all contributions, the energy is found to be
\begin{eqnarray}
  \label{Eresp4}
  E=V\int\dbar^3k\,\left(\frac{1}{4}\omega^{-1}(k)(k^2+m^2)+\frac{1}{4}\omega(k)+\frac{\lambda}{8}\int \dbar^3\ell\,\omega^{-1}(k)\omega^{-1}(\ell-k)\right)
\end{eqnarray}
and it is minimized w.r.t.\ the kernel $\omega$,
\begin{eqnarray}
  \label{gapp4}
  0=\frac{\delta E/V}{\delta\omega^{-1}(k)} = \frac{1}{4}(k^2+m^2)-\frac{1}{4}\omega^2(k)+\frac{\lambda}{4}\int\dbar^3\ell\,\omega^{-1}(\ell)
\end{eqnarray}
which leads to the ''gap equation'':
\begin{eqnarray}
  \label{massgapp4}
  \omega^2(k)=k^2+m^2+\lambda\int\dbar^3\ell\,\omega^{-1}(\ell)
\end{eqnarray}
The term ''gap equation'' is sensible since even if the mass parameter
$m$ was zero to begin with, there would be a mass dynamically
generated by the last term in Eq.\ (\ref{massgapp4}). Hence, it
describes the lowest allowed energy level of the spectrum, i.e.\ a
mass gap. That term stems
from the quartic term in the Hamiltonian and is also present in the YM
Hamiltonian. It is a divergent constant subtracted in our
approach. In the $\phi^4$ theory, Eq.\ (\ref{massgapp4}) must be
solved iteratively and renormalization becomes
necessary. For further reading, see e.g.\ Ref.\ \cite{PabTar88}. If such a term is of any significance for YM theory, then
it will have an effect only on the intermediate regime of
$\omega(k)$. The infrared dominant curvature term from gauge fixing is
of course absent in the $\phi^4$ version (\ref{massgapp4}) of the gap equation.

\chapter{Identities for Gaussian expectation values}
\label{Appgap}

The Gaussian expectation values 
\begin{eqnarray}
  \label{Gaussdefapp}
  \lla \cO \rra_\omega = \int \cD A\,  \tP^*[A] \cO \tP[A]
\end{eqnarray}
with the normalized wave functional,
\begin{eqnarray}
  \label{normPsi}
  1=\int \cD A\,  \left|\tP[A]\right|^2 = \cN^2
\int \cD A\,  \e^{-\int d^d[xy]
  A_i^a(x)\omega_{ij}^{ab}(x,y)A_j^b(y)}\; ,
\end{eqnarray}
can be shown to yield some useful identities. Take a variational
derivative of Eq.\ (\ref{normPsi}) w.r.t.\
$\omega^{ab}_{ij}(x,y)=t_{ij}(x)\omega(x,y)\delta^{ab}$,
\begin{eqnarray}
  \label{lnN2}
  0=\frac{\delta\ln\cN^2}{\delta\omega_{ij}^{ab}(x,y)}-\lla
  A_i^a(x)A_j^b(y)\rra_\omega\; ,
\end{eqnarray}
to find
\begin{eqnarray}
  \label{lnN2res}
  \frac{\delta\ln\cN^2}{\delta\omega_{ij}^{ab}(x,y)}
  =\frac{1}{2}(\omega^{-1})_{ij}^{ab}(x,y)\; .
\end{eqnarray}
For expectation values (\ref{Gaussdefapp}) of operators $\cO=\cO[A]$, a variational
derivative thus gives
\begin{eqnarray}
  \label{cOcovar}
 \frac{\delta\langle\cO\rangle_\omega}{\delta\omega_{ij}^{ab}(x,y)} = \big\langle A_i^a(x)A_j^b(y)
 \big\rangle_\omega \big\langle \cO \big\rangle_\omega - \big\langle A_i^a(x)A_j^b(y) \cO
 \big\rangle_\omega\; ,
\end{eqnarray}
i.e.\  a covariance. Using Wick's theorem, a few manipulations yield the identity
\begin{eqnarray}
  \label{del2O}
  \bigg\langle \frac{\delta^2\cO}{\delta A_i^a(x)\delta A_j^b(y)}\bigg\rangle_\omega = -4\int
  d^d[uv]\omega_{im}^{ac}(x,u)\frac{\delta\lla \cO
    \rra_\omega}{\delta\omega_{mn}^{cd}(u,v)}\omega_{nj}^{db}(v,y)\; .
\end{eqnarray}
For $\cO=H_p[A]$, this leads directly to Eq.\ (\ref{divbyEp}).

A little more effort is required for expectation values of operators
$\cO=\tH[A,\Pi]$ that involve the momentum operator $\Pi$.
The equivalent of Eq.\ (\ref{cOcovar}) is then
\begin{eqnarray}
    \label{tHcovar}
 \frac{\delta\langle\tH\rangle_\omega}{\delta\omega_{ij}^{ab}(x,y)} = \big\langle A_i^a(x)A_j^b(y)
 \big\rangle_\omega \big\langle \tH \big\rangle_\omega - \frac{1}{2}\big\langle\, [A_i^a(x)A_j^b(y), \tH]_+
 \big\rangle_\omega\; ,
\end{eqnarray}
whereby $[\: ,\:]_{\mp}$ denotes the anticommutator ($+$), or the
commutator ($-$), respectively. The anticommutator in Eq.\
(\ref{tHcovar}) can be rewritten as
\begin{eqnarray}
  \label{antibydel2}
  [A_i^a(x)A_j^b(y), \tH]_+ = -\frac{\delta^2\tH}{\delta \Pi_i^a(x)\delta
    \Pi_j^b(y)} + A_i^a(x)\tH A_j^b(y)+A_j^b(y)\tH A_i^a(x)
\end{eqnarray}
using $[A_i , [A_j , \tH]_- ]_- =
-\frac{\delta^2\tH}{\delta\Pi_i\delta\Pi_j}$ with $A_k^a(x)=i\delta /\delta\Pi_k^a(x)$.
With knowledge of the wave
functional,
\begin{eqnarray}
\label{AbyPi}
  A_k^a(x)\tP[A]=-\int
  d^dy\,(\omega^{-1})_{kj}^{ab}(x,y)\frac{\delta}{\delta
    A_j^b(y)}\tP[A]\; ,
\end{eqnarray}
the last two terms in Eq.\ (\ref{antibydel2}) result in
\begin{eqnarray}
  \label{AHA}
  \big\langle A_i^a(x)\tH A_j^b(y) + A_j^b(y)\tH A_i^a(x)\big\rangle_\omega &&\nn\\
 && \hspace{-5cm}=\int d^du\, (\omega^{-1})_{im}^{ac}(x,u)
\lla 
\frac{\delta}{\delta A^c_m(u)} \tH A_j^b(y) 
- A_j^b(y) \tH\frac{\delta}{\delta  A_m^c(u)} \rraw\nn\\
&& \hspace{-5cm}=\int d^du\, (\omega^{-1})_{im}^{ac}(x,u)
\left \{
\bigg\langle \frac{\delta\tH}{\delta  A_m^c(u)} A_j^b(y) \bigg\rangle_\omega
+ \langle \tH \rangle_\omega t_{mj}^{cb}(u,y) \right.\nn\\
&&\hspace{-4cm}-\int d^dv \lla \tH A_j^b(y)A_n^d(v)\rraw
\omega_{mn}^{cd}(u,v) - \int d^dv \lla  A_j^b(y)A_n^d(v)\tH\rraw
\omega_{mn}^{cd}(u,v)
\nn\\
&&\hspace{-4cm}\left.+\bigg\langle  A_j^b(y)\frac{\delta\tH}{\delta  A_m^c(u)} \bigg\rangle_\omega
+ \langle \tH \rangle_\omega\, t_{mj}^{cb}(u,y) \right\}\nn\\
&&\hspace{-5cm}= -\lla \left[A_i^a(x)A_j^b(y), \tH \right]_+ \rraw +
2(\omega^{-1})_{ij}^{ab}(x,y)\langle\tH\rangle_\omega \nn\\
&& \hspace{-1cm}+\int d^du\,
(\omega^{-1})_{im}^{ac}(x,u) \lla\bigg[\frac{\delta\tH}{\delta A_m^c(u)},A_j^b(y)\bigg]_+\rraw
\end{eqnarray}
The anticommutator in the last line can be rewritten as a commutator,
according to the following identity,
\begin{eqnarray}
  \label{commanticomm}
  \big\langle [A,\cO]_\pm \big\rangle_\omega = i\,\omega^{-1} \big\langle [\Pi,\cO]_\mp\big\rangle_\omega
\end{eqnarray}
which holds for Gaussian expectation values, such as (\ref{Gaussdefapp}).

{\it Proof:} Inside the expectation value, the field operator $A$ can be expressed
by a momentum operator $\Pi$, see Eq.\ \ref{AbyPi}. Thus,
\begin{eqnarray}
  \label{proofapp}
  \lla \left[
   A_k^a(x),\cO
   \right]_\pm 
   \rraw &=& 
  \int
  d^dy(\omega^{-1})_{kj}^{ab}(x,y)
  \lla 
   \frac{\delta}{\delta A_j^b(y)}
   \cO \mp \cO 
   \frac{\delta}{\delta A_j^b(y)} 
   \rraw\nn\\
    &=&i\int
  d^dy(\omega^{-1})_{kj}^{ab}(x,y)\lla  [\Pi,\cO]_\mp\rraw\qquad {}_{{}_\square}
\end{eqnarray}
Therefore,
\begin{eqnarray}
  \label{d2HAA}
   \lla\bigg[\frac{\delta\tH}{\delta A_m^c(u)},A_j^b(y)\bigg]_+\rraw
   =  \int
   d^dv\, (\omega^{-1})_{jn}^{bd}(y,v)\bigg\langle\frac{\delta^2\tH}{\delta
     A_m^c(u)\delta A_n^d(v)}\bigg\rangle_\omega\; .
\end{eqnarray}
Plugging Eq.\ (\ref{d2HAA}) into the expression (\ref{AHA}) and the
latter with (\ref{antibydel2}) into Eq.\ (\ref{tHcovar}), some
cancellations occur and we end up with 
\begin{eqnarray}
  \label{gapAP}
  \frac{\delta\langle\tH\rangle_\omega}{\delta\omega_{ij}^{ab}(x,y)} &=&
  \frac{1}{4}\,\bigg\langle\frac{\delta^2\tH}{\delta\Pi_i^a(x)\delta\Pi_j^b(y)}\bigg\rangle_\omega \nn\\
&& - \,\frac{1}{4}\int d^d[uv](\omega^{-1})_{im}^{ac}(x,u)
\bigg\langle\frac{\delta^2\tH}{\delta
     A_m^c(u)\delta A_n^d(v)}\bigg\rangle_\omega
(\omega^{-1})_{nj}^{db}(v,y)\; .
\end{eqnarray}
Such an identity was also derived in Ref.\ \cite{privRei}.

% \include{apps}
%\listoffigures
%\listoftables

%EIGENE PAPERS MIT BUCHSTABEN NUMERIEREN?
\addcontentsline{toc}{chapter}{References}
%\renewcommand{\refname}{Publications}
 % \nocite{*}   
\bibliography{biblio}{}
\bibliographystyle{utcaps}

\clearpage{\pagestyle{empty}\cleardoublepage}
\pagestyle{fancy}
\phantomsection %% wichtig fuer hyperref
\fancyhead[LE]{\scshape Acknowledgments}        % chapter to left
\fancyhead[RO]{\scshape Acknowledgments}       % section to right
\addcontentsline{toc}{chapter}{Acknowledgments}
 %\thispagestyle{empty}
%\selectlanguage{german}

\chapter*{Acknowledgments}

First of all, I would like to thank my supervisor Prof.\ Dr.\ Hugo
Reinhardt. He supported me throughout my PhD thesis unconditionally and I
was always glad to have his experienced advice. His continuous
interest in my work and the brilliant ideas coming from his side were the
driving mechanisms for this thesis to be completed. Moreover, I was
glad to come along with him to the Trento workshop in 2007.

It is a pleasure to recall the great inspiration I received from
numerous people in my undergraduate years that made a way around the
fascinating topic of quantum field theory barely avoidable. I would
like to mention Prof.\ Dr.\ Samson Shatashvili from my Erasmus year in Dublin
as well as Prof.\ Dr.\ Jochen Wambach, Prof.\ Dr.\ Reinhard Alkofer
and Dr.\ Axel Maas from the collaborations during my diploma thesis in
Darmstadt.

During my PhD studies, I profited from discussions with Prof.\
Dr.\ Adam Szczepaniak, Prof.\ Dr.\ Axel Weber, Prof.\ Dr.\ Jeff
Greensite, Prof.\ Dr.\ Daniel Zwanziger and Prof.\ Dr.\ Christian
Fischer. I highly appreciated the opportunity to embrace topics
directly related to my thesis. 

In the institute of theoretical physics in T\"ubingen, the company of
Dr.\ habil.\ Markus Quandt, Dr.\ Claus Feuchter, Dr.\ Peter Watson,
Dr.\ Giuseppe Burgio, Dr.\ Diana Nicmorus, Davide Campagnari, Markus
Leder, Dominik Epple, Burghard Gr\"uter, Hakan Turan, Carina Popovici
and Wolfgang Lutz was very helpful not only for the understanding of
physics issues but also for getting through all this.

For a careful and critical reading of the manuscript, I am indebted to 
Davide Campagnari and Dr.\ Peter Watson.

The European Graduate School Basel--Graz--T\"ubingen ({\textsc{Eurograd}}) supported this thesis both financially and educationally with 
graduate days and workshops. The efforts of the T\"ubingen speakers
Prof.\ Dr.\ Dr.\ h.c.\ mult.\ Amand F\"a\ss{}ler and Prof.\ Dr.\ Josef Jochum in particular are very much appreciated.

My wife, my family and my friends did a great job backing me up,
challenging me, and giving me all I needed, and I will not forget any
of it. 

Finally, I want to emphasize that without the help and support
of my parents, beginning from the first years of my life up till now,
I would have most probably never accomplished this doctoral thesis,
and I am most thankful for that.

\pagestyle{empty}
\cleardoublepage
\chapter*{Publications}
\begin{enumerate}[I.]
\item
{S.J. Cox, S. Neethling, W.R. Rossen, W. Schleifenbaum, P.~Schmidt-Wellenburg,
  and J.J. Cilliers}, ``{A theory of the effective yield stress of foam in
  porous media: the motion of a soap film traversing a three-dimensional
  pore}'', {\em Colloids and Surfaces} {\bf A 245} (2004) 143--151, \\
\href{http://dx.doi.org/10.1016/j.colsurfa.2004.07.004}{{\tt http://dx.doi.org/10.1016/j.colsurfa.2004.07.004}}.
\item
W.~Schleifenbaum, A.~Maas, J.~Wambach, and R.~Alkofer, ``{Infrared behaviour of
  the ghost-gluon vertex in Landau gauge Yang-Mills theory}'', {\em Phys. Rev.}
  {\bf D72} (2005) 014017,\\
  \href{http://dx.doi.org/10.1103/PhysRevD.72.014017}{\tt
    {http://dx.doi.org/10.1103/PhysRevD.72.014017}}, \\
  \href{http://arXiv.org/abs/hep-ph/0411052v2}{\tt {http://arXiv.org/abs/hep-ph/0411052v2}}.
%%CITATION = HEP-PH/0411052;%%.
\item
W.~Schleifenbaum, A.~Maas, J.~Wambach, and R.~Alkofer, ``{The ghost-gluon
  vertex in Landau gauge Yang-Mills theory}'', {\em Proceedings of the 42nd
  International School of Subnuclear Physics in Erice, Italy} (2004), \\
\href{http://arXiv.org/abs/hep-ph/0411060v1}{\tt {http://arXiv.org/abs/hep-ph/0411060v1}}.
%%CITATION = HEP-PH/0411060;%%.
\item
W.~Schleifenbaum, M.~Leder, and H.~Reinhardt, ``Infrared analysis of
  propagators and vertices of Yang-Mills theory in Landau and Coulomb gauge'',
  {\em Phys. Rev.} {\bf D73} (2006) 125019,\\
\href{http://dx.doi.org/10.1103/PhysRevD.73.125019}{\tt {http://dx.doi.org/10.1103/PhysRevD.73.125019}},\\
\href{http://arXiv.org/abs/hep-th/0605115v2}{\tt {http://arXiv.org/abs/hep-th/0605115v2}}.
%%CITATION = HEP-TH/0605115;%%.
\item
H.~Reinhardt, D.~Epple, and W.~Schleifenbaum, ``Hamiltonian approach to
  Yang-Mills theory in Coulomb gauge'', {\em AIP Conf. Proc.} {\bf 892} (2007)
  93--99,\\
\href{http://arXiv.org/abs/hep-th/0610324v1}{\tt {http://arXiv.org/abs/hep-th/0610324v1}}.
%%CITATION = HEP-TH/0610324;%%.
\item
D.~Epple, H.~Reinhardt, and W.~Schleifenbaum, ``Confining Solution of the
  Dyson-Schwinger Equations in Coulomb Gauge'', {\em Phys. Rev.} {\bf D75}
  (2007) 045011,\\
\href{http://dx.doi.org/10.1103/PhysRevD.75.045011}{\tt {http://dx.doi.org/10.1103/PhysRevD.75.045011}},\\
\href{http://arXiv.org/abs/hep-th/0612241v2}{\tt {http://arXiv.org/abs/hep-th/0612241v2}}.
%%CITATION = HEP-TH/0612241;%%.
\item
H.~Reinhardt, W.~Schleifenbaum, D.~Epple, and C.~Feuchter, ``{Hamiltonian
  approach to Coulomb gauge Yang-Mills Theory}'', {\em PoS} {\bf LAT2007}
  (2007), \\
\href{http://arxiv.org/abs/0710.0316v1}{\tt {http://arxiv.org/abs/0710.0316v1}}.
%%CITATION = ARXIV:0710.0316;%%.
\item
D.~Epple, H.~Reinhardt, W.~Schleifenbaum, and A.~P. Szczepaniak, ``{Subcritical
  solution of the Yang-Mills Schroedinger equation in the Coulomb gauge}'',
  {\em Phys. Rev.} {\bf D77} (2008) 085007,\\
\href{http://dx.doi.org/10.1103/PhysRevD.77.085007}{\tt {http://dx.doi.org/10.1103/PhysRevD.77.085007}},\\
\href{http://arxiv.org/abs/0712.3694v1}{\tt {http://arxiv.org/abs/0712.3694v1}}.
%%CITATION = ARXIV:0712.3694;%%.
\end{enumerate}

%\clearpage
%\vspace{-9cm}
%\hspace{-4cm}\includegraphics{pubsEN.pdf}
\end{document}